\documentclass[twocolumn,trackchanges,twocolappendix]{aastex701}
\received{July 17, 2025}

\submitjournal{\apj}

\usepackage{amsmath,environ}
\usepackage{listings}
\usepackage{float}
\usepackage{color}
\usepackage{mathtools}
\usepackage{enumitem}

\usepackage{rotating}

\usepackage{lineno}
\usepackage{hyperref}
\linenumbers

\newcommand{\halfwdth}{0.475\textwidth}

\newcommand{\hii}{H~{\sc ii}}
\newcommand{\hi}{H~{\sc i}}

\newcommand{\ci}[1]{[C~{\sc i}]{#1}} 
\newcommand{\cii}[1]{[C~{\sc ii}]{#1}} 
\newcommand{\nii}[1]{[N~{\sc ii}]{#1}} 
\newcommand{\niii}[1]{[N~{\sc iii}]{#1}} 
\newcommand{\oi}[1]{[O~{\sc i}]{#1}} 
\newcommand{\oiii}[1]{[O~{\sc iii}]{#1}} 
\newcommand{\lcii}[1]{$L_\mathrm{[C~II]{#1}}$} 
\newcommand{\loi}[1]{$L_\mathrm{[O~I]{#1}}$} 
\newcommand{\loiii}[1]{$L_\mathrm{[O~III]{#1}}$} 

\newcommand{\ha}{H$\alpha$}
\newcommand{\lha}{$L_\mathrm{H\alpha}$}
\newcommand{\hb}{H$\beta$}

\newcommand{\haextcorr}{H$\alpha_\mathrm{ext.corr.}$}
\newcommand{\lhaextcorr}{$L_\mathrm{H\alpha,ext.corr.}$}

\newcommand{\lir}{$L_\mathrm{IR}$}
\newcommand{\lfir}{$L_\mathrm{FIR}$}

\newcommand{\lfuv}{$L_\mathrm{FUV}$}
\newcommand{\lline}{$L_\mathrm{line}$}

\newcommand{\hp}{H\textsuperscript{+}}
\newcommand{\cp}{C\textsuperscript{+}}
\newcommand{\op}{O\textsuperscript{+}}
\newcommand{\opp}{O\textsuperscript{2+}}

\newcommand{\oh}{$\log$\,(O/H)}

\newcommand{\edens}{$n_e$}

\newcommand{\cc}{cm\textsuperscript{-3}}
\newcommand{\td}{$T_\mathrm{dust}$}
\newcommand{\te}{$T_e$}

\newcommand{\md}{$M_\mathrm{dust}$}
\newcommand{\lsun}{$L_\odot$}
\newcommand{\msun}{$M_\odot$}
\newcommand{\mhi}{$M_\mathrm{H~I}$}
\newcommand{\mmol}{$M_\mathrm{mol}$}
\newcommand{\mstar}{$M_\star$}
\newcommand{\fciin}{$f_\mathrm{[C~II],neutral}$}
\newcommand{\um}{\micron}

\newcommand{\fnu}[1]{$S${#1}}
\newcommand{\tcolor}{\fnu{60}/\fnu{100}}

\newcommand{\bsf}[1]{\textbf{\textsf{#1}}}
\newcommand{\zz}{\textit{z}}

\newcommand{\pyneb}{\texttt{PyNeb}}

\defcitealias{paperi}{Paper I}
\newcommand{\ppi}{\citetalias{paperi}}
\newcommand{\ppii}{Paper II}

\begin{document}

\title{Fine-structure Line Atlas for Multi-wavelength Extragalactic Study (FLAMES) III: \\\cii{} as Tracer, Crisis of SFR, \oiii{}/\cii{} at High-\zz{}, New Answers and New Questions} 

\shorttitle{FLAMES III: New Answers and New Questions}
\shortauthors{Peng et al.}

\correspondingauthor{Bo Peng}
\email{bp392@cornell.edu}

\author[0000-0002-1605-0032]{Bo Peng}
\affiliation{Max-Planck-Institut für Astrophysik, Garching, D-85748, Germany}
\email{bp392@cornell.edu}

\author{Gordon Stacey}
\affiliation{Department of Astronomy, Cornell University, Ithaca, NY 14853, USA}
\email{stacey@cornell.edu}

\author[0000-0002-4444-8929]{Amit Vishwas}
\affiliation{Cornell Center for Astrophysics and Planetary Science, Cornell University, Ithaca, NY 14853, USA}
\email{vishwas@cornell.edu}

\author[0000-0002-1895-0528]{Catie Ball}
\affiliation{Department of Astronomy, Cornell University, Ithaca, NY 14853, USA}
\email{cjb356@cornell.edu}

\author[0000-0003-1874-7498]{Cody Lamarche}
\affiliation{Department of Physics, Winona State University, Winona, MN 55987, USA}
\email{cody.lamarche@winona.edu}

\author[0000-0002-8513-2971]{Christopher Rooney}
\affiliation{National Institute of Standards and Technology, Boulder, CO 80305, USA}
\email{ctr44@cornell.edu}

\author{Thomas Nikola}
\affiliation{Cornell Center for Astrophysics and Planetary Science, Cornell University, Ithaca, NY 14853, USA}
\email{tn46@cornell.edu}

\author[0000-0001-6266-0213]{Carl Ferkinhoff}
\affiliation{Department of Physics, Winona State University, Winona, MN 55987, USA}
\email{cferkinhoff@winona.edu}

\begin{abstract}

In the final paper of this series, we discuss new perspectives and challenges in the study of interstellar medium (ISM), leveraging comprehensive catalogs and physical insights presented in our previous papers. 
We focus on key questions of far-infrared (FIR) fine-structure lines (FSLs): their origins, diagnostic value, and implications of correlations. 
Our analysis reveals a strong dependence on elemental abundance, so that FSL/\ha{} traces metallicity, \nii{}/\cii{} traces N/O, and $\sim$80\% of \cii{} emission arises from neutral gas without systematic variations. 
We conclude a coherence exists between the emissions from ionized and neutral gases regarding energy sources and distribution. 
We argue that \cii{} is physically a metallicity-dependent star formation rate (SFR) tracer, while its correlations with atomic or molecular gas masses are secondary. 
Crucially, the \cii{} ``deficit'' is only part of a universal ``deficit'' problem that shows in all neutral and ionized gas lines including extinction-corrected \ha{}, caused by infrared (IR) luminosities and characterized by a dichotomy in gas and dust behaviors. 
This universal ``deficit'' marks a breakdown of the obscuration-corrected star-formation rate (SFR) calibration and imperils SFR estimates. 
We argue that it is caused by either IR ``excess'' or ionized gas ``deficit'', and present possible scenarios. 
A renewed picture of ISM structure is needed to reconcile with ionized--neutral gas coherence, metallicity dependence, and gas--dust dichotomy. 
We also discuss differences of FIR FSL at high redshifts: the offset in ``deficit'' trends, the similar \oiii{}/\cii{} in metal-poor galaxies, and elevated \oiii{}/\cii{} in dusty galaxies.

\end{abstract}


\section{Introduction}
\label{sec:implication_intro}

In the first paper of this series \citep[][hereafter \ppi{}]{paperi}, we presented a comprehensive compilation of far-infrared (FIR) and mid-infrared (MIR) fine-structure line (FSL) detections, encompassing the majority of sources known to date. 
Building upon this dataset, the second paper \citep[][hereafter \ppii{}]{paperii} introduced an extensive grid of photoionization models that extend into the neutral gas phase, providing a robust theoretical framework to interpret the observed trends of FSLs in \ppi{}. 
With both a rich observational sample and model based diagnostics in hand, we are now in a position to revisit several long-standing questions concerning the origin and interpretation of FIR FSLs.

As summarized in \ppi{}, earlier studies of FSLs often neglected comparisons with their well-studied optical counterparts. 
This disconnect now poses a significant limitation---particularly in the era of JWST, which is capable of observing rest-frame optical spectra of high-redshift (\zz{}) galaxies, even highly dust-obscured ones, previously accessible only via FIR or submillimeter (sub-mm) observations \citep[e.g.,][]{BJ23,AS24,jones24,ZS24}. 
In \ppi{} and \ppii{}, we demonstrate the concordance of the two wavelength regimes in measuring metallicity and radiation field strength. 
In this work, we will explore the connection of FIR FSLs with the most prominent optical line \ha{}. 

One of the highlights of FSL studies is the interpretation of the \cii{} line, a major coolant of the neutral interstellar medium (ISM). 
Traditionally, \cii{} has been used as a tracer of the global star formation rate (SFR) \citep[e.g.,][]{stacey91,herrera15,S19} and as a diagnostic of photodissociation regions \citep[][]{tielens85,hollenbach91,hollenbach97,wolfire22}. 
However, due to the low ionization potential to turn carbon into \cp{}, \cii{} emission can originate from both neutral and ionized gas phases, complicating its interpretation. 
Observational studies have suggested substantial variation in the relative contributions of these phases across different galaxy populations \citep{B08,C15}. 

Much effort has been devoted to understanding the so-called \cii{} ``deficit''---the observed drop in \lcii{}/\lir{} ratio with increasing IR luminosity---but no consensus has emerged. 
This issue is more than a breakdown of a simple correlation as it calls into question the reliability of both \cii{} and \lir{} as SFR indicators \citep{herrera15,schaerer20}. 
In PDRs, \cii{} emission is powered by photoelectric heating, which is itself closely tied to dust heating and hence to \lir{}. 
Discrepancies between these tracers thus have far-reaching implications for SFR estimation. 

Beyond \cii{}, other FIR lines such as \oiii{} also present interpretive challenges. 
Since the first high-\zz{} detections of \oiii{} \citep{FC10}, numerous studies have reported elevated \oiii{}/\cii{} ratios in high-\zz{} dusty star-forming galaxies (DSFGs) compared to local ultraluminous and luminous infrared galaxies (U/LIRGs) \citep[e.g.,][]{ZZ18,MD18,DC19}. 
This offset, also noted in \ppi{}, suggests evolving physical conditions in the ISM.

Moreover, recent studies of Lyman-break and Ly$\alpha$ emitter galaxies (LBG/LAEs) at \zz{}~\textgreater~5 have suggested that their \oiii{88}/\cii{} ratios lie systematically above local dwarf galaxy scaling relations with SFR \citep[e.g.,][]{HY20}. 
These findings indicate that the high-\zz{} ISM in low-metallicity environment may exhibit enhanced excitation conditions or different ionization structures compared to local ``analogs.''

In this paper, we revisit these open questions in FSL studies using our comprehensive dataset (table FLAMES-low and FLAMES-high, referenced in Appendix~\ref{sec:implication_ref}) in \ppi{} and insights into line diagnostics obtained in \ppii{}. 
Specifically, we will reassess the neutral-ionized phase diagnostics using a nitrogen-corrected \cii{}/\nii{} ratio in Sec.~\ref{sec:implication_neut-ion}; explore the commonality among the ``deficit'' trends in different FIR lines in Sec.~\ref{sec:implication_deficit}; investigate the enhanced oxygen line emission at high redshifts in Sec.~\ref{sec:implication_diff}). 
In Sec.~\ref{sec:implication_discussion}, we will provide a coherent framework for understanding the observed trends of emission lines and their ratios, discuss the implications of the dichotomy between emission lines and IR luminosities, as well as the interpretation of FIR FSLs observed at high-\zz{}. 

The paper adheres to the same plotting style of the series. 
The low-\zz{} galaxies are plotted as small dots, with the color scheme: blue--dwarf galaxies, orange--star-forming (SF) galaxies, red--(ultra-)luminous infrared galaxies (U/LIRGs), cyan--LINERs, black--active galactic nuclei (AGNs), gray--early type galaxies (ETGs), green-yellow--regions. 
All high-\zz{} data points are shown as larger symbols with black outline, in the following colors and symbols: blue star--Lyman break galaxies (LBGs) or Ly$\alpha$ emitters (LAEs), orange diamond--SF, red circle--dusty star-forming galaxies (DSFGs), black square--QSOs or AGNs, gray hexagon--ETGs, green-yellow ``x''--damped Ly$\alpha$ system hosts (DLAs), brown ``+''--proto-clusters. 
The Upper or lower limit is plotted at the 3 $\sigma$ value with the same color and a smaller symbol corresponding to its type, but it is translucent and has an arrow showing the direction of the constraint.

\section{Comparison of Neutral and Ionized Gas Emissions}
\label{sec:implication_neut-ion}

\subsection{\texorpdfstring{\cii{}}{[C II]}/\texorpdfstring{\nii{}}{[N II]}: \texorpdfstring{\cii{}}{[C II]} Origin or N/O Abundance}
\label{sec:implication_fneut}

\begin{figure}[h]
    \centering
    \includegraphics[width=\halfwdth]{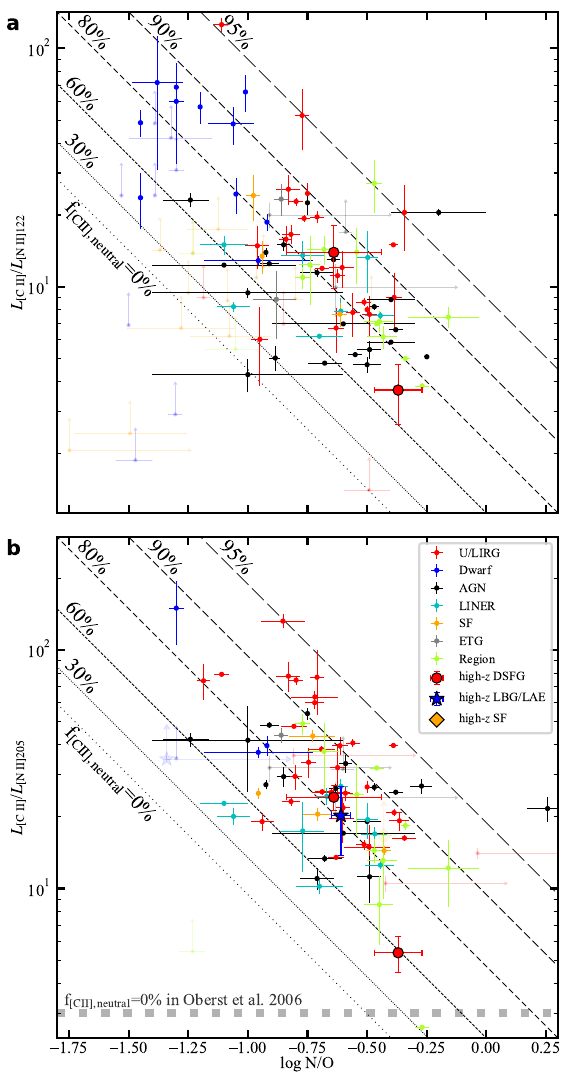}
    \caption[N/O and \fciin{}. ]{N/O vs. \bsf{a}, \cii{}-to-\nii{122}; or \bsf{b}, \cii{}-to-\nii{205}. The increasingly denser and longer segmented diagonal lines represent N/O corrected \cii{} neutral fraction \fciin{} (Eq.~\ref{equ:n_o}) from 0\% to 95\%. The thick short-dashed gray line in (\bsf{b}) represent the \cii{}\textsubscript{ion}/\nii{205} baseline used in \citet{obserst06}.}
    \label{f:data_CII_neutral}
\end{figure}

The fraction of \cii{} emission arising from neutral gas---commonly referred to as the \cii{} neutral fraction (\fciin{})---is a key diagnostic used in studies of PDRs and star formation rates \citep[SFR; e.g.,][]{S19}. 
A widely adopted method to estimate \fciin{} is based on the \cii{}/\nii{205} ratio \citep{obserst06}, based on the fact that \nii{} arises exclusively from ionized gas and that the emissivity ratio of \cii{} and \nii{205} remains relatively invariant across a wide range of electron densities (\edens{}). 

In this framework, any excess in \cii{} flux over the expected \cii{}\textsubscript{ion}/\nii{205} ratio is attributed to additional emission from the neutral ISM. 
However, this approach neglects the impact of elemental abundances---specifically the carbon-to-nitrogen abundance ratio (C/N)---on the interpretation of line ratios. 
On the other hand, nitrogen abundance shows a large variation across galaxies relative to $\alpha$ elements due to an additional secondary production channel in intermediate-mass stars, resulting in a N/O--O/H relation \citep{edmunds78,henry00,pilyugin03}. 
As demonstrated in sec.~3.7 in \ppi{}, the observed \cii{}/\nii{} ratio correlates directly with the nitrogen-to-oxygen abundance ratio (N/O), which is linked to the C/N ratio through the relatively stable C/O ratio \citep{romano20}---both being $\alpha$ elements and originating in supernova events.

The original calibration presented in \citet{obserst06} assumes a fixed C/N ratio, appropriate for localized Galactic ISM environments. 
However, when applied to a diverse galaxy population, this assumption introduces systematic errors---especially given the order-of-magnitude variation in N/O across different galaxy types. 
While early warnings about abundance effects were noted \citep[e.g.,][]{malhotra01}, they have not been sufficiently accounted for in subsequent studies. 
As a result, works using fixed C/N ratios have reported a wide range of \fciin{} values strongly dependent on galaxies types: from low values (down to $\sim$50\%) in metal-rich systems \citep{B08} to high values (exceeding 95\%) in dwarf galaxies \citep{C15}; or correlations with dust temperature and \oiii{88}/\nii{122} ratios \citep{D17}.

\citet{croxall17} found a metallicity dependence of \cii{}/\nii{205} in resolved observations, but attributed the variation primarily to \fciin{} rather than to underlying abundance differences, due to an underestimated nitrogen abundance trend in their adopted C/N–metallicity relation. 

To address this, we revisit the \cii{}/\nii{} diagnostic by explicitly including abundance corrections. 
Fig.~\ref{f:data_CII_neutral} shows \cii{}/\nii{122} and \cii{}/\nii{205} ratios plotted against N/O. 
Diagonal lines correspond to constant \fciin{} values after correcting for abundance effects, following:
\begin{equation}\begin{split}
    f_\mathrm{[C~II],neutral} = 1 - \left(\frac{L_\mathrm{[N~II]}}{L_\mathrm{[C~II]}}\right) \left(\frac{\mathrm{C}}{\mathrm{N}}\right) \left(\frac{\varepsilon_{\mathrm{[C~II],}e^-}}{\varepsilon_\mathrm{[N~II]}}\right)
    \label{equ:n_o}
\end{split}\end{equation}
where C/N is approximated via C/N~=~(C/O)$_\odot$$\times$(O/N), adopting (C/O)$_\odot$ = 10\textsuperscript{-0.26}. 
Emissivities are computed with \pyneb{} \citep{Luridiana15} using \te{}~=~10\textsuperscript{4} K, \edens{}~=~50\,\cc{}, assuming collisional excitation by electron. 
We also plot the original baseline value of \cii{}/\nii{205} calibration used in \citet{obserst06} for comparison. 

After correcting for abundance, the inferred \fciin{} values across galaxy types converge to a narrower range (60--90\%) with a median around 80\%, in contrast to the wide range suggested by fixed-abundance calibrations. 
The use of \nii{122} yields a scatter and a trend similar to \nii{205}, indicating that, in reality, the density effect is minimal---consistent with the small \edens variation found in \ppi{} and \ppii{}. 
Most importantly, we find no intrinsic correlation between \fciin{} and galaxy types---at least when considering integrated galaxy properties. 
Dwarf galaxies that occupy the low N/O part of the parameter space serve as valuable benchmarks, demonstrating how N/O actually causes the variation in \cii{}/\nii{}.

The relatively tight distribution of \fciin{} values implies that large-scale \cii{} emission is generally dominated by neutral gas, and \nii{}/\cii{} is practically a tracer of N/O. 
Residual scatter could also arise from several effects in addition to \fciin{}: observational uncertainties (especially for faint \nii{205} lines), systematic errors in N/O measurements (typically $\gtrsim$0.1 dex), and the assumption of a constant C/O. 
Furthermore, the observed correlation of \fciin{} with other properties in earlier works can now be unified under the framework of N/O-driven line ratio differences. 
For instance, low \fciin{} in metal-rich galaxies and high \fciin{} in dwarfs are naturally explained by their respective differences in N/O. 
Similarly, apparent correlations between \fciin{} and dust temperature or \oiii{88}/\nii{122} may result from underlying trends linking metallicity, ionization parameter $U$, and dust heating (see Sec.~\ref{f:implication_Ha_deficit}).

Nevertheless, we caution that the current data is sparse and subject to large uncertainties, which may obscure subtle residual trends.

\subsection{FIR FSL-to-\texorpdfstring{\ha{}}{H alpha} Ratios}
\label{sec:implication_fir-ha}

\begin{figure*}[h]
    \centering
    \includegraphics[width=\halfwdth]{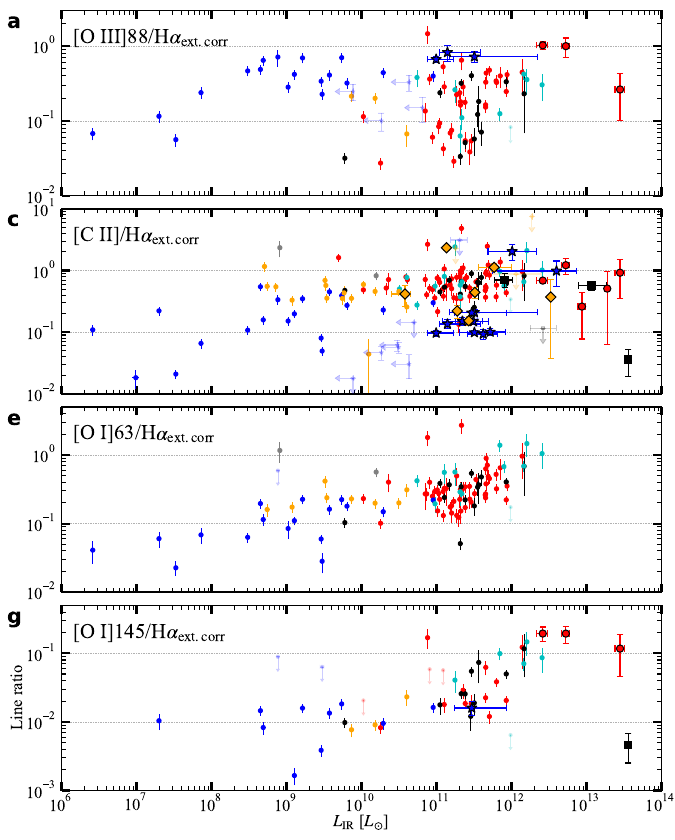}
    \includegraphics[width=0.442\textwidth]{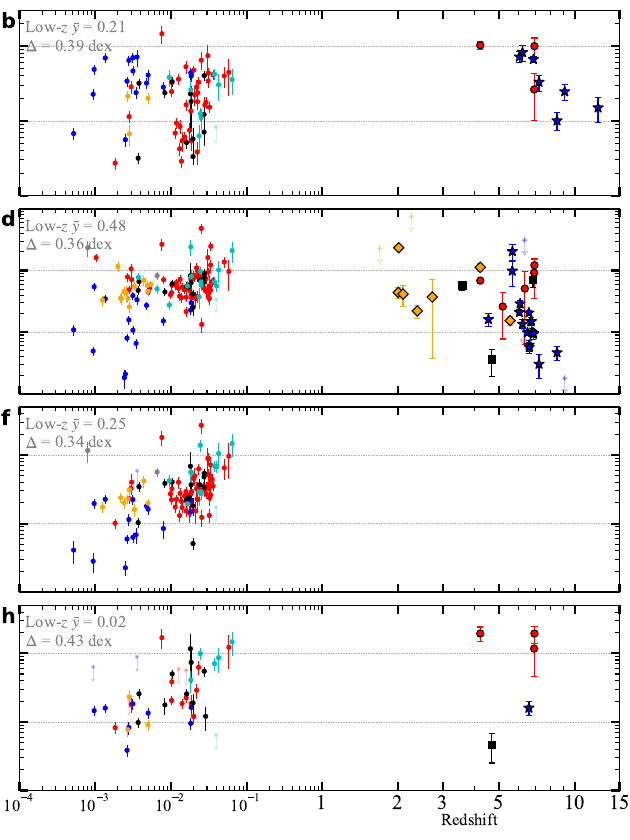}
    \caption[Demographics of line/Halpha. ]{Demographics of the FIR-to-\haextcorr{} discussion in this work. The median value fit on low-z galaxies is printed at the lower right corner in each panel.}
    \label{f:implication_IR-line_Halpha}
\end{figure*}

\begin{figure*}
    \centering
    \includegraphics[width=\textwidth]{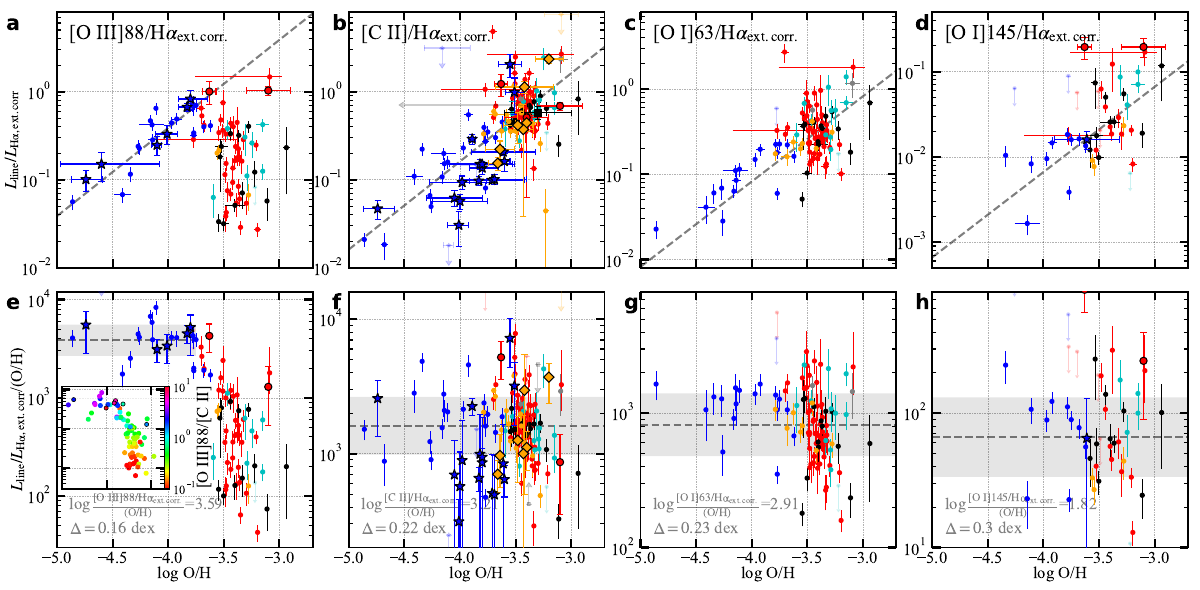}
    \caption[O/H vs. \lline{}/\ha{}.]{Comparison between absolute metallicity O/H and FIR FSL ratio to extinction-corrected \ha{} for \oiii{88} (first column), \cii{} (second column), \oi{63} (third column), \oi{145} (fourth column). The first row shows \oh{} vs. \lline{}/\ha{}, and the second row shows the residual as \lline{}/\ha/(O/H). The gray dashed lines in the first row, and the dashed lines and shades in the second row, represent the linear fitting results printed at the lower left corner in the lower panels. In the inset panel of \oiii{88} comparison, \oiii{88}/\cii{} of each data in the main panel are displayed as color, plotted in the same coordinate as the main panel. The plot style is described in Sec.~\ref{sec:implication_intro}}
    \label{f:implication_O_H-line_Halpha}
\end{figure*}

Due to the lack of a direct tracer of total ionized hydrogen (\hp{}) in the FIR, and the limited availability of free-free radio continuum data, we use extinction-corrected \ha{} (\haextcorr{}) emission as a normalization baseline to investigate the impact of absolute gas-phase metallicity (O/H) on FIR FSL luminosities.

In Fig.~\ref{f:implication_O_H-line_Halpha}, we plot the ratios of \oiii{88}, \cii{}, and the \oi{} doublet relative to \ha{} as a function of \oh{}. 
The bottom panels show the same ratios scaled by O/H to highlight residual trends or scatters.  
All four FIR lines exhibit a clear and approximately linear correlation with metallicity when normalized by \ha{}, indicating a strong abundance dependence. 

For \oiii{88}, the best-fit linear scaling factor is 10\textsuperscript{3.59}, which closely matches the theoretical emissivity ratio $\varepsilon_\mathrm{[O~III]88} / \alpha_\mathrm{B,H\alpha} = 10^{3.57}$ computed at \te{}~=~10\textsuperscript{4}\,K and \edens{}~=~50\,\cc{} with \pyneb{}, and a case-B recombination coefficient $\alpha_\mathrm{B}$ for\ha{} \citep{osterbrock89,draine11}. 
Similarly, the fitted \cii{}/\ha{}/(O/H) ratio of 10\textsuperscript{3.21} is consistent with the expected value, derived using $\varepsilon_{\mathrm{[C~II],}e^-}/\alpha_\mathrm{B,H\alpha}/(1-f_\mathrm{[C~II],neutral})*\mathrm{(C/O)_\odot} = 10^{3.35}$, where $\varepsilon_{e^-}$ denotes the emissivity in ionized gas for \cii{} assuming the aforementioned conditions. 
We adopt \fciin{}~=~0.8 as estimated in the previous section.

These empirical relations show that FIR FSL luminosities scale directly with metallicity, or in words, content of emitting ions.
It also demonstrates the effectiveness of extinction-corrected \ha{} as a normalization for the FIR-based metallicity measurement. 

It is worth noting that the observed trends are primarily driven by dwarf galaxies, which span more than 1.5 dex in \oh{}, while more massive galaxy populations cluster within a narrower metallicity range ($\sim$0.5 dex), thus appearing compressed along the $x$-axis and effectively anchoring the linear-fit. 

A notable deviation from linearity is evident in the \oiii{88}/\ha{} ratio at \oh{}~\textgreater~--3.75. 
This turnover reflects the a decline of the filling factor of doubly ionized oxygen ($ff_{\mathrm{O}^{2+}}$) in metal-rich galaxies, which typically exhibit lower ionization parameters ($U$). 
As $U$ decreases and \oiii{88}/\cii{} falls below $\sim$1.5, the dominant ionization state of oxygen shifts from \opp{} to \op{}, as is also found quantatively in \ppii{}. 
Consequently, both \oiii{88}/\ha{} and \oiii{88}/\cii{} decline by over an order of magnitude, as shown in the inset of Fig.~\ref{f:implication_O_H-line_Halpha}. 
To avoid bias from this ionization effect, only galaxies with \oh{}-$\leq$~--3.75 are included in linear fitting and scatter analysis.

\cii{}/\ha{} plot also shows a hint of deviation, as high-\zz{} LBG/LAEs appear preferentially lower than both the linear fit (dashed line) and the distribution of low-\zz{} dwarf galaxies. 
However, 8 of the 12 high-\zz{} LBG/LAEs below the linear fit are from the same sample \citep[REBELS][]{BR22c,IH22,SL22,RL25,HY25}, and removing the REBELS galaxies would recover a consistent \cii{}/\ha{}--O/H relation, even for other \zz{} \textgreater 6 data points. 
Therefore, we caution that evidence for deviation from this trend is elusive. 
We postulate that the offset of REBELS data points could be a result of intrinsically low \cii{}/\ha{} in \zz{} \textgreater 6 galaxies, or the properties of the galaxies selected in REBELS survey, or the systematic difference in how the metallicity measurement and/or extinction correction is performed in REBELS survey compared to the rest of the FLAMES sample. 

The \oi{145}/\ha{} ratio shows weaker correlation and larger scatter compared to the other lines. 
Despite its theoretical advantage---lower optical depth than \oi{63}---\oi{145} suffers from limited detections and intrinsically lower luminosity (typically $\sim$1/10 of \oi{63}) leading to poorer data quality. 
These factors likely contribute to the observed scatters.

The consistency between the observed \cii{}/\ha{} ratio and its theoretical expectation further supports a nearly constant \cii{} neutral fraction of $\sim$80\%, as shown in Sec.~\ref{sec:implication_fneut}. 
The residual scatter in the \cii{}/\ha{} relation is 0.22 dex, with no discernible dependence on galaxy types. 

These results emphasize the dominant role of elemental abundance in determining FIR FSL luminosities, especially when normalized by a tracer of ionized gas such as \ha{}. 
They also highlight the potential of FIR FSLs---particularly \cii{} and \oiii{88}---as alternative empirical metallicity indicators in high-redshift galaxies where a whole suite of optical lines is not easily accessible, as well as dust-poor UV-selected systems where \ha{} is less attenuated.

\section{Line ``Deficits''}
\label{sec:implication_deficit}

\subsection{Similarities among FIR FSL ``Deficits''}
\label{sec:implication_deficits-deficits}

\begin{figure*}
    \centering
    \includegraphics[width=\textwidth]{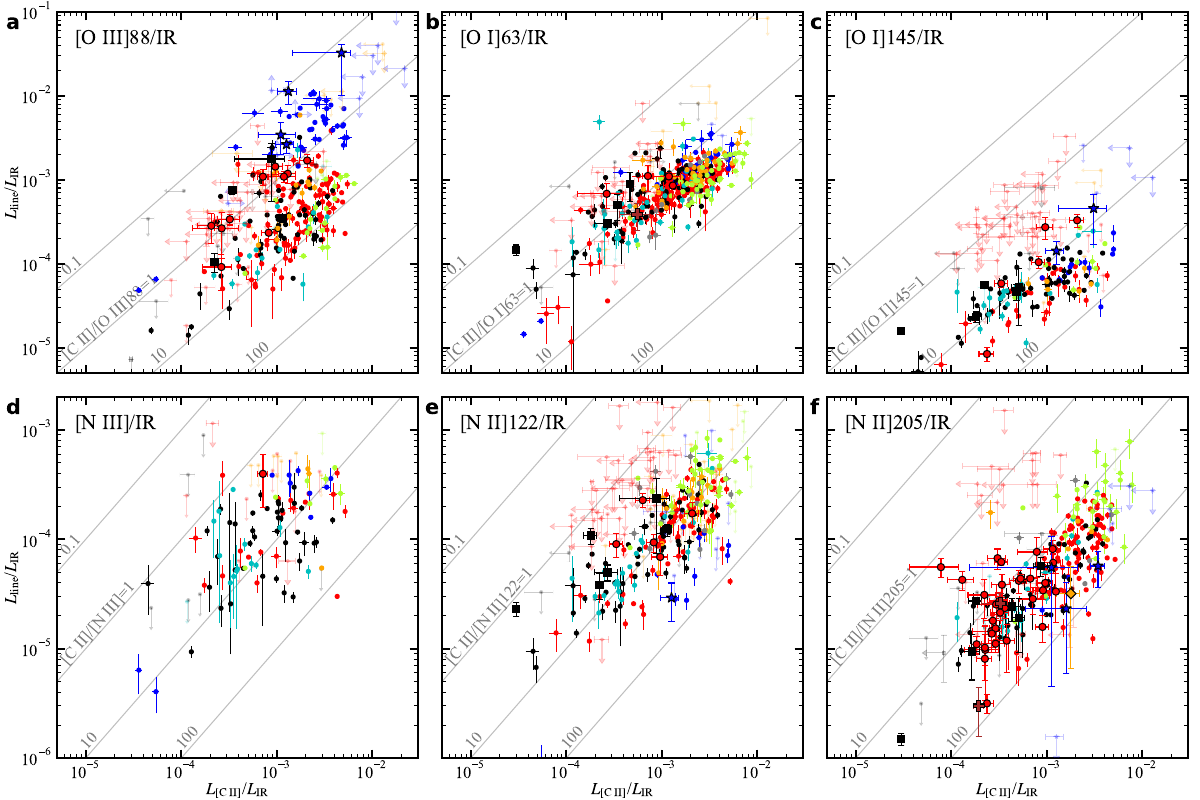}
    \caption[\cii{} ``deficit'' vs. other FIR FSL ``deficits.'']{Comparison of \cii{} ``deficit'' (x-axis) with the ``deficit'' of other FIR FSLs (y-axis). Gray dashed lines indicate constant \cii{}-to-line luminosity ratios. Note that the \oi{145} and \nii{122} panels include a significant number of upper limits.}
    \label{f:implication_deficit-deficit}
\end{figure*}

As highlighted in \ppi{}, the FIR line-to-IR luminosity ``deficit''---most prominently observed in the \cii{} line---is not unique to \cii{}, but is a widespread phenomenon affecting all major FIR fine-structure lines (FSLs). 
To further investigate whether these ``deficits'' arise from a shared physical mechanism, Fig.~\ref{f:implication_deficit-deficit} directly compares the ``deficit'' of \cii{} with that of other FIR FSLs.

Strikingly, all lines except \oiii{88} show strong, nearly linear correlations with \lcii{}/\lir{} over two orders of magnitude throughout the sample. 
These ``deficit-to-deficit'' relations suggest a common origin for the declining line-to-IR luminosity ratios, a conclusion further supported by the fact that high-\zz{} galaxies---despite their offsets in IR-``deficit'' relations---follow the same ``deficit-to-deficit'' trends. 
The apparent universality of these correlations across different environments and redshifts implies that the mechanism reducing line-to-IR luminosity ratios is both fundamental and widespread. 

While the general trends are robust, the degree of scatter and slope deviations vary across lines. 
The \oi{} lines exhibit the tightest correlation with \cii{}, with a scatter below 0.3 dex, although slight deviations from the linear scaling are observed. 
These are likely driven by the correlation between gas density and dust temperature, as will be explored further in Sec.~\ref{sec:implication_ir}. 
The \nii{} lines also show good agreement with the \cii{} ``deficit'' trend, with the exception of dwarf galaxies and LBG/LAEs, which systematically fall below the main relation, reflecting their lower nitrogen abundance. 

In contrast, the \oiii{88} line shows the largest scatter. 
However, within each galaxy class, a linear correlation between \loiii{88}/\lir{} and \lcii{}/\lir{} still persists. 
The total scatter arises from systematic offsets between galaxy populations: local dwarf galaxies and high-\zz{} LBG/LAEs show lower \cii{}/\oiii{88} ratios, and high-\zz{} DSFGs and AGNs are offset to higher values, while low-\zz{} U/LIRGs and AGNs display the highest values of \cii{}/\oiii{88}. 
These offsets are attributed to variations in the ionizing radiation field, as discussed in sec.~3.6 of \ppi{} and quantified in Sec.~3.3 of \ppii{}, and the offset of high-\zz{} dusty galaxies will be further examined in Sec.~\ref{sec:implication_diff}.

\subsection{\texorpdfstring{\ha{}}{H alpha} ``Deficit''}
\label{sec:implication_ha_deficit}

\begin{figure*}
    \centering
    \includegraphics[width=\textwidth]{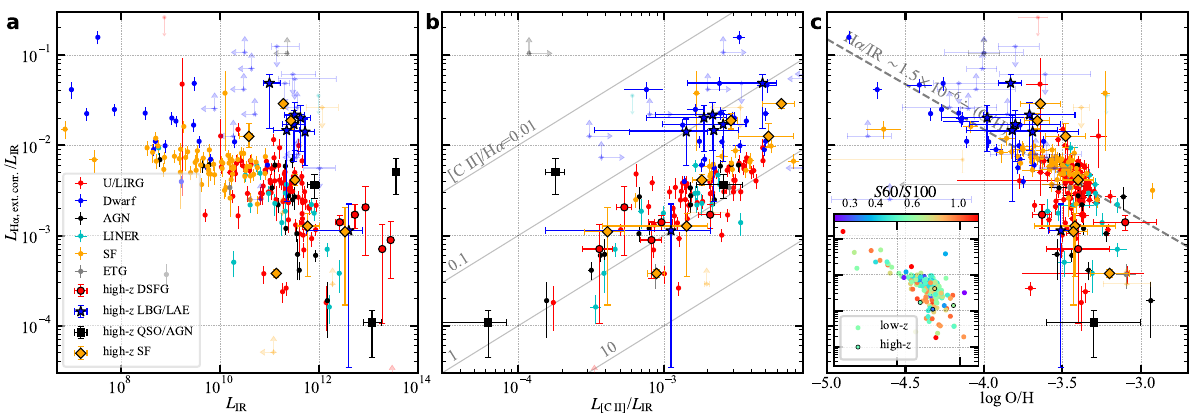}
    \caption[\haextcorr{} vs. IR luminosity and metallicity.]{\bsf{a}, extinction-corrected \lha{}/\lir{} vs. \lir{}, illustrating the \ha{} ``deficit.'' \bsf{b}, Correlation between the ``deficits'' trends of \ha{} and \cii{}. \bsf{c}, \lha{}/\lir{} as a function of absolute gas-phase metallicity O/H, showing a breakdown of the linear trend for dusty galaxies}
    \label{f:implication_Halpha_IR}
\end{figure*}

While most discussions of the line ``deficit'' have focused on FIR lines, this phenomenon is not restricted to FIR. 
Fig.~\ref{f:implication_Halpha_IR} shows that the well-known SFR tracer, the extinction-corrected \ha{} (\haextcorr{}) also exhibits a ``deficit'' at high \lir{}. 
The \lhaextcorr{}/\lir{} ratio declines significantly for \lir{}~$\gtrsim$~10\textsuperscript{11}\,\lsun{} in (\bsf{a}), mirroring the trend seen in \lcii{}/\lir{}. 
Moreover, the two ``deficits'' are tightly correlated in (\bsf{b}), with a scatter of less than 0.2 dex.

This alignment between \ha{} and \cii{} ``deficits'' is particularly revealing. 
As shown in Fig.~\ref{f:implication_O_H-line_Halpha}(\bsf{a}), the \cii{}/\ha{} ratio correlates strongly with metallicity. 
But U/LIRGs and AGNs typically span a narrow range in \oh{}, the observed agreement between their \cii{} and \ha{} ``deficits'' suggests a metallicity-independent mechanism.

Unlike \cii{}, which arises from both neutral and ionized regions and is sensitive to heating efficiency and gas phase, \ha{} originates unambiguously from ionized gas and directly traces the ionizing photon rate. 
Although affected by dust extinction, \ha{} is corrected for attenuation here. 
Therefore, the parallel behavior of \cii{} and \ha{} ``deficits'' rules out explanations based on metallicity, \cii{} excitation conditions, \fciin{}, or extinction effects.

A common factor influencing both tracers is electron density, which affects emissivity. 
However, as shown in Sec.~3.5 of \ppi{}, \edens{} varies only modestly across galaxy populations and redshifts. 
The change in line emissivity due to \edens{} is typically within an order of magnitude—insufficient to explain the 1–2 dex suppression seen in line-to-IR ratios. 
Moreover, no systematic correlation is observed between \edens{} and galaxy type, making density an unlikely primary driver for the trend.

\subsection{Abnormal Behaviour of IR Luminosity}
\label{sec:implication_ir}

\begin{figure*}[ht!]
    \centering
    \includegraphics[width=\textwidth]{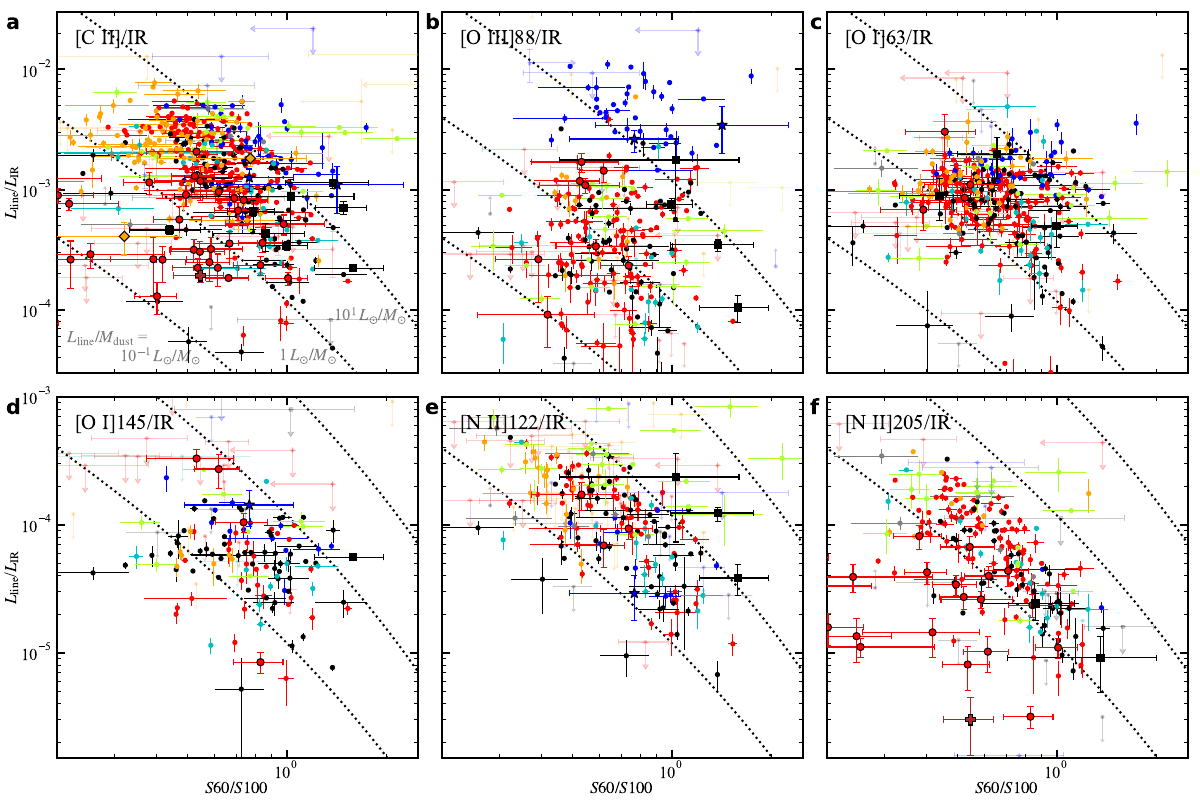}
    \caption[\tcolor{} vs. line ``deficit.'' ]{FIR line ``deficits'' as a function of the dust color temperature \tcolor{}. The dashed lines are the OT-MBB SED model predictions corresponding to constant values of \lline{}/\md{}, as printed in the upper left panel. }
    \label{f:implication_S60_100-deficit}
\end{figure*}

\begin{figure}[h]
    \centering
    \includegraphics[width=\halfwdth]{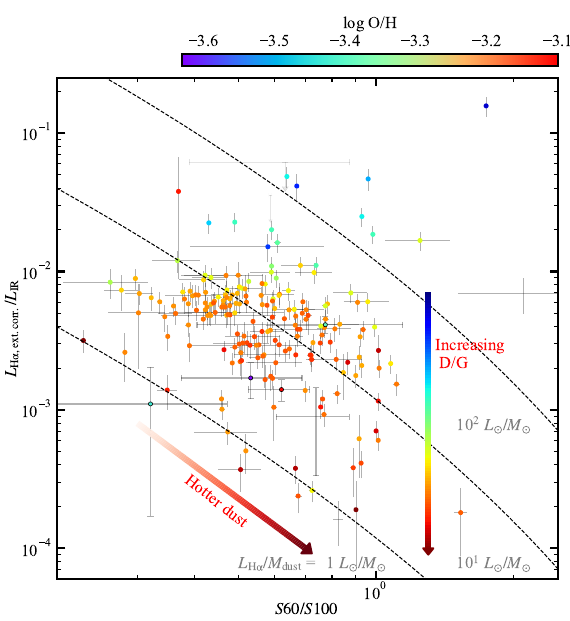}
    \caption[S60/S100 vs. \ha{}/\lir{}.]{Dust color temperature \tcolor{} vs. extinction-corrected \lha{}/\lir{}. Points with measured metallicity are color-coded by \oh{}, otherwise showing only the errorbars. Model lines correspond to constant \lha{}/\md{} ratios derived from OT-MBB SEDs.}
    \label{f:implication_Ha_deficit}
\end{figure}

The tight linear correlations among FIR line ``deficits'' and with \ha{} suggest a common denominator driving the decreasing ratio of line emission to \lir{}. 
Given that the relative line luminosities to each other differ by less than an order of magnitude, while the observed ``deficits'' span over two orders, the variation must originate from \lir{}.

As discussed in sec.~3.2 of \ppi{}, \lir{} is highly sensitive to dust temperature (\td{}), which can be traced by \tcolor{}. 
Fig.~\ref{f:implication_S60_100-deficit} shows that the FIR line ``deficits'' correlate with \tcolor{}. 
The optically thin modified blackbody (OT-MBB) spectral energy distribution (SED) models predict the observed trends under constant \lline{}/\md{}, indicating that increasing \td{} alone---without changing gas excitation or line emissivity or dust-to-gas ratio---can explain the observed ``deficit'' trends.

The same trend holds for \ha{}, as shown in Fig.~\ref{f:implication_Ha_deficit}, reinforcing the hypothesis that elevated dust temperatures enhance \lir{} disproportionately relative to gas emission. 
The correspondence of data points with lines of constant \lline{}/\md{} again suggests that the ``deficit'' is primarily driven by enhanced emissivity in warmer dust.

Scatters remain, particularly for \cii{} and \oiii{88}, due to effects discussed earlier. 
High-\zz{} points are more uncertain, with large error bars and limited statistics. 

\begin{figure}[h]
    \centering
    \includegraphics[width=\halfwdth]{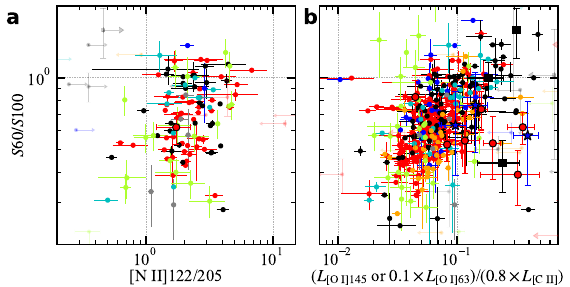}
    \caption[S60/S100 vs. density. ]{Dust color temperature \tcolor{} vs. density diagnostics \bsf{a}, \nii{122/205} (ionized gas density); and \bsf{b}, \oi{145}/\cii{} (neutral gas density proxy).}
    \label{f:implication_n-S60_100}
\end{figure}

We also examine the physical basis of this trend. 
Fig.~\ref{f:implication_n-S60_100} confirms a weak but positive correlation between \tcolor{} and \edens{} traced by \nii{122/205}, consistent with previous findings \citep{herrera16}. 
A tighter correlation is seen between \tcolor{} and \oi{}/\cii{}, though this may partly reflect data quality differences. 
And the interpretation is complicated by degeneracies between density and optical depth.

\begin{figure}[h]
    \centering
    \includegraphics[width=\halfwdth]{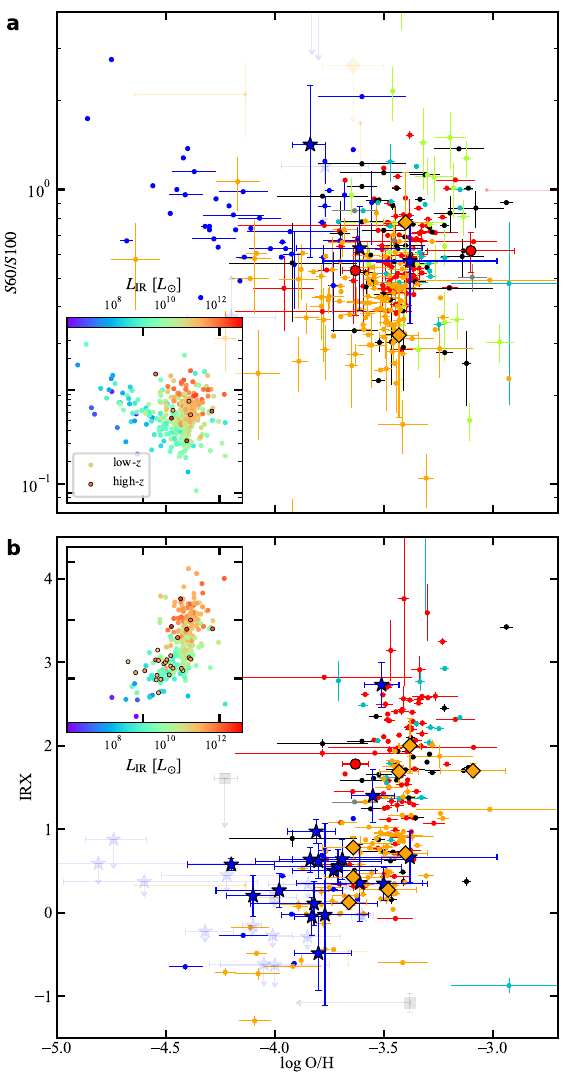}
    \caption[O/H vs. dust color and IRX.]{Metallicity \oh{} vs. \bsf{a}, dust color \tcolor{}; and \bsf{b}, IR excess IRX. Inset panels show data points color-coded by \lir{}. }
    \label{f:implicaiton_O_H-S60_100}
\end{figure}

Finally, Fig.~\ref{f:implicaiton_O_H-S60_100} explores the relationship between metallicity and dust properties. 
A ``V''-shaped trend is evident in the \tcolor{}--\oh{} plot: as metallicity increases to \oh{}~$\sim$~--3.5, dust temperature decreases monotonically, before rising again at higher \lir{}. 
This reflects the turnover of the \td{}--O/H relation found in \citet{engelbracht08}, but with much higher significance and extending to higher \td{}. 
A similar bimodal trend is seen in IRX. 
These transitions correlate with an increase in \lir{}, and the decoupling from metallicity dependence occurs at \lir{}~\textgreater~10\textsuperscript{11}\,\lsun{}, suggesting a shift in dust heating regimes. 
The implications for line ``deficits'' and dust physics are discussed further in the following section.

\section{Discussion}
\label{sec:implication_discussion}

\subsection{The Ionized -- Neutral Emission Coherence}
\label{sec:implication_ionized-neutral}

A recurring theme throughout this study is the strong correlation between the line emissions arising from neutral gas---mainly \cii{} and \oi{}---and those originating in low-ionization ionized gas, such as \nii{} and \haextcorr{}. 
These correlations are particularly evident compared to high-ionization lines (\ppi{}, sec.~3.6) or total infrared luminosity (line ``deficits'' in Sec.~\ref{f:implication_deficit-deficit}). 
In contrast to the strong variations relative to \lir{}, the FIR FSLs from neutral and low-ionization gas are coupled so closely that they often appear interchangeable in diagnostic plots. 

This behavior is physically intuitive: both neutral and ionized gas phases serve as cooling channels for the ISM, and both are primarily heated by ultraviolet (UV) photons emitted by young, massive stars. 
The resulting tight correlations among the corresponding emission lines reflect a shared energy source.

Beyond a shared energy source, the data suggest that neutral and ionized gas phases are not only energetically but also spatially coherent. 
Among all physical parameters examined in \ppi{}, metallicity emerges as the dominant factor that influences the relative strengths of neutral and low-ionization lines (\ppi{}, sec.~3.7), particularly evident in comparisons of neutral gas tracers with \ha{} emission (see Sec.~\ref{sec:implication_fir-ha}). 
These direct dependences on O/H or N/O imply a shared volume, or in other words a close spatial association, between the emitting regions of the two phases.

In principle, differences in ionization energy---EUV photons for ionized gas and far-ultraviolet (FUV) photons for neutral gas---could introduce significant variability due to geometry-dependent photon escape fractions, gas density and pressure, and the heating efficiency of dust grains or gas particles. 
If the emitting regions were physically separated, these factors would lead to observable scatter or decoupling in the line ratios. 
However, the observed tight correlations suggest that these influences are either negligible or highly self-regulated, and that the neutral-to-ionized gas mass ratio in the line-emitting regions remains approximately constant across galaxy types and environments.

This interpretation is further supported by the remarkably consistent \cii{} neutral fraction (\fciin{}), as derived independently from both \cii{}/\nii{} and \cii{}/\ha{} ratios (see Sec.~\ref{sec:implication_fneut} and \ref{sec:implication_fir-ha}).

The coherence between neutral and ionized gas emission has practical implications for ISM studies. 
The strong coupling between these phases allows for the use of neutral gas lines---\cii{} and \oi{}---in conjunction with ionized gas tracers to probe physical conditions like elemental abundances and radiation field strengthes in galaxies. 
This is particularly valuable in high-\zz{} systems, where optical lines are challenging to obtain from the ground and FIR lines may serve as the primary observational window into their ISM.

In summary, the tight correlation between the neutral and low-ionization FIR lines underscores the physical and spatial coherence in their emitting gas. 
This coherence reflects a common UV heating source and a stable mass ratio between the emitting (or irradiated) neutral and ionized gas phases, reinforcing the utility of FIR lines as reliable diagnostics of ISM conditions across cosmic time.

\subsection{Caution on Applying PDR Models on Galaxy Integrated Observations}
\label{sec:implication_pdr}

The coherence between ionized and neutral gas emission discussed in the previous section presents a significant challenge to the classical PDR framework. 
In this section, we critically assess the validity of applying PDR models to galaxy-integrated observations and argue that such an approach may not be appropriate for interpreting FIR FSLs and IR continuum emission on unresolved scales.

The PDR paradigm \citep{tielens85} has served as the foundational theory for studying atomic and molecular ISM for nearly four decades. 
Its success lies in its ability to link atomic and molecular phases through photochemistry and to describe how UV radiation couples to the neutral medium via photoelectric heating on dust grains. 
These models have been successful in reproducing the emission structures and line intensities of resolved PDRs around Galactic \hii{} regions and molecular clouds \citep{hollenbach97,wolfire22}.

However, the application of PDR models has extended far beyond their original domain. 
They are now frequently used to interpret unresolved observations of entire galaxies, including high-\zz{} systems. 
This extrapolation is often made without critical evaluation, relying on implicit assumptions that may not hold at global scales, and frequently lacking validation from independent diagnostics or consistency with other ISM tracers. 

Applying PDR models to unresolved galaxy observations implicitly assumes: 
(1) all observed FIR line emission originates from PDRs;
(2) the observed dust continuum emission is also produced within PDRs;
(3) the ensemble of emitting regions can be represented as a mixture of PDRs described by the model parameter space;
(4) the primary energy sources for both gas and dust are those included in the model (i.e., FUV, X-rays, cosmic rays).
While these assumptions may be reasonable for resolved Galactic regions, they are simplistic and may not hold valid for integrated galaxy measurements. 
For example, in the Milky Way, most FIR dust emission arises from diffuse ISM heated by a mild interstellar radiation field (ISRF), rather than from dense PDRs \citep{caux84,lonsdale87,draine85,li01}. 
Similar findings have been reported in extragalactic studies, where diffuse dust emission dominates the FIR luminosity in normal star-forming systems \citep{draine07}.

Regarding gas emission, about half of the Galactic \cii{} originates in dense PDRs \citep{pineda13}. 
Observations of high-\zz{} galaxies frequently reveal that dust continuum emission is more spatially compact than \cii{} line emission \citep[e.g.,][]{CS18b,FS19,RM20a,MR25,RC25}. 
These findings indicate a spatial disconnect between the two and contradict the assumption of co-spatial origin for FIR continuum and line emission in PDRs. 
Resolution limitations and the blending of multiple ISM phases further complicate the interpretation of unresolved data \citep[see also][]{wolfire22}.

PDR-based interpretations of integrated galaxy data are often the only accessible observational tool but have lacked cross-validation with independent diagnostics. 
Justifications for their use tend to fall into two categories:
\textit{(1) Similarity to Galactic PDR line ratios:} 
Early studies \citep[e.g.,][]{stacey91,kaufman99,malhotra01} noted that normal galaxies and starburst nuclei exhibited FIR line ratios consistent with dense PDR models. 
However, these comparisons were later extrapolated to more extreme systems, such as U/LIRGs and high-\zz{} galaxies, without reevaluating the underlying assumptions. 
In many of these cases, the observed line ratios were interpreted as evidence for extremely high densities and strong ISRF \citep[e.g.,][]{RM19}.

\textit{(2) Self-consistency in parameter space:} 
PDR model applications often involve fitting ratios of a small number of observables (at most five independent observables \lfir{}, \cii{}, \oi{}, low and mid-$J$ CO lines) to a model grid with two primary parameters (typically density $n$ and radiation field strength $G$). 
While fitting four independent ratios to two parameters may appear as strong parameter marginalization and yield a well-constrained solution, in practice the degrees of freedom are often insufficient. 
Typically, only three or four observables are available, and uncertainties in line ratios (up to 0.5 dex) limit their constraining power. 
Several ad hoc corrections---such as adjusting the \cii{} neutral fraction, accounting for \oi{63} self-absorption, CO optical depth, or contributions from shock or x-ray-dominated ragions (XDRs)---are selectively invoked to tune a fit, undermining the physical robustness of the model inference. 

Beyond a lack of validation, PDR-based interpretations can contradict other well-established observational diagnostics using ionized gas. 
Observations consistently show that metal-poor dwarf galaxies exhibit higher $U$ than metal-rich systems like ULIRGs, evidenced by stronger high-ionization lines such as \oiii{} \citep[\ppi{}][]{kewley19}. 
Since no significant difference in electron density (\edens{}) is observed between these populations, $U$ should scale directly with the ISRF strength $G$ \citep{C19}. 
Photoionization models that couple \hii{} regions with PDRs do predict a correlation between increasing $U$ and rising \oiii{} nd \niii{} over low-ionization lines or \lfir{} \citep[\ppii{}][]{abel05,abel09}. 
The high $G$ in dwarf galaxies is further backed by their high \td{}. 

However, PDR-based fits to \lcii{}/\lfir{} or \loi{}/\lfir{} ratios yield the opposite conclusion: ULIRGs are inferred to have the highest $G$, while dwarf galaxies are interpreted to have mild ISRF due to their \textit{lack of} a \cii{} ``deficit.'' 
This contradicts the intense radiation field expected in dwarf galaxies, not only because of high $U$, but also the low dust-to-gas ratio (D/G) that further increases FUV photo-to-dust grain contrast and lowers heating efficiency. 
To reconcile this contradiction, another degree of freedom, PDR porosity, had to be introduced to explain \lcii{}/\lir{} in dwarf galaxies \citep{C19}. 

The PDR framework also struggles to explain the tight correlation between neutral and ionized gas lines identified in this study (Sec.~\ref{sec:implication_ionized-neutral}). 
It appears as if the ionization structure extends beyond the classical ionization front, making low-ionization lines only reflect the amount of the photons that creates \cp{}, so that \oiii{88}/\cii{} measures $U$, while \cii{}/\ha{} is a function of metallicity. 
However, this contradicts the PDR paradigm where neutral gas is heated independently by all the incidence of FUV photons in regions offset from ionized gas. 
Although D/G is affected by metallicity, it mainly changes the physical thickness of PDR, while PDR strucutre should remain similar as a funciton of $A_\mathrm{V}$. 
As a result, total heating in PDR reflected in \cii{} and IR luminosity should not depend on elemental abundance. 

Furthermore, the PDR framework faces an intrinsic limitation in explaining the observed line ``deficit,'' particularly for ionized gas emission. 
PDRs are, by definition, powered by non-ionizing FUV ($6\leq h\nu \leq 13\,\mathrm{eV}$) photons that escape from the surrounding ionized regions and deposit energy downstream in the neutral medium. 
As such, any mechanism proposed within the PDR paradigm to explain the \cii{} ``deficit''---such as reduced photoelectric heating efficiency, suppression of \cii{} emissivity due to thermalization at high densities, self-absorption, or dust extinction---can only affect the neutral gas phase. 
These mechanisms cannot account for the simultaneous and equally significant suppression of ionized gas emission lines like \nii{} or \ha{}, which are powered by ionizing (EUV) photons and arise upstream in the energy flow. 

To understand all these, we may need to re-evaluate our conceptual model of the neutral ISM heating and distribution, or develop additional theories that can explain the observations. 

This critique is not intended to undermine the value of PDR models, which remain indispensable for interpreting detailed, resolved observations of galactic clouds. 
PDRs undoubtedly exist in all galaxies and could contribute significantly to FIR line and continuum emission. 
However, interpreting galaxy-integrated data---particularly in IR-luminous systems showing strong line ``deficits''---using this prescriptive framework often leads to inferred physical conditions that are extreme and inconsistent with other ISM tracers.

We therefore urge caution when applying PDR models to unresolved extragalactic observations. 
Conclusions drawn from such modeling should be critically evaluated in light of: 
(1) the assumptions about emission origins; 
(2) the lack of external validation from independent tracers;
(3) the apparent contradictions with ionized gas diagnostics;
(4) the inability to self-consistently explain the line ``deficit'' phenomenon across gas phases.
Without appropriately addressing these issues---through multiphase modeling, improved spatial resolution, or more comprehensive diagnostics---the interpretation of FIR lines in galaxy-integrated data should not rely solely on classical PDR models.

\subsection{Ionized \& Neutral Gas -- Dust Dichotomy}
\label{sec:implication_dichotomy}

\begin{figure}[]
    \centering
    \includegraphics[width=\halfwdth]{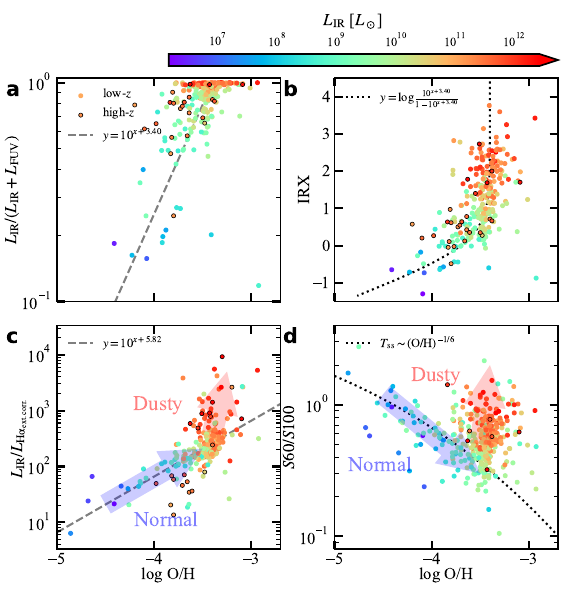}
    \caption{Metallicity \oh{} v.s. dust properties, color-coded by IR luminosity \lir{}, according to the color bar displayed on the top right. \bsf{a}, \oh{} v.s. the dust obscuration fraction $f_\mathrm{IR}$ = \lir{}/(\lir{}+\lfuv{}), the gray dashed line illustrates a linear fit of the low $f_\mathrm{IR}$ part of the data. \bsf{b}, \oh{} v.s. IRX, same as Fig.~\ref{f:implicaiton_O_H-S60_100}(\bsf{b}). The black dotted line represents $\mathrm{IRX} = \log f_\mathrm{IR}/(1-f_\mathrm{IR})$ using the linear fit of $f_\mathrm{IR}$ in (\bsf{a}). \bsf{c}, \oh{} v.s. \lir{}/\lhaextcorr{}, another measure of obscuration fraction that is normalized by extinction-corrected \ha{}, same as Fig.~\ref{f:implication_Halpha_IR}(\bsf{c}). The gray dashed line shows a linear fit to the low-metallicity data, the faint blue and red arrows denote the trends of data points along increasing \lir{}, for normal and dusty galaxies, respectively. \bsf{d}, \oh{} v.s. dust color temperature \tcolor{}, same as Fig.~\ref{f:implicaiton_O_H-S60_100}(\bsf{a}). The black dotted line shows the predicted scaling between the steady state temperature $T_\mathrm{ss}$ and metallicity, adopting OT-MBB SED. The arrows denotes the trends of data points distribution along \lir{}. }
    \label{f:implication_O_H-dust}
\end{figure}

In Sec.~\ref{sec:implication_deficits-deficits} and \ref{sec:implication_ha_deficit}, we demonstrate that the line ``deficit'' is a universal phenomenom and is primarily driven by the wildly varying \lir{}. 
Sec.~\ref{sec:implication_ir} shows that the line ``deficit'' is mainly caused by the fact that dust becomes warmer, instead of just a higher dust mass-to-line luminosity ratio. 
In fact, this trend of increasing dust temperature corresponds to the abrupt deviation from the \td{}-O/H relation seen in normal galaxies (Fig.~\ref{f:implicaiton_O_H-S60_100} and \ref{f:implication_O_H-dust}). 
Here, we argue that this ``deficit'' reflects a deeper phenomenon: a decoupling between dust emission and gas spectral line emission in dusty galaxies, which we refer to as the \textit{gas–dust dichotomy}. 
We will summarize the observed characteristics of the anomalous behavior of dust emission in these galaxies, and propose several potential scenarios in the next section. 

We begin by examining the behavior of dust emission in normal galaxies. 

As shown in Fig.~\ref{f:implication_O_H-dust}, across the range from low-metallicity dwarfs to SF galaxies, increasing metallicity (\oh{} from --5.0 to --3.5) is accompanied by a steady rise in \lir{}. 
Simultaneously, the dust-obscuration fraction---quantified either as $f_\mathrm{IR} = L_\mathrm{IR} / (L_\mathrm{IR} + L_\mathrm{FUV})$, obscuration of the bolometric luminosity, or as $L_\mathrm{IR} / L_{\mathrm{H}\alpha,\mathrm{ext.corr.}}$, which tracks the fraction of ionizing radiation reprocessed by dust---increases linearly with metallicity (Fig.~\ref{f:implication_O_H-dust}(\bsf{a}) and (\bsf{c})). 
These trends can be approximated as:
\begin{equation}\begin{split}
    f_\mathrm{IR} \sim 10^{3.40} \cdot (\mathrm{O/H}),~~
L_\mathrm{IR} / L_{\mathrm{H}\alpha,\mathrm{ext.corr.}} \sim 10^{5.82} \cdot (\mathrm{O/H}).
\end{split}\end{equation}
This linear relation in $f_\mathrm{IR}$ also reproduces the curved trend of IRX versus metallicity (Fig.~\ref{f:implicaiton_O_H-S60_100}(\bsf{b})), shown as the dotted line in Fig.~\ref{f:implication_O_H-dust}(\bsf{b}).

The linear scaling between $f_\mathrm{IR}$ and metallicity is non-trivial, given that the D/G has been shown to correlate more steeply with \oh{} \citep{D19}. 
If the dust absorption cross-section remains constant, one would expect the obscuration fraction to drop more rapidly than metallicity in more meta-poor environments due to the super linear D/G. 
The observed linear trend therefore implies the presence of an additional compensating factor: increased dust temperature.

In fact, as metallicity decreases from --3.5 to --5.0, \tcolor{} rises (Fig.~\ref{f:implicaiton_O_H-S60_100}(\bsf{a}) and \ref{f:implication_O_H-dust}(\bsf{d})), suggesting enhanced heating efficiency. 
A likely explanation is that the ISRF becomes more intense in metal-poor galaxies. 
This is consistent with the well-established anti-correlation between metallicity and the ionization parameter $U$ \citep[see sec.~3.7 in \ppi{}, sec.~4.1 in \ppii{}, and][]{kewley19}.

Assuming a scaling relation $U \propto (\mathrm{O/H})^{-1}$, as estimated in \ppii{}, and extrapolating to the ISRF strength $G$, we can estimate the equilibrium dust temperature using eq.~24.18 in \citet{draine11} as $T_\mathrm{ss} \approx \left(\langle Q_\mathrm{abs}\rangle_\star/Q_0\right)^{1/(4+\beta)}\left(G/G_0\right)^{1/(4+\beta)}$. 
Adopting $\beta = 2$, and normalizing to $T_\mathrm{ss} = 25$ K at $\log(\mathrm{O/H}) = -3.31$, we derive $T_\mathrm{ss}\sim \mathrm{(O/H)}^{-1/6}$, shown as the dotted line in Fig.~\ref{f:implication_O_H-dust}(\bsf{d}). 
The observed trend aligns reasonably well with this scaling, though the data suggest a slightly steeper slope, implying either a stronger increase in $G$ or an enhanced dust absorption cross-section in low-metallicity environments.

Additionally, the fact that $f_\mathrm{IR}$ scales linearly with the total metal content---rather than the dust mass---compensated by higher dust temperatures in low-metallicity galaxies, as well as a similar linear O/H dependence of \cii{}/\ha{}, may point to a gas-, instead of dust-, regulated heating–cooling balance in neutral gas. 
These behaviors are consistent with gas heating through a component that scales linearly with O/H, e.g., photoionization of carbon \citep{werner70}.
However, exploring this possibility in detail is beyond the scope of this work.

IR-luminous dusty galaxies deviate significantly from the trends described above. 
As metallicity increases to $\sim$--3.4, \lir{} surpasses 10\textsuperscript{11}\,\lsun{}, and the obscuration fraction $f_\mathrm{IR}$ saturates near 100\% (Fig.~\ref{f:implication_O_H-dust}(\bsf{a})). 
However, \lir{} continues to increase, and \lir{}/\haextcorr{} rises well beyond the linear trend (Fig.~\ref{f:implication_O_H-dust}(\bsf{c})), despite little or no change in metallicity (also supported by Fig.~\ref{f:implication_Halpha_IR}). 
This results in an over an order-of-magnitude increase in the obscuration fraction of ionizing radiation---manifesting as the \ha{} ``deficit.'' 

This deviation is driven by a sharp increase in dust temperature. 
In Fig.~\ref{f:implication_O_H-dust}(\bsf{d}), \tcolor{} for dusty galaxies rises steeply and becomes uncorrelated with metallicity, forming a distinct ``V''-shaped trend. 
Sec.~\ref{sec:implication_ir} further demonstrates that this rapid increase in \td{}, rather than D/G or line emissivity, is manly responsible for the observed line ``deficit.''

This abnormal behavior indicates a fundamental change in the dust heating mechanism. 
While gas heating continues to ``trace'' the ionizing radiation field---evidenced by the strong correlation between gas line emission across phases (see Sec.~\ref{sec:implication_ionized-neutral})---dust heating becomes increasingly decoupled. 

The breakdown of the \td{}–\oh{} relation in dusty galaxies marks the onset of a \textit{gas--dust dichotomy}: a physical decoupling between the luminosities of gas-phase spectral lines emission and the cold dust continuum. 
While gas emission remains anchored to the ionizing photon budget, dust emission becomes disproportionately enhanced, leading to the systematic suppression of line-to-continuum ratios observed as line ``deficits.''

\subsection{The \texorpdfstring{\cii{}}{[C II]} ``Deficit'' Problem: Ionized Gas ``Deficit'' or Dust IR ``Excess''}
\label{sec:implication_cii_deficit}

\begin{figure}[]
    \centering
    \includegraphics[width=\halfwdth]{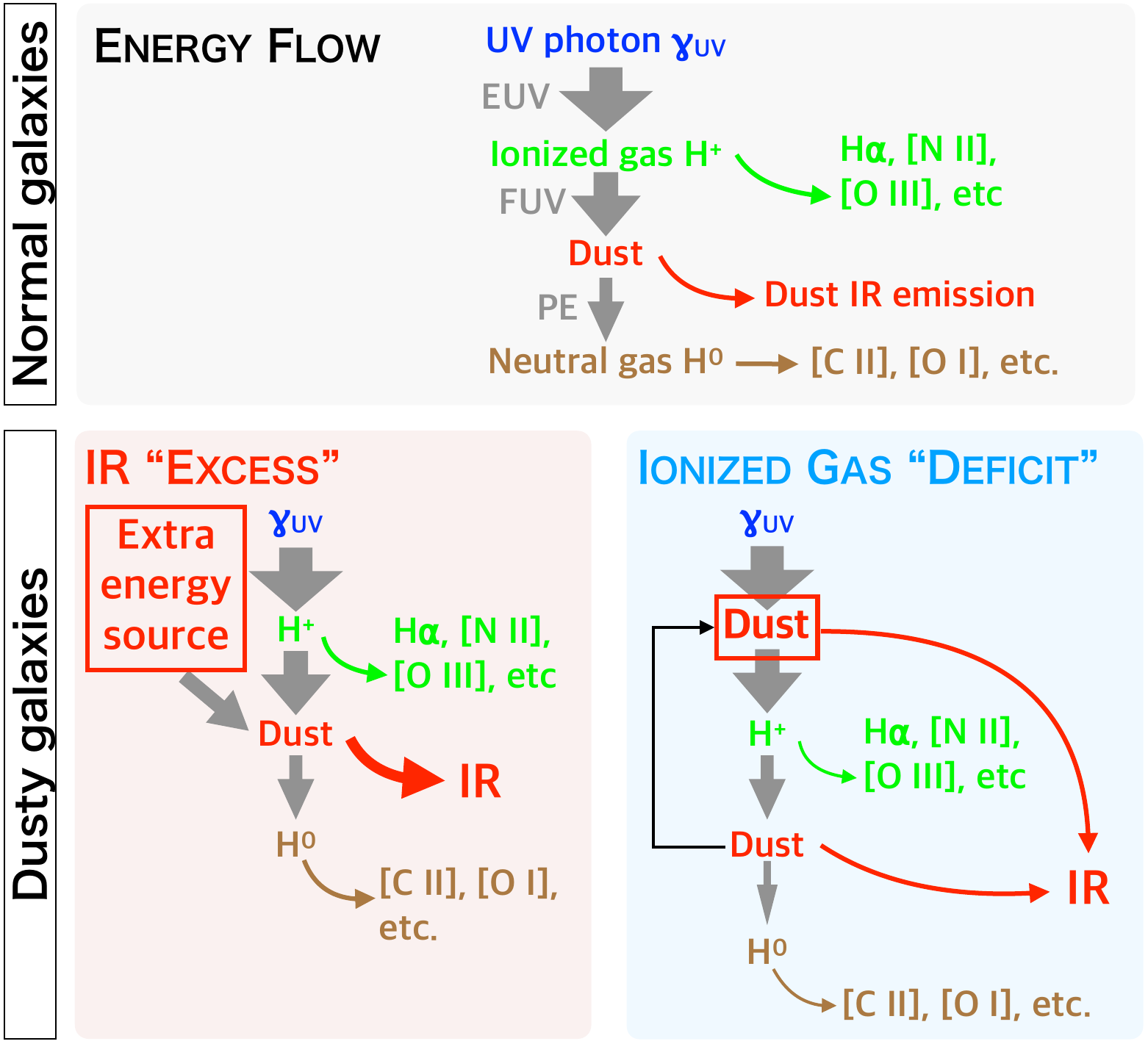}
    \caption[Energy flow diagram. ]{Energy flow diagram for the IR ``excess'' and ionized gas ``deficit'' scenarios}
    \label{f:flow}
\end{figure}

In this section, we present a new perspective on the well-known \cii{} ``deficit'' problem. 
Before exploring theoretical explanations, we first summarize the key observational features of galaxies that fall along the ``deficit'' branch:
\begin{enumerate}
    \item The \cii{} ``deficit'' is defined as a significant drop in the \cii{}/\lir{} ratio below 10\textsuperscript{-3}, observed in heavily dust-obscured environments. This typically occurs at \lir{} \textgreater 3 \texttimes 10\textsuperscript{11} \lsun{} for low-\zz{} galaxies, and \lir{} \textgreater 3 \texttimes 10\textsuperscript{12} \lsun{} for high-\zz{} galaxies.
    \item Along the ``deficit'' sequence, \lir{} increases by more than two orders of magnitude, while \lcii{} increases by less than an order of magnitude.
    \item These galaxies exhibit tight correlations among low-ionization lines from both ionized (\nii{122,205}, \ha{}) and neutral gas (\cii{}, \oi{}), after correcting for elemental abundances. 
    \item All low-ionization lines including \haextcorr{} display line ``deficit.'' Even high-ionization lines like \oiii{88} follow the same ``deficit'' trend within each galaxy population.
    \item After correcting for nitrogen abundance effects, dusty galaxies show similar \fciin{} values to those of galaxies without a \cii{} ``deficit.'' 
    \item The decline in \lline{}/\lir{} is primarily driven by an increase in \td{}, with only a modest change ($\sim$0.5 dex) in the line luminosity to dust mass ratio ($L_\mathrm{line}/M_\mathrm{dust}$).
    \item The rise in \td{} and \lir{} marks a break from the \td{}--O/H relation observed in normal galaxies. Dusty galaxies show little variation in metallicity, and based on diagnostics from \ppi{} and \ppii{}, they also exhibit the lowest $U$ and \te{} among star-forming systems.
    \item Dust temperature (\td{}) shows a mild correlation with ionized gas density as traced by \nii{122}/\nii{205}, and a strong correlation with neutral gas density inferred from \oi{}/\cii{}.
\end{enumerate}
Any explanation of the \cii{} or general line ``deficit'' must account for all of these constraints.

With these observational constraints in mind, we revisit the previous explanations for the \cii{} ``deficit'':
\begin{enumerate}[]
    \item \label{itm:pdr} Reduced photoelectric heating in PDR due to intense ISRF or charged dust \citep{kaufman99,malhotra01,herrera18b}. 
    \item \label{itm:fciin} Reduced \fciin{} leading to lower \lcii{} \citep{malhotra01,croxall17}.
    \item \label{itm:optical_thick} Optically thick \cii{} \citep{abel07,gerin15}. 
    \item \label{itm:attenuation} \cii{} attenuated by dust \citep{papadopoulos10}. 
    \item \label{itm:cii_ionized} Carbon mostly in higher ionization state \citep{langer15}. 
    \item \label{itm:dense} Thermalization in dense environments \citep{luhman03,D17,sutter21}. 
    \item \label{itm:dusty_hii} IR from dust in \hii{} region \citep{luhman03}.
    \item \label{itm:non_fuv} Dust heated by non-FUV radiation \citep{luhman03,kaufman99}. 
    \item \label{itm:agn} AGN activity contributes significantly to dust luminosity \citep{sargsyan12}.
\end{enumerate}

Several of the hypotheses are inconsistent with observations.

\begin{itemize}[label=-]
    \item Despite popularity, the PDR-based explanation \ref{itm:pdr} is limited by the fact that PDRs sit \textit{downstream} of ionized gas in the radiation energy flow, and therefore cannot account for the observed ``deficits'' in ionized gas lines like \nii{}, \ha{}, or \oiii{}. We discuss this further below.

    \item Explanation \ref{itm:fciin} is ruled out by our findings in Sec.~\ref{sec:implication_fneut}, which show that \fciin{} remains consistent across galaxy types when corrected for elemental abundance, and cannot explain the simultaneous ``deficit'' in both ionized and neutral gas lines.

    \item Explanations \ref{itm:optical_thick} and \ref{itm:attenuation} are unlikely because they would affect different lines to varying degrees, depending on wavelengths and optical depths. Instead, we observe similar ``deficit'' trends across lines with very different properties.

    \item Explanation \ref{itm:cii_ionized} is incompatible with the observed \ha{} and \oiii{} ``deficits,'' particularly since the strongest ``deficits'' occur in ULIRGs with the lowest inferred ionization states, while metal-poor dwarf galaxies with the strongest radiation fields do not show ``deficits.''

    \item Explanation \ref{itm:dense} does not align with measured gas densities. The typical electron density from \nii{122}/\nii{205} is $\sim$50 cm\textsuperscript{–3}, and the neutral gas density inferred from \oi{}/\cii{} is $\sim$3000 cm\textsuperscript{–3}, both near the critical density for \cii{}, and insufficient for strong thermalization effects. This explanation also fails to account for the similar ``deficit'' observed in \oi{} lines, which have even higher critical densities. 
\end{itemize}

To understand the root of the ``deficit,'' we consider the energy flow in the ISM. 
As shown in Fig.~\ref{f:flow}, stellar radiation originates from young massive stars; ionizing photons create \hii{} regions and are fully absorbed locally, while non-ionizing FUV photons propagate farther into the PDRs, where they heat dust and neutral gas. 

In this framework, both dust and neutral gas lie \textit{downstream} of the ionized gas in terms of energy input. 
Therefore, explanations that reduce the heating or excitation efficiencies in neutral gas---such as \ref{itm:pdr} and \ref{itm:dense}---cannot account for the observed suppression of ionized gas lines. 
Nor can they explain how the dust continuum can increase significantly without affecting the neutral-to-ionized gas line ratios. 

If \lhaextcorr{} is used as a proxy for the total ionizing luminosity, then deviations in the \lir{}/\lhaextcorr{}--\oh{} relation imply either a substantial \textit{upstream} depletion of ionizing photon radiation, or an additional energy source not traced by EUV and FUV radiations.
To resolve this, we propose two conceptual categories of explanations (Fig.~\ref{f:flow}).

\begin{itemize}
    \item \textbf{Ionized Gas ``Deficit''}: Dust is located \textit{upstream} in the energy flow, intercepting radiation before it reaches the gas. This reduces the energy available to ionize or excite the gas, suppressing line emission.
    \item \textbf{Dust IR ``Excess''}: An additional energy source preferentially heats dust, enhancing IR continuum without proportionally increasing gas line emission.
\end{itemize}

Among previous theories, explanation \ref{itm:dusty_hii}---IR emission from dust in \hii{} regions---fits the ionized gas ``deficit'' scenario, while \ref{itm:non_fuv}---non-FUV heating, and \ref{itm:agn}---AGN heated dust, fit the IR ``excess'' scenario.

Inspired by this framework, we propose two additional scenarios:
\begin{enumerate}[resume]
    \item \label{itm:diffuse} Emission from diffuse ISM dominates both dust and gas line luminosities, leading to a geometrical or energetic decoupling.
    \item \label{itm:shock} Dust is heated by non-radiative energy sources, such as turbulence or low-velocity shocks.
\end{enumerate}

The scenario \ref{itm:diffuse} aligns with the ionized gas ``deficit'' model, while \ref{itm:shock} is compatible with the IR ``excess'' framework.

\begin{figure*}
    \centering
    \includegraphics[width=\textwidth]{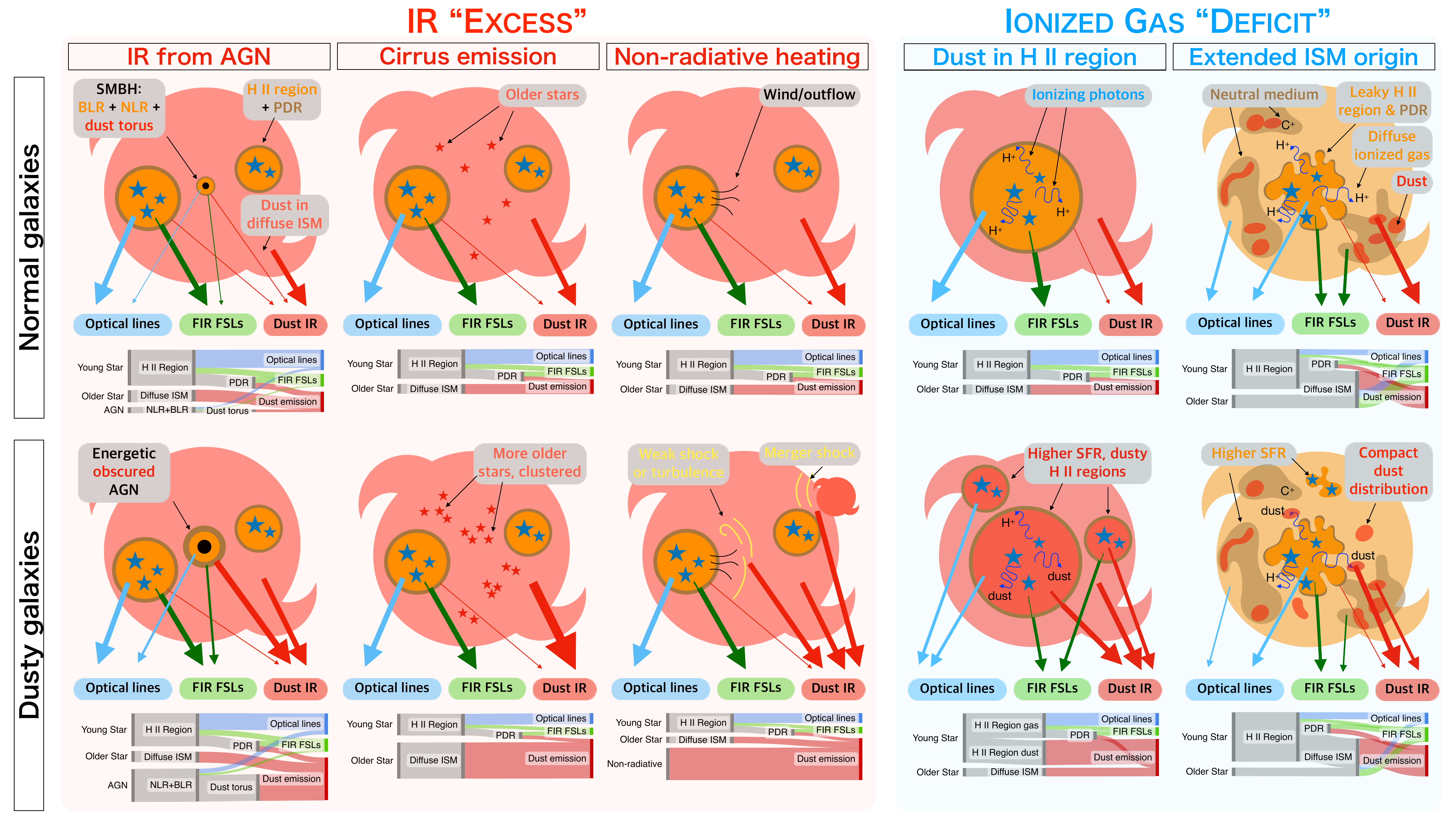}
    \caption[Illustration of the five models for ``deficit.'' ]{Illustration of the five models considered in the text that involved either IR ``excess'' (left three scenarios) or ionized gas ``deficit'' (right two scenarios) to explain the line ``deficit'' and dust--gas dichotomy seen in dusty galaxies.}
    \label{f:data_model}
\end{figure*}

These five scenarios---two in the ionized gas ``deficit'' group and three in the IR ``excess'' group---are illustrated in Fig.~\ref{f:data_model}. 
Although none of them can fully explain all observations on their own, they offer heuristic frameworks to guide future studies. 
We do not claim that these are definitive solutions; other explanations may exist beyond those currently considered.

This discussion has focused on ionized and neutral atomic gas, and does not include molecular gas. 
However, cross-matched CO observations from \citet{israel15,R15,K16,L17a} reveal strong correlations between CO and \ci{} line excitation and dust temperature, similar to the results in \citet{R15}. 
This places CO and \ci{} on the ``dust side'' in the gas--dust dichotomy, suggesting they behave more like dust tracers in these systems. 
A full treatment of molecular gas in this context is deferred to a forthcoming study.

\subsubsection{IR Emission from AGN}

AGN contributions to \lfir{} are often dismissed, as MIR emission---tracing warm dust---has been shown to correlate more closely with AGN activity. 
Nevertheless, we include this scenario for completeness and because it cannot be ignored as many high-\zz{} quasars exhibit \lir{}~$\gtrsim$~10\textsuperscript{13}\,\lsun{}, and the hottest low-\zz{} galaxies often host AGNs \citep[see fig.~1 and 3 in \ppi{}, and ][]{sargsyan12}.

While AGN hosts do exhibit enhanced 24\,\um{} emission (fig.~31 in \ppi{}), the bulk of their IR luminosity still originates from the FIR (i.e., cold dust). 
This high \lfir{} is often attributed to either starbursts triggered by AGN feedback or the co-evolution scenarios where cold gas fuels both star formation and AGN accretion. 
However, AGNs could also act as an independent, powerful energy source that heats dust and produces an IR ``excess,'' thereby increasing \td{} without proportionally enhancing gas line emission. 

Recent studies suggest that AGNs may heat dust on kpc scales and potentially dominate even FIR emission in some systems \citep{symeonidis18,mckinney21}, making this scenario physically plausible. 
Conversely, other observations argue that AGN bolometric luminosities derived from X-ray data are insufficient to account for the observed FIR output \citep[e.g.,][]{uematsu23}. 
Thus, the prevalence and energetic significance of AGNs in powering cold dust emission remains an open question that requires further observational evidence.

\subsubsection{Dust heated by Cirrus Emission}

An alternative explanation for the IR ``excess'' is increased dust heating from non-FUV photons, particularly from older stellar populations. 
This radiation--often referred to as the ``cirrus'' component--is the dominant heating source for FIR-emitting dust in normal galaxies like Milky Way \citep{lonsdale87,caux84,draine85,draine07}. 
Enhancing either the fraction of this cirrus component or its effective obscuration can boost FIR emission without impacting gas excitation, thereby decoupling dust and gas luminosities.

This scenario could be tested by modeling the extremely dusty galaxy SED under the constraint of energy conservation between stellar attenuation and FIR re-emission. 
However, such modeling is beyond the scope of this work.

The origin of the enhanced cirrus heating remains uncertain. 
One possibility is a top-light initial mass function (IMF), suppressing the formation of massive stars and thus the ionizing photon budget. 
Another possibility involves geometric changes, such as more compact star clusters, higher internal dust obscuration, or more enshrouded stellar populations---any of which could simultaneously increase \td{} and reduce gas excitation.

A significant concern with this scenario is its energetic feasibility. 
To explain \lir{}~$\gtrsim$~10\textsuperscript{14}\,\lsun{} in high-$z$ galaxies via old stellar populations, the total bolometric luminosity of evolved stars would need to exceed this number, implying an extraordinarily large mass of old stellar populations at \zz{}~\textgreater~3---a requirement that would challenge current models of galaxy formation.

\subsubsection{Dust Emission Powered by Non-Radiative Processes}

All explanations thus far assume that dust IR emission is powered by stellar radiation. 
A more unconventional but physically motivated alternative involves non-radiative energy sources---specifically, mechanical heating through turbulence or weak shocks in the ISM.

This hypothesis requires mechanisms that can heat dust significantly (up to two orders of magnitude in energy) without simultaneously exciting gas-phase emission lines. 
This constraint rules out energetic cosmic rays, magnetic fields, and strong radiative shocks, which would also boost line emission. 
Instead, moderate mechanical processes such as turbulence or low-velocity shocks in dense neutral or molecular gas remain viable candidates.

In this scenario, dust in these dense regions is heated to preferentially higher temperatures, and its thermal emission---scaling as $T_\mathrm{dust}^4$---acts as the primary cooling channel, matching the mechanical energy input. 
Meanwhile, FIR fine-structure lines (FSLs), which are more sensitive to gas density than temperature, would be affected to lesser extend---explaining the apparent gas–-dust decoupling. 
This mechanism aligns with the observed correlation between \td{} and \oi{}/\cii{}, and with the frequent association of IR-luminous systems with AGNs and/or mergers. 

Observational support includes examples of luminous dust emission powered in part by shocks, such as in radio galaxies \citep{ogle07,ogle10,villa24} and U/LIRGs \citep{beirao09,chandar23}. 
However, this hypothesis faces substantial observational and theoretical challenges.

Observationally, the proposed mechanical heating must be moderate---unlike the strong shocks commonly studied in molecular gas---and would need to heat cold dust without exciting gas lines. 
The resulting observables---enhanced cold dust luminosity and \td{}, slightly elevated gas densities, boosted low-excitation molecular lines (e.g., CO), and spatial overlap with regions of AGN-driven outflows or merger-induced turbulence---are subtle and difficult to detect.

Theoretically, dust destruction via sputtering in shocks poses a problem, as does the limited efficiency of gas-to-dust energy transfer, which is only effective in dense regions and for small grains. 
These issues suggest that this scenario may require a revised understanding of dust grain populations and ISM structures in dusty galaxies.

\subsubsection{Dust Emission in \texorpdfstring{\hii{}}{H II} Regions}

Another potential explanation involves enhanced FIR emission from dust within \hii{} regions. 
This has long been a topic of research and debate \citep[e.g.,][]{petrosian74,draine11b,jones17}. 
In this scenario, dust coexists with ionized gas and absorbs a significant fraction of EUV and FUV photons, thereby reducing the energy available for gas heating and line emission, resulting in both the ionized and neutral gas line ``deficits.''

This model naturally explains the increase in \td{} and the shift in energy balance towards dust. 
It also preserves the neutral–ionized line coherence seen in observations. 
However, it faces some major shortcomings:
First, theoretical models and observations suggest that dust within \hii{} regions typically reaches temperatures of 100–200\,K---much higher than the \td{} inferred from FIR SEDs (20–50\,K). 
Although some studies have found cold dust ($\sim$25–75\,K) within \hii{} regions \citep[e.g.,][]{ochsendorf15}, such cases are rare and are not representative of the bulk FIR emission.
Second, radiation pressure and dust opacity limit the fraction of ionizing photons that can be absorbed by dust to about 80\% \citep{draine11b,inoue02}, which corresponds to a maximum ``deficit'' factor of $\sim$0.25--insufficient to explain the observed trends. 
Additionally, cooler large grains are expected to migrate outward from ionized regions, making it difficult to reconcile this model with the observed FIR emission \citep{ishiki18}.
Furthermore, explaining the large variations in \ha{} obscuration fractions in Sec.~\ref{sec:implication_dichotomy} without major changes in metallicity or D/G may require complex dust distributions or star–dust geometries.

Nevertheless, these constraints are based primarily on Galactic \hii{} regions, and more extreme dust geometries or obscuration conditions may exist in extragalactic environments.

\subsubsection{Line Emission from Diffuse ISM structures}

The final hypothesis challenges the conventional picture of line emission originating from compact, hierarchical structures (i.e., OB stars surrounded by \hii{} regions and dense PDRs). 
Instead, it posits that a significant fraction of gas emission arises from extended, diffuse ISM phases in normal galaxies in the first place. 

For this to occur, most of the EUV and FUV photons must escape from their natal clouds into the surrounding diffuse medium. 
In dusty galaxies with high star formation rates, either (1) dust in the diffuse ISM becomes more effective at intercepting these photons, or (2) the \hii{} regions become more porous or density-bounded, increasing the photon escape fraction. 
In either case, more radiation is absorbed by dust rather than gas, resulting in suppressed gas line emission and elevated FIR continuum---a manifestation of the ionized gas ``deficit.''

This model is motivated by our growing understanding of the diffuse ISM. 
Observations show that diffuse ionized gas (DIG) can contribute up to $\sim$50\% of total \lha{} \citep{ferguson96,belfiore22}, and that a significant fraction of \cii{} emission in the Galactic plane originates from diffuse, non-PDR gas \citep{madden93,pineda13,gerin15,heyer22,madden23}.

However, this scenario also faces pressures and density constraints. 
The electron densities and gas pressures derived in sec.~3.5 and 3.8 in \ppi{} ($P/k_\mathrm{B} \sim 3 \times 10^5\,\mathrm{K\,cm^{-3}}$) are more than an order of magnitude higher than those of typical DIG \citep{ferguson96,della20}, bringing the emission closer to the regime of normal \hii{} regions or molecular cloud envelopes. 

We therefore speculate that the emitting gas may represent a transitional phase---spanning moderately dense ionized gas, diffuse PDRs, and molecular envelopes---similar to the extended warm ionized medium or warm dark clouds described by \citet{langer10,langer15b,langer21}. 
Confirming this scenario observationally is challenging due to the low surface brightness and need for high spatial resolution (\textless100\,pc) to separate extended structures from compact sources. 

In addition, this hypothesis cannot explain the correlation with dust temperatures, and predicts an anti-correlation between ``deficit'' trend with gas density, opposite to the observations. 
The implied changes in the photon escape fraction and dust geometry also require further investigation.

\subsection{Crisis of SFR Measurements in Dusty Galaxies}
\label{sec:implication_sfr}

\begin{figure}[h]
    \centering
    \includegraphics[width=\halfwdth]{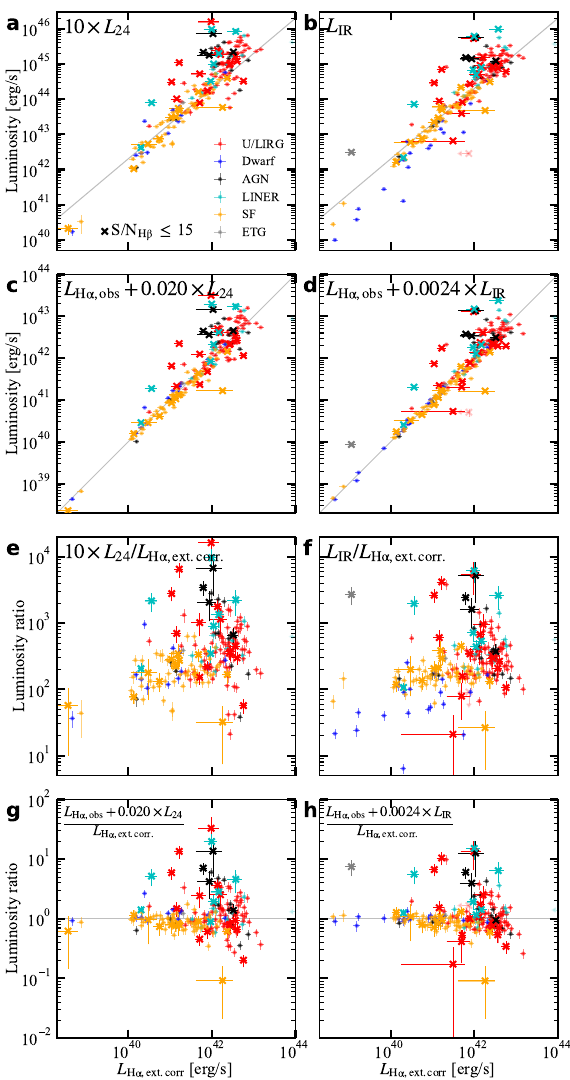}
    \caption[Reproduction of fig.~6–8 in \citet{kennicutt09}.]{Reproduction of fig.~6–8 from \citet{kennicutt09}, using the same data sources but without imposing their sample selection criteria. Galaxies that would be excluded by the \hb{} S/N cut of 15 (as described in their sec. 2.1) are marked with cross symbols.}
    \label{f:implication_Ha_corr_SFR}
\end{figure}

The significance of the \cii{} ``deficit'' problem lies in the fact that \cii{} has been widely proposed as a extinction-insensitive tracer of star formation rate (SFR). 
However, the breakdown of its linear correlation with \lir{} or $L_{24}$ challenges this usage, especially since the \lir{} and mid-IR luminosities are themselves established SFR indicators and are used to calibrate other tracers \citep{murphy11}. 
Consequently, some studies such as \citet{delooze14} suggested that \cii{} may not be a reliable SFR indicator.

As discussed in previous sections, all low-ionization emission lines---whether from neutral or ionized gas---exhibit a similar ``deficit'' relative to \lir{}. 
This fundamentally undermines the foundation of widely used SFR tracers, particularly in dusty galaxies. 
To illustrate this, we revisit one of the most influential obscuration-corrected SFR calibrations: the IR-corrected \ha{} method from \citet{kennicutt09}. 

In fig.~7 and 8 of \citet{kennicutt09}, \lir{} is compared against the extinction-corrected \ha{}, but no ``deficit'' trend (i.e.,  \lha{}/\lir{}~\textless~10\textsuperscript{-2.5}) is observed---unlike what we find in our Fig.~\ref{f:implication_Halpha_IR}, despite both studies using the same IR and optical datasets. 

Upon re-examination, we realize that this difference arises from a key aspect of sample selection. 
As described in sec.~2.1 of \citet{kennicutt09}, a signal-to-noise (S/N) requirement is imposed on the Balmer decrement, specifically through a combination of formal S/N cuts and visual inspection of the continuum-subtracted \ha{} spectra. 
The authors noted that this selection roughly corresponds to an S/N cut of 15 at \hb{}. 

In Fig.~\ref{f:implication_Ha_corr_SFR}, we reproduce the analysis of \citet{kennicutt09} following their methodology but highlight galaxies with $\mathrm{S/N(H\beta)} \leq 15$ using a cross marker. 
Comparing their sample to ours, this S/N criterion effectively excludes a large number of IR-luminous galaxies (\lir{}~\textgreater~10\textsuperscript{45}\,erg/s; see (\bsf{b})) and removes most of the outliers in the IR-corrected \ha{} versus \haextcorr{} comparisons ((\bsf{g}) \& (\bsf{h})). 
While the simple S/N cut yields 157 galaxies, \citet{kennicutt09} selected 147 out of 410 galaxies from \citet{moustakas06}, effectively excluding most outliers. 

Thus, the \haextcorr{} ``deficit'' was already present in earlier datasets but was dismissed as a possible result of poor extinction correction, rather than recognized as an intrinsic discrepancy. 
More importantly, for IR-bright galaxies---including U/LIRGs, AGN hosts, high-\zz{} DSFGs, and QSOs---the dust-corrected \ha{} luminosity was never directly calibrated against other SFR tracers, but rather inferred under assumptions that they follow the same calibration. 
The limitation in sample range and ''anomalously large \lir{}/\ha{}'' were already noted in sec.~6.2 of \citet{kennicutt09}. 
These caveats on applicability limits were, however, ignored by most of the later studies that make use of these calibrations. 

Another widely referenced study is \citet{murphy11}, which provides a more comprehensive framework for IR-based SFR measurements. 
It acknowledged that IR emission may arise from ionizing photons absorbed by dust before interacting with gas---a mechanism consistent with the ionized gas ``deficit'' scenario \ref{itm:dusty_hii}. 
However, \citet{murphy11} did not discuss the observational or theoretical implications of this mode of obscuration.

Similar discrepancies also appear in other hydrogen recombination lines, such as Paschen and Brackett lines, which are less sensitive to dust attenuation due to their near-IR wavelengths. 
Notably, integrated SFRs derived from Paschen lines also underestimate the total SFR compared to IR-based estimates at the highest SFR end, as shown in fig.~5 of \citet{piqueras16} and figs.~14–15 of \citet{gimenez22}.

\begin{figure}[]
    \centering
    \includegraphics[width=\halfwdth{}]{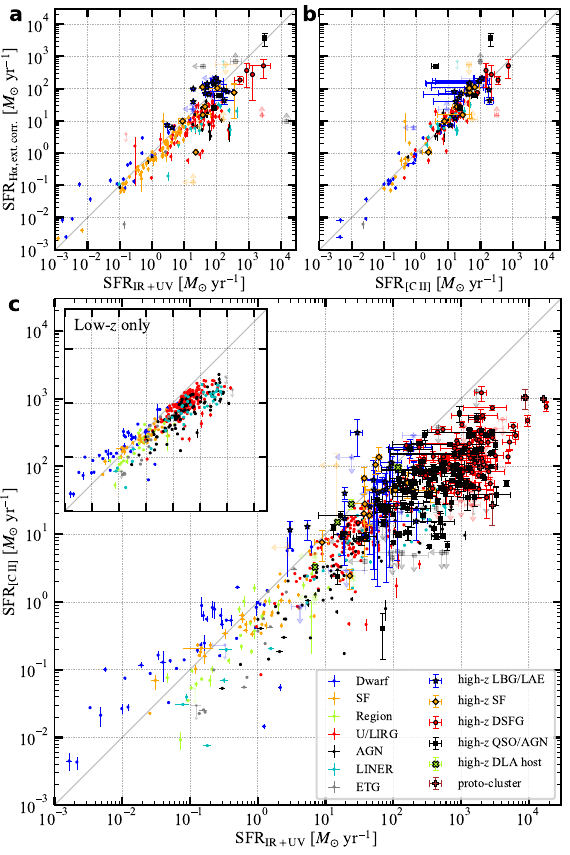}
    \caption[SFR Comparisons. ]{Comparison of different SFR indicators. Extinction-corrected \ha{} based SFR\textsubscript{H$\alpha$,ext.corr.} compared to \bsf{a}, \lir{} and \lfuv{} based totol SFR\textsubscript{IR+UV}; and \bsf{b}, \lcii{} based SFR\textsubscript{[C~II]}. \bsf{c}, SFR\textsubscript{IR+UV} compared to SFR\textsubscript{[C~II]}. The gray diagonal line in each panel denotes a unity relation. The inset in (\bsf{c}) show the same plot but include only low-\zz{} data points for clarity. Note that for DSFGs and QSOs, only \lir{} is used to compute SFR\textsubscript{IR+UV}.}
    \label{f:implication_sfr-sfr}
\end{figure}

All of this evidence suggests that the ``deficits'' of \haextcorr{} and other recombination lines have been present in SFR studies, though previously attributed to poor extinction corrections rather than recognized as intrinsic deviations. 
As discussed earlier, the interpretation of this discrepancy depends on the physical scenarios: (1) In the case of an \textit{ionized gas ``deficit''}, \lir{} remains a valid SFR tracer, as dust absorbs ionizing photons before they can heat or ionize gas. (2) In the case of an \textit{IR ``excess''}, \lir{} significantly overestimates the true SFR, and extinction-corrected \ha{} is a more faithful tracer.

Extending the discussion on gas--dust dichotomy, we argue that this discrepancy is not limited to recombination lines, but extends to FIR fine-structure lines (FSLs). 
In this view, the SFRs derived from neutral or ionized gas lines are consistently lower than those inferred from \lir{}.

This discrepancy has profound implications. 
In Fig.~\ref{f:implication_sfr-sfr}, we compare the total SFR estimated from \lir{} and \lfuv{} \citep[SFR\textsubscript{IR+UV}; following][; high-\zz{} DSFGs and QSOs only use \lir{}
]{murphy11} to the SFR derived from \cii{} emission, which we calibrate in Sec.~\ref{sec:implication_cii_origin}. 
In a scenario where IR ``excess'' dominates, SFRs would need to be revised downward by factors up to 100, and merely any galaxy (excluding proto-clusters) exhibits SFRs exceeding 1000\,\msun/yr---well below the sometimes-quoted 10\textsuperscript{4}\,\msun/yr for DSFGs.

This calls into question the validity of previously accepted dust-corrected SFR calibrations and reveals a potential order-of-magnitude uncertainty in SFR estimates for dusty galaxies. 
We therefore face a serious \textit{crisis in SFR measurement} in the most luminous and dust-obscured systems.

\subsection{Puzzles of FSL Emitting Region}
\label{sec:implication_where}

As discussed in previous sections, the line ``deficit'' challenges the classical picture of the ISM, where star formation is embedded in compact star-forming regions with orion-like \hii{} regions surrounded by dense PDRs \citep[e.g.,][]{RM20a}. 
Several other properties of FIR FSLs also disfavor this picture.

One notable characteristic of FSLs is that they appear to arise from ionized gas with universally low densities. 
As discussed in \ppi{} and \ppii{}, the typical electron density derived from the \nii{122}/\nii{205} ratio is $\sim$50\,\cc{}. 
This is at the low end of the density distribution for classical \hii{} regions \citep{hunt09}, which typically exhibit densities ranging from 100 to 10\textsuperscript{5}\,\cc{}, especially in compact, high-luminosity \hii{} regions such as Orion. 
On the other hand, such low densities are common for more extended or evolved \hii{} regions like the Carina Nebula or for emission-line clusters in the Antennae Galaxy \citep{gilbert07}. 
They are also higher than the densities found in extragalactic giant \hii{} regions studied in integrated spectra of nearby galaxies \citep{kennicutt84}. 
Based on the \hii{} region size--density scaling relation, an electron density of 50 cm\,\cc{} corresponds to an \hii{} region with a typical size of $\sim$50\,pc.

Given that FSL emission is density-weighted up to the line’s critical density ($n_\mathrm{crit}$), this low \edens{} value---especially for \nii{} with $n_{e\mathrm{,crit}}$~$\sim$~300\,\cc{}---suggests that the bulk of the emission arises from low-density gas (see discussion in sec.~4.1 in \ppi{}). 
Other ionized gas density tracers also place the emission in their low-density limits. 
This implies that dense, compact star-forming regions like Orion contribute negligibly to the total FSL luminosity integrated over galaxies, and that more diffuse gas in and around normal to giant \hii{} regions dominate the emission. 

Another puzzling observation is the prevalence of extended emission in neutral gas lines, particularly \cii{} and \oi{}. 
Multiple studies on cloud scales \citep[e.g.,][]{cigan16,heyer22,madden23} and galactic scales \citep{madden93,pineda13,gerin15} report that \cii{} and \oi{} emission often extend well beyond the narrow PDR layers, with high filling factors and distributions that sometimes follow \hi{} 21\,cm emission. 

At high \zz{}, this phenomenon is even more pronounced. 
Observations frequently show that \cii{} emission is more spatially extended than the compact FIR continuum, across a range of systems including LBG/LAEs \citep[e.g.,][]{CS18b,FS19}, QSOs \citep[e.g.,][]{VB20,MR25}, and DSFGs \citep[e.g.,][]{RM19,RC25}. 
The origin of this extended emission remains uncertain, with possible contributors including CO-dark molecular gas clouds, diffuse PDRs, cold neutral medium, or even circumgalactic medium. 

The contribution of extended neutral gas emission in galaxies beyond the Milky Way remains poorly constrained due to the lack of facilities capable of high-resolution and deep low-surface brightness observations. 
Reconciling this extended structure with the observed constancy of \fciin{} and the presence of the ionized gas ``deficit'' is particularly challenging. 
It suggests either that the fractional luminosity of the extended emission is negligible in most galaxies (unlikely given the observations), or more plausibly, that both neutral and ionized gas lines originate from similarly extended, physically associated structures in the diffuse ISM. 

The ionized--neutral gas coherence (Sec.~\ref{sec:implication_ionized-neutral}) further complicates the traditional interpretation of FIR FSLs. 
As is already discussed in Sec.~\ref{sec:implication_pdr}, in a traditional PDR framework, the neutral gas line luminosities are controlled mainly by incident FUV strength and photoelectric heating efficiency, rather than metallicity. 
But observationally, the luminosities of both major cooling lines in PDR (\cii{} and \oi{}) and IR luminosity are primarily correlated with ionized gas tracers \haextcorr{}, after accounting for the linear O/H abundance dependence. 

Combining discussion in  Sec.~\ref{sec:implication_ir} and \ref{sec:implication_dichotomy}, we argue that these observations support a scenario in which both ionized and neutral FIR FSLs arise from radiation-bounded, physically coherent regions—possibly extended envelopes of normal-to-giant \hii{} regions or mixed diffuse ionized/neutral structures \citep[see also][]{werner70}. 
And the metallicity dependence likely arises from gas heating through agents that's directly proportional to O/H. 
Since the emission discussed here is spatially integrated over entire galaxies, this suggests that the dominant sources of FIR FSLs differ from the classical picture of dust photo-electric heating powered dense PDRs layers.

\subsection{What Does \texorpdfstring{\cii{}}{[C II]} Actually Trace?}
\label{sec:implication_cii_origin}

\cii{} is one of the brightest FIR fine-structure lines and is often the only accessible spectral feature in high-\zz{} galaxies. 
For this reason, it has been widely used as a tracer of various physical quantities, including SFR, atomic hydrogen mass (\mhi{}), molecular gas mass (\mmol{}), and more. 
However, many of these applications implicitly assume fixed relationships between physical properties such as molecular gas fraction or star formation efficiency (SFE). 
In this section, we critically assess what \cii{} truly traces, based on empirical trends and physical interpretation.

\subsubsection{SFR? Yes, with Caveats.}

\begin{figure}[]
    \centering
    \includegraphics[width=\halfwdth{}]{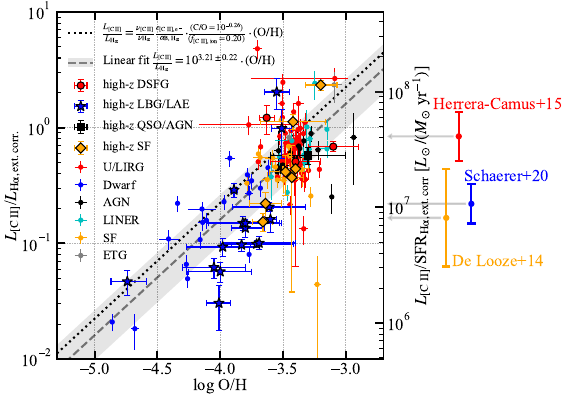}
    \caption[\oh{} vs. \lcii{}/\lhaextcorr{}. ]{\oh{} vs. \lcii{}/\lhaextcorr{}, showing only detections. The right axis converts \lcii{}/\lha{} to \lcii{}/SFR using the \ha-to-SFR conversion from \citet{murphy11}. Literature \cii-to-SFR calibrations and scatters from \citet{herrera15}, \citet{schaerer20}, and \citet{delooze14} are shown for comparison.}
    \label{f:implication_o_h-cii_ha}
\end{figure}

\cii{} has frequently been used as a tracer of SFR, particularly for galaxies where traditional optical or UV tracers are obscured by dust. 
However, as discussed earlier, the validity of this approach varies between normal galaxies and dusty systems showing FIR line ``deficits.''

For galaxies with normal dust heating, a tight correlation exists between \lcii{} and \lhaextcorr{}, modulated by metallicity (see Sec.~\ref{sec:implication_fir-ha} and Fig.~\ref{f:implication_O_H-line_Halpha}(\bsf{a})). 
Converting \lhaextcorr{} to SFR using the calibration from \citet{murphy11}, we derive a metallicity-dependent \cii-to-SFR relation:
\begin{equation}\begin{split}
    \mathrm{SFR_{[C~II]}} = 2.59 \left(\frac{L_\mathrm{[C~II]}}{10^8\,L_\odot}\right)\left(\frac{\mathrm{O/H}}{10^{-3.31}}\right)^{-1} \,M_\odot\,\mathrm{yr^{-1}}
\label{equ:cii_sfr}
\end{split}\end{equation}
This relation, shown in Fig.~\ref{f:implication_o_h-cii_ha}, has a scatter of 0.22 dex. 
It can be approximated by combining the emissivity ratio (\pyneb) of \cii{} and \ha{} in ionized gas (assumed \edens{}~=~50\,\cc{}, (C/O)~=~10\textsuperscript{--0.26}) with an ionized gas \cii{} fraction $f_\mathrm{[C~II],ion} = 20\%$ (i.e., \fciin{} = 80\%), shown as the dotted line in Fig.~\ref{f:implication_o_h-cii_ha}. 
Comments on the seemingly offset of high-\zz{} LBG/LAEs can be found in Sec.~\ref{sec:implication_fir-ha}. 

This metallicity dependent SFR calibration clarifies the different \cii{}-to-SFR conversion factors identified in previous studies (right side in Fig.~\ref{f:implication_o_h-cii_ha}), which primarily depend on the metallicity of galaxies in each sample. 
For instance, studies of high-\zz{} LBGs \citep[e.g.,][ALPINE survey]{schaerer20} which are often metal-poor, or studies that include dwarf galaxies \citep[e.g.,][]{delooze14}, show lower \cii{}/SFR ratios compared to studies focusing on star-forming galaxies \citep{herrera15}. 
The comprehensive sample in \citet{delooze14}, which spans a wide metallicity range, also exhibits the largest scatter. 
Besides, this calibration offers an explanation for the weak or non-detection of \cii{} in galaxies in the early universe \citep{carniani20,BC21,PG23,SS25a} without requiring adjustments to the carbon-to-oxygen abundance ratio or IMF \citep{katz22}. 

This metallicity dependence also limits the utility of \cii{} as a universal SFR tracer. A single \cii{}--SFR relation cannot be extrapolated to galaxies with different metallicities or redshifts. 
Since both the star formation main sequence (\mstar--SFR) and mass-metallicity relations (\mstar--O/H) evolve with cosmic time, the \cii{}--SFR relation must be adjusted accordingly. 
This explains the offset seen between the \cii{}--SFR relations in \citet{herrera15} and \citet{schaerer20}, despite both targeting star-forming galaxies at their respective epochs.

This limitation has critical implications for line-intensity mapping (LIM) studies aimed at probing cosmic star formation history. 
\cii{} LIM at \zz{}$\sim$6 is designed to capture emission from low-mass low-luminosity galaxies that potentially dominate the SFR density and ionizing photon budget during the epoch of reionization. 
However, such low-mass galaxies are typically metal-poor and thus faint in \cii{}, diminishing their expected contribution to the total intensity signal \citep[e.g.,][]{karoumpis22}.

In practice, \cii{} is less reliable than \ha{} as an SFR tracer, especially in low-metallicity galaxies such as dwarfs and high-\zz{} main-sequence galaxies. 
The abundance dependence (reducing \lcii{}/SFR by up to 1.5 dex) severely limits its utility, compounded by the relatively low dust attenuation in these systems. 

For dusty galaxies showing FIR line ``deficits'' (see Sec.~\ref{sec:implication_cii_deficit}), the choice of SFR tracer depends on the physical scenario: (1) Under the \textit{IR ``excess''} scenario, emission lines better estimate SFR, and Eq.~\ref{equ:cii_sfr} can be applied—possibly simplified by assuming a fixed metallicity (e.g., \oh{}~$\sim$~--3.4). (2) Under the \textit{ionized gas ``deficit''} scenario, only \lir{} probes the total SFR, and \cii{} serves only as a rough proxy with an empirical ``deficit'' correction. 

In summary, \cii{} can be used to estimate SFR for its correlation to \haextcorr{}, but only with careful attention to metallicity and the understanding of the line-to-IR ``deficits.''

\subsubsection{Ionized Gas Mass? No.}

Given the empirical correlation between \cii{} and \ha{}, one might infer that \cii{} traces ionized gas mass. 
However, this is incorrect. 
Approximately 80\% of \cii{} emission originates from neutral gas, as shown in Sec.~\ref{sec:implication_fneut}. 
Therefore, while \lcii{} correlates with \lha{}, it does not physically ``trace'' the ionized gas mass. 

This apparent contradiction is resolved by recognizing that \ha{} itself is not a direct tracer of total ionized gas mass. 
Rather, it measures the total heating of ionized gas, primarily in irradiated \hii{} regions. 
In contrast, most of the ionized gas mass resides in high-filling factor diffuse phases (e.g., DIG) with relatively small contribution to the \ha{} luminosity. 
Thus, \lha{} traces the energy input from young stars, not the total ionized gas mass. 

The ionized--neutral gas coherence implies that \cii{} and \ha{} both ``trace'' the energy budget of the corresponding gas phases, not the total mass their respective phases. 
\textbf{As a cooling line, \cii{} primarily traces the energy incident on neutral gas, modulated by carbon abundance. }
If converted to a mass, \lcii{} corresponds to the mass of \cp{} in the irradiated neutral gas, not the total ionized gas mass.

\subsubsection{Neutral Gas Mass? No.}

\begin{figure}[h]
    \centering
    \includegraphics[width=\halfwdth]{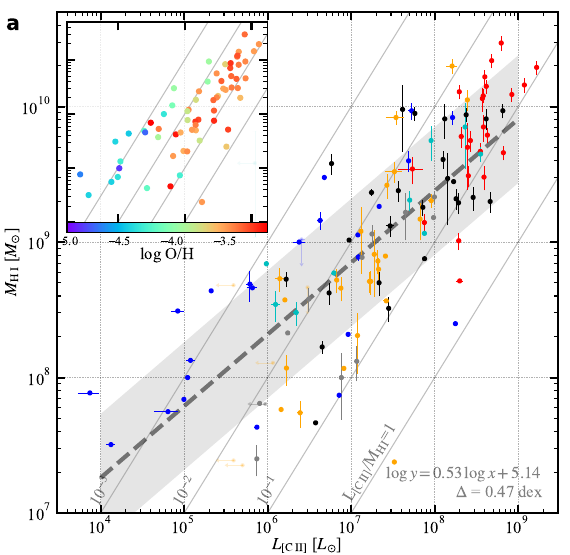}
    \includegraphics[width=\halfwdth]{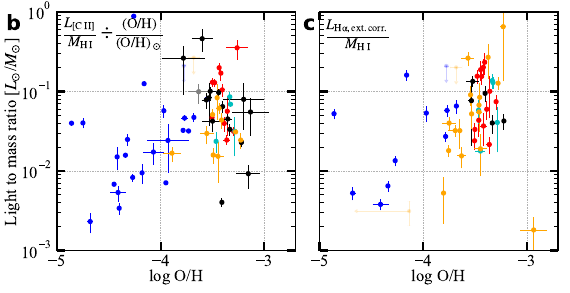}
    \caption[\lcii{}-to-\mhi{} calibration. ]{\bsf{a}, total \hi{} mass \mhi{} vs. \cii{} luminosity \lcii{}. The gray dashed line and shade denote the power-law fit result, which is printed at the lower right corner. The dotted lines correspond to constant \cii{} light-to-\hi{} mass values. The inset panel shows the absolute metallicity of every data point in color, plotted in the same coordinate. Lower row: light-to-\hi{} mass \lline/\mhi{} ratio of \bsf{b}, \cii{}; and \bsf{c}, extinction-corrected \ha{}. Note that the plotted \cii{} L/M is further corrected for the absolute metallicity O/H.}
    \label{f:data_HI}
\end{figure}

Because \cii{} predominantly arises from neutral gas, it has been proposed as a tracer of neutral gas mass (\mhi{}), traditionally measured via the 21-cm line. 
This idea has been explored in gamma-ray burst (GRB) sightlines \citep{heintz21} and simulations \citep{vizgan22,casavecchia24}, but lacks validation from local galaxy observations.

Fig.~\ref{f:data_HI}(\bsf{a}) shows a sublinear correlation between \lcii{} and \mhi{}, with a power-law index of $\sim$0.6 and very substantial scatter. 
The \lcii{}/\mhi{} ratio strongly depends on metallicity, as seen in the inset. 
Even after correcting for metallicity (\bsf{b}), a residual trend remains, similar to that seen in \lha{}/\mhi{} (\bsf{c}), pointing to a broader variation in ionized-to-neutral gas ratios across galaxy types. 

This trend likely reflects an increase in molecular gas fractions in metal-rich galaxies, due to more abundant dust as catalysts and metals as coolents. 
The result is a higher efficiency of converting \hi{} into H\textsubscript{2}, and thus stronger ionized line emission per unit \mhi{}. 
Consequently, \lcii{}/\mhi{} depends quadratically on metallicity, as \lcii{}~$\sim$~(O/H)\textsuperscript{2}$\cdot$\mhi{}, contrasting with the shallower dependence found in GRB studies. 
Beacause the \hi{} disk is much more extended than \cii{} \citep{deblok16}, and the data is often integrated over larger areas, we limit the comparison in Fig.~\ref{f:data_HI} to galaxies with a \hi{} size estimated to be less than 200\arcsec{} by using the mass-size relation \citep{wang16}. 
Although we warn that the mismatch in field of views might still add to the scatters and outliers seen in the figure. 

A recent study by \citet{wilson23} attempted to calibrate \loi{}/\mhi{}, also using high-$z$ GRBs. 
However, our local galaxy data show a steeper metallicity dependence, concluding that \loi{} and \lcii{} do not robustly ``trace'' total \hi{} mass in diverse galaxy populations.

\subsubsection{Molecular Gas Mass? No.}

\cii{} has also been proposed as a tracer of CO-dark molecular gas \citep{stacey91,madden20}, and even as a proxy for the total molecular gas mass \citep{ZA18}. 
While the former has physical merit and requires detailed modeling that we defer to future work, the latter can be discussed on empirical grounds. 

First, the correlation between SFR and molecular gas mass \citep[Kennicutt–Schmidt law;][and references therein]{kennicutt12} naturally leads to a correlation between the star-formation powered \lcii{} and \mmol{}. 
This correlation does not mean that \cii{} physically ``traces'' molecular gas mass. 

Second, accurate calibration of \cii{}--\mmol{} requires reliable \mmol{} measurements. 
But \mmol{} is often estimated from SFR (circular logic) or from CO with assumed CO-to-H\textsubscript{2} conversion factors ($\alpha_\mathrm{CO}$), which are metallicity and environment dependent. 
For example, in \citet{ZA18}, many galaxies used to calibrate the \cii{}–\mmol{} relation have \mmol{} from either SFR or CO with a uniform or weakly (power index --1.5) metallicity-dependent $\alpha_\mathrm{CO}$. 
However, for low-metallicity dwarf galaxies, the true $\alpha_\mathrm{CO}$ may be much steeper---$\alpha_\mathrm{CO}~\sim~(\mathrm{O/H})^{-3.39}$ as found in \citet{madden20}---which would increase their \mmol{} by over an order of magnitude and break the claimed linear \cii{}--\mmol{} relation.

This also leads to a paradox: if \mmol{} estimates are correct based on CO-only observations, there is little need for invoking \cii{} to trace CO-dark gas; if they are incorrect, then there is no reliable reference to calibrate \cii{} as a molecular gas tracer. 
Without independent, accurate \mmol{} measurements, \cii{} cannot be reliably calibrate to ``trace'' total molecular gas mass. \\

In summary, \cii{} is a major cooling line in neutral gas, its luminosity mainly traces the energy input from star formation on neutral gas.
In practice, it is a metallicity-dependent tracer for SFR. 
\cii{} luminosity does not physically ``trace'' the total gas mass of any phase, except for the mass of \cp{} in \textit{irradiated} neutral gas. 
Although observationally, \lcii{} shows correlations with \mhi{} and \mmol{}, these correlations reflect a convolution of many effects: (1) ``bigger-things-brighter,'' (2) the Kennicutt–Schmidt law for atomic and molecular gas, and (3) the metallicity-dependent \cii{}–\ha{} relation. 
These empirical correlations should not be mistaken for physical tracers.

\subsection{High-\texorpdfstring{\zz{}}{z} Difference}
\label{sec:implication_diff}

As discuseed in sec.~4.4 of \ppi{}, only a limited number of ISM diagnostics display clear, systematic differences between low- and high-\zz{} galaxies when grouped by galaxy types. 
These include: (1) an offset in the FIR line ``deficit'' trend, (2) enhanced \oiii{} ratios in high-\zz{} dusty galaxies, and (3) elevated \oi{}/\cii{} ratios. 
In this section, we investigate the underlying causes and implications of these differences in the context of the ionized and neutral gas properties and dust emission behavior.

\subsubsection{Evolution in Line ``Deficit''}

\begin{figure}[h]
    \centering
    \includegraphics[width=\halfwdth]{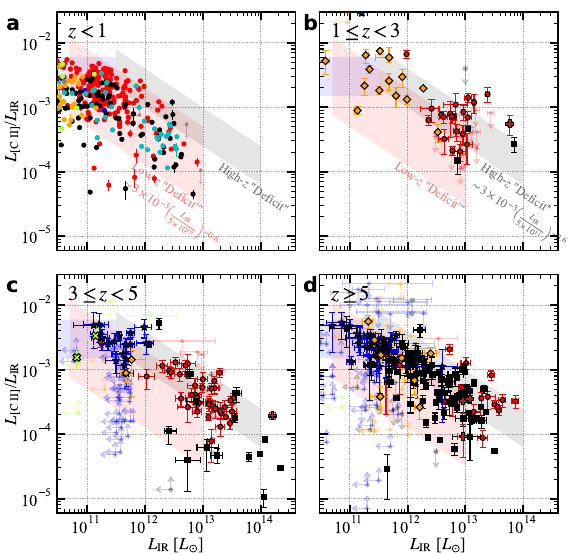}
    \caption[\cii{} ``deficit'' along \zz{}. ]{\cii{} ``deficit'' at Low- and high-\zz{} in four redshift bins: \bsf{a}, \zz{}~\textless~1 (low-\zz{}); \bsf{b}, 1~$\leq$~\zz{}~\textless~3; \bsf{c}, 3~$\leq$~\zz{}~\textless~5; \bsf{d}, \zz{}~\textgreater~5. The none-``deficit'' branch, low-\zz{} \cii{} ``deficit'' and total high-\zz{} ``deficit'' branches are shown as blue, red, and gray translucent shades, respectively. }
    \label{f:z-deficit}
\end{figure}

One of the most prominent differences between low- and high-\zz{} galaxies is the systematic offset in the FIR FSL ``deficit'' trends. 
As shown in Fig.~\ref{f:z-deficit}, the turnover point of the ``deficit'' branch for \cii{} shifts toward one order of magnitude higher \lir{} at higher redshifts. 
This trend is not unique to \cii{}--a similar offset is observed in other FIR lines, as shown in figs.~7 of \ppi{}.

Notably, while the turnover point shifts, the slope of the ``deficit'' trend remains largely unchanged. 
Furthermore, there is no strong evidence for evolution in the ``deficit'' slope or normalization across different high-\zz{} bins: (\bsf{b}), 1~$\leq$~\zz{}~\textless~3 (Cosmic Noon); (\bsf{c}), 3~$\leq$~\zz{}~\textless~5; and (\bsf{d}), \zz{}~\textgreater~5 (Epoch of Reionization).

This redshift-dependent offset means that, for galaxies with the same \lir{}, high-\zz{} dusty galaxies exhibit higher IR line luminosities than their low-\zz{} counterparts. 
Interpreting this result depends on the physical origin of the line ``deficit,'' which we discussed in Sec.~\ref{sec:implication_cii_deficit} under two main scenarios: \textit{IR ``excess''} and \textit{ionized gas ``deficit''}.

Under the \textit{IR ``excess''} scenario, line luminosities is related to SFR, while \lir{} reflects both star formation and additional energy input. 
The offset suggests that, at high \zz{}, a given level of this additional energy source corresponds to a higher SFR. 
This could imply that either AGN growth lags behind stellar mass assembly at high \zz{} (AGN heated dust), or high-\zz{} galaxies have a lower fraction of evolved stellar populations (cirrus emission to dust), or feedback processes are less efficient or merger-triggered starbursts are more prominent (non-radiatively heated dust), in the three scenarios. 
Each interpretation offers valuable insight into the co-evolution of galaxies, AGN, and the ISM, but all require further study using detailed modeling and high-resolution data.

In contrast, under the \textit{ionized gas ``deficit''} scenario, the offset suggests that a smaller fraction of ionizing photons are intercepted by dust at high redshift, either within \hii{} regions or along escape channels into the diffuse ISM. 
This could reflect more porous ISM geometries, less dust per unit gas mass, or evolutionary differences in dust composition. 
It also implies that upstream dust heating---responsible for suppressing gas line emission---only becomes significant at higher SFRs in high-\zz{} galaxies.

Both scenarios are consistent with current theoretical expectations for high-\zz{} galaxies, including lower dust content, higher gas fractions, and more turbulent and porous ISM structures. 
However, the implications are far-reaching and highlight the need for comprehensive, multi-wavelength modeling to understand the evolution of galaxy-scale energy budgets.

A small but notable deviation is observed among high-\zz{} QSOs, particularly in (\bsf{d}). 
These sources lie slightly ($\sim$0.1–0.2\,dex) below the composite high-\zz{} ``deficit'' trend (gray region), as well as below DSFGs at similar redshifts. 
This may indicate a systematically lower line-to-IR ratio in QSOs compared to starburst-dominated DSFGs. 
However, this discrepancy could also result from systematics in the measurement of \lir{} for high-\zz{} QSOs, which often rely on limited photometric data and template fitting. 
Given these uncertainties, we refrain from making claims about a population- or redshift-dependent divergence in the FIR line ``deficit'' among high-\zz{} QSOs.

\subsubsection{Invariant \texorpdfstring{\oiii{}}{[O III]}/\texorpdfstring{\cii{}}{[C II]} in High-\texorpdfstring{\zz}{z} Normal Galaxies---False Analogy of Low-\texorpdfstring{\zz}{z} ``Analogs''}

\begin{figure}[]
    \centering
    \includegraphics[width=\halfwdth]{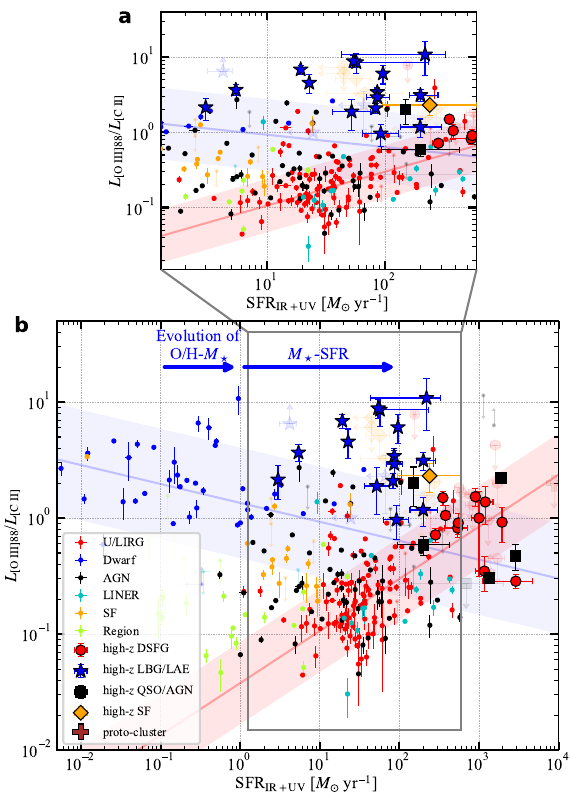}
    \caption[\oiii{88}/\cii{} vs. SFR for high-\zz{} normal galaxies.]{\loiii{88}/\lcii{} vs. total star formation rate SFR\textsubscript{IR+UV}. \textbf{a}, axis limits match fig.~5 in \citet{HY20}. The blue and red shaded regions represent the \oiii/\cii{}–SFR relations derived by dividing the \oiii{}–SFR and \cii{}–SFR fits from \citet{delooze14} for dwarf and starburst galaxies, respectively. \textbf{b}, zoomed-out view to show the full parameter space of both the \oiii/\cii{} and SFR. The gray box indicates the region shown in panel (\bsf{a}). Arrows indicate the shifts in SFR due to redshift evolution in the mass–metallicity (O/H-\mstar{}) and star-forming main sequence (\mstar{}-SFR) relations.}
    \label{f:implication_oiii_cii-SFR}
\end{figure}

The line ratio \oiii{}/\cii{} in high-\zz{} galaxies has often been a focus of study, particularly in two contexts: (1) normal metal-poor galaxies like LBG/LAEs, and (2) dusty galaxies including DSFGs and QSOs. 
In this section, we focus on the first category.

Several studies have reported ``elevated'' \oiii{}/\cii{} ratios in high-\zz{} LBG/LAEs compared to local dwarf galaxies---often referred to as ``low-\zz{} analogs'' \citep[e.g.,][]{LN19,HY20}. 
In some cases, the observed low \cii{} luminosities were mis-characterized as a ``deficit,'' despite the true \cii{} ``deficit'' phenomenon refers to dusty galaxies. 
This claim has led to various hypotheses, including altered C/O ratios, top-heavy IMFs, harder radiation fields, low PDR covering fractions, and extended emission regions \citep{carniani20,HY20,katz22}.

However, we argue that the \oiii{}/\cii{} ratios in high-\zz{} LBG/LAEs are not systematically higher than in local dwarfs. 
The apparent discrepancy is the product of how SFR is used as the baseline for comparison.

Fig.~\ref{f:implication_oiii_cii-SFR}(\bsf{a}) reproduces the common \oiii{}/\cii{}--SFR plot from the literature \citep[e.g.,][]{HY20}. 
The high-\zz{} LBG/LAE data appear to lie above the dwarf galaxy trend (blue line), suggesting elevated \oiii{}/\cii{}. 
However, this comparison is misleading for several reasons. 
First, very few dwarf galaxies are actually plotted in this figure. 
Second, the \oiii{88}/\cii{}–SFR relation is constructed by dividing two SFR calibration fits (\oiii{88}--SFR and \cii{}--SFR) from \citet{delooze14}, which are not intended to define a physically meaningful line ratio of \oiii{}/\cii{}, and the very shallow slope (power index 0.16) hints at non-correlation. 
Third, the shaded regions result from a quadratic sum of the fit scatters, without proper propagation of uncertainties that accounts for the slope.
Fourth, the use of SFR on the x-axis introduces a systematic offset between LBG/LAEs and dwarfs due to galaxy evolution, not intrinsic line ratio differences.

Fig.~\ref{f:implication_oiii_cii-SFR}(\bsf{b}) expands the axis range to include the full distribution of low-\zz{} dwarf galaxies. 
This more complete view shows that both LBG/LAEs and dwarfs occupy the same range of \oiii{}/\cii{} ($\sim$1–10) with no clear trend with SFR. 
The perceived offset in (\bsf{a}) is entirely driven by the use of SFR as x-axis, misunderstanding a \oiii{88}/\cii{}-SFR relation, as well as the evolution in the global galaxy property, SFR.

Here we provide a more accurate interpretation: at fixed \oiii{88}/\cii{}, high-\zz{} galaxies have significantly higher SFR than their low-\zz{} counterparts. 
This increase in SFR is expected from the evolution of two well-established scaling relations: 
(1) the mass--metallicity relation (O/H--\mstar{}), such that \oh{}~=~-–4.0 corresponds to $\sim$ 10\textsuperscript{8} \msun{} at \zz{}~$\sim$~0 \citep{andrews03,lee06} and $\sim$ 10\textsuperscript{9} \msun{} \zz{}~\textgreater~5 \citep{NK23,curti24}; 
(2) the star-forming main sequence (\mstar{}--SFR), which at fixed mass shows a $\sim$2 dex increase in SFR from \zz{}~$\sim$~0 to \zz{}~$\sim$~6 \citep{speagle14,schreiber20,popesso23,curti24}. 
These evolutionary shifts explain the $\sim$3 dex increase in SFR at fixed \oiii{88}/\cii{} seen in Fig.~\ref{f:implication_oiii_cii-SFR}, assuming \oiii{88}/\cii{} probes $U$, which is correlated with O/H (sec.~3.7 in \ppi{} and sec.~4.1 in \ppii{}).

This issue highlights a broader problem in the use of low-\zz{} ``analogs’’ to study high-\zz{} galaxies. 
ISM properties such as metallicity, gas density, and radiation field are governed by local physical processes and can be estimated via line ratios. 
In contrast, global properties such as stellar mass, SFR, and size scale with galaxy mass and evolve over time. 
Attempting to match both types of properties simultaneously is often impossible due to the evolution of scaling relations. 
While low-\zz{} dwarf galaxies may mimic the ISM conditions of high-\zz{} galaxies, they differ dramatically in mass, SFR, and compactness \citep{NK23,sanders24,curti24,sarkar25,popesso23,ward24}. 
Therefore, any use of local ``analogs'' must be explicit in which properties are being matched, and avoid conflating gas-phase similarity with global morphological or evolutionary equivalence.

\subsubsection{Enhanced \texorpdfstring{\oiii{}}{[O III]} and \texorpdfstring{\oi{}}{[O I]} in High-\texorpdfstring{\zz}{z} Dusty Galaxies}

\begin{figure}[h]
    \centering
    \includegraphics[width=\halfwdth]{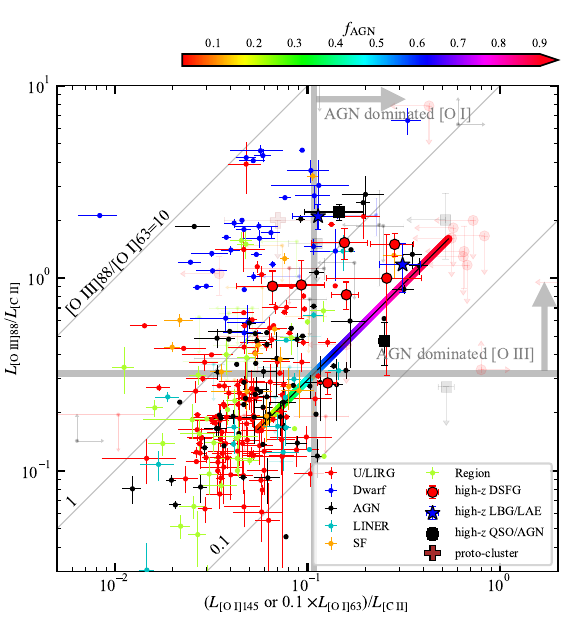}
    \caption[High-z AGN. ]{\oi{}/\cii{} vs. \oiii{88}/\cii{}, the diagonal dashed lines represent different \oiii{88}/\oi{63} values. The colored track shows the expected trajectory as AGN fraction increases, based on decomposition models from \ppi{}. Horizontal and vertical shaded bands indicate the 50\% AGN contribution levels for \oiii{} and \oi{}, respectively.}
    \label{f:implication_agn-oi}
\end{figure}

Although \oiii{88}/\cii{} shows no systematic redshift evolution in normal metal-poor galaxies, it behaves differently in dusty systems. 
In \ppi{}, we show that both \oiii{88}/\cii{} and \oiii{88}/\nii{} are elevated by $\sim$0.5 dex in high-\zz{} DSFGs and QSOs compared to their low-\zz{} counterparts. 
Fig.~\ref{f:implication_IR-line_Halpha} also shows elevated \oiii{}/\haextcorr{}, though the sample of high-\zz{} dusty galaxies with both measurements is limited.

The ionized gas density does not evolve significantly across redshifts (sec.~3.5 of \ppi{}), ruling it out as the cause. 
The coherence between low-ionization and neutral lines also remains at high \zz{}, suggesting that the elevated \loiii{88} caueses the difference. 

Interestingly, this enhancement also extends to neutral oxygen lines. 
The \oi{} doublet show elevated luminosities at high redshift, relative to lines like \cii{}, \nii{}, and \haextcorr{}. 
This simultaneous increase in both low- and high-ionization oxygen lines cannot be easily explained by a stronger radiation field alone.

Two viable explanations remain, either an increase in oxygen abundance relative to other metals, or an additional emission source that enhances both \oiii{} and \oi{}---most plausibly, AGN activity (exemplified in \ppi{} sec.~4.2). 

While some high-\zz{} studies have invoked a low C/O ratio to explain ``elevated'' \oiii{}/\cii{} \citep[e.g.,][]{HY20}, those arguments are only relevant to low-mass low-metallicity galaxies. 
For chemically mature dusty galaxies, an oxygen overabundance of 0.5 dex relative to all other elements is not theoretically expected. 
And such abundance anomaly should be detectable in their descendant at low redshift, namely elliptical galaxies. 
On the other hand, carbon and nitrogen abundances, despite their different nucleosynthetic origins, show little evolution in \nii{}/\cii{} ratios between low- and high-\zz{} dusty galaxies. 
Moreover, changes in abundance of carbon or nitrogen cannot explain enhanced \oiii{}/\ha{} and \oi{}/\ha{}, which are normalized to hydrogen.
Thus, an abundance-driven explanation is inconsistent with the observations.

As suggested in \citet{DC19}, AGN activity provides a more plausible explanation. 
In Fig.~\ref{f:implication_agn-oi}, we plot the effectss of increasing AGN contributions on both \oiii{88} and \oi{} using decomposition recipe from \ppi{} sec.~4.2. 
The colored trajectory follows an increasing AGN fraction ($f_\mathrm{AGN}$), and the majority of high-\zz{} dusty galaxies lie in the quadrant where AGN dominates both lines (i.e., $f_\mathrm{AGN}>50\%$). 
This makes AGN activity a quantitatively viable explanation for the enhanced oxygen lines.

However, verifying this scenario is challenging due to the lack of robust AGN diagnostics in heavily obscured systems. 
In particular, DSFGs at high \zz{} often lack X-ray or MIR AGN observations that are deep enough, making indirect line diagnostics such as those shown here essential for future studies.

\subsection{New Questions for Future FIR FSL Studies}

The increasing availability of FIR FSL observations has significantly advanced our understanding of the ISM conditions and the physical parameters that control these lines. 
However, serious discrepancies and uncertainties remain. 
Long-standing assumptions about the origin, diagnostic power, and calibration of FIR continuum and FSLs---particularly \cii{}, \oi{}, and \oiii{}---require critical revision. 
The findings presented in this work raise several open questions that define the priorities for future FIR FSL studies. 

The empirical relations discussed in this work are derived from unresolved, galaxy-integrated measurements. 
However, galaxies are composed of diverse ISM phases and structures, and integrated quantities can conflate physically distinct emission regions. 
Future studies need to explore and validate these relations on resolved spatial scales, both in nearby and high-redshift galaxies. 
Mapping FIR FSLs alongside multi-wavelength data (e.g., \ha{}, CO, dust continuum) will help disentangle the dominant emission zones and determine how reliably global line ratios reflect local physical conditions. 

As discussed in Sec.~\ref{sec:implication_cii_deficit}, the interpretation of the line ``deficit'' problem has critical implications for whether \lir{} overestimates or spectral line emission underestimates SFR. 
Disentangling this requires line-independent SFR measurements. 
One promising approach is the use of core-collapse supernova (SN) rates as direct tracers of massive star formation. 
Upcoming time-domain observatories, such as the Nancy Grace Roman Space Telescope and the Vera C. Rubin Observatory, will offer new opportunities to apply this method, particularly in nearby galaxies with rich archival data.

The physical origin of FIR FSLs---whether from compact \hii{} regions, dense PDRs, diffuse ionized gas, or extended neutral media---remains a core uncertainty. 
While cloud-scale studies in the Milky Way and nearby systems provide valuable insight, most extragalactic FIR FSL measurements remain spatially unresolved. 
Future instruments with enhanced sensitivity to low-surface brightness emission, such as balloon-borne or space-based FIR telescopes and next-generation interferometers, will be essential to detect and characterize the diffuse, extended components of FIR line emission. 
A critical goal is to quantify the filling factors and spatial associations of FIR lines with their ionizing or heating sources.

The spatial relationship between FIR continuum emission and FIR FSLs remains poorly understood, especially at high redshifts. 
Although the dust continuum emission often appears compact, FIR lines such as \cii{} can be significantly more extended. 
Understanding this mismatch is essential for interpreting the line ``deficit'' problem and for constructing physically motivated models of global ISM structures and energy balance. 
High-angular-resolution observations from ALMA and future FIR facilities will be needed to map both line and continuum emission and evaluate their spatial correlations and morphological differences. 

As shown in sec.~4.2 in \ppi{}, AGN activity can significantly enhance FIR FSL emission, particularly for \oiii{} and \oi{} at high \zz{} (Sec.~\ref{f:implication_agn-oi}). 
Disentangling AGN and star formation contributions remains challenging, especially in dusty systems where traditional AGN tracers (e.g., BPTdiagram, X-ray emission, MIR colors) are obscured. 
Future studies should refine AGN-sensitive diagnostics using FIR line ratios and develop robust decomposition methods that isolate AGN-related line components. 
Combining FIR spectroscopy with JWST/NIRSpec data and future FIR, MIR and X-ray observations will be critical to establishing a comprehensive picture of obscured AGN populations at high redshift and their influence on FIR FSLs.

\section{Summary}
\label{sec:implication_summary}

In this work, we investigate the origins, diagnostic power, and empirical correlations of FIR FSLs in galaxies, based on the comprehensive catalogs compiled in \ppi{} and physical insights obtained in \ppii{}. 
By combining FIR data with ancillary multi-wavelength data, we reassess long-standing questions and propose a new perspective for FIR FSLs and ISM structures.

\noindent\textbf{New Answers}
\begin{itemize}
    \item a common fraction of $\sim$80\% \cii{} emission arise from neutral gas, and no systematic differences exist in \fciin{} between galaxy populations when using the N/O-corrected \nii{}/\cii{} calibrations. 
    \item FIR FSL/\haextcorr{} practically measures metallicity. 
    \item FIR FSLs from low-ionized state gas are strongly coupled to those of neutral gas with abundance bring the sole modulation, showing an \textit{ionized--neutral gas coherence} in terms of both energy source and spatial structure. 
    \item \lcii{} as a cooling line is physically a metallicity-dependent tracer of SFR. 
    \item \lcii{} also display poor correlation to atomic gas mass (\mhi{}), with quadratic dependence on metallicity. 
    \item \cii{} is correlated to \mhi{} and \mmol{} through secondary relations, rather than being physical tracers. 
    \item Extra justifications are needed for the applicability of PDR models on galaxy-integrated data. 
    \item The \cii{} ``deficit'' is only a manifestation of a universal phenomenon affecting ionized and neutral gas lines, even including extinction-corrected \ha{}. 
    \item The obscuration-corrected SFR was not calibrated on dusty galaxies, and comparison with FIR FSL ``deficits'' shows that the calibration breaks down for dusty galaxies. 
    \item Line-based calibrations suggest a crisis in SFR estimations in dusty galaxies, and a potential down-revised SFR by factors up to 100. 
    \item Line ``deficits'' are tightly correlated with each other, caused by the common denominator \lir{}. 
    \item Metallicity only controls dust properties in normal galaxies, linearly to obscuration fraciton and a --1/6 power to dust temperature. 
    \item Line ``deficits'' correspond to a \textit{gas--dust dichotomy}, characterzied by deviation of dust temperature from the normal metallicity dependence and an increase in density. 
    \item Offset in line ``deficits'' trends in high-\zz{} galaxies have profound implications in feedback and galaxy evolution, depending on scenarios. 
    \item High-\zz{} LBG/LAEs do not have elevated \oiii{}/\cii{}, and the offset in SFR can be readily explained by redshift evolution of the galaxy scaling relations. 
    \item Extra justifications are needed for invoking low-\zz{} ``analogs''. 
\end{itemize}

\noindent\textbf{New Questions}
\begin{itemize}
    \item The line ``deficit'' and \textit{``gas–-dust dichotomy''} reflect a decoupling of dust heating from gas, caused by either an \textit{IR ``excess''} or intrinsic \textit{ionized gas ``deficit''}. We discuss five possible scenarios. 
    \item The spatial origins of FIR FSLs and their positions in ISM structures remain elusive, complicated by low density, \textit{ionized--neutral gas coherence}, and metallicity dependence. 
    \item The true SFR tracers in dusty galaxies rely on the understanding of the ``deficit'' problem, and may require significant revision. 
    \item Obscured AGN may be prevalent in high-\zz{} DSFGs and cause the enhanced oxygen emission, though data of extinction-free AGN tracers are missing for these galaxies. 
    \item To reconcile various questions presented in this work, further understanding of FIR FSLs requires a new theoretical picture of ISM structures, supplemented by resolved, deep low-surface brightness observations. 
\end{itemize}


\begin{acknowledgments}

We thank Padelis Papadopoulos for the discussion on dust and gas heating. 
We thank Iker Millan Irigoyen for the discussion on IMF. 
B.P. acknowledges the support of NRAO SOS 1519126. 
Support for this work at Cornell was provided in part by NASA grant NNX17AF37G, NSF grants AST-1716229 and AST-1910107 and NASA/SOFIA grant NNA17BF53C (SOF09-0185).  

\software{Astropy \citep{astropy13,astropy18}, PyNeb \citep{Luridiana15}}
\end{acknowledgments}



\appendix

\section{Data Sources and References}
\label{sec:implication_ref}

This work uses the data in the tables FLAMES-low and FLAMES-high in \ppi{}. 

The plotted low-\zz{} data are sourced from \ppi{}, SHINING \citep{H18a}, DGS \citep{M13,C15,C19}, HERCULES \citep{R15}, HERUS \citep{F13}, \citet{D17}, \citet{F16}, \citet{Z16}, \citet{B08}, \citet{F14}, \citet{F07}, \citet{H10}, \citet{K16}, \citet{P21}, \citet{S22}, \citet{C22}, \citet{S15}, \citet{D15}, \citet{L17a}, \citet{L17b}, \citet{S19}, \citet{M06}, DustPedia \citep{D19,C18}, ALFALFA \citep{H18b,D20}, HIPASS \citep{K04}.

The plotted high-\zz{} data are source from \ppi{}, \citet{ZJ15}, \citet{DD99a}, \citet{KM23}, \citet{EA06}, \citet{MI21}, \citet{AH25}, \citet{KK23}, \citet{ZS24}, \citet{PE25}, \citet{IT19}, \citet{WJ12}, \citet{TK19}, \citet{BC06}, \citet{VG24}, \citet{IR10a}, \citet{BM25}, \citet{MI25}, \citet{MR05}, \citet{VB17a}, \citet{SD15c}, \citet{WD15}, \citet{UB16}, \citet{DT21}, \citet{TR23a}, \citet{BE15}, \citet{PB23}, \citet{DC11}, \citet{PA24}, \citet{FC11}, \citet{VE11}, \citet{DC14}, \citet{IA20}, \citet{SM25a}, \citet{HA24}, \citet{VV24}, \citet{GB18}, \citet{FD99}, \citet{BM20}, \citet{LD07}, \citet{MC24}, \citet{MY19}, \citet{KN15}, \citet{LK22}, \citet{CK12}, \citet{SM16}, \citet{SC19}, \citet{MS17}, \citet{CA11}, \citet{KT25}, \citet{TS23}, \citet{SV12}, \citet{HK09}, \citet{HT18}, \citet{RF21}, \citet{WF21a}, \citet{RL24}, \citet{RS21}, \citet{IE21}, \citet{VB17b}, \citet{DT18}, \citet{SY25}, \citet{MY16}, \citet{SS23}, \citet{SF24}, \citet{GG23}, \citet{HA23}, \citet{WR19}, \citet{AY08}, \citet{MI24}, \citet{RD20}, \citet{HY25}, \citet{RL25}, \citet{PJ04}, \citet{YJ19b}, \citet{WW25}, \citet{SS25a}, \citet{IH22}, \citet{CD20}, \citet{PA23a}, \citet{AJ24}, \citet{SV15}, \citet{FS25}, \citet{VB19}, \citet{KK17}, \citet{GC24}, \citet{SY17}, \citet{SC18}, \citet{RM23}, \citet{KD23}, \citet{TY23}, \citet{SG10}, \citet{UH24a}, \citet{WF12}, \citet{BM24a}, \citet{LM21}, \citet{PA23b}, \citet{FR24}, \citet{HT19b}, \citet{MH14}, \citet{YM15}, \citet{IT21a}, \citet{AH24b}, \citet{TT22}, \citet{GB15}, \citet{NM19c}, \citet{ID06}, \citet{CF12}, \citet{KS12}, \citet{LJ20}, \citet{OY12}, \citet{LO20}, \citet{JG25}, \citet{DC19}, \citet{FS24a}, \citet{BC21}, \citet{ZJ18}, \citet{DR22}, \citet{PR19}, \citet{IA16}, \citet{MR25}, \citet{LC18}, \citet{WE12b}, \citet{ZA24}, \citet{OP16}, \citet{MT23}, \citet{MY18b}, \citet{RD14}, \citet{SY21}, \citet{RJ11a}, \citet{WA03}, \citet{OK14}, \citet{SY22}, \citet{FT19}, \citet{ZZ18}, \citet{HK16}, \citet{WA13}, \citet{BT23}, \citet{DD99b}, \citet{KN13}, \citet{SE10}, \citet{LJ22}, \citet{NK23}, \citet{BL12}, \citet{BT24a}, \citet{AH24a}, \citet{PL11}, \citet{LN21a}, \citet{CC20}, \citet{HR22}, \citet{AM16}, \citet{PR18}, \citet{CP15}, \citet{MC25a}, \citet{ZJ24}, \citet{RY23}, \citet{IT18}, \citet{SM12}, \citet{SM24}, \citet{AH22}, \citet{CC15}, \citet{EA23}, \citet{RJ11b}, \citet{DT16}, \citet{BI17}, \citet{FY21}, \citet{PA17}, \citet{CS17}, \citet{UH23}, \citet{BR18}, \citet{RM20a}, \citet{VB16}, \citet{SA13}, \citet{WF22}, \citet{AA25}, \citet{MS22}, \citet{WJ17}, \citet{NT12}, \citet{NM19a}, \citet{AJ23}, \citet{SD15a}, \citet{YW14}, \citet{HS10}, \citet{RM20b}, \citet{PA21}, \citet{BM17}, \citet{BE24}, \citet{LF21}, \citet{MR12}, \citet{HK23}, \citet{YM21}, \citet{WJ13}, \citet{HT23}, \citet{ID16}, \citet{FY24b}, \citet{FS21}, \citet{AS24}, \citet{MR15}, \citet{MZ24}, \citet{WC07}, \citet{TK22}, \citet{IR16}, \citet{GK24}, \citet{FY24a}, \citet{JG24b}, \citet{BE18b}, \citet{FS24b}, \citet{VF22}, \citet{SM25b}, \citet{WR13}, \citet{NM17}, \citet{SR18}, \citet{NM19b}, \citet{DR14}, \citet{FY25}, \citet{VB12}, \citet{EA20}, \citet{WC15b}, \citet{RJ08}, \citet{FF21}, \citet{YJ21}, \citet{JX24}, \citet{PR16}, \citet{IT21b}, \citet{LT23}, \citet{DM15}, \citet{PL16}, \citet{GM23}, \citet{WF09}, \citet{WY22b}, \citet{GR13}, \citet{RW12}, \citet{WR16}, \citet{MR22}, \citet{WA07}, \citet{TH00}, \citet{BM16}, \citet{RM19}, \citet{WF24a}, \citet{LN18}, \citet{HT25}, \citet{SC21}, \citet{KK16}, \citet{RC20}, \citet{CA23}, \citet{MR09}, \citet{MY18a}, \citet{SA12}, \citet{FS09}, \citet{SJ25}, \citet{GC18}, \citet{ZJ25}, \citet{BJ23}, \citet{HR21}, \citet{VB20}, \citet{LH09}, \citet{LN21b}, \citet{MM23}, \citet{FA20}, \citet{CM24}, \citet{DR17}, \citet{MC17}, \citet{WY22a}, \citet{MT18}, \citet{SF25}, \citet{BM18}, \citet{BE16}, \citet{DR23}, \citet{ER23}, \citet{PG23}, \citet{CS24}, \citet{LN17a}, \citet{FS22}, \citet{RF23}, \citet{IR13}, \citet{WF18}, \citet{CL11}, \citet{YJ19a}, \citet{MJ19}, \citet{IR25}, \citet{RD18}, \citet{SS22}, \citet{RC25}, \citet{RF20}, \citet{WJ22}, \citet{JG24a}, \citet{SM23}, \citet{DR18}, \citet{OS09}, \citet{BR22c}, \citet{CP02}, \citet{WC13}, \citet{FH13}, \citet{RT14}, \citet{OM13}, \citet{WF21b}, \citet{HT19a}, \citet{AM08}, \citet{BT21}, \citet{SY19}, \citet{CP11a}, \citet{WF19b}, \citet{UH17}, \citet{WF19a}, \citet{WW24}, \citet{SJ20}, \citet{CS18a}, \citet{BE25}, \citet{VI11}, \citet{GM20}, \citet{FS19}, \citet{OA01}, \citet{LK23}, \citet{WJ10}, \citet{SM17}, \citet{RD13}, \citet{OI16}, \citet{BR22b}, \citet{BE18a}, \citet{MD18}, \citet{YJ20}, \citet{TY20}, \citet{JG17}, \citet{JE20}, \citet{GS12}, \citet{GJ16}, \citet{FC15}, \citet{SL22}, \citet{SS25b}, \citet{BS24}, \citet{LD19}, \citet{WE12a}, \citet{TR22}, \citet{NR14}, \citet{WC15a}, \citet{ZA18}, \citet{vI24}, \citet{IN25}, \citet{CP11b}, \citet{CS13}, \citet{BT20}, \citet{UH21}, \citet{LJ24}, \citet{LF24}, \citet{MG14}, \citet{BA23}, \citet{WC17}, \citet{MJ17}, \citet{MC25b}, \citet{HY20}, \citet{BE21}, \citet{XM24}, \citet{LN19}, \citet{CC13}, \citet{KY22}, \citet{NM20}, \citet{WY21}, \citet{LK19}, \citet{FC10}, \citet{FR10}, \citet{LM19}, \citet{BA06}, \citet{CS18b}, \citet{SD15b}, \citet{TR23b}, \citet{LK24}, \citet{SH18}, \citet{TT23}, \citet{KS25}.



\bibliography{main,flames-low,flames-high}{}

\begin{thebibliography}{}
\expandafter\ifx\csname natexlab\endcsname\relax\def\natexlab#1{#1}\fi
\providecommand{\url}[1]{\href{#1}{#1}}
\providecommand{\dodoi}[1]{doi:~\href{http://doi.org/#1}{\nolinkurl{#1}}}
\providecommand{\doeprint}[1]{\href{http://ascl.net/#1}{\nolinkurl{http://ascl.net/#1}}}
\providecommand{\doarXiv}[1]{\href{https://arxiv.org/abs/#1}{\nolinkurl{https://arxiv.org/abs/#1}}}

\bibitem[{{Abel} {et~al.}(2009){Abel}, {Dudley}, {Fischer}, {Satyapal}, \& {van
  Hoof}}]{abel09}
{Abel}, N.~P., {Dudley}, C., {Fischer}, J., {Satyapal}, S., \& {van Hoof},
  P.~A.~M. 2009, \apj, 701, 1147, \dodoi{10.1088/0004-637X/701/2/1147}

\bibitem[{{Abel} {et~al.}(2005){Abel}, {Ferland}, {Shaw}, \& {van
  Hoof}}]{abel05}
{Abel}, N.~P., {Ferland}, G.~J., {Shaw}, G., \& {van Hoof}, P.~A.~M. 2005,
  \apjs, 161, 65, \dodoi{10.1086/432913}

\bibitem[{{Abel} {et~al.}(2007){Abel}, {Sarma}, {Troland}, \&
  {Ferland}}]{abel07}
{Abel}, N.~P., {Sarma}, A.~P., {Troland}, T.~H., \& {Ferland}, G.~J. 2007,
  \apj, 662, 1024, \dodoi{10.1086/517987}

\bibitem[{{Akins} {et~al.}(2022){Akins}, {Fujimoto}, {Finlator}, {Watson},
  {Knudsen}, {Richard}, {Bakx}, {Hashimoto}, {Inoue}, {Matsuo},
  {Micha{\l}owski}, \& {Tamura}}]{AH22}
{Akins}, H.~B., {Fujimoto}, S., {Finlator}, K., {et~al.} 2022, \apj, 934, 64,
  \dodoi{10.3847/1538-4357/ac795b}

\bibitem[{{Algera} {et~al.}(2025){Algera}, {Rowland}, {Stefanon}, {Palla},
  {Sommovigo}, {Inami}, {Bouwens}, {Aravena}, {Bowler}, {Dayal}, {De Looze},
  {Ferrara}, {Fisher}, {Graziani}, {Gulis}, {Heintz}, {Hodge}, {van Leeuwen},
  {Pallottini}, {Phillips}, {Schouws}, {Smit}, {Stark}, \& {van der
  Werf}}]{AH25}
{Algera}, H., {Rowland}, L., {Stefanon}, M., {et~al.} 2025, arXiv e-prints,
  arXiv:2501.10508, \dodoi{10.48550/arXiv.2501.10508}

\bibitem[{{Algera} {et~al.}(2024{\natexlab{a}}){Algera}, {Inami}, {De Looze},
  {Ferrara}, {Hirashita}, {Aravena}, {Bakx}, {Bouwens}, {Bowler}, {Da Cunha},
  {Dayal}, {Fudamoto}, {Hodge}, {Hygate}, {van Leeuwen}, {Nanayakkara},
  {Palla}, {Pallottini}, {Rowland}, {Smit}, {Sommovigo}, {Stefanon}, {Vijayan},
  \& {van der Werf}}]{AH24b}
{Algera}, H. S.~B., {Inami}, H., {De Looze}, I., {et~al.} 2024{\natexlab{a}},
  \mnras, 533, 3098, \dodoi{10.1093/mnras/stae1994}

\bibitem[{{Algera} {et~al.}(2024{\natexlab{b}}){Algera}, {Inami}, {Sommovigo},
  {Fudamoto}, {Schneider}, {Graziani}, {Dayal}, {Bouwens}, {Aravena}, {da
  Cunha}, {Ferrara}, {Hygate}, {van Leeuwen}, {De Looze}, {Palla},
  {Pallottini}, {Smit}, {Stefanon}, {Topping}, \& {van der Werf}}]{AH24a}
{Algera}, H. S.~B., {Inami}, H., {Sommovigo}, L., {et~al.} 2024{\natexlab{b}},
  \mnras, 527, 6867, \dodoi{10.1093/mnras/stad3111}

\bibitem[{{{\'A}lvarez-M{\'a}rquez} {et~al.}(2023){{\'A}lvarez-M{\'a}rquez},
  {Crespo G{\'o}mez}, {Colina}, {Neeleman}, {Walter}, {Labiano},
  {P{\'e}rez-Gonz{\'a}lez}, {Bik}, {Noorgaard-Nielsen}, {Ostlin}, {Wright},
  {Alonso-Herrero}, {Azollini}, {Caputi}, {Eckart}, {Le F{\`e}vre},
  {Garc{\'\i}a-Mar{\'\i}n}, {Greve}, {Hjorth}, {Ilbert}, {Kendrew}, {Pye},
  {Tikkanen}, {Topinka}, {van der Werf}, {Ward}, {van Dishoeck}, {G{\"u}del},
  {Henning}, {Lagage}, {Ray}, \& {Waelkens}}]{AJ23}
{{\'A}lvarez-M{\'a}rquez}, J., {Crespo G{\'o}mez}, A., {Colina}, L., {et~al.}
  2023, \aap, 671, A105, \dodoi{10.1051/0004-6361/202245400}

\bibitem[{{{\'A}lvarez-M{\'a}rquez} {et~al.}(2024){{\'A}lvarez-M{\'a}rquez},
  {Colina}, {Crespo G{\'o}mez}, {Rinaldi}, {Melinder}, {{\"O}stlin},
  {Annunziatella}, {Labiano}, {Bik}, {Bosman}, {Greve}, {Wright},
  {Alonso-Herrero}, {Boogaard}, {Azollini}, {Caputi}, {Costantin}, {Eckart},
  {Garc{\'\i}a-Mar{\'\i}n}, {Gillman}, {Hjorth}, {Iani}, {Ilbert}, {Jermann},
  {Langeroodi}, {Meyer}, {Pei{\ss}ker}, {P{\'e}rez-Gonz{\'a}lez}, {Pye},
  {Tikkanen}, {Topinka}, {van der Werf}, {Walter}, {Henning}, \& {Ray}}]{AJ24}
{{\'A}lvarez-M{\'a}rquez}, J., {Colina}, L., {Crespo G{\'o}mez}, A., {et~al.}
  2024, \aap, 686, A85, \dodoi{10.1051/0004-6361/202347946}

\bibitem[{{Amvrosiadis} {et~al.}(2025){Amvrosiadis}, {Lange}, {Nightingale},
  {He}, {Frenk}, {Oman}, {Smail}, {Swinbank}, {Fragkoudi}, {Gadotti}, {Cole},
  {Borsato}, {Robertson}, {Massey}, {Cao}, \& {Li}}]{AA25}
{Amvrosiadis}, A., {Lange}, S., {Nightingale}, J.~W., {et~al.} 2025, \mnras,
  537, 1163, \dodoi{10.1093/mnras/staf048}

\bibitem[{{Andrews} \& {Martini}(2013)}]{andrews03}
{Andrews}, B.~H., \& {Martini}, P. 2013, \apj, 765, 140,
  \dodoi{10.1088/0004-637X/765/2/140}

\bibitem[{{Ao} {et~al.}(2008){Ao}, {Wei{\ss}}, {Downes}, {Walter}, {Henkel}, \&
  {Menten}}]{AY08}
{Ao}, Y., {Wei{\ss}}, A., {Downes}, D., {et~al.} 2008, \aap, 491, 747,
  \dodoi{10.1051/0004-6361:200810482}

\bibitem[{{Aravena} {et~al.}(2008){Aravena}, {Bertoldi}, {Schinnerer}, {Weiss},
  {Jahnke}, {Carilli}, {Frayer}, {Henkel}, {Brusa}, {Menten}, {Salvato}, \&
  {Smolcic}}]{AM08}
{Aravena}, M., {Bertoldi}, F., {Schinnerer}, E., {et~al.} 2008, \aap, 491, 173,
  \dodoi{10.1051/0004-6361:200810628}

\bibitem[{{Aravena} {et~al.}(2016){Aravena}, {Spilker}, {Bethermin},
  {Bothwell}, {Chapman}, {de Breuck}, {Furstenau}, {G{\'o}nzalez-L{\'o}pez},
  {Greve}, {Litke}, {Ma}, {Malkan}, {Marrone}, {Murphy}, {Stark}, {Strandet},
  {Vieira}, {Weiss}, {Welikala}, {Wong}, \& {Collier}}]{AM16}
{Aravena}, M., {Spilker}, J.~S., {Bethermin}, M., {et~al.} 2016, \mnras, 457,
  4406, \dodoi{10.1093/mnras/stw275}

\bibitem[{{Arribas} {et~al.}(2024){Arribas}, {Perna}, {Rodr{\'\i}guez Del
  Pino}, {Lamperti}, {D'Eugenio}, {P{\'e}rez-Gonz{\'a}lez}, {Jones}, {Crespo
  G{\'o}mez}, {Curti}, {Lim}, {{\'A}lvarez-M{\'a}rquez}, {Bunker}, {Carniani},
  {Charlot}, {Jakobsen}, {Maiolino}, {{\"U}bler}, {Willott}, {B{\"o}ker},
  {Chevallard}, {Circosta}, {Cresci}, {Kumari}, {Parlanti}, {Scholtz},
  {Venturi}, \& {Witstok}}]{AS24}
{Arribas}, S., {Perna}, M., {Rodr{\'\i}guez Del Pino}, B., {et~al.} 2024, \aap,
  688, A146, \dodoi{10.1051/0004-6361/202348824}

\bibitem[{{Astropy Collaboration} {et~al.}(2013){Astropy Collaboration},
  {Robitaille}, {Tollerud}, {Greenfield}, {Droettboom}, {Bray}, {Aldcroft},
  {Davis}, {Ginsburg}, {Price-Whelan}, {Kerzendorf}, {Conley}, {Crighton},
  {Barbary}, {Muna}, {Ferguson}, {Grollier}, {Parikh}, {Nair}, {Unther},
  {Deil}, {Woillez}, {Conseil}, {Kramer}, {Turner}, {Singer}, {Fox}, {Weaver},
  {Zabalza}, {Edwards}, {Azalee Bostroem}, {Burke}, {Casey}, {Crawford},
  {Dencheva}, {Ely}, {Jenness}, {Labrie}, {Lim}, {Pierfederici}, {Pontzen},
  {Ptak}, {Refsdal}, {Servillat}, \& {Streicher}}]{astropy13}
{Astropy Collaboration}, {Robitaille}, T.~P., {Tollerud}, E.~J., {et~al.} 2013,
  \aap, 558, A33, \dodoi{10.1051/0004-6361/201322068}

\bibitem[{{Astropy Collaboration} {et~al.}(2018){Astropy Collaboration},
  {Price-Whelan}, {Sip{\H{o}}cz}, {G{\"u}nther}, {Lim}, {Crawford}, {Conseil},
  {Shupe}, {Craig}, {Dencheva}, {Ginsburg}, {VanderPlas}, {Bradley},
  {P{\'e}rez-Su{\'a}rez}, {de Val-Borro}, {Aldcroft}, {Cruz}, {Robitaille},
  {Tollerud}, {Ardelean}, {Babej}, {Bach}, {Bachetti}, {Bakanov}, {Bamford},
  {Barentsen}, {Barmby}, {Baumbach}, {Berry}, {Biscani}, {Boquien}, {Bostroem},
  {Bouma}, {Brammer}, {Bray}, {Breytenbach}, {Buddelmeijer}, {Burke},
  {Calderone}, {Cano Rodr{\'\i}guez}, {Cara}, {Cardoso}, {Cheedella}, {Copin},
  {Corrales}, {Crichton}, {D'Avella}, {Deil}, {Depagne}, {Dietrich}, {Donath},
  {Droettboom}, {Earl}, {Erben}, {Fabbro}, {Ferreira}, {Finethy}, {Fox},
  {Garrison}, {Gibbons}, {Goldstein}, {Gommers}, {Greco}, {Greenfield},
  {Groener}, {Grollier}, {Hagen}, {Hirst}, {Homeier}, {Horton}, {Hosseinzadeh},
  {Hu}, {Hunkeler}, {Ivezi{\'c}}, {Jain}, {Jenness}, {Kanarek}, {Kendrew},
  {Kern}, {Kerzendorf}, {Khvalko}, {King}, {Kirkby}, {Kulkarni}, {Kumar},
  {Lee}, {Lenz}, {Littlefair}, {Ma}, {Macleod}, {Mastropietro}, {McCully},
  {Montagnac}, {Morris}, {Mueller}, {Mumford}, {Muna}, {Murphy}, {Nelson},
  {Nguyen}, {Ninan}, {N{\"o}the}, {Ogaz}, {Oh}, {Parejko}, {Parley}, {Pascual},
  {Patil}, {Patil}, {Plunkett}, {Prochaska}, {Rastogi}, {Reddy Janga},
  {Sabater}, {Sakurikar}, {Seifert}, {Sherbert}, {Sherwood-Taylor}, {Shih},
  {Sick}, {Silbiger}, {Singanamalla}, {Singer}, {Sladen}, {Sooley},
  {Sornarajah}, {Streicher}, {Teuben}, {Thomas}, {Tremblay}, {Turner},
  {Terr{\'o}n}, {van Kerkwijk}, {de la Vega}, {Watkins}, {Weaver}, {Whitmore},
  {Woillez}, {Zabalza}, \& {Astropy Contributors}}]{astropy18}
{Astropy Collaboration}, {Price-Whelan}, A.~M., {Sip{\H{o}}cz}, B.~M., {et~al.}
  2018, \aj, 156, 123, \dodoi{10.3847/1538-3881/aabc4f}

\bibitem[{{Ba{\~n}ados} {et~al.}(2018{\natexlab{a}}){Ba{\~n}ados}, {Carilli},
  {Walter}, {Momjian}, {Decarli}, {Farina}, {Mazzucchelli}, \&
  {Venemans}}]{BE18b}
{Ba{\~n}ados}, E., {Carilli}, C., {Walter}, F., {et~al.} 2018{\natexlab{a}},
  \apjl, 861, L14, \dodoi{10.3847/2041-8213/aac511}

\bibitem[{{Ba{\~n}ados} {et~al.}(2015){Ba{\~n}ados}, {Decarli}, {Walter},
  {Venemans}, {Farina}, \& {Fan}}]{BE15}
{Ba{\~n}ados}, E., {Decarli}, R., {Walter}, F., {et~al.} 2015, \apjl, 805, L8,
  \dodoi{10.1088/2041-8205/805/1/L8}

\bibitem[{{Ba{\~n}ados} {et~al.}(2016){Ba{\~n}ados}, {Venemans}, {Decarli},
  {Farina}, {Mazzucchelli}, {Walter}, {Fan}, {Stern}, {Schlafly}, {Chambers},
  {Rix}, {Jiang}, {McGreer}, {Simcoe}, {Wang}, {Yang}, {Morganson}, {De Rosa},
  {Greiner}, {Balokovi{\'c}}, {Burgett}, {Cooper}, {Draper}, {Flewelling},
  {Hodapp}, {Jun}, {Kaiser}, {Kudritzki}, {Magnier}, {Metcalfe}, {Miller},
  {Schindler}, {Tonry}, {Wainscoat}, {Waters}, \& {Yang}}]{BE16}
{Ba{\~n}ados}, E., {Venemans}, B.~P., {Decarli}, R., {et~al.} 2016, \apjs, 227,
  11, \dodoi{10.3847/0067-0049/227/1/11}

\bibitem[{{Ba{\~n}ados} {et~al.}(2018{\natexlab{b}}){Ba{\~n}ados}, {Venemans},
  {Mazzucchelli}, {Farina}, {Walter}, {Wang}, {Decarli}, {Stern}, {Fan},
  {Davies}, {Hennawi}, {Simcoe}, {Turner}, {Rix}, {Yang}, {Kelson}, {Rudie}, \&
  {Winters}}]{BE18a}
{Ba{\~n}ados}, E., {Venemans}, B.~P., {Mazzucchelli}, C., {et~al.}
  2018{\natexlab{b}}, \nat, 553, 473, \dodoi{10.1038/nature25180}

\bibitem[{{Ba{\~n}ados} {et~al.}(2021){Ba{\~n}ados}, {Mazzucchelli}, {Momjian},
  {Eilers}, {Wang}, {Schindler}, {Connor}, {Andika}, {Barth}, {Carilli},
  {Davies}, {Decarli}, {Fan}, {Farina}, {Hennawi}, {Pensabene}, {Stern},
  {Venemans}, {Wenzl}, \& {Yang}}]{BE21}
{Ba{\~n}ados}, E., {Mazzucchelli}, C., {Momjian}, E., {et~al.} 2021, \apj, 909,
  80, \dodoi{10.3847/1538-4357/abe239}

\bibitem[{{Ba{\~n}ados} {et~al.}(2025){Ba{\~n}ados}, {Momjian}, {Connor},
  {Belladitta}, {Decarli}, {Mazzucchelli}, {Venemans}, {Walter}, {Wang}, {Xie},
  {Barth}, {Eilers}, {Fan}, {Khusanova}, {Schindler}, {Stern}, {Yang},
  {Andika}, {Carilli}, {Farina}, {Fabian}, {Hennawi}, {Pensabene}, \&
  {Rojas-Ruiz}}]{BE25}
{Ba{\~n}ados}, E., {Momjian}, E., {Connor}, T., {et~al.} 2025, Nature
  Astronomy, 9, 293, \dodoi{10.1038/s41550-024-02431-4}

\bibitem[{{Bakx} {et~al.}(2020){Bakx}, {Tamura}, {Hashimoto}, {Inoue}, {Lee},
  {Mawatari}, {Ota}, {Umehata}, {Zackrisson}, {Hatsukade}, {Kohno}, {Matsuda},
  {Matsuo}, {Okamoto}, {Shibuya}, {Shimizu}, {Taniguchi}, \& {Yoshida}}]{BT20}
{Bakx}, T. J.~L.~C., {Tamura}, Y., {Hashimoto}, T., {et~al.} 2020, \mnras, 493,
  4294, \dodoi{10.1093/mnras/staa509}

\bibitem[{{Bakx} {et~al.}(2021){Bakx}, {Sommovigo}, {Carniani}, {Ferrara},
  {Akins}, {Fujimoto}, {Hagimoto}, {Knudsen}, {Pallottini}, {Tamura}, \&
  {Watson}}]{BT21}
{Bakx}, T. J.~L.~C., {Sommovigo}, L., {Carniani}, S., {et~al.} 2021, \mnras,
  508, L58, \dodoi{10.1093/mnrasl/slab104}

\bibitem[{{Bakx} {et~al.}(2023){Bakx}, {Zavala}, {Mitsuhashi}, {Treu},
  {Fontana}, {Tadaki}, {Casey}, {Castellano}, {Glazebrook}, {Hagimoto},
  {Ikeda}, {Jones}, {Leethochawalit}, {Mason}, {Morishita}, {Nanayakkara},
  {Pentericci}, {Roberts-Borsani}, {Santini}, {Serjeant}, {Tamura}, {Trenti},
  \& {Vanzella}}]{BT23}
{Bakx}, T. J.~L.~C., {Zavala}, J.~A., {Mitsuhashi}, I., {et~al.} 2023, \mnras,
  519, 5076, \dodoi{10.1093/mnras/stac3723}

\bibitem[{{Bakx} {et~al.}(2024){Bakx}, {Algera}, {Venemans}, {Sommovigo},
  {Fujimoto}, {Carniani}, {Hagimoto}, {Hashimoto}, {Inoue}, {Salak},
  {Serjeant}, {Vallini}, {Eales}, {Ferrara}, {Fudamoto}, {Imamura}, {Inoue},
  {Knudsen}, {Matsuo}, {Sugahara}, {Tamura}, {Taniguchi}, \&
  {Yamanaka}}]{BT24a}
{Bakx}, T. J.~L.~C., {Algera}, H. S.~B., {Venemans}, B., {et~al.} 2024, \mnras,
  532, 2270, \dodoi{10.1093/mnras/stae1613}

\bibitem[{{Banados} {et~al.}(2024){Banados}, {Khusanova}, {Decarli}, {Momjian},
  {Walter}, {Connor}, {Carilli}, {Mazzucchelli}, {Rojas-Ruiz}, \&
  {Venemans}}]{BE24}
{Banados}, E., {Khusanova}, Y., {Decarli}, R., {et~al.} 2024, arXiv e-prints,
  arXiv:2408.12299, \dodoi{10.48550/arXiv.2408.12299}

\bibitem[{{Barisic} {et~al.}(2017){Barisic}, {Faisst}, {Capak}, {Pavesi},
  {Riechers}, {Scoville}, {Cooke}, {Kartaltepe}, {Casey}, \& {Smolcic}}]{BI17}
{Barisic}, I., {Faisst}, A.~L., {Capak}, P.~L., {et~al.} 2017, \apj, 845, 41,
  \dodoi{10.3847/1538-4357/aa7eda}

\bibitem[{{Beelen} {et~al.}(2006){Beelen}, {Cox}, {Benford}, {Dowell},
  {Kov{\'a}cs}, {Bertoldi}, {Omont}, \& {Carilli}}]{BA06}
{Beelen}, A., {Cox}, P., {Benford}, D.~J., {et~al.} 2006, \apj, 642, 694,
  \dodoi{10.1086/500636}

\bibitem[{{Beir{\~a}o} {et~al.}(2009){Beir{\~a}o}, {Appleton}, {Brandl},
  {Seibert}, {Jarrett}, \& {Houck}}]{beirao09}
{Beir{\~a}o}, P., {Appleton}, P.~N., {Brandl}, B.~R., {et~al.} 2009, \apj, 693,
  1650, \dodoi{10.1088/0004-637X/693/2/1650}

\bibitem[{{Belfiore} {et~al.}(2022){Belfiore}, {Santoro}, {Groves},
  {Schinnerer}, {Kreckel}, {Glover}, {Klessen}, {Emsellem}, {Blanc}, {Congiu},
  {Barnes}, {Boquien}, {Chevance}, {Dale}, {Kruijssen}, {Leroy}, {Pan},
  {Pessa}, {Schruba}, \& {Williams}}]{belfiore22}
{Belfiore}, F., {Santoro}, F., {Groves}, B., {et~al.} 2022, \aap, 659, A26,
  \dodoi{10.1051/0004-6361/202141859}

\bibitem[{{B{\'e}thermin} {et~al.}(2016){B{\'e}thermin}, {De Breuck},
  {Gullberg}, {Aravena}, {Bothwell}, {Chapman}, {Gonzalez}, {Greve}, {Litke},
  {Ma}, {Malkan}, {Marrone}, {Murphy}, {Spilker}, {Stark}, {Strandet},
  {Vieira}, {Wei{\ss}}, \& {Welikala}}]{BM16}
{B{\'e}thermin}, M., {De Breuck}, C., {Gullberg}, B., {et~al.} 2016, \aap, 586,
  L7, \dodoi{10.1051/0004-6361/201527739}

\bibitem[{{B{\'e}thermin} {et~al.}(2020){B{\'e}thermin}, {Fudamoto}, {Ginolfi},
  {Loiacono}, {Khusanova}, {Capak}, {Cassata}, {Faisst}, {Le F{\`e}vre},
  {Schaerer}, {Silverman}, {Yan}, {Amorin}, {Bardelli}, {Boquien}, {Cimatti},
  {Davidzon}, {Dessauges-Zavadsky}, {Fujimoto}, {Gruppioni}, {Hathi}, {Ibar},
  {Jones}, {Koekemoer}, {Lagache}, {Lemaux}, {Moreau}, {Oesch}, {Pozzi},
  {Riechers}, {Talia}, {Toft}, {Vallini}, {Vergani}, {Zamorani}, \&
  {Zucca}}]{BM20}
{B{\'e}thermin}, M., {Fudamoto}, Y., {Ginolfi}, M., {et~al.} 2020, \aap, 643,
  A2, \dodoi{10.1051/0004-6361/202037649}

\bibitem[{{Binggeli} {et~al.}(2021){Binggeli}, {Inoue}, {Hashimoto}, {Toribio},
  {Zackrisson}, {Ramstedt}, {Mawatari}, {Harikane}, {Matsuo}, {Okamoto}, {Ota},
  {Shimizu}, {Tamura}, {Taniguchi}, \& {Umehata}}]{BC21}
{Binggeli}, C., {Inoue}, A.~K., {Hashimoto}, T., {et~al.} 2021, \aap, 646, A26,
  \dodoi{10.1051/0004-6361/202038180}

\bibitem[{{Birkin} {et~al.}(2023){Birkin}, {Hutchison}, {Welch}, {Spilker},
  {Aravena}, {Bayliss}, {Cathey}, {Chapman}, {Gonzalez}, {Gururajan},
  {Hayward}, {Khullar}, {Kim}, {Mahler}, {Malkan}, {Narayanan}, {Olivier},
  {Phadke}, {Reuter}, {Rigby}, {Smith}, {Solimano}, {Sulzenauer}, {Vieira},
  {Vizgan}, \& {Weiss}}]{BJ23}
{Birkin}, J.~E., {Hutchison}, T.~A., {Welch}, B., {et~al.} 2023, \apj, 958, 64,
  \dodoi{10.3847/1538-4357/acf712}

\bibitem[{{Bischetti} {et~al.}(2025){Bischetti}, {Feruglio}, {Carniani},
  {D'Odorico}, {Salvestrini}, \& {Fiore}}]{BM25}
{Bischetti}, M., {Feruglio}, C., {Carniani}, S., {et~al.} 2025, arXiv e-prints,
  arXiv:2504.15357, \dodoi{10.48550/arXiv.2504.15357}

\bibitem[{{Bischetti} {et~al.}(2018){Bischetti}, {Piconcelli}, {Feruglio},
  {Duras}, {Bongiorno}, {Carniani}, {Marconi}, {Pappalardo}, {Schneider},
  {Travascio}, {Valiante}, {Vietri}, {Zappacosta}, \& {Fiore}}]{BM18}
{Bischetti}, M., {Piconcelli}, E., {Feruglio}, C., {et~al.} 2018, \aap, 617,
  A82, \dodoi{10.1051/0004-6361/201833249}

\bibitem[{{Bischetti} {et~al.}(2024){Bischetti}, {Choi}, {Fiore}, {Feruglio},
  {Carniani}, {D'Odorico}, {Ba{\~n}ados}, {Chen}, {Decarli}, {Gallerani},
  {Hlavacek-Larrondo}, {Lai}, {Leighly}, {Mazzucchelli}, {Perreault-Levasseur},
  {Tripodi}, {Walter}, {Wang}, {Yang}, {Zanchettin}, \& {Zhu}}]{BM24a}
{Bischetti}, M., {Choi}, H., {Fiore}, F., {et~al.} 2024, \apj, 970, 9,
  \dodoi{10.3847/1538-4357/ad4a77}

\bibitem[{{Borys} {et~al.}(2006){Borys}, {Blain}, {Dey}, {Le Floc'h},
  {Jannuzi}, {Barnard}, {Bian}, {Brodwin}, {Men{\'e}ndez-Delmestre},
  {Thompson}, {Brand}, {Brown}, {Dowell}, {Eisenhardt}, {Farrah}, {Frayer},
  {Higdon}, {Higdon}, {Phillips}, {Soifer}, {Stern}, \& {Weedman}}]{BC06}
{Borys}, C., {Blain}, A.~W., {Dey}, A., {et~al.} 2006, \apj, 636, 134,
  \dodoi{10.1086/497983}

\bibitem[{{Bosman} {et~al.}(2024){Bosman}, {{\'A}lvarez-M{\'a}rquez}, {Colina},
  {Walter}, {Alonso-Herrero}, {Ward}, {{\~A}-stlin}, {Greve}, {Wright}, {Bik},
  {Boogaard}, {Caputi}, {Costantin}, {Eckart}, {Garc{\'\i}a-Mar{\'\i}n},
  {Gillman}, {Hjorth}, {Iani}, {Ilbert}, {Jermann}, {Labiano}, {Langeroodi},
  {Pei{\ss}ker}, {Rinaldi}, {Topinka}, {van der Werf}, {G{\"u}del}, {Henning},
  {Lagage}, {Ray}, {van Dishoeck}, \& {Vandenbussche}}]{BS24}
{Bosman}, S. E.~I., {{\'A}lvarez-M{\'a}rquez}, J., {Colina}, L., {et~al.} 2024,
  Nature Astronomy, 8, 1054, \dodoi{10.1038/s41550-024-02273-0}

\bibitem[{{Bouwens} {et~al.}(2022){Bouwens}, {Smit}, {Schouws}, {Stefanon},
  {Bowler}, {Endsley}, {Gonzalez}, {Inami}, {Stark}, {Oesch}, {Hodge},
  {Aravena}, {da Cunha}, {Dayal}, {de Looze}, {Ferrara}, {Fudamoto},
  {Graziani}, {Li}, {Nanayakkara}, {Pallottini}, {Schneider}, {Sommovigo},
  {Topping}, {van der Werf}, {Algera}, {Barrufet}, {Hygate}, {Labb{\'e}},
  {Riechers}, \& {Witstok}}]{BR22c}
{Bouwens}, R.~J., {Smit}, R., {Schouws}, S., {et~al.} 2022, \apj, 931, 160,
  \dodoi{10.3847/1538-4357/ac5a4a}

\bibitem[{{Bowler} {et~al.}(2018){Bowler}, {Bourne}, {Dunlop}, {McLure}, \&
  {McLeod}}]{BR18}
{Bowler}, R.~A.~A., {Bourne}, N., {Dunlop}, J.~S., {McLure}, R.~J., \&
  {McLeod}, D.~J. 2018, \mnras, 481, 1631, \dodoi{10.1093/mnras/sty2368}

\bibitem[{{Bowler} {et~al.}(2022){Bowler}, {Cullen}, {McLure}, {Dunlop}, \&
  {Avison}}]{BR22b}
{Bowler}, R.~A.~A., {Cullen}, F., {McLure}, R.~J., {Dunlop}, J.~S., \&
  {Avison}, A. 2022, \mnras, 510, 5088, \dodoi{10.1093/mnras/stab3744}

\bibitem[{{Brada{\v{c}}} {et~al.}(2017){Brada{\v{c}}}, {Garcia-Appadoo},
  {Huang}, {Vallini}, {Quinn Finney}, {Hoag}, {Lemaux}, {Borello Schmidt},
  {Treu}, {Carilli}, {Dijkstra}, {Ferrara}, {Fontana}, {Jones}, {Ryan}, {Wagg},
  \& {Gonzalez}}]{BM17}
{Brada{\v{c}}}, M., {Garcia-Appadoo}, D., {Huang}, K.-H., {et~al.} 2017, \apjl,
  836, L2, \dodoi{10.3847/2041-8213/836/1/L2}

\bibitem[{{Bradley} {et~al.}(2012){Bradley}, {Bouwens}, {Zitrin}, {Smit},
  {Coe}, {Ford}, {Zheng}, {Illingworth}, {Ben{\'\i}tez}, \&
  {Broadhurst}}]{BL12}
{Bradley}, L.~D., {Bouwens}, R.~J., {Zitrin}, A., {et~al.} 2012, \apj, 747, 3,
  \dodoi{10.1088/0004-637X/747/1/3}

\bibitem[{{Brauher} {et~al.}(2008){Brauher}, {Dale}, \& {Helou}}]{B08}
{Brauher}, J.~R., {Dale}, D.~A., \& {Helou}, G. 2008, \apjs, 178, 280,
  \dodoi{10.1086/590249}

\bibitem[{{Bunker} {et~al.}(2023){Bunker}, {Saxena}, {Cameron}, {Willott},
  {Curtis-Lake}, {Jakobsen}, {Carniani}, {Smit}, {Maiolino}, {Witstok},
  {Curti}, {D'Eugenio}, {Jones}, {Ferruit}, {Arribas}, {Charlot}, {Chevallard},
  {Giardino}, {de Graaff}, {Looser}, {L{\"u}tzgendorf}, {Maseda}, {Rawle},
  {Rix}, {Del Pino}, {Alberts}, {Egami}, {Eisenstein}, {Endsley}, {Hainline},
  {Hausen}, {Johnson}, {Rieke}, {Rieke}, {Robertson}, {Shivaei}, {Stark},
  {Sun}, {Tacchella}, {Tang}, {Williams}, {Willmer}, {Baker}, {Baum},
  {Bhatawdekar}, {Bowler}, {Boyett}, {Chen}, {Circosta}, {Helton}, {Ji},
  {Kumari}, {Lyu}, {Nelson}, {Parlanti}, {Perna}, {Sandles}, {Scholtz},
  {Suess}, {Topping}, {{\"U}bler}, {Wallace}, \& {Whitler}}]{BA23}
{Bunker}, A.~J., {Saxena}, A., {Cameron}, A.~J., {et~al.} 2023, \aap, 677, A88,
  \dodoi{10.1051/0004-6361/202346159}

\bibitem[{{Cameron} {et~al.}(2023){Cameron}, {Katz}, {Rey}, \& {Saxena}}]{CA23}
{Cameron}, A.~J., {Katz}, H., {Rey}, M.~P., \& {Saxena}, A. 2023, \mnras, 523,
  3516, \dodoi{10.1093/mnras/stad1579}

\bibitem[{{Capak} {et~al.}(2011){Capak}, {Riechers}, {Scoville}, {Carilli},
  {Cox}, {Neri}, {Robertson}, {Salvato}, {Schinnerer}, {Yan}, {Wilson}, {Yun},
  {Civano}, {Elvis}, {Karim}, {Mobasher}, \& {Staguhn}}]{CP11b}
{Capak}, P.~L., {Riechers}, D., {Scoville}, N.~Z., {et~al.} 2011, \nat, 470,
  233, \dodoi{10.1038/nature09681}

\bibitem[{{Capak} {et~al.}(2015){Capak}, {Carilli}, {Jones}, {Casey},
  {Riechers}, {Sheth}, {Carollo}, {Ilbert}, {Karim}, {Lefevre}, {Lilly},
  {Scoville}, {Smolcic}, \& {Yan}}]{CP15}
{Capak}, P.~L., {Carilli}, C., {Jones}, G., {et~al.} 2015, \nat, 522, 455,
  \dodoi{10.1038/nature14500}

\bibitem[{{Carilli} {et~al.}(2013){Carilli}, {Riechers}, {Walter}, {Maiolino},
  {Wagg}, {Lentati}, {McMahon}, \& {Wolfe}}]{CC13}
{Carilli}, C.~L., {Riechers}, D., {Walter}, F., {et~al.} 2013, \apj, 763, 120,
  \dodoi{10.1088/0004-637X/763/2/120}

\bibitem[{{Carniani} {et~al.}(2018{\natexlab{a}}){Carniani}, {Maiolino},
  {Smit}, \& {Amor{\'\i}n}}]{CS18b}
{Carniani}, S., {Maiolino}, R., {Smit}, R., \& {Amor{\'\i}n}, R.
  2018{\natexlab{a}}, \apjl, 854, L7, \dodoi{10.3847/2041-8213/aaab45}

\bibitem[{{Carniani} {et~al.}(2013){Carniani}, {Marconi}, {Biggs}, {Cresci},
  {Cupani}, {D'Odorico}, {Humphreys}, {Maiolino}, {Mannucci}, {Molaro},
  {Nagao}, {Testi}, \& {Zwaan}}]{CS13}
{Carniani}, S., {Marconi}, A., {Biggs}, A., {et~al.} 2013, \aap, 559, A29,
  \dodoi{10.1051/0004-6361/201322320}

\bibitem[{{Carniani} {et~al.}(2017){Carniani}, {Maiolino}, {Pallottini},
  {Vallini}, {Pentericci}, {Ferrara}, {Castellano}, {Vanzella}, {Grazian},
  {Gallerani}, {Santini}, {Wagg}, \& {Fontana}}]{CS17}
{Carniani}, S., {Maiolino}, R., {Pallottini}, A., {et~al.} 2017, \aap, 605,
  A42, \dodoi{10.1051/0004-6361/201630366}

\bibitem[{{Carniani} {et~al.}(2018{\natexlab{b}}){Carniani}, {Maiolino},
  {Amorin}, {Pentericci}, {Pallottini}, {Ferrara}, {Willott}, {Smit},
  {Matthee}, {Sobral}, {Santini}, {Castellano}, {De Barros}, {Fontana},
  {Grazian}, \& {Guaita}}]{CS18a}
{Carniani}, S., {Maiolino}, R., {Amorin}, R., {et~al.} 2018{\natexlab{b}},
  \mnras, 478, 1170, \dodoi{10.1093/mnras/sty1088}

\bibitem[{{Carniani} {et~al.}(2020){Carniani}, {Ferrara}, {Maiolino},
  {Castellano}, {Gallerani}, {Fontana}, {Kohandel}, {Lupi}, {Pallottini},
  {Pentericci}, {Vallini}, \& {Vanzella}}]{carniani20}
{Carniani}, S., {Ferrara}, A., {Maiolino}, R., {et~al.} 2020, \mnras, 499,
  5136, \dodoi{10.1093/mnras/staa3178}

\bibitem[{{Carniani} {et~al.}(2024){Carniani}, {Hainline}, {D'Eugenio},
  {Eisenstein}, {Jakobsen}, {Witstok}, {Johnson}, {Chevallard}, {Maiolino},
  {Helton}, {Willott}, {Robertson}, {Alberts}, {Arribas}, {Baker},
  {Bhatawdekar}, {Boyett}, {Bunker}, {Cameron}, {Cargile}, {Charlot}, {Curti},
  {Curtis-Lake}, {Egami}, {Giardino}, {Isaak}, {Ji}, {Jones}, {Kumari},
  {Maseda}, {Parlanti}, {P{\'e}rez-Gonz{\'a}lez}, {Rawle}, {Rieke}, {Rieke},
  {Del Pino}, {Saxena}, {Scholtz}, {Smit}, {Sun}, {Tacchella}, {{\"U}bler},
  {Venturi}, {Williams}, \& {Willmer}}]{CS24}
{Carniani}, S., {Hainline}, K., {D'Eugenio}, F., {et~al.} 2024, \nat, 633, 318,
  \dodoi{10.1038/s41586-024-07860-9}

\bibitem[{{Casavecchia} {et~al.}(2024){Casavecchia}, {Maio}, {P{\'e}roux}, \&
  {Ciardi}}]{casavecchia24}
{Casavecchia}, B., {Maio}, U., {P{\'e}roux}, C., \& {Ciardi}, B. 2024, \aap,
  689, A106, \dodoi{10.1051/0004-6361/202450332}

\bibitem[{{Castellano} {et~al.}(2024){Castellano}, {Napolitano}, {Fontana},
  {Roberts-Borsani}, {Treu}, {Vanzella}, {Zavala}, {Arrabal Haro},
  {Calabr{\`o}}, {Llerena}, {Mascia}, {Merlin}, {Paris}, {Pentericci},
  {Santini}, {Bakx}, {Bergamini}, {Cupani}, {Dickinson}, {Filippenko},
  {Glazebrook}, {Grillo}, {Kelly}, {Malkan}, {Mason}, {Morishita},
  {Nanayakkara}, {Rosati}, {Sani}, {Wang}, \& {Yoon}}]{CM24}
{Castellano}, M., {Napolitano}, L., {Fontana}, A., {et~al.} 2024, \apj, 972,
  143, \dodoi{10.3847/1538-4357/ad5f88}

\bibitem[{{Caux} {et~al.}(1984){Caux}, {Serra}, {Gispert}, {Puget}, {Ryter}, \&
  {Coron}}]{caux84}
{Caux}, E., {Serra}, G., {Gispert}, R., {et~al.} 1984, \aap, 137, 1

\bibitem[{{Chandar} {et~al.}(2023){Chandar}, {Caputo}, {Linden}, {Mok},
  {Whitmore}, {Calzetti}, {Elmegreen}, {Lee}, {Ubeda}, {White}, \&
  {Cook}}]{chandar23}
{Chandar}, R., {Caputo}, M., {Linden}, S., {et~al.} 2023, \apj, 943, 142,
  \dodoi{10.3847/1538-4357/acac96}

\bibitem[{{Chartab} {et~al.}(2022){Chartab}, {Cooray}, {Ma}, {Nayyeri},
  {Zilliot}, {Lopez}, {Fadda}, {Herrera-Camus}, {Malkan}, {Rigopoulou},
  {Sheth}, \& {Wardlow}}]{C22}
{Chartab}, N., {Cooray}, A., {Ma}, J., {et~al.} 2022, Nature Astronomy, 6, 844,
  \dodoi{10.1038/s41550-022-01679-y}

\bibitem[{{Cheng} {et~al.}(2020){Cheng}, {Cao}, {Lu}, {Li}, {Yang},
  {Rigopoulou}, {Charmandaris}, {Gao}, {Xu}, {van der Werf}, {Diaz Santos},
  {Privon}, {Zhao}, {Cao}, {Dai}, {Huang}, {Sanders}, {Wang}, {Wang}, \&
  {Zhu}}]{CC20}
{Cheng}, C., {Cao}, X., {Lu}, N., {et~al.} 2020, \apj, 898, 33,
  \dodoi{10.3847/1538-4357/ab980b}

\bibitem[{{Cicone} {et~al.}(2015){Cicone}, {Maiolino}, {Gallerani}, {Neri},
  {Ferrara}, {Sturm}, {Fiore}, {Piconcelli}, \& {Feruglio}}]{CC15}
{Cicone}, C., {Maiolino}, R., {Gallerani}, S., {et~al.} 2015, \aap, 574, A14,
  \dodoi{10.1051/0004-6361/201424980}

\bibitem[{{Cigan} {et~al.}(2016){Cigan}, {Young}, {Cormier}, {Lebouteiller},
  {Madden}, {Hunter}, {Brinks}, {Elmegreen}, {Schruba}, {Heesen}, \& {the <SPAN
  CLASS=''sml''>Little Things</SPAN> Team}}]{cigan16}
{Cigan}, P., {Young}, L., {Cormier}, D., {et~al.} 2016, \aj, 151, 14,
  \dodoi{10.3847/0004-6256/151/1/14}

\bibitem[{{Clark} {et~al.}(2018){Clark}, {Verstocken}, {Bianchi}, {Fritz},
  {Viaene}, {Smith}, {Baes}, {Casasola}, {Cassara}, {Davies}, {De Looze}, {De
  Vis}, {Evans}, {Galametz}, {Jones}, {Lianou}, {Madden}, {Mosenkov}, \&
  {Xilouris}}]{C18}
{Clark}, C.~J.~R., {Verstocken}, S., {Bianchi}, S., {et~al.} 2018, \aap, 609,
  A37, \dodoi{10.1051/0004-6361/201731419}

\bibitem[{{Combes} {et~al.}(2012){Combes}, {Rex}, {Rawle}, {Egami}, {Boone},
  {Smail}, {Richard}, {Ivison}, {Gurwell}, {Casey}, {Omont}, {Berciano Alba},
  {Dessauges-Zavadsky}, {Edge}, {Fazio}, {Kneib}, {Okabe}, {Pell{\'o}},
  {P{\'e}rez-Gonz{\'a}lez}, {Schaerer}, {Smith}, {Swinbank}, \& {van der
  Werf}}]{CF12}
{Combes}, F., {Rex}, M., {Rawle}, T.~D., {et~al.} 2012, \aap, 538, L4,
  \dodoi{10.1051/0004-6361/201118750}

\bibitem[{{Conley} {et~al.}(2011){Conley}, {Cooray}, {Vieira}, {Gonz{\'a}lez
  Solares}, {Kim}, {Aguirre}, {Amblard}, {Auld}, {Baker}, {Beelen}, {Blain},
  {Blundell}, {Bock}, {Bradford}, {Bridge}, {Brisbin}, {Burgarella},
  {Carpenter}, {Chanial}, {Chapin}, {Christopher}, {Clements}, {Cox},
  {Djorgovski}, {Dowell}, {Eales}, {Earle}, {Ellsworth-Bowers}, {Farrah},
  {Franceschini}, {Frayer}, {Fu}, {Gavazzi}, {Glenn}, {Griffin}, {Gurwell},
  {Halpern}, {Ibar}, {Ivison}, {Jarvis}, {Kamenetzky}, {Krips}, {Levenson},
  {Lupu}, {Mahabal}, {Maloney}, {Maraston}, {Marchetti}, {Marsden},
  {Matsuhara}, {Mortier}, {Murphy}, {Naylor}, {Neri}, {Nguyen}, {Oliver},
  {Omont}, {Page}, {Papageorgiou}, {Pearson}, {P{\'e}rez-Fournon}, {Pohlen},
  {Rangwala}, {Rawlings}, {Raymond}, {Riechers}, {Rodighiero}, {Roseboom},
  {Rowan-Robinson}, {Schulz}, {Scott}, {Scott}, {Serra}, {Seymour}, {Shupe},
  {Smith}, {Symeonidis}, {Tugwell}, {Vaccari}, {Valiante}, {Valtchanov},
  {Verma}, {Viero}, {Vigroux}, {Wang}, {Wiebe}, {Wright}, {Xu}, {Zeimann},
  {Zemcov}, \& {Zmuidzinas}}]{CA11}
{Conley}, A., {Cooray}, A., {Vieira}, J.~D., {et~al.} 2011, \apjl, 732, L35,
  \dodoi{10.1088/2041-8205/732/2/L35}

\bibitem[{{Coppin} {et~al.}(2012){Coppin}, {Danielson}, {Geach}, {Hodge},
  {Swinbank}, {Wardlow}, {Bertoldi}, {Biggs}, {Brandt}, {Caselli}, {Chapman},
  {Dannerbauer}, {Dunlop}, {Greve}, {Hamann}, {Ivison}, {Karim}, {Knudsen},
  {Menten}, {Schinnerer}, {Smail}, {Spaans}, {Walter}, {Webb}, \& {van der
  Werf}}]{CK12}
{Coppin}, K.~E.~K., {Danielson}, A.~L.~R., {Geach}, J.~E., {et~al.} 2012,
  \mnras, 427, 520, \dodoi{10.1111/j.1365-2966.2012.21977.x}

\bibitem[{{Cormier} {et~al.}(2015){Cormier}, {Madden}, {Lebouteiller}, {Abel},
  {Hony}, {Galliano}, {R{\'e}my-Ruyer}, {Bigiel}, {Baes}, {Boselli},
  {Chevance}, {Cooray}, {De Looze}, {Doublier}, {Galametz}, {Hughes},
  {Karczewski}, {Lee}, {Lu}, \& {Spinoglio}}]{C15}
{Cormier}, D., {Madden}, S.~C., {Lebouteiller}, V., {et~al.} 2015, \aap, 578,
  A53, \dodoi{10.1051/0004-6361/201425207}

\bibitem[{{Cormier} {et~al.}(2019){Cormier}, {Abel}, {Hony}, {Lebouteiller},
  {Madden}, {Polles}, {Galliano}, {De Looze}, {Galametz}, \&
  {Lambert-Huyghe}}]{C19}
{Cormier}, D., {Abel}, N.~P., {Hony}, S., {et~al.} 2019, \aap, 626, A23,
  \dodoi{10.1051/0004-6361/201834457}

\bibitem[{{Cowie} {et~al.}(2011){Cowie}, {Hu}, \& {Songaila}}]{CL11}
{Cowie}, L.~L., {Hu}, E.~M., \& {Songaila}, A. 2011, \apjl, 735, L38,
  \dodoi{10.1088/2041-8205/735/2/L38}

\bibitem[{{Cox} {et~al.}(2002){Cox}, {Omont}, {Djorgovski}, {Bertoldi}, {Pety},
  {Carilli}, {Isaak}, {Beelen}, {McMahon}, \& {Castro}}]{CP02}
{Cox}, P., {Omont}, A., {Djorgovski}, S.~G., {et~al.} 2002, \aap, 387, 406,
  \dodoi{10.1051/0004-6361:20020382}

\bibitem[{{Cox} {et~al.}(2011){Cox}, {Krips}, {Neri}, {Omont}, {G{\"u}sten},
  {Menten}, {Wyrowski}, {Wei{\ss}}, {Beelen}, {Gurwell}, {Dannerbauer},
  {Ivison}, {Negrello}, {Aretxaga}, {Hughes}, {Auld}, {Baes}, {Blundell},
  {Buttiglione}, {Cava}, {Cooray}, {Dariush}, {Dunne}, {Dye}, {Eales},
  {Frayer}, {Fritz}, {Gavazzi}, {Hopwood}, {Ibar}, {Jarvis}, {Maddox},
  {Micha{\l}owski}, {Pascale}, {Pohlen}, {Rigby}, {Smith}, {Swinbank}, {Temi},
  {Valtchanov}, {van der Werf}, \& {de Zotti}}]{CP11a}
{Cox}, P., {Krips}, M., {Neri}, R., {et~al.} 2011, \apj, 740, 63,
  \dodoi{10.1088/0004-637X/740/2/63}

\bibitem[{{Croxall} {et~al.}(2017){Croxall}, {Smith}, {Pellegrini}, {Groves},
  {Bolatto}, {Herrera-Camus}, {Sandstrom}, {Draine}, {Wolfire}, {Armus},
  {Boquien}, {Brandl}, {Dale}, {Galametz}, {Hunt}, {Kennicutt}, {Kreckel},
  {Rigopoulou}, {van der Werf}, \& {Wilson}}]{croxall17}
{Croxall}, K.~V., {Smith}, J.~D., {Pellegrini}, E., {et~al.} 2017, \apj, 845,
  96, \dodoi{10.3847/1538-4357/aa8035}

\bibitem[{{Cunningham} {et~al.}(2020){Cunningham}, {Chapman}, {Aravena}, {De
  Breuck}, {B{\'e}thermin}, {Chen}, {Dong}, {Gonzalez}, {Greve}, {Litke}, {Ma},
  {Malkan}, {Marrone}, {Miller}, {Phadke}, {Reuter}, {Rotermund}, {Spilker},
  {Stark}, {Strandet}, {Vieira}, \& {Wei{\ss}}}]{CD20}
{Cunningham}, D.~J.~M., {Chapman}, S.~C., {Aravena}, M., {et~al.} 2020, \mnras,
  494, 4090, \dodoi{10.1093/mnras/staa820}

\bibitem[{{Curti} {et~al.}(2024){Curti}, {Maiolino}, {Curtis-Lake},
  {Chevallard}, {Carniani}, {D'Eugenio}, {Looser}, {Scholtz}, {Charlot},
  {Cameron}, {{\"U}bler}, {Witstok}, {Boyett}, {Laseter}, {Sandles}, {Arribas},
  {Bunker}, {Giardino}, {Maseda}, {Rawle}, {Rodr{\'\i}guez Del Pino}, {Smit},
  {Willott}, {Eisenstein}, {Hausen}, {Johnson}, {Rieke}, {Robertson},
  {Tacchella}, {Williams}, {Willmer}, {Baker}, {Bhatawdekar}, {Egami},
  {Helton}, {Ji}, {Kumari}, {Perna}, {Shivaei}, \& {Sun}}]{curti24}
{Curti}, M., {Maiolino}, R., {Curtis-Lake}, E., {et~al.} 2024, \aap, 684, A75,
  \dodoi{10.1051/0004-6361/202346698}

\bibitem[{{de Blok} {et~al.}(2016){de Blok}, {Walter}, {Smith},
  {Herrera-Camus}, {Bolatto}, {Requena-Torres}, {Crocker}, {Croxall},
  {Kennicutt}, {Koda}, {Armus}, {Boquien}, {Dale}, {Kreckel}, \&
  {Meidt}}]{deblok16}
{de Blok}, W.~J.~G., {Walter}, F., {Smith}, J. D.~T., {et~al.} 2016, \aj, 152,
  51, \dodoi{10.3847/0004-6256/152/2/51}

\bibitem[{{De Breuck} {et~al.}(2011){De Breuck}, {Maiolino}, {Caselli},
  {Coppin}, {Hailey-Dunsheath}, \& {Nagao}}]{DC11}
{De Breuck}, C., {Maiolino}, R., {Caselli}, P., {et~al.} 2011, \aap, 530, L8,
  \dodoi{10.1051/0004-6361/201116868}

\bibitem[{{De Breuck} {et~al.}(2014){De Breuck}, {Williams}, {Swinbank},
  {Caselli}, {Coppin}, {Davis}, {Maiolino}, {Nagao}, {Smail}, {Walter},
  {Wei{\ss}}, \& {Zwaan}}]{DC14}
{De Breuck}, C., {Williams}, R.~J., {Swinbank}, M., {et~al.} 2014, \aap, 565,
  A59, \dodoi{10.1051/0004-6361/201323331}

\bibitem[{{De Breuck} {et~al.}(2019){De Breuck}, {Wei{\ss}}, {B{\'e}thermin},
  {Cunningham}, {Apostolovski}, {Aravena}, {Archipley}, {Chapman}, {Chen},
  {Fu}, {Jarugula}, {Malkan}, {Mangian}, {Phadke}, {Reuter}, {Stacey},
  {Strandet}, {Vieira}, \& {Vishwas}}]{DC19}
{De Breuck}, C., {Wei{\ss}}, A., {B{\'e}thermin}, M., {et~al.} 2019, \aap, 631,
  A167, \dodoi{10.1051/0004-6361/201936169}

\bibitem[{{De Looze} {et~al.}(2014){De Looze}, {Cormier}, {Lebouteiller},
  {Madden}, {Baes}, {Bendo}, {Boquien}, {Boselli}, {Clements}, {Cortese},
  {Cooray}, {Galametz}, {Galliano}, {Graci{\'a}-Carpio}, {Isaak}, {Karczewski},
  {Parkin}, {Pellegrini}, {R{\'e}my-Ruyer}, {Spinoglio}, {Smith}, \&
  {Sturm}}]{delooze14}
{De Looze}, I., {Cormier}, D., {Lebouteiller}, V., {et~al.} 2014, \aap, 568,
  A62, \dodoi{10.1051/0004-6361/201322489}

\bibitem[{{De Vis} {et~al.}(2019){De Vis}, {Jones}, {Viaene}, {Casasola},
  {Clark}, {Baes}, {Bianchi}, {Cassara}, {Davies}, {De Looze}, {Galametz},
  {Galliano}, {Lianou}, {Madden}, {Manilla-Robles}, {Mosenkov}, {Nersesian},
  {Roychowdhury}, {Xilouris}, \& {Ysard}}]{D19}
{De Vis}, P., {Jones}, A., {Viaene}, S., {et~al.} 2019, \aap, 623, A5,
  \dodoi{10.1051/0004-6361/201834444}

\bibitem[{{Decarli} {et~al.}(2014){Decarli}, {Walter}, {Carilli}, {Bertoldi},
  {Cox}, {Ferkinhoff}, {Groves}, {Maiolino}, {Neri}, {Riechers}, \&
  {Weiss}}]{DR14}
{Decarli}, R., {Walter}, F., {Carilli}, C., {et~al.} 2014, \apjl, 782, L17,
  \dodoi{10.1088/2041-8205/782/2/L17}

\bibitem[{{Decarli} {et~al.}(2017){Decarli}, {Walter}, {Venemans},
  {Ba{\~n}ados}, {Bertoldi}, {Carilli}, {Fan}, {Farina}, {Mazzucchelli},
  {Riechers}, {Rix}, {Strauss}, {Wang}, \& {Yang}}]{DR17}
{Decarli}, R., {Walter}, F., {Venemans}, B.~P., {et~al.} 2017, \nat, 545, 457,
  \dodoi{10.1038/nature22358}

\bibitem[{{Decarli} {et~al.}(2018){Decarli}, {Walter}, {Venemans},
  {Ba{\~n}ados}, {Bertoldi}, {Carilli}, {Fan}, {Farina}, {Mazzucchelli},
  {Riechers}, {Rix}, {Strauss}, {Wang}, \& {Yang}}]{DR18}
---. 2018, \apj, 854, 97, \dodoi{10.3847/1538-4357/aaa5aa}

\bibitem[{{Decarli} {et~al.}(2022){Decarli}, {Pensabene}, {Venemans}, {Walter},
  {Ba{\~n}ados}, {Bertoldi}, {Carilli}, {Cox}, {Fan}, {Farina}, {Ferkinhoff},
  {Groves}, {Li}, {Mazzucchelli}, {Neri}, {Riechers}, {Uzgil}, {Wang}, {Wang},
  {Weiss}, {Winters}, \& {Yang}}]{DR22}
{Decarli}, R., {Pensabene}, A., {Venemans}, B., {et~al.} 2022, \aap, 662, A60,
  \dodoi{10.1051/0004-6361/202142871}

\bibitem[{{Decarli} {et~al.}(2023){Decarli}, {Pensabene}, {Diaz-Santos},
  {Ferkinhoff}, {Strauss}, {Venemans}, {Walter}, {Ba{\~n}ados}, {Bertoldi},
  {Fan}, {Farina}, {Riechers}, {Rix}, \& {Wang}}]{DR23}
{Decarli}, R., {Pensabene}, A., {Diaz-Santos}, T., {et~al.} 2023, \aap, 673,
  A157, \dodoi{10.1051/0004-6361/202245674}

\bibitem[{{Della Bruna} {et~al.}(2020){Della Bruna}, {Adamo}, {Bik},
  {Fumagalli}, {Walterbos}, {{\"O}stlin}, {Bruzual}, {Calzetti}, {Charlot},
  {Grasha}, {Smith}, {Thilker}, \& {Wofford}}]{della20}
{Della Bruna}, L., {Adamo}, A., {Bik}, A., {et~al.} 2020, \aap, 635, A134,
  \dodoi{10.1051/0004-6361/201937173}

\bibitem[{{Dessauges-Zavadsky} {et~al.}(2015){Dessauges-Zavadsky}, {Zamojski},
  {Schaerer}, {Combes}, {Egami}, {Swinbank}, {Richard}, {Sklias}, {Rawle},
  {Rex}, {Kneib}, {Boone}, \& {Blain}}]{DM15}
{Dessauges-Zavadsky}, M., {Zamojski}, M., {Schaerer}, D., {et~al.} 2015, \aap,
  577, A50, \dodoi{10.1051/0004-6361/201424661}

\bibitem[{{D{\'\i}az-Santos} {et~al.}(2016){D{\'\i}az-Santos}, {Assef},
  {Blain}, {Tsai}, {Aravena}, {Eisenhardt}, {Wu}, {Stern}, \& {Bridge}}]{DT16}
{D{\'\i}az-Santos}, T., {Assef}, R.~J., {Blain}, A.~W., {et~al.} 2016, \apjl,
  816, L6, \dodoi{10.3847/2041-8205/816/1/L6}

\bibitem[{{D{\'\i}az-Santos} {et~al.}(2017){D{\'\i}az-Santos}, {Armus},
  {Charmandaris}, {Lu}, {Stierwalt}, {Stacey}, {Malhotra}, {van der Werf},
  {Howell}, {Privon}, {Mazzarella}, {Goldsmith}, {Murphy}, {Barcos-Mu{\~n}oz},
  {Linden}, {Inami}, {Larson}, {Evans}, {Appleton}, {Iwasawa}, {Lord},
  {Sanders}, \& {Surace}}]{D17}
{D{\'\i}az-Santos}, T., {Armus}, L., {Charmandaris}, V., {et~al.} 2017, \apj,
  846, 32, \dodoi{10.3847/1538-4357/aa81d7}

\bibitem[{{D{\'\i}az-Santos} {et~al.}(2018){D{\'\i}az-Santos}, {Assef},
  {Blain}, {Aravena}, {Stern}, {Tsai}, {Eisenhardt}, {Wu}, {Jun}, {Dibert},
  {Inami}, {Lansbury}, \& {Leclercq}}]{DT18}
{D{\'\i}az-Santos}, T., {Assef}, R.~J., {Blain}, A.~W., {et~al.} 2018, Science,
  362, 1034, \dodoi{10.1126/science.aap7605}

\bibitem[{{D{\'\i}az-Santos} {et~al.}(2021){D{\'\i}az-Santos}, {Assef},
  {Eisenhardt}, {Jun}, {Jones}, {Blain}, {Stern}, {Aravena}, {Tsai}, {Lake},
  {Wu}, \& {Gonz{\'a}lez-L{\'o}pez}}]{DT21}
{D{\'\i}az-Santos}, T., {Assef}, R.~J., {Eisenhardt}, P. R.~M., {et~al.} 2021,
  \aap, 654, A37, \dodoi{10.1051/0004-6361/202140455}

\bibitem[{{Dimaratos} {et~al.}(2015){Dimaratos}, {Cormier}, {Bigiel}, \&
  {Madden}}]{D15}
{Dimaratos}, A., {Cormier}, D., {Bigiel}, F., \& {Madden}, S.~C. 2015, \aap,
  580, A135, \dodoi{10.1051/0004-6361/201526447}

\bibitem[{{Downes} {et~al.}(1999{\natexlab{a}}){Downes}, {Neri}, {Wiklind},
  {Wilner}, \& {Shaver}}]{DD99b}
{Downes}, D., {Neri}, R., {Wiklind}, T., {Wilner}, D.~J., \& {Shaver}, P.~A.
  1999{\natexlab{a}}, \apjl, 513, L1, \dodoi{10.1086/311896}

\bibitem[{{Downes} {et~al.}(1999{\natexlab{b}}){Downes}, {Neri}, {Greve},
  {Guilloteau}, {Casoli}, {Hughes}, {Lutz}, {Menten}, {Wilner}, {Andreani},
  {Bertoldi}, {Carilli}, {Dunlop}, {Genzel}, {Gueth}, {Ivison}, {Mann},
  {Mellier}, {Oliver}, {Peacock}, {Rigopoulou}, {Rowan-Robinson}, {Schilke},
  {Serjeant}, {Tacconi}, \& {Wright}}]{DD99a}
{Downes}, D., {Neri}, R., {Greve}, A., {et~al.} 1999{\natexlab{b}}, \aap, 347,
  809, \dodoi{10.48550/arXiv.astro-ph/9907139}

\bibitem[{{Draine}(2011{\natexlab{a}})}]{draine11}
{Draine}, B.~T. 2011{\natexlab{a}}, {Physics of the Interstellar and
  Intergalactic Medium}

\bibitem[{{Draine}(2011{\natexlab{b}})}]{draine11b}
---. 2011{\natexlab{b}}, \apj, 732, 100, \dodoi{10.1088/0004-637X/732/2/100}

\bibitem[{{Draine} \& {Anderson}(1985)}]{draine85}
{Draine}, B.~T., \& {Anderson}, N. 1985, \apj, 292, 494, \dodoi{10.1086/163181}

\bibitem[{{Draine} {et~al.}(2007){Draine}, {Dale}, {Bendo}, {Gordon}, {Smith},
  {Armus}, {Engelbracht}, {Helou}, {Kennicutt}, {Li}, {Roussel}, {Walter},
  {Calzetti}, {Moustakas}, {Murphy}, {Rieke}, {Bot}, {Hollenbach}, {Sheth}, \&
  {Teplitz}}]{draine07}
{Draine}, B.~T., {Dale}, D.~A., {Bendo}, G., {et~al.} 2007, \apj, 663, 866,
  \dodoi{10.1086/518306}

\bibitem[{{Durbala} {et~al.}(2020){Durbala}, {Finn}, {Crone Odekon}, {Haynes},
  {Koopmann}, \& {O'Donoghue}}]{D20}
{Durbala}, A., {Finn}, R.~A., {Crone Odekon}, M., {et~al.} 2020, \aj, 160, 271,
  \dodoi{10.3847/1538-3881/abc018}

\bibitem[{{Edmunds} \& {Pagel}(1978)}]{edmunds78}
{Edmunds}, M.~G., \& {Pagel}, B.~E.~J. 1978, \mnras, 185, 77P,
  \dodoi{10.1093/mnras/185.1.77P}

\bibitem[{{Eilers} {et~al.}(2020){Eilers}, {Hennawi}, {Decarli}, {Davies},
  {Venemans}, {Walter}, {Ba{\~n}ados}, {Fan}, {Farina}, {Mazzucchelli},
  {Novak}, {Schindler}, {Simcoe}, {Wang}, \& {Yang}}]{EA20}
{Eilers}, A.-C., {Hennawi}, J.~F., {Decarli}, R., {et~al.} 2020, \apj, 900, 37,
  \dodoi{10.3847/1538-4357/aba52e}

\bibitem[{{Eilers} {et~al.}(2023){Eilers}, {Simcoe}, {Yue}, {Mackenzie},
  {Matthee}, {{\v{D}}urov{\v{c}}{\'\i}kov{\'a}}, {Kashino}, {Bordoloi}, \&
  {Lilly}}]{EA23}
{Eilers}, A.-C., {Simcoe}, R.~A., {Yue}, M., {et~al.} 2023, \apj, 950, 68,
  \dodoi{10.3847/1538-4357/acd776}

\bibitem[{{Endsley} {et~al.}(2023){Endsley}, {Stark}, {Lyu}, {Wang}, {Yang},
  {Fan}, {Smit}, {Bouwens}, {Hainline}, \& {Schouws}}]{ER23}
{Endsley}, R., {Stark}, D.~P., {Lyu}, J., {et~al.} 2023, \mnras, 520, 4609,
  \dodoi{10.1093/mnras/stad266}

\bibitem[{{Engelbracht} {et~al.}(2008){Engelbracht}, {Rieke}, {Gordon},
  {Smith}, {Werner}, {Moustakas}, {Willmer}, \& {Vanzi}}]{engelbracht08}
{Engelbracht}, C.~W., {Rieke}, G.~H., {Gordon}, K.~D., {et~al.} 2008, \apj,
  678, 804, \dodoi{10.1086/529513}

\bibitem[{{Evans} {et~al.}(2006){Evans}, {Solomon}, {Tacconi}, {Vavilkin}, \&
  {Downes}}]{EA06}
{Evans}, A.~S., {Solomon}, P.~M., {Tacconi}, L.~J., {Vavilkin}, T., \&
  {Downes}, D. 2006, \aj, 132, 2398, \dodoi{10.1086/508416}

\bibitem[{{Fadely} {et~al.}(2010){Fadely}, {Allam}, {Baker}, {Lin}, {Lutz},
  {Shapley}, {Shin}, {Allyn Smith}, {Strauss}, \& {Tucker}}]{FR10}
{Fadely}, R., {Allam}, S.~S., {Baker}, A.~J., {et~al.} 2010, \apj, 723, 729,
  \dodoi{10.1088/0004-637X/723/1/729}

\bibitem[{{Faisst} {et~al.}(2020){Faisst}, {Fudamoto}, {Oesch}, {Scoville},
  {Riechers}, {Pavesi}, \& {Capak}}]{FA20}
{Faisst}, A.~L., {Fudamoto}, Y., {Oesch}, P.~A., {et~al.} 2020, \mnras, 498,
  4192, \dodoi{10.1093/mnras/staa2545}

\bibitem[{{Falkendal} {et~al.}(2019){Falkendal}, {De Breuck}, {Lehnert},
  {Drouart}, {Vernet}, {Emonts}, {Lee}, {Nesvadba}, {Seymour}, {B{\'e}thermin},
  {Kolwa}, {Gullberg}, \& {Wylezalek}}]{FT19}
{Falkendal}, T., {De Breuck}, C., {Lehnert}, M.~D., {et~al.} 2019, \aap, 621,
  A27, \dodoi{10.1051/0004-6361/201732485}

\bibitem[{{Farrah} {et~al.}(2007){Farrah}, {Bernard-Salas}, {Spoon}, {Soifer},
  {Armus}, {Brandl}, {Charmandaris}, {Desai}, {Higdon}, {Devost}, \&
  {Houck}}]{F07}
{Farrah}, D., {Bernard-Salas}, J., {Spoon}, H.~W.~W., {et~al.} 2007, \apj, 667,
  149, \dodoi{10.1086/520834}

\bibitem[{{Farrah} {et~al.}(2013){Farrah}, {Lebouteiller}, {Spoon},
  {Bernard-Salas}, {Pearson}, {Rigopoulou}, {Smith}, {Gonz{\'a}lez-Alfonso},
  {Clements}, {Efstathiou}, {Cormier}, {Afonso}, {Petty}, {Harris}, {Hurley},
  {Borys}, {Verma}, {Cooray}, \& {Salvatelli}}]{F13}
{Farrah}, D., {Lebouteiller}, V., {Spoon}, H.~W.~W., {et~al.} 2013, \apj, 776,
  38, \dodoi{10.1088/0004-637X/776/1/38}

\bibitem[{{Ferguson} {et~al.}(1996){Ferguson}, {Wyse}, {Gallagher}, \&
  {Hunter}}]{ferguson96}
{Ferguson}, A. M.~N., {Wyse}, R. F.~G., {Gallagher}, J.~S., I., \& {Hunter},
  D.~A. 1996, \aj, 111, 2265, \dodoi{10.1086/117961}

\bibitem[{{Ferkinhoff} {et~al.}(2015){Ferkinhoff}, {Brisbin}, {Nikola},
  {Stacey}, {Sheth}, {Hailey-Dunsheath}, \& {Falgarone}}]{FC15}
{Ferkinhoff}, C., {Brisbin}, D., {Nikola}, T., {et~al.} 2015, \apj, 806, 260,
  \dodoi{10.1088/0004-637X/806/2/260}

\bibitem[{{Ferkinhoff} {et~al.}(2010){Ferkinhoff}, {Hailey-Dunsheath},
  {Nikola}, {Parshley}, {Stacey}, {Benford}, \& {Staguhn}}]{FC10}
{Ferkinhoff}, C., {Hailey-Dunsheath}, S., {Nikola}, T., {et~al.} 2010, \apjl,
  714, L147, \dodoi{10.1088/2041-8205/714/1/L147}

\bibitem[{{Ferkinhoff} {et~al.}(2011){Ferkinhoff}, {Brisbin}, {Nikola},
  {Parshley}, {Stacey}, {Phillips}, {Falgarone}, {Benford}, {Staguhn}, \&
  {Tucker}}]{FC11}
{Ferkinhoff}, C., {Brisbin}, D., {Nikola}, T., {et~al.} 2011, \apjl, 740, L29,
  \dodoi{10.1088/2041-8205/740/1/L29}

\bibitem[{{Fern{\'a}ndez Aranda} {et~al.}(2024){Fern{\'a}ndez Aranda},
  {D{\'\i}az Santos}, {Hatziminaoglou}, {Assef}, {Aravena}, {Eisenhardt},
  {Ferkinhoff}, {Pensabene}, {Nikola}, {Andreani}, {Vishwas}, {Stacey},
  {Decarli}, {Blain}, {Brisbin}, {Charmandaris}, {Jun}, {Li}, {Liao}, {Martin},
  {Stern}, {Tsai}, {Wu}, \& {Zewdie}}]{FR24}
{Fern{\'a}ndez Aranda}, R., {D{\'\i}az Santos}, T., {Hatziminaoglou}, E.,
  {et~al.} 2024, \aap, 682, A166, \dodoi{10.1051/0004-6361/202347869}

\bibitem[{{Fern{\'a}ndez-Ontiveros} {et~al.}(2016){Fern{\'a}ndez-Ontiveros},
  {Spinoglio}, {Pereira-Santaella}, {Malkan}, {Andreani}, \& {Dasyra}}]{F16}
{Fern{\'a}ndez-Ontiveros}, J.~A., {Spinoglio}, L., {Pereira-Santaella}, M.,
  {et~al.} 2016, \apjs, 226, 19, \dodoi{10.3847/0067-0049/226/2/19}

\bibitem[{{Finkelstein} {et~al.}(2009){Finkelstein}, {Papovich}, {Rudnick},
  {Egami}, {Le Floc'h}, {Rieke}, {Rigby}, \& {Willmer}}]{FS09}
{Finkelstein}, S.~L., {Papovich}, C., {Rudnick}, G., {et~al.} 2009, \apj, 700,
  376, \dodoi{10.1088/0004-637X/700/1/376}

\bibitem[{{Fischer} {et~al.}(2014){Fischer}, {Abel}, {Gonz{\'a}lez-Alfonso},
  {Dudley}, {Satyapal}, \& {van Hoof}}]{F14}
{Fischer}, J., {Abel}, N.~P., {Gonz{\'a}lez-Alfonso}, E., {et~al.} 2014, \apj,
  795, 117, \dodoi{10.1088/0004-637X/795/2/117}

\bibitem[{{F{\"o}rster Schreiber} \& {Wuyts}(2020)}]{schreiber20}
{F{\"o}rster Schreiber}, N.~M., \& {Wuyts}, S. 2020, \araa, 58, 661,
  \dodoi{10.1146/annurev-astro-032620-021910}

\bibitem[{{Fraternali} {et~al.}(2021){Fraternali}, {Karim}, {Magnelli},
  {G{\'o}mez-Guijarro}, {Jim{\'e}nez-Andrade}, \& {Posses}}]{FF21}
{Fraternali}, F., {Karim}, A., {Magnelli}, B., {et~al.} 2021, \aap, 647, A194,
  \dodoi{10.1051/0004-6361/202039807}

\bibitem[{{Frayer} {et~al.}(1999){Frayer}, {Ivison}, {Scoville}, {Evans},
  {Yun}, {Smail}, {Barger}, {Blain}, \& {Kneib}}]{FD99}
{Frayer}, D.~T., {Ivison}, R.~J., {Scoville}, N.~Z., {et~al.} 1999, \apjl, 514,
  L13, \dodoi{10.1086/311940}

\bibitem[{{Fu} {et~al.}(2013){Fu}, {Cooray}, {Feruglio}, {Ivison}, {Riechers},
  {Gurwell}, {Bussmann}, {Harris}, {Altieri}, {Aussel}, {Baker}, {Bock},
  {Boylan-Kolchin}, {Bridge}, {Calanog}, {Casey}, {Cava}, {Chapman},
  {Clements}, {Conley}, {Cox}, {Farrah}, {Frayer}, {Hopwood}, {Jia}, {Magdis},
  {Marsden}, {Mart{\'\i}nez-Navajas}, {Negrello}, {Neri}, {Oliver}, {Omont},
  {Page}, {P{\'e}rez-Fournon}, {Schulz}, {Scott}, {Smith}, {Vaccari},
  {Valtchanov}, {Vieira}, {Viero}, {Wang}, {Wardlow}, \& {Zemcov}}]{FH13}
{Fu}, H., {Cooray}, A., {Feruglio}, C., {et~al.} 2013, \nat, 498, 338,
  \dodoi{10.1038/nature12184}

\bibitem[{{Fudamoto} {et~al.}(2021){Fudamoto}, {Oesch}, {Schouws}, {Stefanon},
  {Smit}, {Bouwens}, {Bowler}, {Endsley}, {Gonzalez}, {Inami}, {Labbe},
  {Stark}, {Aravena}, {Barrufet}, {da Cunha}, {Dayal}, {Ferrara}, {Graziani},
  {Hodge}, {Hutter}, {Li}, {De Looze}, {Nanayakkara}, {Pallottini}, {Riechers},
  {Schneider}, {Ucci}, {van der Werf}, \& {White}}]{FY21}
{Fudamoto}, Y., {Oesch}, P.~A., {Schouws}, S., {et~al.} 2021, \nat, 597, 489,
  \dodoi{10.1038/s41586-021-03846-z}

\bibitem[{{Fudamoto} {et~al.}(2024{\natexlab{a}}){Fudamoto}, {Oesch}, {Walter},
  {Decarli}, {Carilli}, {Ferrara}, {Barrufet}, {Bouwens}, {Dessauges-Zavadsky},
  {Nelson}, {Dannerbauer}, {Illingworth}, {Inoue}, {Marques-Chaves},
  {P{\'e}rez-Fournon}, {Riechers}, {Schaerer}, {Smit}, {Sugahara}, \& {van der
  Werf}}]{FY24b}
{Fudamoto}, Y., {Oesch}, P.~A., {Walter}, F., {et~al.} 2024{\natexlab{a}},
  \mnras, 530, 340, \dodoi{10.1093/mnras/stae556}

\bibitem[{{Fudamoto} {et~al.}(2024{\natexlab{b}}){Fudamoto}, {Inoue}, {Coe},
  {Welch}, {Acebron}, {Ricotti}, {Mandelker}, {Windhorst}, {Xu}, {Sugahara},
  {Bauer}, {Brada{\v{c}}}, {Bradley}, {Diego}, {Florian}, {Frye}, {Fujimoto},
  {Hashimoto}, {Henry}, {Mahler}, {Oesch}, {Ravindranath}, {Rigby}, {Sharon},
  {Strait}, {Tamura}, {Trenti}, {Vanzella}, {Zackrisson}, \& {Zitrin}}]{FY24a}
{Fudamoto}, Y., {Inoue}, A.~K., {Coe}, D., {et~al.} 2024{\natexlab{b}}, \apj,
  961, 71, \dodoi{10.3847/1538-4357/ad0f95}

\bibitem[{{Fudamoto} {et~al.}(2025){Fudamoto}, {Inoue}, {Bouwens}, {Inami},
  {Smit}, {Stark}, {Aravena}, {Pallottini}, {Hashimoto}, {Oguri}, {Bowler}, {da
  Cunha}, {Dayal}, {Ferrara}, {Fujimoto}, {Heintz}, {Hygate}, {van Leeuwen},
  {De Looze}, {Rowland}, {Stefanon}, {Sugahara}, {Witstok}, \& {van der
  Werf}}]{FY25}
{Fudamoto}, Y., {Inoue}, A.~K., {Bouwens}, R., {et~al.} 2025, arXiv e-prints,
  arXiv:2504.03831, \dodoi{10.48550/arXiv.2504.03831}

\bibitem[{{Fujimoto} {et~al.}(2019){Fujimoto}, {Ouchi}, {Ferrara},
  {Pallottini}, {Ivison}, {Behrens}, {Gallerani}, {Arata}, {Yajima}, \&
  {Nagamine}}]{FS19}
{Fujimoto}, S., {Ouchi}, M., {Ferrara}, A., {et~al.} 2019, \apj, 887, 107,
  \dodoi{10.3847/1538-4357/ab480f}

\bibitem[{{Fujimoto} {et~al.}(2021){Fujimoto}, {Oguri}, {Brammer}, {Yoshimura},
  {Laporte}, {Gonz{\'a}lez-L{\'o}pez}, {Caminha}, {Kohno}, {Zitrin}, {Richard},
  {Ouchi}, {Bauer}, {Smail}, {Hatsukade}, {Ono}, {Kokorev}, {Umehata},
  {Schaerer}, {Knudsen}, {Sun}, {Magdis}, {Valentino}, {Ao}, {Toft},
  {Dessauges-Zavadsky}, {Shimasaku}, {Caputi}, {Kusakabe}, {Morokuma-Matsui},
  {Shotaro}, {Egami}, {Lee}, {Rawle}, \& {Espada}}]{FS21}
{Fujimoto}, S., {Oguri}, M., {Brammer}, G., {et~al.} 2021, \apj, 911, 99,
  \dodoi{10.3847/1538-4357/abd7ec}

\bibitem[{{Fujimoto} {et~al.}(2022){Fujimoto}, {Brammer}, {Watson}, {Magdis},
  {Kokorev}, {Greve}, {Toft}, {Walter}, {Valiante}, {Ginolfi}, {Schneider},
  {Valentino}, {Colina}, {Vestergaard}, {Marques-Chaves}, {Fynbo}, {Krips},
  {Steinhardt}, {Cortzen}, {Rizzo}, \& {Oesch}}]{FS22}
{Fujimoto}, S., {Brammer}, G.~B., {Watson}, D., {et~al.} 2022, \nat, 604, 261,
  \dodoi{10.1038/s41586-022-04454-1}

\bibitem[{{Fujimoto} {et~al.}(2024{\natexlab{a}}){Fujimoto}, {Ouchi},
  {Nakajima}, {Harikane}, {Isobe}, {Brammer}, {Oguri}, {Gim{\'e}nez-Arteaga},
  {Heintz}, {Kokorev}, {Bauer}, {Ferrara}, {Kojima}, {Lagos}, {Laura},
  {Schaerer}, {Shimasaku}, {Hatsukade}, {Kohno}, {Sun}, {Valentino}, {Watson},
  {Fudamoto}, {Inoue}, {Gonz{\'a}lez-L{\'o}pez}, {Koekemoer}, {Knudsen}, {Lee},
  {Magdis}, {Richard}, {Strait}, {Sugahara}, {Tamura}, {Toft}, {Umehata}, \&
  {Walth}}]{FS24a}
{Fujimoto}, S., {Ouchi}, M., {Nakajima}, K., {et~al.} 2024{\natexlab{a}}, \apj,
  964, 146, \dodoi{10.3847/1538-4357/ad235c}

\bibitem[{{Fujimoto} {et~al.}(2024{\natexlab{b}}){Fujimoto}, {Ouchi}, {Kohno},
  {Valentino}, {Gim\textbackslash'enez-Arteaga}, {Brammer}, {Furtak},
  {Kohandel}, {Oguri}, {Pallottini}, {Richard}, {Zitrin}, {Bauer},
  {Boylan-Kolchin}, {Dessauges-Zavadsky}, {Egami}, {Finkelstein}, {Ma},
  {Smail}, {Watson}, {Hutchison}, {Rigby}, {Welch}, {Ao}, {Bradley}, {Caminha},
  {Caputi}, {Espada}, {Endsley}, {Fudamoto},
  {Gonz\textbackslash'alez-L\textbackslash'opez}, {Hatsukade}, {Koekemoer},
  {Kokorev}, {Laporte}, {Lee}, {Magdis}, {Ono}, {Rizzo}, {Shibuya},
  {Shimasaku}, {Sun}, {Toft}, {Umehata}, {Wang}, \& {Yajima}}]{FS24b}
{Fujimoto}, S., {Ouchi}, M., {Kohno}, K., {et~al.} 2024{\natexlab{b}}, arXiv
  e-prints, arXiv:2402.18543, \dodoi{10.48550/arXiv.2402.18543}

\bibitem[{{Fujimoto} {et~al.}(2025){Fujimoto}, {Bezanson}, {Labbe}, {Brammer},
  {Price}, {Wang}, {Weaver}, {Fudamoto}, {Oesch}, {Williams}, {Dayal},
  {Feldmann}, {Greene}, {Leja}, {Whitaker}, {Zitrin}, {Cutler}, {Furtak},
  {Pan}, {Chemerynska}, {Kokorev}, {Miller}, {Atek}, {van Dokkum}, {Juneau},
  {Kassin}, {Khullar}, {Marchesini}, {Maseda}, {Nelson}, {Setton}, \&
  {Smit}}]{FS25}
{Fujimoto}, S., {Bezanson}, R., {Labbe}, I., {et~al.} 2025, \apjs, 278, 45,
  \dodoi{10.3847/1538-4365/adc677}

\bibitem[{{Gallerani} {et~al.}(2012){Gallerani}, {Neri}, {Maiolino},
  {Mart{\'\i}n}, {De Breuck}, {Walter}, {Caselli}, {Krips}, {Meneghetti},
  {Nagao}, {Wagg}, \& {Walmsley}}]{GS12}
{Gallerani}, S., {Neri}, R., {Maiolino}, R., {et~al.} 2012, \aap, 543, A114,
  \dodoi{10.1051/0004-6361/201118705}

\bibitem[{{Geach} {et~al.}(2016){Geach}, {Narayanan}, {Matsuda}, {Hayes},
  {Mas-Ribas}, {Dijkstra}, {Steidel}, {Chapman}, {Feldmann}, {Avison},
  {Agertz}, {Ao}, {Birkinshaw}, {Bremer}, {Clements}, {Dannerbauer}, {Farrah},
  {Harrison}, {Kubo}, {Micha{\l}owski}, {Scott}, {Smith}, {Spaans}, {Simpson},
  {Swinbank}, {Taniguchi}, {van der Werf}, {Verma}, \& {Yamada}}]{GJ16}
{Geach}, J.~E., {Narayanan}, D., {Matsuda}, Y., {et~al.} 2016, \apj, 832, 37,
  \dodoi{10.3847/0004-637X/832/1/37}

\bibitem[{{George} {et~al.}(2013){George}, {Ivison}, {Hopwood}, {Riechers},
  {Bussmann}, {Cox}, {Dye}, {Krips}, {Negrello}, {Neri}, {Serjeant},
  {Valtchanov}, {Baes}, {Bourne}, {Clements}, {de Zotti}, {Dunne}, {Eales},
  {Ibar}, {Maddox}, {Smith}, {Valiante}, \& {van der Werf}}]{GR13}
{George}, R.~D., {Ivison}, R.~J., {Hopwood}, R., {et~al.} 2013, \mnras, 436,
  L99, \dodoi{10.1093/mnrasl/slt122}

\bibitem[{{Gerin} {et~al.}(2015){Gerin}, {Ruaud}, {Goicoechea}, {Gusdorf},
  {Godard}, {de Luca}, {Falgarone}, {Goldsmith}, {Lis}, {Menten}, {Neufeld},
  {Phillips}, \& {Liszt}}]{gerin15}
{Gerin}, M., {Ruaud}, M., {Goicoechea}, J.~R., {et~al.} 2015, \aap, 573, A30,
  \dodoi{10.1051/0004-6361/201424349}

\bibitem[{{Gilbert} \& {Graham}(2007)}]{gilbert07}
{Gilbert}, A.~M., \& {Graham}, J.~R. 2007, \apj, 668, 168,
  \dodoi{10.1086/520910}

\bibitem[{{Gim{\'e}nez-Arteaga} {et~al.}(2022){Gim{\'e}nez-Arteaga}, {Brammer},
  {Marchesini}, {Colina}, {Bajaj}, {Brinch}, {Calzetti}, {Lange-Vagle},
  {Murphy}, {Perna}, {Piqueras-L{\'o}pez}, \& {Snyder}}]{gimenez22}
{Gim{\'e}nez-Arteaga}, C., {Brammer}, G.~B., {Marchesini}, D., {et~al.} 2022,
  \apjs, 263, 17, \dodoi{10.3847/1538-4365/ac958c}

\bibitem[{{Gim{\'e}nez-Arteaga} {et~al.}(2024){Gim{\'e}nez-Arteaga},
  {Fujimoto}, {Valentino}, {Brammer}, {Mason}, {Rizzo}, {Rusakov}, {Colina},
  {Prieto-Lyon}, {Oesch}, {Espada}, {Heintz}, {Knudsen}, {Dessauges-Zavadsky},
  {Laporte}, {Lee}, {Magdis}, {Ono}, {Ao}, {Ouchi}, {Kohno}, \&
  {Koekemoer}}]{GC24}
{Gim{\'e}nez-Arteaga}, C., {Fujimoto}, S., {Valentino}, F., {et~al.} 2024,
  \aap, 686, A63, \dodoi{10.1051/0004-6361/202349135}

\bibitem[{{Ginolfi} {et~al.}(2020){Ginolfi}, {Jones}, {B{\'e}thermin},
  {Faisst}, {Lemaux}, {Schaerer}, {Fudamoto}, {Oesch}, {Dessauges-Zavadsky},
  {Fujimoto}, {Carniani}, {Le F{\`e}vre}, {Cassata}, {Silverman}, {Capak},
  {Yan}, {Bardelli}, {Cucciati}, {Gal}, {Gruppioni}, {Hathi}, {Lubin},
  {Maiolino}, {Morselli}, {Pelliccia}, {Talia}, {Vergani}, \&
  {Zamorani}}]{GM20}
{Ginolfi}, M., {Jones}, G.~C., {B{\'e}thermin}, M., {et~al.} 2020, \aap, 643,
  A7, \dodoi{10.1051/0004-6361/202038284}

\bibitem[{{Giulietti} {et~al.}(2023){Giulietti}, {Lapi}, {Massardi}, {Behiri},
  {Torsello}, {D'Amato}, {Ronconi}, {Perrotta}, \& {Bressan}}]{GM23}
{Giulietti}, M., {Lapi}, A., {Massardi}, M., {et~al.} 2023, \apj, 943, 151,
  \dodoi{10.3847/1538-4357/aca53f}

\bibitem[{{Glazer} {et~al.}(2024){Glazer}, {Brad{\u{a}}c}, {Sanders},
  {Fujimoto}, {Bolan}, {Ferrara}, {Strait}, {Jones}, {Lemaux}, {Vallini}, \&
  {Ryan}}]{GK24}
{Glazer}, K., {Brad{\u{a}}c}, M., {Sanders}, R.~L., {et~al.} 2024, \mnras, 531,
  945, \dodoi{10.1093/mnras/stae1178}

\bibitem[{{G{\'o}mez-Guijarro} {et~al.}(2018){G{\'o}mez-Guijarro}, {Toft},
  {Karim}, {Magnelli}, {Magdis}, {Jim{\'e}nez-Andrade}, {Capak}, {Fraternali},
  {Fujimoto}, {Riechers}, {Schinnerer}, {Smol{\v{c}}i{\'c}}, {Aravena},
  {Bertoldi}, {Cortzen}, {Hasinger}, {Hu}, {Jones}, {Koekemoer}, {Lee},
  {McCracken}, {Micha{\l}owski}, {Navarrete}, {Povi{\'c}}, {Puglisi},
  {Romano-D{\'\i}az}, {Sheth}, {Silverman}, {Staguhn}, {Steinhardt},
  {Stockmann}, {Tanaka}, {Valentino}, {van Kampen}, \& {Zirm}}]{GC18}
{G{\'o}mez-Guijarro}, C., {Toft}, S., {Karim}, A., {et~al.} 2018, \apj, 856,
  121, \dodoi{10.3847/1538-4357/aab206}

\bibitem[{{Gullberg} {et~al.}(2015){Gullberg}, {De Breuck}, {Vieira},
  {Wei{\ss}}, {Aguirre}, {Aravena}, {B{\'e}thermin}, {Bradford}, {Bothwell},
  {Carlstrom}, {Chapman}, {Fassnacht}, {Gonzalez}, {Greve}, {Hezaveh},
  {Holzapfel}, {Husband}, {Ma}, {Malkan}, {Marrone}, {Menten}, {Murphy},
  {Reichardt}, {Spilker}, {Stark}, {Strandet}, \& {Welikala}}]{GB15}
{Gullberg}, B., {De Breuck}, C., {Vieira}, J.~D., {et~al.} 2015, \mnras, 449,
  2883, \dodoi{10.1093/mnras/stv372}

\bibitem[{{Gullberg} {et~al.}(2018){Gullberg}, {Swinbank}, {Smail}, {Biggs},
  {Bertoldi}, {De Breuck}, {Chapman}, {Chen}, {Cooke}, {Coppin}, {Cox},
  {Dannerbauer}, {Dunlop}, {Edge}, {Farrah}, {Geach}, {Greve}, {Hodge}, {Ibar},
  {Ivison}, {Karim}, {Schinnerer}, {Scott}, {Simpson}, {Stach}, {Thomson}, {van
  der Werf}, {Walter}, {Wardlow}, \& {Weiss}}]{GB18}
{Gullberg}, B., {Swinbank}, A.~M., {Smail}, I., {et~al.} 2018, \apj, 859, 12,
  \dodoi{10.3847/1538-4357/aabe8c}

\bibitem[{{Gururajan} {et~al.}(2023){Gururajan}, {Bethermin}, {Sulzenauer},
  {Theul{\'e}}, {Spilker}, {Aravena}, {Chapman}, {Gonzalez}, {Greve},
  {Narayanan}, {Reuter}, {Vieira}, \& {Weiss}}]{GG23}
{Gururajan}, G., {Bethermin}, M., {Sulzenauer}, N., {et~al.} 2023, \aap, 676,
  A89, \dodoi{10.1051/0004-6361/202346449}

\bibitem[{{Hailey-Dunsheath} {et~al.}(2010){Hailey-Dunsheath}, {Nikola},
  {Stacey}, {Oberst}, {Parshley}, {Benford}, {Staguhn}, \& {Tucker}}]{HS10}
{Hailey-Dunsheath}, S., {Nikola}, T., {Stacey}, G.~J., {et~al.} 2010, \apjl,
  714, L162, \dodoi{10.1088/2041-8205/714/1/L162}

\bibitem[{{Hainline} {et~al.}(2009){Hainline}, {Shapley}, {Kornei}, {Pettini},
  {Buckley-Geer}, {Allam}, \& {Tucker}}]{HK09}
{Hainline}, K.~N., {Shapley}, A.~E., {Kornei}, K.~A., {et~al.} 2009, \apj, 701,
  52, \dodoi{10.1088/0004-637X/701/1/52}

\bibitem[{{Harikane} {et~al.}(2020){Harikane}, {Ouchi}, {Inoue}, {Matsuoka},
  {Tamura}, {Bakx}, {Fujimoto}, {Moriwaki}, {Ono}, {Nagao}, {Tadaki}, {Kojima},
  {Shibuya}, {Egami}, {Ferrara}, {Gallerani}, {Hashimoto}, {Kohno}, {Matsuda},
  {Matsuo}, {Pallottini}, {Sugahara}, \& {Vallini}}]{HY20}
{Harikane}, Y., {Ouchi}, M., {Inoue}, A.~K., {et~al.} 2020, \apj, 896, 93,
  \dodoi{10.3847/1538-4357/ab94bd}

\bibitem[{{Harikane} {et~al.}(2025){Harikane}, {Sanders}, {Ellis}, {Jones},
  {Ouchi}, {Laporte}, {Roberts-Borsani}, {Katz}, {Nakajima}, {Ono}, \&
  {Gupta}}]{HY25}
{Harikane}, Y., {Sanders}, R.~L., {Ellis}, R., {et~al.} 2025, arXiv e-prints,
  arXiv:2505.09186, \dodoi{10.48550/arXiv.2505.09186}

\bibitem[{{Harshan} {et~al.}(2024){Harshan}, {Tripodi}, {Martis},
  {Rihtar{\v{s}}i{\v{c}}}, {Brada{\v{c}}}, {Asada}, {Brammer}, {Desprez},
  {Estrada-Carpenter}, {Matharu}, {Markov}, {Muzzin}, {Mowla}, {Noirot},
  {Sarrouh}, {Sawicki}, {Strait}, \& {Willott}}]{HA24}
{Harshan}, A., {Tripodi}, R., {Martis}, N.~S., {et~al.} 2024, \apjl, 977, L36,
  \dodoi{10.3847/2041-8213/ad9741}

\bibitem[{{Hashimoto} {et~al.}(2019{\natexlab{a}}){Hashimoto}, {Inoue},
  {Tamura}, {Matsuo}, {Mawatari}, \& {Yamaguchi}}]{HT19b}
{Hashimoto}, T., {Inoue}, A.~K., {Tamura}, Y., {et~al.} 2019{\natexlab{a}},
  \pasj, 71, 109, \dodoi{10.1093/pasj/psz094}

\bibitem[{{Hashimoto} {et~al.}(2018){Hashimoto}, {Laporte}, {Mawatari},
  {Ellis}, {Inoue}, {Zackrisson}, {Roberts-Borsani}, {Zheng}, {Tamura},
  {Bauer}, {Fletcher}, {Harikane}, {Hatsukade}, {Hayatsu}, {Matsuda}, {Matsuo},
  {Okamoto}, {Ouchi}, {Pell{\'o}}, {Rydberg}, {Shimizu}, {Taniguchi},
  {Umehata}, \& {Yoshida}}]{HT18}
{Hashimoto}, T., {Laporte}, N., {Mawatari}, K., {et~al.} 2018, \nat, 557, 392,
  \dodoi{10.1038/s41586-018-0117-z}

\bibitem[{{Hashimoto} {et~al.}(2019{\natexlab{b}}){Hashimoto}, {Inoue},
  {Mawatari}, {Tamura}, {Matsuo}, {Furusawa}, {Harikane}, {Shibuya}, {Knudsen},
  {Kohno}, {Ono}, {Zackrisson}, {Okamoto}, {Kashikawa}, {Oesch}, {Ouchi},
  {Ota}, {Shimizu}, {Taniguchi}, {Umehata}, \& {Watson}}]{HT19a}
{Hashimoto}, T., {Inoue}, A.~K., {Mawatari}, K., {et~al.} 2019{\natexlab{b}},
  \pasj, 71, 71, \dodoi{10.1093/pasj/psz049}

\bibitem[{{Hashimoto} {et~al.}(2023){Hashimoto}, {{\'A}lvarez-M{\'a}rquez},
  {Fudamoto}, {Colina}, {Inoue}, {Nakazato}, {Ceverino}, {Yoshida},
  {Costantin}, {Sugahara}, {G{\'o}mez}, {Blanco-Prieto}, {Mawatari}, {Arribas},
  {Marques-Chaves}, {Pereira-Santaella}, {Bakx}, {Hagimoto}, {Hashigaya},
  {Matsuo}, {Tamura}, {Usui}, \& {Ren}}]{HT23}
{Hashimoto}, T., {{\'A}lvarez-M{\'a}rquez}, J., {Fudamoto}, Y., {et~al.} 2023,
  \apjl, 955, L2, \dodoi{10.3847/2041-8213/acf57c}

\bibitem[{{Haynes} {et~al.}(2018){Haynes}, {Giovanelli}, {Kent}, {Adams},
  {Balonek}, {Craig}, {Fertig}, {Finn}, {Giovanardi}, {Hallenbeck}, {Hess},
  {Hoffman}, {Huang}, {Jones}, {Koopmann}, {Kornreich}, {Leisman}, {Miller},
  {Moorman}, {O'Connor}, {O'Donoghue}, {Papastergis}, {Troischt}, {Stark}, \&
  {Xiao}}]{H18b}
{Haynes}, M.~P., {Giovanelli}, R., {Kent}, B.~R., {et~al.} 2018, \apj, 861, 49,
  \dodoi{10.3847/1538-4357/aac956}

\bibitem[{{Heintz} {et~al.}(2021){Heintz}, {Watson}, {Oesch}, {Narayanan}, \&
  {Madden}}]{heintz21}
{Heintz}, K.~E., {Watson}, D., {Oesch}, P.~A., {Narayanan}, D., \& {Madden},
  S.~C. 2021, \apj, 922, 147, \dodoi{10.3847/1538-4357/ac2231}

\bibitem[{{Heintz} {et~al.}(2023){Heintz}, {Gim{\'e}nez-Arteaga}, {Fujimoto},
  {Brammer}, {Espada}, {Gillman}, {Gonz{\'a}lez-L{\'o}pez}, {Greve},
  {Harikane}, {Hatsukade}, {Knudsen}, {Koekemoer}, {Kohno}, {Kokorev}, {Lee},
  {Magdis}, {Nelson}, {Rizzo}, {Sanders}, {Schaerer}, {Shapley}, {Strait},
  {Toft}, {Valentino}, {van der Wel}, {Vijayan}, {Watson}, {Bauer},
  {Christiansen}, \& {Wilson}}]{HK23}
{Heintz}, K.~E., {Gim{\'e}nez-Arteaga}, C., {Fujimoto}, S., {et~al.} 2023,
  \apjl, 944, L30, \dodoi{10.3847/2041-8213/acb2cf}

\bibitem[{{Henry} {et~al.}(2000){Henry}, {Edmunds}, \& {K{\"o}ppen}}]{henry00}
{Henry}, R.~B.~C., {Edmunds}, M.~G., \& {K{\"o}ppen}, J. 2000, \apj, 541, 660,
  \dodoi{10.1086/309471}

\bibitem[{{Herard-Demanche} {et~al.}(2025){Herard-Demanche}, {Bouwens},
  {Oesch}, {Naidu}, {Decarli}, {Nelson}, {Brammer}, {Weibel}, {Xiao},
  {Stefanon}, {Walter}, {Matthee}, {Meyer}, {Wuyts}, {Reddy}, {Rowland}, {van
  Leeuwen}, {Haro}, {Dannerbauer}, {Shapley}, {Chisholm}, {van Dokkum},
  {Labbe}, {Illingworth}, {Schaerer}, \& {Shivaei}}]{HT25}
{Herard-Demanche}, T., {Bouwens}, R.~J., {Oesch}, P.~A., {et~al.} 2025, \mnras,
  537, 788, \dodoi{10.1093/mnras/staf030}

\bibitem[{{Herrera-Camus} {et~al.}(2015){Herrera-Camus}, {Bolatto}, {Wolfire},
  {Smith}, {Croxall}, {Kennicutt}, {Calzetti}, {Helou}, {Walter}, {Leroy},
  {Draine}, {Brandl}, {Armus}, {Sandstrom}, {Dale}, {Aniano}, {Meidt},
  {Boquien}, {Hunt}, {Galametz}, {Tabatabaei}, {Murphy}, {Appleton}, {Roussel},
  {Engelbracht}, \& {Beirao}}]{herrera15}
{Herrera-Camus}, R., {Bolatto}, A.~D., {Wolfire}, M.~G., {et~al.} 2015, \apj,
  800, 1, \dodoi{10.1088/0004-637X/800/1/1}

\bibitem[{{Herrera-Camus} {et~al.}(2016){Herrera-Camus}, {Bolatto}, {Smith},
  {Draine}, {Pellegrini}, {Wolfire}, {Croxall}, {de Looze}, {Calzetti},
  {Kennicutt}, {Crocker}, {Armus}, {van der Werf}, {Sandstrom}, {Galametz},
  {Brandl}, {Groves}, {Rigopoulou}, {Walter}, {Leroy}, {Boquien}, {Tabatabaei},
  \& {Beirao}}]{herrera16}
{Herrera-Camus}, R., {Bolatto}, A., {Smith}, J.~D., {et~al.} 2016, \apj, 826,
  175, \dodoi{10.3847/0004-637X/826/2/175}

\bibitem[{{Herrera-Camus} {et~al.}(2018{\natexlab{a}}){Herrera-Camus}, {Sturm},
  {Graci{\'a}-Carpio}, {Lutz}, {Contursi}, {Veilleux}, {Fischer},
  {Gonz{\'a}lez-Alfonso}, {Poglitsch}, {Tacconi}, {Genzel}, {Maiolino},
  {Sternberg}, {Davies}, \& {Verma}}]{herrera18b}
{Herrera-Camus}, R., {Sturm}, E., {Graci{\'a}-Carpio}, J., {et~al.}
  2018{\natexlab{a}}, \apj, 861, 95, \dodoi{10.3847/1538-4357/aac0f9}

\bibitem[{{Herrera-Camus} {et~al.}(2018{\natexlab{b}}){Herrera-Camus}, {Sturm},
  {Graci{\'a}-Carpio}, {Lutz}, {Contursi}, {Veilleux}, {Fischer},
  {Gonz{\'a}lez-Alfonso}, {Poglitsch}, {Tacconi}, {Genzel}, {Maiolino},
  {Sternberg}, {Davies}, \& {Verma}}]{H18a}
---. 2018{\natexlab{b}}, \apj, 861, 94, \dodoi{10.3847/1538-4357/aac0f6}

\bibitem[{{Herrera-Camus} {et~al.}(2021){Herrera-Camus}, {F{\"o}rster
  Schreiber}, {Genzel}, {Tacconi}, {Bolatto}, {Davies}, {Fisher}, {Lutz},
  {Naab}, {Shimizu}, {Tadaki}, \& {{\"U}bler}}]{HR21}
{Herrera-Camus}, R., {F{\"o}rster Schreiber}, N., {Genzel}, R., {et~al.} 2021,
  \aap, 649, A31, \dodoi{10.1051/0004-6361/202039704}

\bibitem[{{Herrera-Camus} {et~al.}(2022){Herrera-Camus}, {F{\"o}rster
  Schreiber}, {Price}, {{\"U}bler}, {Bolatto}, {Davies}, {Fisher}, {Genzel},
  {Lutz}, {Naab}, {Nestor}, {Shimizu}, {Sternberg}, {Tacconi}, \&
  {Tadaki}}]{HR22}
{Herrera-Camus}, R., {F{\"o}rster Schreiber}, N.~M., {Price}, S.~H., {et~al.}
  2022, \aap, 665, L8, \dodoi{10.1051/0004-6361/202142562}

\bibitem[{{Heyer} {et~al.}(2022){Heyer}, {Goldsmith}, {Simon}, {Aladro}, \&
  {Ricken}}]{heyer22}
{Heyer}, M., {Goldsmith}, P.~F., {Simon}, R., {Aladro}, R., \& {Ricken}, O.
  2022, \apj, 941, 62, \dodoi{10.3847/1538-4357/aca097}

\bibitem[{{Hollenbach} {et~al.}(1991){Hollenbach}, {Takahashi}, \&
  {Tielens}}]{hollenbach91}
{Hollenbach}, D.~J., {Takahashi}, T., \& {Tielens}, A.~G.~G.~M. 1991, \apj,
  377, 192, \dodoi{10.1086/170347}

\bibitem[{{Hollenbach} \& {Tielens}(1997)}]{hollenbach97}
{Hollenbach}, D.~J., \& {Tielens}, A.~G.~G.~M. 1997, \araa, 35, 179,
  \dodoi{10.1146/annurev.astro.35.1.179}

\bibitem[{{Howell} {et~al.}(2010){Howell}, {Armus}, {Mazzarella}, {Evans},
  {Surace}, {Sanders}, {Petric}, {Appleton}, {Bothun}, {Bridge}, {Chan},
  {Charmandaris}, {Frayer}, {Haan}, {Inami}, {Kim}, {Lord}, {Madore},
  {Melbourne}, {Schulz}, {U}, {Vavilkin}, {Veilleux}, \& {Xu}}]{H10}
{Howell}, J.~H., {Armus}, L., {Mazzarella}, J.~M., {et~al.} 2010, \apj, 715,
  572, \dodoi{10.1088/0004-637X/715/1/572}

\bibitem[{{Huang} {et~al.}(2016){Huang}, {Brada{\v{c}}}, {Lemaux}, {Ryan},
  {Hoag}, {Castellano}, {Amor{\'\i}n}, {Fontana}, {Brammer}, {Cain}, {Lubin},
  {Merlin}, {Schmidt}, {Schrabback}, {Treu}, {Gonzalez}, {von der Linden}, \&
  {Knight}}]{HK16}
{Huang}, K.-H., {Brada{\v{c}}}, M., {Lemaux}, B.~C., {et~al.} 2016, \apj, 817,
  11, \dodoi{10.3847/0004-637X/817/1/11}

\bibitem[{{Hunt} \& {Hirashita}(2009)}]{hunt09}
{Hunt}, L.~K., \& {Hirashita}, H. 2009, \aap, 507, 1327,
  \dodoi{10.1051/0004-6361/200912020}

\bibitem[{{Hygate} {et~al.}(2023){Hygate}, {Hodge}, {da Cunha}, {Rybak},
  {Schouws}, {Inami}, {Stefanon}, {Graziani}, {Schneider}, {Dayal}, {Bouwens},
  {Smit}, {Bowler}, {Endsley}, {Gonzalez}, {Oesch}, {Stark}, {Algera},
  {Aravena}, {Barrufet}, {Ferrara}, {Fudamoto}, {Hilhorst}, {De Looze},
  {Nanayakkara}, {Pallottini}, {Riechers}, {Sommovigo}, {Topping}, \& {van der
  Werf}}]{HA23}
{Hygate}, A.~P.~S., {Hodge}, J.~A., {da Cunha}, E., {et~al.} 2023, \mnras, 524,
  1775, \dodoi{10.1093/mnras/stad1212}

\bibitem[{{Iani} {et~al.}(2021){Iani}, {Zanella}, {Vernet}, {Richard},
  {Gronke}, {Harrison}, {Arrigoni-Battaia}, {Rodighiero}, {Burkert},
  {Behrendt}, {Chen}, {Emsellem}, {Fensch}, {Hibon}, {Hilker}, {Le Floc'h},
  {Mainieri}, {Swinbank}, {Valentino}, {Vanzella}, \& {Zwaan}}]{IE21}
{Iani}, E., {Zanella}, A., {Vernet}, J., {et~al.} 2021, \mnras, 507, 3830,
  \dodoi{10.1093/mnras/stab2376}

\bibitem[{{Ikeda} {et~al.}(2025){Ikeda}, {Tadaki}, {Mitsuhashi}, {Aravena}, {De
  Looze}, {F{\"o}rster Schreiber}, {Gonz{\'a}lez-L{\'o}pez}, {Herrera-Camus},
  {Spilker}, {Barcos-Mu{\~n}oz}, {Bowler}, {Calistro Rivera}, {da Cunha},
  {Davies}, {D{\'\i}az-Santos}, {Ferrara}, {Killi}, {Lee}, {Li}, {Lutz},
  {Posses}, {Smit}, {Solimano}, {Telikova}, {{\"U}bler}, {Veilleux}, \&
  {Villanueva}}]{IR25}
{Ikeda}, R., {Tadaki}, K.-i., {Mitsuhashi}, I., {et~al.} 2025, \aap, 693, A237,
  \dodoi{10.1051/0004-6361/202451811}

\bibitem[{{Inami} {et~al.}(2022){Inami}, {Algera}, {Schouws}, {Sommovigo},
  {Bouwens}, {Smit}, {Stefanon}, {Bowler}, {Endsley}, {Ferrara}, {Oesch},
  {Stark}, {Aravena}, {Barrufet}, {da Cunha}, {Dayal}, {De Looze}, {Fudamoto},
  {Gonzalez}, {Graziani}, {Hodge}, {Hygate}, {Nanayakkara}, {Pallottini},
  {Riechers}, {Schneider}, {Topping}, \& {van der Werf}}]{IH22}
{Inami}, H., {Algera}, H. S.~B., {Schouws}, S., {et~al.} 2022, \mnras, 515,
  3126, \dodoi{10.1093/mnras/stac1779}

\bibitem[{{Inoue}(2002)}]{inoue02}
{Inoue}, A.~K. 2002, \apj, 570, 688, \dodoi{10.1086/339788}

\bibitem[{{Inoue} {et~al.}(2020){Inoue}, {Hashimoto}, {Chihara}, \&
  {Koike}}]{IA20}
{Inoue}, A.~K., {Hashimoto}, T., {Chihara}, H., \& {Koike}, C. 2020, \mnras,
  495, 1577, \dodoi{10.1093/mnras/staa1203}

\bibitem[{{Inoue} {et~al.}(2016){Inoue}, {Tamura}, {Matsuo}, {Mawatari},
  {Shimizu}, {Shibuya}, {Ota}, {Yoshida}, {Zackrisson}, {Kashikawa}, {Kohno},
  {Umehata}, {Hatsukade}, {Iye}, {Matsuda}, {Okamoto}, \& {Yamaguchi}}]{IA16}
{Inoue}, A.~K., {Tamura}, Y., {Matsuo}, H., {et~al.} 2016, Science, 352, 1559,
  \dodoi{10.1126/science.aaf0714}

\bibitem[{{Iono} {et~al.}(2006){Iono}, {Yun}, {Elvis}, {Peck}, {Ho}, {Wilner},
  {Hunter}, {Matsushita}, \& {Muller}}]{ID06}
{Iono}, D., {Yun}, M.~S., {Elvis}, M., {et~al.} 2006, \apjl, 645, L97,
  \dodoi{10.1086/506344}

\bibitem[{{Iono} {et~al.}(2016){Iono}, {Yun}, {Aretxaga}, {Hatsukade},
  {Hughes}, {Ikarashi}, {Izumi}, {Kawabe}, {Kohno}, {Lee}, {Matsuda},
  {Nakanishi}, {Saito}, {Tamura}, {Ueda}, {Umehata}, {Wilson}, {Michiyama}, \&
  {Ando}}]{ID16}
{Iono}, D., {Yun}, M.~S., {Aretxaga}, I., {et~al.} 2016, \apjl, 829, L10,
  \dodoi{10.3847/2041-8205/829/1/L10}

\bibitem[{{Ishii} {et~al.}(2025){Ishii}, {Hashimoto}, {Ferkinhoff}, {Rybak},
  {Inoue}, {Michiyama}, {Donevski}, {Fujimoto}, {Salak}, {Kuno}, {Matsuo},
  {Mawatari}, {Tamura}, {Izumi}, {Nagao}, {Nakazato}, {Osone}, {Sugahara},
  {Usui}, {Wakasugi}, {Yajima}, {Bakx}, {Fudamoto}, {Meyer}, {Walter}, \&
  {Yoshida}}]{IN25}
{Ishii}, N., {Hashimoto}, T., {Ferkinhoff}, C., {et~al.} 2025, \pasj, 77, 139,
  \dodoi{10.1093/pasj/psae105}

\bibitem[{{Ishiki} {et~al.}(2018){Ishiki}, {Okamoto}, \& {Inoue}}]{ishiki18}
{Ishiki}, S., {Okamoto}, T., \& {Inoue}, A.~K. 2018, \mnras, 474, 1935,
  \dodoi{10.1093/mnras/stx2833}

\bibitem[{{Israel} {et~al.}(2015){Israel}, {Rosenberg}, \& {van der
  Werf}}]{israel15}
{Israel}, F.~P., {Rosenberg}, M.~J.~F., \& {van der Werf}, P. 2015, \aap, 578,
  A95, \dodoi{10.1051/0004-6361/201425175}

\bibitem[{{Ivison} {et~al.}(2010){Ivison}, {Swinbank}, {Swinyard}, {Smail},
  {Pearson}, {Rigopoulou}, {Polehampton}, {Baluteau}, {Barlow}, {Blain},
  {Bock}, {Clements}, {Coppin}, {Cooray}, {Danielson}, {Dwek}, {Edge},
  {Franceschini}, {Fulton}, {Glenn}, {Griffin}, {Isaak}, {Leeks}, {Lim},
  {Naylor}, {Oliver}, {Page}, {P{\'e}rez Fournon}, {Rowan-Robinson}, {Savini},
  {Scott}, {Spencer}, {Valtchanov}, {Vigroux}, \& {Wright}}]{IR10a}
{Ivison}, R.~J., {Swinbank}, A.~M., {Swinyard}, B., {et~al.} 2010, \aap, 518,
  L35, \dodoi{10.1051/0004-6361/201014548}

\bibitem[{{Ivison} {et~al.}(2013){Ivison}, {Swinbank}, {Smail}, {Harris},
  {Bussmann}, {Cooray}, {Cox}, {Fu}, {Kov{\'a}cs}, {Krips}, {Narayanan},
  {Negrello}, {Neri}, {Pe{\~n}arrubia}, {Richard}, {Riechers}, {Rowlands},
  {Staguhn}, {Targett}, {Amber}, {Baker}, {Bourne}, {Bertoldi}, {Bremer},
  {Calanog}, {Clements}, {Dannerbauer}, {Dariush}, {De Zotti}, {Dunne},
  {Eales}, {Farrah}, {Fleuren}, {Franceschini}, {Geach}, {George}, {Helly},
  {Hopwood}, {Ibar}, {Jarvis}, {Kneib}, {Maddox}, {Omont}, {Scott}, {Serjeant},
  {Smith}, {Thompson}, {Valiante}, {Valtchanov}, {Vieira}, \& {van der
  Werf}}]{IR13}
{Ivison}, R.~J., {Swinbank}, A.~M., {Smail}, I., {et~al.} 2013, \apj, 772, 137,
  \dodoi{10.1088/0004-637X/772/2/137}

\bibitem[{{Ivison} {et~al.}(2016){Ivison}, {Lewis}, {Weiss}, {Arumugam},
  {Simpson}, {Holland}, {Maddox}, {Dunne}, {Valiante}, {van der Werf}, {Omont},
  {Dannerbauer}, {Smail}, {Bertoldi}, {Bremer}, {Bussmann}, {Cai}, {Clements},
  {Cooray}, {De Zotti}, {Eales}, {Fuller}, {Gonzalez-Nuevo}, {Ibar},
  {Negrello}, {Oteo}, {P{\'e}rez-Fournon}, {Riechers}, {Stevens}, {Swinbank},
  \& {Wardlow}}]{IR16}
{Ivison}, R.~J., {Lewis}, A.~J.~R., {Weiss}, A., {et~al.} 2016, \apj, 832, 78,
  \dodoi{10.3847/0004-637X/832/1/78}

\bibitem[{{Izumi} {et~al.}(2018){Izumi}, {Onoue}, {Shirakata}, {Nagao},
  {Kohno}, {Matsuoka}, {Imanishi}, {Strauss}, {Kashikawa}, {Schulze},
  {Silverman}, {Fujimoto}, {Harikane}, {Toba}, {Umehata}, {Nakanishi},
  {Greene}, {Tamura}, {Taniguchi}, {Yamaguchi}, {Goto}, {Hashimoto},
  {Ikarashi}, {Iono}, {Iwasawa}, {Lee}, {Makiya}, {Minezaki}, \& {Tang}}]{IT18}
{Izumi}, T., {Onoue}, M., {Shirakata}, H., {et~al.} 2018, \pasj, 70, 36,
  \dodoi{10.1093/pasj/psy026}

\bibitem[{{Izumi} {et~al.}(2019){Izumi}, {Onoue}, {Matsuoka}, {Nagao},
  {Strauss}, {Imanishi}, {Kashikawa}, {Fujimoto}, {Kohno}, {Toba}, {Umehata},
  {Goto}, {Ueda}, {Shirakata}, {Silverman}, {Greene}, {Harikane}, {Hashimoto},
  {Ikarashi}, {Iono}, {Iwasawa}, {Lee}, {Minezaki}, {Nakanishi}, {Tamura},
  {Tang}, \& {Taniguchi}}]{IT19}
{Izumi}, T., {Onoue}, M., {Matsuoka}, Y., {et~al.} 2019, \pasj, 71, 111,
  \dodoi{10.1093/pasj/psz096}

\bibitem[{{Izumi} {et~al.}(2021{\natexlab{a}}){Izumi}, {Onoue}, {Matsuoka},
  {Strauss}, {Fujimoto}, {Umehata}, {Imanishi}, {Kawamuro}, {Nagao}, {Toba},
  {Kohno}, {Kashikawa}, {Inayoshi}, {Kawaguchi}, {Iwasawa}, {Inoue}, {Goto},
  {Baba}, {Schramm}, {Suh}, {Harikane}, {Ueda}, {Silverman}, {Hashimoto},
  {Hashimoto}, {Ikarashi}, {Iono}, {Lee}, {Lee}, {Minezaki}, {Nakanishi},
  {Nakano}, {Tamura}, \& {Tang}}]{IT21a}
---. 2021{\natexlab{a}}, \apj, 908, 235, \dodoi{10.3847/1538-4357/abd7ef}

\bibitem[{{Izumi} {et~al.}(2021{\natexlab{b}}){Izumi}, {Matsuoka}, {Fujimoto},
  {Onoue}, {Strauss}, {Umehata}, {Imanishi}, {Kohno}, {Kawaguchi}, {Kawamuro},
  {Baba}, {Nagao}, {Toba}, {Inayoshi}, {Silverman}, {Inoue}, {Ikarashi},
  {Iwasawa}, {Kashikawa}, {Hashimoto}, {Nakanishi}, {Ueda}, {Schramm}, {Lee},
  \& {Suh}}]{IT21b}
{Izumi}, T., {Matsuoka}, Y., {Fujimoto}, S., {et~al.} 2021{\natexlab{b}}, \apj,
  914, 36, \dodoi{10.3847/1538-4357/abf6dc}

\bibitem[{{Ji} {et~al.}(2024){Ji}, {{\"U}bler}, {Maiolino}, {D'Eugenio},
  {Arribas}, {Bunker}, {Charlot}, {Perna}, {Rodr{\'\i}guez Del Pino},
  {B{\"o}ker}, {Cresci}, {Curti}, {Kumari}, \& {Lamperti}}]{JX24}
{Ji}, X., {{\"U}bler}, H., {Maiolino}, R., {et~al.} 2024, \mnras, 535, 881,
  \dodoi{10.1093/mnras/stae2375}

\bibitem[{{Jim{\'e}nez-Andrade} {et~al.}(2020){Jim{\'e}nez-Andrade}, {Zavala},
  {Magnelli}, {Casey}, {Liu}, {Romano-D{\'\i}az}, {Schinnerer}, {Harrington},
  {Aretxaga}, {Karim}, {Staguhn}, {Burnham}, {Monta{\~n}a},
  {Smol{\v{c}}i{\'c}}, {Yun}, {Bertoldi}, \& {Hughes}}]{JE20}
{Jim{\'e}nez-Andrade}, E.~F., {Zavala}, J.~A., {Magnelli}, B., {et~al.} 2020,
  \apj, 890, 171, \dodoi{10.3847/1538-4357/ab6dec}

\bibitem[{{Jones} {et~al.}(2017{\natexlab{a}}){Jones}, {K{\"o}hler}, {Ysard},
  {Bocchio}, \& {Verstraete}}]{jones17}
{Jones}, A.~P., {K{\"o}hler}, M., {Ysard}, N., {Bocchio}, M., \& {Verstraete},
  L. 2017{\natexlab{a}}, \aap, 602, A46, \dodoi{10.1051/0004-6361/201630225}

\bibitem[{{Jones} {et~al.}(2017{\natexlab{b}}){Jones}, {Willott}, {Carilli},
  {Ferrara}, {Wang}, \& {Wagg}}]{JG17}
{Jones}, G.~C., {Willott}, C.~J., {Carilli}, C.~L., {et~al.}
  2017{\natexlab{b}}, \apj, 845, 175, \dodoi{10.3847/1538-4357/aa7d0d}

\bibitem[{{Jones} {et~al.}(2024{\natexlab{a}}){Jones}, {{\"U}bler}, {Perna},
  {Arribas}, {Bunker}, {Carniani}, {Charlot}, {Maiolino}, {Del Pino},
  {Willott}, {Bowler}, {B{\"o}ker}, {Cameron}, {Chevallard}, {Cresci}, {Curti},
  {D'Eugenio}, {Kumari}, {Saxena}, {Scholtz}, {Venturi}, \&
  {Witstok}}]{jones24}
{Jones}, G.~C., {{\"U}bler}, H., {Perna}, M., {et~al.} 2024{\natexlab{a}},
  \aap, 682, A122, \dodoi{10.1051/0004-6361/202347838}

\bibitem[{{Jones} {et~al.}(2024{\natexlab{b}}){Jones}, {Bowler}, {Bunker},
  {Arribas}, {Carniani}, {Charlot}, {Perna}, {Rodr{\'\i}guez Del Pino},
  {{\"U}bler}, {Willott}, {Chevallard}, {Cresci}, {Parlanti}, {Scholtz}, \&
  {Venturi}}]{JG24b}
{Jones}, G.~C., {Bowler}, R., {Bunker}, A.~J., {et~al.} 2024{\natexlab{b}},
  arXiv e-prints, arXiv:2412.15027, \dodoi{10.48550/arXiv.2412.15027}

\bibitem[{{Jones} {et~al.}(2024{\natexlab{c}}){Jones}, {{\"U}bler}, {Perna},
  {Arribas}, {Bunker}, {Carniani}, {Charlot}, {Maiolino}, {Del Pino},
  {Willott}, {Bowler}, {B{\"o}ker}, {Cameron}, {Chevallard}, {Cresci}, {Curti},
  {D'Eugenio}, {Kumari}, {Saxena}, {Scholtz}, {Venturi}, \& {Witstok}}]{JG24a}
{Jones}, G.~C., {{\"U}bler}, H., {Perna}, M., {et~al.} 2024{\natexlab{c}},
  \aap, 682, A122, \dodoi{10.1051/0004-6361/202347838}

\bibitem[{{Jones} {et~al.}(2025){Jones}, {Bunker}, {Telikova}, {Arribas},
  {Carniani}, {Charlot}, {D'Eugenio}, {Maiolino}, {Perna}, {Rodr{\'\i}guez Del
  Pino}, {{\"U}bler}, {Willott}, {Aravena}, {B{\"o}ker}, {Cresci}, {Curti},
  {Gonz{\'a}lez-L{\'o}pez}, {Herrera-Camus}, {Lamperti}, {Parlanti},
  {P{\'e}rez-Gonz{\'a}lez}, \& {Villanueva}}]{JG25}
{Jones}, G.~C., {Bunker}, A.~J., {Telikova}, K., {et~al.} 2025, \mnras, 540,
  3311, \dodoi{10.1093/mnras/staf899}

\bibitem[{{Kade} {et~al.}(2023){Kade}, {Knudsen}, {Vlemmings}, {Stanley},
  {Gullberg}, \& {K{\"o}nig}}]{KK23}
{Kade}, K., {Knudsen}, K.~K., {Vlemmings}, W., {et~al.} 2023, \aap, 673, A116,
  \dodoi{10.1051/0004-6361/202141839}

\bibitem[{{Kamenetzky} {et~al.}(2016){Kamenetzky}, {Rangwala}, {Glenn},
  {Maloney}, \& {Conley}}]{K16}
{Kamenetzky}, J., {Rangwala}, N., {Glenn}, J., {Maloney}, P.~R., \& {Conley},
  A. 2016, \apj, 829, 93, \dodoi{10.3847/0004-637X/829/2/93}

\bibitem[{{Kanekar} {et~al.}(2013){Kanekar}, {Wagg}, {Chary}, \&
  {Carilli}}]{KN13}
{Kanekar}, N., {Wagg}, J., {Chary}, R.~R., \& {Carilli}, C.~L. 2013, \apjl,
  771, L20, \dodoi{10.1088/2041-8205/771/2/L20}

\bibitem[{{Karoumpis} {et~al.}(2022){Karoumpis}, {Magnelli},
  {Romano-D{\'\i}az}, {Haslbauer}, \& {Bertoldi}}]{karoumpis22}
{Karoumpis}, C., {Magnelli}, B., {Romano-D{\'\i}az}, E., {Haslbauer}, M., \&
  {Bertoldi}, F. 2022, \aap, 659, A12, \dodoi{10.1051/0004-6361/202141293}

\bibitem[{{Kashikawa} {et~al.}(2015){Kashikawa}, {Ishizaki}, {Willott},
  {Onoue}, {Im}, {Furusawa}, {Toshikawa}, {Ishikawa}, {Niino}, {Shimasaku},
  {Ouchi}, \& {Hibon}}]{KN15}
{Kashikawa}, N., {Ishizaki}, Y., {Willott}, C.~J., {et~al.} 2015, \apj, 798,
  28, \dodoi{10.1088/0004-637X/798/1/28}

\bibitem[{{Kashino} {et~al.}(2023){Kashino}, {Lilly}, {Simcoe}, {Bordoloi},
  {Mackenzie}, {Matthee}, \& {Eilers}}]{KD23}
{Kashino}, D., {Lilly}, S.~J., {Simcoe}, R.~A., {et~al.} 2023, \nat, 617, 261,
  \dodoi{10.1038/s41586-023-05901-3}

\bibitem[{{Katz} {et~al.}(2022){Katz}, {Rosdahl}, {Kimm}, {Garel}, {Blaizot},
  {Haehnelt}, {Michel-Dansac}, {Martin-Alvarez}, {Devriendt}, {Slyz},
  {Teyssier}, {Ocvirk}, {Laporte}, \& {Ellis}}]{katz22}
{Katz}, H., {Rosdahl}, J., {Kimm}, T., {et~al.} 2022, \mnras, 510, 5603,
  \dodoi{10.1093/mnras/stac028}

\bibitem[{{Kaufman} {et~al.}(1999){Kaufman}, {Wolfire}, {Hollenbach}, \&
  {Luhman}}]{kaufman99}
{Kaufman}, M.~J., {Wolfire}, M.~G., {Hollenbach}, D.~J., \& {Luhman}, M.~L.
  1999, \apj, 527, 795, \dodoi{10.1086/308102}

\bibitem[{{Kennicutt}(1984)}]{kennicutt84}
{Kennicutt}, R.~C., J. 1984, \apj, 287, 116, \dodoi{10.1086/162669}

\bibitem[{{Kennicutt} {et~al.}(2009){Kennicutt}, {Hao}, {Calzetti},
  {Moustakas}, {Dale}, {Bendo}, {Engelbracht}, {Johnson}, \&
  {Lee}}]{kennicutt09}
{Kennicutt}, Robert~C., J., {Hao}, C.-N., {Calzetti}, D., {et~al.} 2009, \apj,
  703, 1672, \dodoi{10.1088/0004-637X/703/2/1672}

\bibitem[{{Kennicutt} \& {Evans}(2012)}]{kennicutt12}
{Kennicutt}, R.~C., \& {Evans}, N.~J. 2012, \araa, 50, 531,
  \dodoi{10.1146/annurev-astro-081811-125610}

\bibitem[{{Kewley} {et~al.}(2019){Kewley}, {Nicholls}, \&
  {Sutherland}}]{kewley19}
{Kewley}, L.~J., {Nicholls}, D.~C., \& {Sutherland}, R.~S. 2019, \araa, 57,
  511, \dodoi{10.1146/annurev-astro-081817-051832}

\bibitem[{{Khusanova} {et~al.}(2022){Khusanova}, {Ba{\~n}ados}, {Mazzucchelli},
  {Rojas-Ruiz}, {Momjian}, {Walter}, {Decarli}, {Venemans}, {Farina}, {Meyer},
  {Wang}, \& {Yang}}]{KY22}
{Khusanova}, Y., {Ba{\~n}ados}, E., {Mazzucchelli}, C., {et~al.} 2022, \aap,
  664, A39, \dodoi{10.1051/0004-6361/202243660}

\bibitem[{{Killi} {et~al.}(2023){Killi}, {Watson}, {Fujimoto}, {Akins},
  {Knudsen}, {Richard}, {Harikane}, {Rigopoulou}, {Rizzo}, {Ginolfi},
  {Popping}, \& {Kokorev}}]{KM23}
{Killi}, M., {Watson}, D., {Fujimoto}, S., {et~al.} 2023, \mnras, 521, 2526,
  \dodoi{10.1093/mnras/stad687}

\bibitem[{{Kiyota} {et~al.}(2025){Kiyota}, {Ouchi}, {Xu}, {Nakazato}, {Soga},
  {Yajima}, {Fujimoto}, {Harikane}, {Nakajima}, {Ono}, {Sun}, {Kusakabe},
  {Ceverino}, {Hatsukade}, {Iono}, {Kohno}, \& {Nakanishi}}]{KT25}
{Kiyota}, T., {Ouchi}, M., {Xu}, Y., {et~al.} 2025, arXiv e-prints,
  arXiv:2504.03156, \dodoi{10.48550/arXiv.2504.03156}

\bibitem[{{Knudsen} {et~al.}(2016){Knudsen}, {Richard}, {Kneib}, {Jauzac},
  {Cl{\'e}ment}, {Drouart}, {Egami}, \& {Lindroos}}]{KK16}
{Knudsen}, K.~K., {Richard}, J., {Kneib}, J.-P., {et~al.} 2016, \mnras, 462,
  L6, \dodoi{10.1093/mnrasl/slw114}

\bibitem[{{Knudsen} {et~al.}(2017){Knudsen}, {Watson}, {Frayer}, {Christensen},
  {Gallazzi}, {Micha{\l}owski}, {Richard}, \& {Zavala}}]{KK17}
{Knudsen}, K.~K., {Watson}, D., {Frayer}, D., {et~al.} 2017, \mnras, 466, 138,
  \dodoi{10.1093/mnras/stw3066}

\bibitem[{{Kolupuri} {et~al.}(2025){Kolupuri}, {Decarli}, {Neri}, {Cox},
  {Ferkinhoff}, {Bertoldi}, {Weiss}, {Venemans}, {Riechers}, {Paolo Farina}, \&
  {Walter}}]{KS25}
{Kolupuri}, S., {Decarli}, R., {Neri}, R., {et~al.} 2025, \aap, 695, A201,
  \dodoi{10.1051/0004-6361/202452374}

\bibitem[{{K{\"o}nig} {et~al.}(2012){K{\"o}nig}, {Greve}, {Seymour},
  {Rawlings}, {Papadopoulos}, {Ivison}, {De Breuck}, {Stevens}, {Smail}, \&
  {Kovacs}}]{KS12}
{K{\"o}nig}, S., {Greve}, T.~R., {Seymour}, N., {et~al.} 2012, in Journal of
  Physics Conference Series, Vol. 372, Journal of Physics Conference Series
  (IOP), 012064, \dodoi{10.1088/1742-6596/372/1/012064}

\bibitem[{{Koribalski} {et~al.}(2004){Koribalski}, {Staveley-Smith}, {Kilborn},
  {Ryder}, {Kraan-Korteweg}, {Ryan-Weber}, {Ekers}, {Jerjen}, {Henning},
  {Putman}, {Zwaan}, {de Blok}, {Calabretta}, {Disney}, {Minchin}, {Bhathal},
  {Boyce}, {Drinkwater}, {Freeman}, {Gibson}, {Green}, {Haynes}, {Juraszek},
  {Kesteven}, {Knezek}, {Mader}, {Marquarding}, {Meyer}, {Mould}, {Oosterloo},
  {O'Brien}, {Price}, {Sadler}, {Schr{\"o}der}, {Stewart}, {Stootman}, {Waugh},
  {Warren}, {Webster}, \& {Wright}}]{K04}
{Koribalski}, B.~S., {Staveley-Smith}, L., {Kilborn}, V.~A., {et~al.} 2004,
  \aj, 128, 16, \dodoi{10.1086/421744}

\bibitem[{{Lamarche} {et~al.}(2018){Lamarche}, {Verma}, {Vishwas}, {Stacey},
  {Brisbin}, {Ferkinhoff}, {Nikola}, {Higdon}, {Higdon}, \& {Tecza}}]{LC18}
{Lamarche}, C., {Verma}, A., {Vishwas}, A., {et~al.} 2018, \apj, 867, 140,
  \dodoi{10.3847/1538-4357/aae394}

\bibitem[{{Lambert} {et~al.}(2023){Lambert}, {Posses}, {Aravena},
  {G{\'o}nzalez-L{\'o}pez}, {Assef}, {D{\'\i}az-Santos}, {Brisbin}, {Decarli},
  {Herrera-Camus}, {Mej{\'\i}a}, \& {Ricci}}]{LT23}
{Lambert}, T.~S., {Posses}, A., {Aravena}, M., {et~al.} 2023, \mnras, 518,
  3183, \dodoi{10.1093/mnras/stac3016}

\bibitem[{{Langer} {et~al.}(2015){Langer}, {Goldsmith}, {Pineda}, {Velusamy},
  {Requena-Torres}, \& {Wiesemeyer}}]{langer15b}
{Langer}, W.~D., {Goldsmith}, P.~F., {Pineda}, J.~L., {et~al.} 2015, \aap, 576,
  A1, \dodoi{10.1051/0004-6361/201425360}

\bibitem[{{Langer} \& {Pineda}(2015)}]{langer15}
{Langer}, W.~D., \& {Pineda}, J.~L. 2015, \aap, 580, A5,
  \dodoi{10.1051/0004-6361/201525950}

\bibitem[{{Langer} {et~al.}(2010){Langer}, {Velusamy}, {Pineda}, {Goldsmith},
  {Li}, \& {Yorke}}]{langer10}
{Langer}, W.~D., {Velusamy}, T., {Pineda}, J.~L., {et~al.} 2010, \aap, 521,
  L17, \dodoi{10.1051/0004-6361/201015088}

\bibitem[{{Langer} {et~al.}(2021){Langer}, {Pineda}, {Goldsmith}, {Chambers},
  {Riquelme}, {Anderson}, {Luisi}, {Justen}, \& {Buchbender}}]{langer21}
{Langer}, W.~D., {Pineda}, J.~L., {Goldsmith}, P.~F., {et~al.} 2021, \aap, 651,
  A59, \dodoi{10.1051/0004-6361/202040223}

\bibitem[{{Lapham} {et~al.}(2017){Lapham}, {Young}, \& {Crocker}}]{L17b}
{Lapham}, R.~C., {Young}, L.~M., \& {Crocker}, A. 2017, \apj, 840, 51,
  \dodoi{10.3847/1538-4357/aa6d83}

\bibitem[{{Laporte} {et~al.}(2021{\natexlab{a}}){Laporte}, {Meyer}, {Ellis},
  {Robertson}, {Chisholm}, \& {Roberts-Borsani}}]{LN21b}
{Laporte}, N., {Meyer}, R.~A., {Ellis}, R.~S., {et~al.} 2021{\natexlab{a}},
  \mnras, 505, 3336, \dodoi{10.1093/mnras/stab1239}

\bibitem[{{Laporte} {et~al.}(2019){Laporte}, {Katz}, {Ellis}, {Lagache},
  {Bauer}, {Boone}, {Inoue}, {Hashimoto}, {Matsuo}, {Mawatari}, \&
  {Tamura}}]{LN19}
{Laporte}, N., {Katz}, H., {Ellis}, R.~S., {et~al.} 2019, \mnras, 487, L81,
  \dodoi{10.1093/mnrasl/slz094}

\bibitem[{{Laporte} {et~al.}(2021{\natexlab{b}}){Laporte}, {Zitrin}, {Ellis},
  {Fujimoto}, {Brammer}, {Richard}, {Oguri}, {Caminha}, {Kohno}, {Yoshimura},
  {Ao}, {Bauer}, {Caputi}, {Egami}, {Espada}, {Gonz{\'a}lez-L{\'o}pez},
  {Hatsukade}, {Knudsen}, {Lee}, {Magdis}, {Ouchi}, {Valentino}, \&
  {Wang}}]{LN21a}
{Laporte}, N., {Zitrin}, A., {Ellis}, R.~S., {et~al.} 2021{\natexlab{b}},
  \mnras, 505, 4838, \dodoi{10.1093/mnras/stab191}

\bibitem[{{Le F{\`e}vre} {et~al.}(2020){Le F{\`e}vre}, {B{\'e}thermin},
  {Faisst}, {Jones}, {Capak}, {Cassata}, {Silverman}, {Schaerer}, {Yan},
  {Amorin}, {Bardelli}, {Boquien}, {Cimatti}, {Dessauges-Zavadsky},
  {Giavalisco}, {Hathi}, {Fudamoto}, {Fujimoto}, {Ginolfi}, {Gruppioni},
  {Hemmati}, {Ibar}, {Koekemoer}, {Khusanova}, {Lagache}, {Lemaux}, {Loiacono},
  {Maiolino}, {Mancini}, {Narayanan}, {Morselli}, {M{\'e}ndez-Hern{\`a}ndez},
  {Oesch}, {Pozzi}, {Romano}, {Riechers}, {Scoville}, {Talia}, {Tasca},
  {Thomas}, {Toft}, {Vallini}, {Vergani}, {Walter}, {Zamorani}, \&
  {Zucca}}]{LO20}
{Le F{\`e}vre}, O., {B{\'e}thermin}, M., {Faisst}, A., {et~al.} 2020, \aap,
  643, A1, \dodoi{10.1051/0004-6361/201936965}

\bibitem[{{Lee} {et~al.}(2006){Lee}, {Skillman}, {Cannon}, {Jackson}, {Gehrz},
  {Polomski}, \& {Woodward}}]{lee06}
{Lee}, H., {Skillman}, E.~D., {Cannon}, J.~M., {et~al.} 2006, \apj, 647, 970,
  \dodoi{10.1086/505573}

\bibitem[{{Lee} {et~al.}(2024){Lee}, {Akiyama}, {Kohno}, {Iono}, {Imanishi},
  {Hatsukade}, {Umehata}, {Nagao}, {Toba}, {Chen}, {Egusa}, {Ichikawa},
  {Izumi}, {Matsumoto}, {Schramm}, \& {Matsuoka}}]{LK24}
{Lee}, K., {Akiyama}, M., {Kohno}, K., {et~al.} 2024, \apj, 972, 111,
  \dodoi{10.3847/1538-4357/ad5be5}

\bibitem[{{Lee} {et~al.}(2019){Lee}, {Nagao}, {De Breuck}, {Carniani},
  {Cresci}, {Hatsukade}, {Kawabe}, {Kohno}, {Maiolino}, {Mannucci}, {Marconi},
  {Nakanishi}, {Saito}, {Tamura}, {Troncoso}, {Umehata}, \& {Yun}}]{LM19}
{Lee}, M.~M., {Nagao}, T., {De Breuck}, C., {et~al.} 2019, \apjl, 883, L29,
  \dodoi{10.3847/2041-8213/ab412e}

\bibitem[{{Lee} {et~al.}(2021){Lee}, {Nagao}, {De Breuck}, {Carniani},
  {Cresci}, {Hatsukade}, {Kawabe}, {Kohno}, {Maiolino}, {Mannucci}, {Marconi},
  {Nakanishi}, {Troncoso}, \& {Umehata}}]{LM21}
---. 2021, \apj, 913, 41, \dodoi{10.3847/1538-4357/abe7ea}

\bibitem[{{Lelli} {et~al.}(2021){Lelli}, {Di Teodoro}, {Fraternali}, {Man},
  {Zhang}, {De Breuck}, {Davis}, \& {Maiolino}}]{LF21}
{Lelli}, F., {Di Teodoro}, E.~M., {Fraternali}, F., {et~al.} 2021, Science,
  371, 713, \dodoi{10.1126/science.abc1893}

\bibitem[{{Li} \& {Draine}(2001)}]{li01}
{Li}, A., \& {Draine}, B.~T. 2001, \apj, 554, 778, \dodoi{10.1086/323147}

\bibitem[{{Li} {et~al.}(2022){Li}, {Venemans}, {Walter}, {Decarli}, {Wang}, \&
  {Cai}}]{LJ22}
{Li}, J., {Venemans}, B.~P., {Walter}, F., {et~al.} 2022, \apj, 930, 27,
  \dodoi{10.3847/1538-4357/ac61d7}

\bibitem[{{Li} {et~al.}(2020){Li}, {Wang}, {Cox}, {Gao}, {Walter}, {Wagg},
  {Menten}, {Bertoldi}, {Shao}, {Venemans}, {Decarli}, {Riechers}, {Neri},
  {Fan}, {Omont}, \& {Narayanan}}]{LJ20}
{Li}, J., {Wang}, R., {Cox}, P., {et~al.} 2020, \apj, 900, 131,
  \dodoi{10.3847/1538-4357/ababac}

\bibitem[{{Li} {et~al.}(2024){Li}, {Da Cunha}, {Gonz{\'a}lez-L{\'o}pez},
  {Aravena}, {De Looze}, {F{\"o}rster Schreiber}, {Herrera-Camus}, {Spilker},
  {Tadaki}, {Barcos-Munoz}, {Battisti}, {Birkin}, {Bowler}, {Davies},
  {D{\'\i}az-Santos}, {Ferrara}, {Fisher}, {Hodge}, {Ikeda}, {Killi}, {Lee},
  {Liu}, {Lutz}, {Mitsuhashi}, {Naab}, {Posses}, {Rela{\~n}o}, {Solimano},
  {{\"U}bler}, {van der Giessen}, \& {Villanueva}}]{LJ24}
{Li}, J., {Da Cunha}, E., {Gonz{\'a}lez-L{\'o}pez}, J., {et~al.} 2024, \apj,
  976, 70, \dodoi{10.3847/1538-4357/ad7fee}

\bibitem[{{Lin} {et~al.}(2009){Lin}, {Buckley-Geer}, {Allam}, {Tucker},
  {Diehl}, {Kubik}, {Kubo}, {Annis}, {Frieman}, {Oguri}, \& {Inada}}]{LH09}
{Lin}, H., {Buckley-Geer}, E., {Allam}, S.~S., {et~al.} 2009, \apj, 699, 1242,
  \dodoi{10.1088/0004-637X/699/2/1242}

\bibitem[{{Litke} {et~al.}(2019){Litke}, {Marrone}, {Spilker}, {Aravena},
  {B{\'e}thermin}, {Chapman}, {Chen}, {de Breuck}, {Dong}, {Gonzalez}, {Greve},
  {Hayward}, {Hezaveh}, {Jarugula}, {Ma}, {Morningstar}, {Narayanan}, {Phadke},
  {Reuter}, {Vieira}, \& {Weiss}}]{LK19}
{Litke}, K.~C., {Marrone}, D.~P., {Spilker}, J.~S., {et~al.} 2019, \apj, 870,
  80, \dodoi{10.3847/1538-4357/aaf057}

\bibitem[{{Litke} {et~al.}(2022){Litke}, {Marrone}, {Aravena}, {B{\'e}thermin},
  {Chapman}, {Dong}, {Hayward}, {Hill}, {Jarugula}, {Malkan}, {Narayanan},
  {Reuter}, {Spilker}, {Sulzenauer}, {Vieira}, \& {Wei{\ss}}}]{LK22}
{Litke}, K.~C., {Marrone}, D.~P., {Aravena}, M., {et~al.} 2022, \apj, 928, 179,
  \dodoi{10.3847/1538-4357/ac58f9}

\bibitem[{{Litke} {et~al.}(2023){Litke}, {Marrone}, {Aravena}, {Archipley},
  {B{\'e}thermin}, {Burgoyne}, {Cathey}, {Chapman}, {Gonzalez}, {Greve},
  {Gururajan}, {Hayward}, {Malkan}, {Phadke}, {Reuter}, {Rotermund}, {Spilker},
  {Stark}, {Sulzenauer}, {Vieira}, {Vizgan}, \& {Wei{\ss}}}]{LK23}
---. 2023, \apj, 949, 87, \dodoi{10.3847/1538-4357/acc93a}

\bibitem[{{Liu} {et~al.}(2019){Liu}, {Lang}, {Magnelli}, {Schinnerer},
  {Leslie}, {Fudamoto}, {Bondi}, {Groves}, {Jim{\'e}nez-Andrade}, {Harrington},
  {Karim}, {Oesch}, {Sargent}, {Vardoulaki}, {B{\v{a}}descu}, {Moser},
  {Bertoldi}, {Battisti}, {da Cunha}, {Zavala}, {Vaccari}, {Davidzon},
  {Riechers}, \& {Aravena}}]{LD19}
{Liu}, D., {Lang}, P., {Magnelli}, B., {et~al.} 2019, \apjs, 244, 40,
  \dodoi{10.3847/1538-4365/ab42da}

\bibitem[{{Loiacono} {et~al.}(2024){Loiacono}, {Decarli}, {Mignoli}, {Farina},
  {Ba{\~n}ados}, {Bosman}, {Eilers}, {Schindler}, {Strauss}, {Vestergaard},
  {Wang}, {Blecha}, {Carilli}, {Comastri}, {Connor}, {Costa}, {Dotti}, {Fan},
  {Gilli}, {Jun}, {Liu}, {Lupi}, {Marshall}, {Mazzucchelli}, {Meyer},
  {Neeleman}, {Overzier}, {Pensabene}, {Riechers}, {Trakhtenbrot}, {Trebitsch},
  {Venemans}, {Walter}, \& {Yang}}]{LF24}
{Loiacono}, F., {Decarli}, R., {Mignoli}, M., {et~al.} 2024, \aap, 685, A121,
  \dodoi{10.1051/0004-6361/202348535}

\bibitem[{{Lonsdale Persson} \& {Helou}(1987)}]{lonsdale87}
{Lonsdale Persson}, C.~J., \& {Helou}, G. 1987, \apj, 314, 513,
  \dodoi{10.1086/165082}

\bibitem[{{Lu} {et~al.}(2017{\natexlab{a}}){Lu}, {Zhao}, {D{\'\i}az-Santos},
  {Xu}, {Gao}, {Armus}, {Isaak}, {Mazzarella}, {van der Werf}, {Appleton},
  {Charmandaris}, {Evans}, {Howell}, {Iwasawa}, {Leech}, {Lord}, {Petric},
  {Privon}, {Sanders}, {Schulz}, \& {Surace}}]{L17a}
{Lu}, N., {Zhao}, Y., {D{\'\i}az-Santos}, T., {et~al.} 2017{\natexlab{a}},
  \apjs, 230, 1, \dodoi{10.3847/1538-4365/aa6476}

\bibitem[{{Lu} {et~al.}(2017{\natexlab{b}}){Lu}, {Zhao}, {D{\'\i}az-Santos},
  {Xu}, {Charmandaris}, {Gao}, {van der Werf}, {Privon}, {Inami}, {Rigopoulou},
  {Sanders}, \& {Zhu}}]{LN17a}
---. 2017{\natexlab{b}}, \apjl, 842, L16, \dodoi{10.3847/2041-8213/aa77fc}

\bibitem[{{Lu} {et~al.}(2018){Lu}, {Cao}, {D{\'\i}az-Santos}, {Zhao}, {Privon},
  {Cheng}, {Gao}, {Xu}, {Charmandaris}, {Rigopoulou}, {van der Werf}, {Huang},
  {Wang}, {Evans}, \& {Sanders}}]{LN18}
{Lu}, N., {Cao}, T., {D{\'\i}az-Santos}, T., {et~al.} 2018, \apj, 864, 38,
  \dodoi{10.3847/1538-4357/aad3c9}

\bibitem[{{Luhman} {et~al.}(2003){Luhman}, {Satyapal}, {Fischer}, {Wolfire},
  {Sturm}, {Dudley}, {Lutz}, \& {Genzel}}]{luhman03}
{Luhman}, M.~L., {Satyapal}, S., {Fischer}, J., {et~al.} 2003, \apj, 594, 758,
  \dodoi{10.1086/376965}

\bibitem[{{Luridiana} {et~al.}(2015){Luridiana}, {Morisset}, \&
  {Shaw}}]{Luridiana15}
{Luridiana}, V., {Morisset}, C., \& {Shaw}, R.~A. 2015, \aap, 573, A42,
  \dodoi{10.1051/0004-6361/201323152}

\bibitem[{{Lutz} {et~al.}(2007){Lutz}, {Sturm}, {Tacconi}, {Valiante},
  {Schweitzer}, {Netzer}, {Maiolino}, {Andreani}, {Shemmer}, \&
  {Veilleux}}]{LD07}
{Lutz}, D., {Sturm}, E., {Tacconi}, L.~J., {et~al.} 2007, \apjl, 661, L25,
  \dodoi{10.1086/518537}

\bibitem[{{Ma} {et~al.}(2024){Ma}, {Sun}, {Cheng}, {Yan}, {Ling}, {Sun}, {Foo},
  {Egami}, {Diego}, {Cohen}, {Jansen}, {Summers}, {Windhorst}, {D'Silva},
  {Koekemoer}, {Coe}, {Conselice}, {Driver}, {Frye}, {Grogin}, {Marshall},
  {Nonino}, {Ortiz}, {Pirzkal}, {Robotham}, {Ryan}, {Willmer}, {Adams},
  {Hathi}, {Dole}, {Willner}, {Espada}, {Furtak}, {Hsiao}, {Li}, {Chen},
  {Jolly}, \& {Chen}}]{MZ24}
{Ma}, Z., {Sun}, B., {Cheng}, C., {et~al.} 2024, \apj, 975, 87,
  \dodoi{10.3847/1538-4357/ad7b32}

\bibitem[{{Madden} {et~al.}(1993){Madden}, {Geis}, {Genzel}, {Herrmann},
  {Jackson}, {Poglitsch}, {Stacey}, \& {Townes}}]{madden93}
{Madden}, S.~C., {Geis}, N., {Genzel}, R., {et~al.} 1993, \apj, 407, 579,
  \dodoi{10.1086/172539}

\bibitem[{{Madden} \& {LMC+ Consortium}(2023)}]{madden23}
{Madden}, S.~C., \& {LMC+ Consortium}. 2023, in Physics and Chemistry of Star
  Formation: The Dynamical ISM Across Time and Spatial Scales, 60,
  \dodoi{10.48550/arXiv.2303.00120}

\bibitem[{{Madden} {et~al.}(2013){Madden}, {R{\'e}my-Ruyer}, {Galametz},
  {Cormier}, {Lebouteiller}, {Galliano}, {Hony}, {Bendo}, {Smith}, {Pohlen},
  {Roussel}, {Sauvage}, {Wu}, {Sturm}, {Poglitsch}, {Contursi}, {Doublier},
  {Baes}, {Barlow}, {Boselli}, {Boquien}, {Carlson}, {Ciesla}, {Cooray},
  {Cortese}, {de Looze}, {Irwin}, {Isaak}, {Kamenetzky}, {Karczewski}, {Lu},
  {MacHattie}, {O'Halloran}, {Parkin}, {Rangwala}, {Schirm}, {Schulz},
  {Spinoglio}, {Vaccari}, {Wilson}, \& {Wozniak}}]{M13}
{Madden}, S.~C., {R{\'e}my-Ruyer}, A., {Galametz}, M., {et~al.} 2013, \pasp,
  125, 600, \dodoi{10.1086/671138}

\bibitem[{{Madden} {et~al.}(2020){Madden}, {Cormier}, {Hony}, {Lebouteiller},
  {Abel}, {Galametz}, {De Looze}, {Chevance}, {Polles}, {Lee}, {Galliano},
  {Lambert-Huyghe}, {Hu}, \& {Ramambason}}]{madden20}
{Madden}, S.~C., {Cormier}, D., {Hony}, S., {et~al.} 2020, \aap, 643, A141,
  \dodoi{10.1051/0004-6361/202038860}

\bibitem[{{Magdis} {et~al.}(2014){Magdis}, {Rigopoulou}, {Hopwood}, {Huang},
  {Farrah}, {Pearson}, {Alonso-Herrero}, {Bock}, {Clements}, {Cooray},
  {Griffin}, {Oliver}, {Perez Fournon}, {Riechers}, {Swinyard}, {Scott},
  {Thatte}, {Valtchanov}, \& {Vaccari}}]{MG14}
{Magdis}, G.~E., {Rigopoulou}, D., {Hopwood}, R., {et~al.} 2014, \apj, 796, 63,
  \dodoi{10.1088/0004-637X/796/1/63}

\bibitem[{{Maiolino} {et~al.}(2009){Maiolino}, {Caselli}, {Nagao}, {Walmsley},
  {De Breuck}, \& {Meneghetti}}]{MR09}
{Maiolino}, R., {Caselli}, P., {Nagao}, T., {et~al.} 2009, \aap, 500, L1,
  \dodoi{10.1051/0004-6361/200912265}

\bibitem[{{Maiolino} {et~al.}(2005){Maiolino}, {Cox}, {Caselli}, {Beelen},
  {Bertoldi}, {Carilli}, {Kaufman}, {Menten}, {Nagao}, {Omont}, {Wei{\ss}},
  {Walmsley}, \& {Walter}}]{MR05}
{Maiolino}, R., {Cox}, P., {Caselli}, P., {et~al.} 2005, \aap, 440, L51,
  \dodoi{10.1051/0004-6361:200500165}

\bibitem[{{Maiolino} {et~al.}(2012){Maiolino}, {Gallerani}, {Neri}, {Cicone},
  {Ferrara}, {Genzel}, {Lutz}, {Sturm}, {Tacconi}, {Walter}, {Feruglio},
  {Fiore}, \& {Piconcelli}}]{MR12}
{Maiolino}, R., {Gallerani}, S., {Neri}, R., {et~al.} 2012, \mnras, 425, L66,
  \dodoi{10.1111/j.1745-3933.2012.01303.x}

\bibitem[{{Maiolino} {et~al.}(2015){Maiolino}, {Carniani}, {Fontana},
  {Vallini}, {Pentericci}, {Ferrara}, {Vanzella}, {Grazian}, {Gallerani},
  {Castellano}, {Cristiani}, {Brammer}, {Santini}, {Wagg}, \&
  {Williams}}]{MR15}
{Maiolino}, R., {Carniani}, S., {Fontana}, A., {et~al.} 2015, \mnras, 452, 54,
  \dodoi{10.1093/mnras/stv1194}

\bibitem[{{Malhotra} {et~al.}(2001){Malhotra}, {Kaufman}, {Hollenbach},
  {Helou}, {Rubin}, {Brauher}, {Dale}, {Lu}, {Lord}, {Stacey}, {Contursi},
  {Hunter}, \& {Dinerstein}}]{malhotra01}
{Malhotra}, S., {Kaufman}, M.~J., {Hollenbach}, D., {et~al.} 2001, \apj, 561,
  766, \dodoi{10.1086/323046}

\bibitem[{{Malhotra} {et~al.}(2017){Malhotra}, {Rhoads}, {Finkelstein}, {Yang},
  {Carilli}, {Combes}, {Dassas}, {Finkelstein}, {Frye}, {Gerin}, {Guillard},
  {Nesvadba}, {Rigby}, {Shin}, {Spaans}, {Strauss}, \& {Papovich}}]{MS17}
{Malhotra}, S., {Rhoads}, J.~E., {Finkelstein}, K., {et~al.} 2017, \apj, 835,
  110, \dodoi{10.3847/1538-4357/835/1/110}

\bibitem[{{Marconcini} {et~al.}(2024){Marconcini}, {D'Eugenio}, {Maiolino},
  {Arribas}, {Bunker}, {Carniani}, {Charlot}, {Perna}, {Rodr{\'\i}guez Del
  Pino}, {{\"U}bler}, {Willott}, {B{\"o}ker}, {Cresci}, {Curti}, {Jones},
  {Lamperti}, {Parlanti}, \& {Venturi}}]{MC24}
{Marconcini}, C., {D'Eugenio}, F., {Maiolino}, R., {et~al.} 2024, \mnras, 533,
  2488, \dodoi{10.1093/mnras/stae1971}

\bibitem[{{Marconcini} {et~al.}(2025){Marconcini}, {D'Eugenio}, {Maiolino},
  {Arribas}, {Bunker}, {Carniani}, {Charlot}, {Perna}, {Rodr{\'\i}guez Del
  Pino}, {{\"U}bler}, {P{\'e}rez-Gonz{\'a}lez}, {Willott}, {B{\"o}ker},
  {Cresci}, {Curti}, {Lamperti}, {Scholtz}, {Parlanti}, \& {Venturi}}]{MC25b}
---. 2025, \aap, 699, A154, \dodoi{10.1051/0004-6361/202452994}

\bibitem[{{Marrone} {et~al.}(2018){Marrone}, {Spilker}, {Hayward}, {Vieira},
  {Aravena}, {Ashby}, {Bayliss}, {B{\'e}thermin}, {Brodwin}, {Bothwell},
  {Carlstrom}, {Chapman}, {Chen}, {Crawford}, {Cunningham}, {De Breuck},
  {Fassnacht}, {Gonzalez}, {Greve}, {Hezaveh}, {Lacaille}, {Litke}, {Lower},
  {Ma}, {Malkan}, {Miller}, {Morningstar}, {Murphy}, {Narayanan}, {Phadke},
  {Rotermund}, {Sreevani}, {Stalder}, {Stark}, {Strandet}, {Tang}, \&
  {Wei{\ss}}}]{MD18}
{Marrone}, D.~P., {Spilker}, J.~S., {Hayward}, C.~C., {et~al.} 2018, \nat, 553,
  51, \dodoi{10.1038/nature24629}

\bibitem[{{Marshall} {et~al.}(2023){Marshall}, {Perna}, {Willott}, {Maiolino},
  {Scholtz}, {{\"U}bler}, {Carniani}, {Arribas}, {L{\"u}tzgendorf}, {Bunker},
  {Charlot}, {Ferruit}, {Jakobsen}, {Rix}, {Rodr{\'\i}guez Del Pino},
  {B{\"o}ker}, {Cameron}, {Cresci}, {Curtis-Lake}, {Jones}, {Kumari},
  {P{\'e}rez-Gonz{\'a}lez}, \& {Reed}}]{MM23}
{Marshall}, M.~A., {Perna}, M., {Willott}, C.~J., {et~al.} 2023, \aap, 678,
  A191, \dodoi{10.1051/0004-6361/202346113}

\bibitem[{{Matsuoka} {et~al.}(2016){Matsuoka}, {Onoue}, {Kashikawa}, {Iwasawa},
  {Strauss}, {Nagao}, {Imanishi}, {Niida}, {Toba}, {Akiyama}, {Asami}, {Bosch},
  {Foucaud}, {Furusawa}, {Goto}, {Gunn}, {Harikane}, {Ikeda}, {Kawaguchi},
  {Kikuta}, {Komiyama}, {Lupton}, {Minezaki}, {Miyazaki}, {Morokuma},
  {Murayama}, {Nishizawa}, {Ono}, {Ouchi}, {Price}, {Sameshima}, {Silverman},
  {Sugiyama}, {Tait}, {Takada}, {Takata}, {Tanaka}, {Tang}, \& {Utsumi}}]{MY16}
{Matsuoka}, Y., {Onoue}, M., {Kashikawa}, N., {et~al.} 2016, \apj, 828, 26,
  \dodoi{10.3847/0004-637X/828/1/26}

\bibitem[{{Matsuoka} {et~al.}(2018{\natexlab{a}}){Matsuoka}, {Iwasawa},
  {Onoue}, {Kashikawa}, {Strauss}, {Lee}, {Imanishi}, {Nagao}, {Akiyama},
  {Asami}, {Bosch}, {Furusawa}, {Goto}, {Gunn}, {Harikane}, {Ikeda}, {Izumi},
  {Kawaguchi}, {Kato}, {Kikuta}, {Kohno}, {Komiyama}, {Lupton}, {Minezaki},
  {Miyazaki}, {Morokuma}, {Murayama}, {Niida}, {Nishizawa}, {Oguri}, {Ono},
  {Ouchi}, {Price}, {Sameshima}, {Schulze}, {Shirakata}, {Silverman},
  {Sugiyama}, {Tait}, {Takada}, {Takata}, {Tanaka}, {Tang}, {Toba}, {Utsumi},
  {Wang}, \& {Yamashita}}]{MY18b}
{Matsuoka}, Y., {Iwasawa}, K., {Onoue}, M., {et~al.} 2018{\natexlab{a}}, \apjs,
  237, 5, \dodoi{10.3847/1538-4365/aac724}

\bibitem[{{Matsuoka} {et~al.}(2018{\natexlab{b}}){Matsuoka}, {Onoue},
  {Kashikawa}, {Iwasawa}, {Strauss}, {Nagao}, {Imanishi}, {Lee}, {Akiyama},
  {Asami}, {Bosch}, {Foucaud}, {Furusawa}, {Goto}, {Gunn}, {Harikane}, {Ikeda},
  {Izumi}, {Kawaguchi}, {Kikuta}, {Kohno}, {Komiyama}, {Lupton}, {Minezaki},
  {Miyazaki}, {Morokuma}, {Murayama}, {Niida}, {Nishizawa}, {Oguri}, {Ono},
  {Ouchi}, {Price}, {Sameshima}, {Schulze}, {Shirakata}, {Silverman},
  {Sugiyama}, {Tait}, {Takada}, {Takata}, {Tanaka}, {Tang}, {Toba}, {Utsumi},
  \& {Wang}}]{MY18a}
{Matsuoka}, Y., {Onoue}, M., {Kashikawa}, N., {et~al.} 2018{\natexlab{b}},
  \pasj, 70, S35, \dodoi{10.1093/pasj/psx046}

\bibitem[{{Matsuoka} {et~al.}(2019){Matsuoka}, {Onoue}, {Kashikawa}, {Strauss},
  {Iwasawa}, {Lee}, {Imanishi}, {Nagao}, {Akiyama}, {Asami}, {Bosch},
  {Furusawa}, {Goto}, {Gunn}, {Harikane}, {Ikeda}, {Izumi}, {Kawaguchi},
  {Kato}, {Kikuta}, {Kohno}, {Komiyama}, {Koyama}, {Lupton}, {Minezaki},
  {Miyazaki}, {Murayama}, {Niida}, {Nishizawa}, {Noboriguchi}, {Oguri}, {Ono},
  {Ouchi}, {Price}, {Sameshima}, {Schulze}, {Shirakata}, {Silverman},
  {Sugiyama}, {Tait}, {Takada}, {Takata}, {Tanaka}, {Tang}, {Toba}, {Utsumi},
  {Wang}, \& {Yamashita}}]{MY19}
---. 2019, \apjl, 872, L2, \dodoi{10.3847/2041-8213/ab0216}

\bibitem[{{Matthee} {et~al.}(2017){Matthee}, {Sobral}, {Boone},
  {R{\"o}ttgering}, {Schaerer}, {Girard}, {Pallottini}, {Vallini}, {Ferrara},
  {Darvish}, \& {Mobasher}}]{MJ17}
{Matthee}, J., {Sobral}, D., {Boone}, F., {et~al.} 2017, \apj, 851, 145,
  \dodoi{10.3847/1538-4357/aa9931}

\bibitem[{{Matthee} {et~al.}(2019){Matthee}, {Sobral}, {Boogaard},
  {R{\"o}ttgering}, {Vallini}, {Ferrara}, {Paulino-Afonso}, {Boone},
  {Schaerer}, \& {Mobasher}}]{MJ19}
{Matthee}, J., {Sobral}, D., {Boogaard}, L.~A., {et~al.} 2019, \apj, 881, 124,
  \dodoi{10.3847/1538-4357/ab2f81}

\bibitem[{{Mazzucchelli} {et~al.}(2017){Mazzucchelli}, {Ba{\~n}ados},
  {Venemans}, {Decarli}, {Farina}, {Walter}, {Eilers}, {Rix}, {Simcoe},
  {Stern}, {Fan}, {Schlafly}, {De Rosa}, {Hennawi}, {Chambers}, {Greiner},
  {Burgett}, {Draper}, {Kaiser}, {Kudritzki}, {Magnier}, {Metcalfe}, {Waters},
  \& {Wainscoat}}]{MC17}
{Mazzucchelli}, C., {Ba{\~n}ados}, E., {Venemans}, B.~P., {et~al.} 2017, \apj,
  849, 91, \dodoi{10.3847/1538-4357/aa9185}

\bibitem[{{Mazzucchelli} {et~al.}(2025){Mazzucchelli}, {Decarli}, {Belladitta},
  {Ba{\~n}ados}, {Meyer}, {Connor}, {Momjian}, {Rojas-Ruiz}, {Eilers},
  {Khusanova}, {Farina}, {Drake}, {Walter}, {Wang}, {Onoue}, \&
  {Venemans}}]{MC25a}
{Mazzucchelli}, C., {Decarli}, R., {Belladitta}, S., {et~al.} 2025, \aap, 694,
  A171, \dodoi{10.1051/0004-6361/202451290}

\bibitem[{{McKinney} {et~al.}(2021){McKinney}, {Hayward}, {Rosenthal},
  {Mart{\'\i}nez-Galarza}, {Pope}, {Sajina}, \& {Smith}}]{mckinney21}
{McKinney}, J., {Hayward}, C.~C., {Rosenthal}, L.~J., {et~al.} 2021, \apj, 921,
  55, \dodoi{10.3847/1538-4357/ac185f}

\bibitem[{{Messias} {et~al.}(2014){Messias}, {Dye}, {Nagar}, {Orellana},
  {Bussmann}, {Calanog}, {Dannerbauer}, {Fu}, {Ibar}, {Inohara}, {Ivison},
  {Negrello}, {Riechers}, {Sheen}, {Aguirre}, {Amber}, {Birkinshaw}, {Bourne},
  {Bradford}, {Clements}, {Cooray}, {De Zotti}, {Demarco}, {Dunne}, {Eales},
  {Fleuren}, {Kamenetzky}, {Lupu}, {Maddox}, {Marrone}, {Micha{\l}owski},
  {Murphy}, {Nguyen}, {Omont}, {Rowlands}, {Smith}, {Smith}, {Valiante}, \&
  {Vieira}}]{MH14}
{Messias}, H., {Dye}, S., {Nagar}, N., {et~al.} 2014, \aap, 568, A92,
  \dodoi{10.1051/0004-6361/201424410}

\bibitem[{{Meyer} {et~al.}(2025){Meyer}, {Venemans}, {Neeleman}, {Decarli}, \&
  {Walter}}]{MR25}
{Meyer}, R.~A., {Venemans}, B., {Neeleman}, M., {Decarli}, R., \& {Walter}, F.
  2025, \apj, 980, 20, \dodoi{10.3847/1538-4357/ada351}

\bibitem[{{Meyer} {et~al.}(2022){Meyer}, {Walter}, {Cicone}, {Cox}, {Decarli},
  {Neri}, {Novak}, {Pensabene}, {Riechers}, \& {Weiss}}]{MR22}
{Meyer}, R.~A., {Walter}, F., {Cicone}, C., {et~al.} 2022, \apj, 927, 152,
  \dodoi{10.3847/1538-4357/ac4e94}

\bibitem[{{Miller} {et~al.}(2018){Miller}, {Chapman}, {Aravena}, {Ashby},
  {Hayward}, {Vieira}, {Wei{\ss}}, {Babul}, {B{\'e}thermin}, {Bradford},
  {Brodwin}, {Carlstrom}, {Chen}, {Cunningham}, {De Breuck}, {Gonzalez},
  {Greve}, {Harnett}, {Hezaveh}, {Lacaille}, {Litke}, {Ma}, {Malkan},
  {Marrone}, {Morningstar}, {Murphy}, {Narayanan}, {Pass}, {Perry}, {Phadke},
  {Rennehan}, {Rotermund}, {Simpson}, {Spilker}, {Sreevani}, {Stark},
  {Strandet}, \& {Strom}}]{MT18}
{Miller}, T.~B., {Chapman}, S.~C., {Aravena}, M., {et~al.} 2018, \nat, 556,
  469, \dodoi{10.1038/s41586-018-0025-2}

\bibitem[{{Mitsuhashi} {et~al.}(2021){Mitsuhashi}, {Matsuda}, {Smail},
  {Hayatsu}, {Simpson}, {Swinbank}, {Umehata},
  {Dudzevi{\v{c}}i{\={u}}t{\.{e}}}, {Birkin}, {Ikarashi}, {Chen}, {Tadaki},
  {Yajima}, {Harikane}, {Inami}, {Chapman}, {Hatsukade}, {Iono}, {Bunker},
  {Ao}, {Saito}, {Ueda}, \& {Sakamoto}}]{MI21}
{Mitsuhashi}, I., {Matsuda}, Y., {Smail}, I., {et~al.} 2021, \apj, 907, 122,
  \dodoi{10.3847/1538-4357/abcc72}

\bibitem[{{Mitsuhashi} {et~al.}(2024){Mitsuhashi}, {Tadaki}, {Ikeda},
  {Herrera-Camus}, {Aravena}, {De Looze}, {F{\"o}rster Schreiber},
  {Gonz{\'a}lez-L{\'o}pez}, {Spilker}, {Assef}, {Bouwens}, {Barcos-Munoz},
  {Birkin}, {Bowler}, {Calistro Rivera}, {Davies}, {Da Cunha},
  {D{\'\i}az-Santos}, {Ferrara}, {Fisher}, {Lee}, {Li}, {Lutz}, {Rela{\~n}o},
  {Naab}, {Palla}, {Posses}, {Solimano}, {Tacconi}, {{\"U}bler}, {van der
  Giessen}, \& {Veilleux}}]{MI24}
{Mitsuhashi}, I., {Tadaki}, K.-i., {Ikeda}, R., {et~al.} 2024, \aap, 690, A197,
  \dodoi{10.1051/0004-6361/202348782}

\bibitem[{{Mitsuhashi} {et~al.}(2025){Mitsuhashi}, {Zavala}, {Bakx}, {Inoue},
  {Castellano}, {Calabro}, {Casey}, {Franco}, {Hatsukade}, {Hathi}, {Ikeda},
  {Koekemoer}, {Kartaltepe}, {Knudsen}, {Santini}, {Saito}, {Terlevich}, \&
  {Terlevich}}]{MI25}
{Mitsuhashi}, I., {Zavala}, J.~A., {Bakx}, T. J.~L.~C., {et~al.} 2025, arXiv
  e-prints, arXiv:2501.19384, \dodoi{10.48550/arXiv.2501.19384}

\bibitem[{{Molyneux} {et~al.}(2022){Molyneux}, {Smit}, {Schaerer}, {Bouwens},
  {Bradley}, {Hodge}, {Longmore}, {Schouws}, {van der Werf}, {Zitrin}, \&
  {Phillips}}]{MS22}
{Molyneux}, S.~J., {Smit}, R., {Schaerer}, D., {et~al.} 2022, \mnras, 512, 535,
  \dodoi{10.1093/mnras/stac557}

\bibitem[{{Morishita} {et~al.}(2023){Morishita}, {Roberts-Borsani}, {Treu},
  {Brammer}, {Mason}, {Trenti}, {Vulcani}, {Wang}, {Acebron}, {Bah{\'e}},
  {Bergamini}, {Boyett}, {Bradac}, {Calabr{\`o}}, {Castellano}, {Chen}, {De
  Lucia}, {Filippenko}, {Fontana}, {Glazebrook}, {Grillo}, {Henry}, {Jones},
  {Kelly}, {Koekemoer}, {Leethochawalit}, {Lu}, {Marchesini}, {Mascia},
  {Mercurio}, {Merlin}, {Metha}, {Nanayakkara}, {Nonino}, {Paris},
  {Pentericci}, {Rosati}, {Santini}, {Strait}, {Vanzella}, {Windhorst}, \&
  {Xie}}]{MT23}
{Morishita}, T., {Roberts-Borsani}, G., {Treu}, T., {et~al.} 2023, \apjl, 947,
  L24, \dodoi{10.3847/2041-8213/acb99e}

\bibitem[{{Moustakas} \& {Kennicutt}(2006{\natexlab{a}})}]{moustakas06}
{Moustakas}, J., \& {Kennicutt}, Robert~C., J. 2006{\natexlab{a}}, \apjs, 164,
  81, \dodoi{10.1086/500971}

\bibitem[{{Moustakas} \& {Kennicutt}(2006{\natexlab{b}})}]{M06}
{Moustakas}, J., \& {Kennicutt}, Jr., R.~C. 2006{\natexlab{b}}, \apjs, 164, 81,
  \dodoi{10.1086/500971}

\bibitem[{{Murphy} {et~al.}(2011){Murphy}, {Condon}, {Schinnerer}, {Kennicutt},
  {Calzetti}, {Armus}, {Helou}, {Turner}, {Aniano}, {Beir{\~a}o}, {Bolatto},
  {Brandl}, {Croxall}, {Dale}, {Donovan Meyer}, {Draine}, {Engelbracht},
  {Hunt}, {Hao}, {Koda}, {Roussel}, {Skibba}, \& {Smith}}]{murphy11}
{Murphy}, E.~J., {Condon}, J.~J., {Schinnerer}, E., {et~al.} 2011, \apj, 737,
  67, \dodoi{10.1088/0004-637X/737/2/67}

\bibitem[{{Nagao} {et~al.}(2012){Nagao}, {Maiolino}, {De Breuck}, {Caselli},
  {Hatsukade}, \& {Saigo}}]{NT12}
{Nagao}, T., {Maiolino}, R., {De Breuck}, C., {et~al.} 2012, \aap, 542, L34,
  \dodoi{10.1051/0004-6361/201219518}

\bibitem[{{Nakajima} {et~al.}(2023){Nakajima}, {Ouchi}, {Isobe}, {Harikane},
  {Zhang}, {Ono}, {Umeda}, \& {Oguri}}]{NK23}
{Nakajima}, K., {Ouchi}, M., {Isobe}, Y., {et~al.} 2023, \apjs, 269, 33,
  \dodoi{10.3847/1538-4365/acd556}

\bibitem[{{Neeleman} {et~al.}(2017){Neeleman}, {Kanekar}, {Prochaska},
  {Rafelski}, {Carilli}, \& {Wolfe}}]{NM17}
{Neeleman}, M., {Kanekar}, N., {Prochaska}, J.~X., {et~al.} 2017, Science, 355,
  1285, \dodoi{10.1126/science.aal1737}

\bibitem[{{Neeleman} {et~al.}(2019{\natexlab{a}}){Neeleman}, {Kanekar},
  {Prochaska}, {Rafelski}, \& {Carilli}}]{NM19b}
{Neeleman}, M., {Kanekar}, N., {Prochaska}, J.~X., {Rafelski}, M.~A., \&
  {Carilli}, C.~L. 2019{\natexlab{a}}, \apjl, 870, L19,
  \dodoi{10.3847/2041-8213/aaf871}

\bibitem[{{Neeleman} {et~al.}(2020){Neeleman}, {Prochaska}, {Kanekar}, \&
  {Rafelski}}]{NM20}
{Neeleman}, M., {Prochaska}, J.~X., {Kanekar}, N., \& {Rafelski}, M. 2020,
  \nat, 581, 269, \dodoi{10.1038/s41586-020-2276-y}

\bibitem[{{Neeleman} {et~al.}(2019{\natexlab{b}}){Neeleman}, {Ba{\~n}ados},
  {Walter}, {Decarli}, {Venemans}, {Carilli}, {Fan}, {Farina}, {Mazzucchelli},
  {Novak}, {Riechers}, {Rix}, \& {Wang}}]{NM19a}
{Neeleman}, M., {Ba{\~n}ados}, E., {Walter}, F., {et~al.} 2019{\natexlab{b}},
  \apj, 882, 10, \dodoi{10.3847/1538-4357/ab2ed3}

\bibitem[{{Neri} {et~al.}(2014){Neri}, {Downes}, {Cox}, \& {Walter}}]{NR14}
{Neri}, R., {Downes}, D., {Cox}, P., \& {Walter}, F. 2014, \aap, 562, A35,
  \dodoi{10.1051/0004-6361/201322528}

\bibitem[{{Novak} {et~al.}(2019){Novak}, {Ba{\~n}ados}, {Decarli}, {Walter},
  {Venemans}, {Neeleman}, {Farina}, {Mazzucchelli}, {Carilli}, {Fan}, {Rix}, \&
  {Wang}}]{NM19c}
{Novak}, M., {Ba{\~n}ados}, E., {Decarli}, R., {et~al.} 2019, \apj, 881, 63,
  \dodoi{10.3847/1538-4357/ab2beb}

\bibitem[{{Oberst} {et~al.}(2006){Oberst}, {Parshley}, {Stacey}, {Nikola},
  {L{\"o}hr}, {Harnett}, {Tothill}, {Lane}, {Stark}, \& {Tucker}}]{obserst06}
{Oberst}, T.~E., {Parshley}, S.~C., {Stacey}, G.~J., {et~al.} 2006, \apjl, 652,
  L125, \dodoi{10.1086/510289}

\bibitem[{{Ochsendorf} \& {Tielens}(2015)}]{ochsendorf15}
{Ochsendorf}, B.~B., \& {Tielens}, A.~G.~G.~M. 2015, \aap, 576, A2,
  \dodoi{10.1051/0004-6361/201424799}

\bibitem[{{Oesch} {et~al.}(2016){Oesch}, {Brammer}, {van Dokkum},
  {Illingworth}, {Bouwens}, {Labb{\'e}}, {Franx}, {Momcheva}, {Ashby}, {Fazio},
  {Gonzalez}, {Holden}, {Magee}, {Skelton}, {Smit}, {Spitler}, {Trenti}, \&
  {Willner}}]{OP16}
{Oesch}, P.~A., {Brammer}, G., {van Dokkum}, P.~G., {et~al.} 2016, \apj, 819,
  129, \dodoi{10.3847/0004-637X/819/2/129}

\bibitem[{{Ogle} {et~al.}(2007){Ogle}, {Antonucci}, {Appleton}, \&
  {Whysong}}]{ogle07}
{Ogle}, P., {Antonucci}, R., {Appleton}, P.~N., \& {Whysong}, D. 2007, \apj,
  668, 699, \dodoi{10.1086/521334}

\bibitem[{{Ogle} {et~al.}(2010){Ogle}, {Boulanger}, {Guillard}, {Evans},
  {Antonucci}, {Appleton}, {Nesvadba}, \& {Leipski}}]{ogle10}
{Ogle}, P., {Boulanger}, F., {Guillard}, P., {et~al.} 2010, \apj, 724, 1193,
  \dodoi{10.1088/0004-637X/724/2/1193}

\bibitem[{{Omont} {et~al.}(2001){Omont}, {Cox}, {Bertoldi}, {McMahon},
  {Carilli}, \& {Isaak}}]{OA01}
{Omont}, A., {Cox}, P., {Bertoldi}, F., {et~al.} 2001, \aap, 374, 371,
  \dodoi{10.1051/0004-6361:20010721}

\bibitem[{{Ono} {et~al.}(2012){Ono}, {Ouchi}, {Mobasher}, {Dickinson},
  {Penner}, {Shimasaku}, {Weiner}, {Kartaltepe}, {Nakajima}, {Nayyeri},
  {Stern}, {Kashikawa}, \& {Spinrad}}]{OY12}
{Ono}, Y., {Ouchi}, M., {Mobasher}, B., {et~al.} 2012, \apj, 744, 83,
  \dodoi{10.1088/0004-637X/744/2/83}

\bibitem[{{Osterbrock}(1989)}]{osterbrock89}
{Osterbrock}, D.~E. 1989, {Astrophysics of gaseous nebulae and active galactic
  nuclei}

\bibitem[{{Ota} {et~al.}(2014){Ota}, {Walter}, {Ohta}, {Hatsukade}, {Carilli},
  {da Cunha}, {Gonz{\'a}lez-L{\'o}pez}, {Decarli}, {Hodge}, {Nagai}, {Egami},
  {Jiang}, {Iye}, {Kashikawa}, {Riechers}, {Bertoldi}, {Cox}, {Neri}, \&
  {Weiss}}]{OK14}
{Ota}, K., {Walter}, F., {Ohta}, K., {et~al.} 2014, \apj, 792, 34,
  \dodoi{10.1088/0004-637X/792/1/34}

\bibitem[{{Oteo} {et~al.}(2016){Oteo}, {Ivison}, {Dunne}, {Smail}, {Swinbank},
  {Zhang}, {Lewis}, {Maddox}, {Riechers}, {Serjeant}, {Van der Werf}, {Biggs},
  {Bremer}, {Cigan}, {Clements}, {Cooray}, {Dannerbauer}, {Eales}, {Ibar},
  {Messias}, {Micha{\l}owski}, {P{\'e}rez-Fournon}, \& {van Kampen}}]{OI16}
{Oteo}, I., {Ivison}, R.~J., {Dunne}, L., {et~al.} 2016, \apj, 827, 34,
  \dodoi{10.3847/0004-637X/827/1/34}

\bibitem[{{Ouchi} {et~al.}(2013){Ouchi}, {Ellis}, {Ono}, {Nakanishi}, {Kohno},
  {Momose}, {Kurono}, {Ashby}, {Shimasaku}, {Willner}, {Fazio}, {Tamura}, \&
  {Iono}}]{OM13}
{Ouchi}, M., {Ellis}, R., {Ono}, Y., {et~al.} 2013, \apj, 778, 102,
  \dodoi{10.1088/0004-637X/778/2/102}

\bibitem[{{Oyabu} {et~al.}(2009){Oyabu}, {Kawara}, {Tsuzuki}, {Matsuoka},
  {Sameshima}, {Asami}, \& {Ohyama}}]{OS09}
{Oyabu}, S., {Kawara}, K., {Tsuzuki}, Y., {et~al.} 2009, \apj, 697, 452,
  \dodoi{10.1088/0004-637X/697/1/452}

\bibitem[{{Papadopoulos} {et~al.}(2010){Papadopoulos}, {van der Werf}, {Isaak},
  \& {Xilouris}}]{papadopoulos10}
{Papadopoulos}, P.~P., {van der Werf}, P., {Isaak}, K., \& {Xilouris}, E.~M.
  2010, \apj, 715, 775, \dodoi{10.1088/0004-637X/715/2/775}

\bibitem[{{Parlanti} {et~al.}(2025){Parlanti}, {Carniani}, {Venturi},
  {Herrera-Camus}, {Arribas}, {Bunker}, {Charlot}, {D'Eugenio}, {Maiolino},
  {Perna}, {{\"U}bler}, {B{\"o}ker}, {Cresci}, {Curti}, {Jones}, {Lamperti},
  {P{\'e}rez-Gonz{\'a}lez}, {Del Pino}, \& {Zamora}}]{PE25}
{Parlanti}, E., {Carniani}, S., {Venturi}, G., {et~al.} 2025, \aap, 695, A6,
  \dodoi{10.1051/0004-6361/202451692}

\bibitem[{{Pavesi} {et~al.}(2019){Pavesi}, {Riechers}, {Faisst}, {Stacey}, \&
  {Capak}}]{PR19}
{Pavesi}, R., {Riechers}, D.~A., {Faisst}, A.~L., {Stacey}, G.~J., \& {Capak},
  P.~L. 2019, \apj, 882, 168, \dodoi{10.3847/1538-4357/ab3a46}

\bibitem[{{Pavesi} {et~al.}(2016){Pavesi}, {Riechers}, {Capak}, {Carilli},
  {Sharon}, {Stacey}, {Karim}, {Scoville}, \& {Smol{\v{c}}i{\'c}}}]{PR16}
{Pavesi}, R., {Riechers}, D.~A., {Capak}, P.~L., {et~al.} 2016, \apj, 832, 151,
  \dodoi{10.3847/0004-637X/832/2/151}

\bibitem[{{Pavesi} {et~al.}(2018){Pavesi}, {Riechers}, {Sharon},
  {Smol{\v{c}}i{\'c}}, {Faisst}, {Schinnerer}, {Carilli}, {Capak}, {Scoville},
  \& {Stacey}}]{PR18}
{Pavesi}, R., {Riechers}, D.~A., {Sharon}, C.~E., {et~al.} 2018, \apj, 861, 43,
  \dodoi{10.3847/1538-4357/aac6b6}

\bibitem[{{Peng} {et~al.}(2025){Peng}, {Lamarche}, {Ball}, {Vishwas}, {Stacey},
  {Rooney}, {Nikola}, \& {Ferkinhoff}}]{paperi}
{Peng}, B., {Lamarche}, C., {Ball}, C., {et~al.} 2025, arXiv e-prints,
  arXiv:2507.10702.
\newblock \doarXiv{2507.10702}

\bibitem[{Peng {et~al.}(2025)Peng, Vishwas, Lamarche, Stacey, Ball, Rooney,
  Nikola, \& Ferkinhoff}]{paperii}
Peng, B., Vishwas, A., Lamarche, C., {et~al.} 2025, Fine-structure Line Atlas
  for Multi-wavelength Extragalactic Study (FLAMES) II: Photoionization Model
  View of Ionized to Neutral Gas Emission.
\newblock \doarXiv{2507.11829}

\bibitem[{{Peng} {et~al.}(2021){Peng}, {Lamarche}, {Stacey}, {Nikola},
  {Vishwas}, {Ferkinhoff}, {Rooney}, {Ball}, {Brisbin}, {Higdon}, \&
  {Higdon}}]{P21}
{Peng}, B., {Lamarche}, C., {Stacey}, G.~J., {et~al.} 2021, \apj, 908, 166,
  \dodoi{10.3847/1538-4357/abd4e2}

\bibitem[{{Peng} {et~al.}(2023){Peng}, {Vishwas}, {Stacey}, {Nikola},
  {Lamarche}, {Rooney}, {Ball}, {Ferkinhoff}, \& {Spoon}}]{PB23}
{Peng}, B., {Vishwas}, A., {Stacey}, G., {et~al.} 2023, \apjl, 944, L36,
  \dodoi{10.3847/2041-8213/acb59c}

\bibitem[{{Pensabene} {et~al.}(2021){Pensabene}, {Decarli}, {Ba{\~n}ados},
  {Venemans}, {Walter}, {Bertoldi}, {Fan}, {Farina}, {Li}, {Mazzucchelli},
  {Novak}, {Riechers}, {Rix}, {Strauss}, {Wang}, {Wei{\ss}}, {Yang}, \&
  {Yang}}]{PA21}
{Pensabene}, A., {Decarli}, R., {Ba{\~n}ados}, E., {et~al.} 2021, \aap, 652,
  A66, \dodoi{10.1051/0004-6361/202039696}

\bibitem[{{Pentericci} {et~al.}(2011){Pentericci}, {Fontana}, {Vanzella},
  {Castellano}, {Grazian}, {Dijkstra}, {Boutsia}, {Cristiani}, {Dickinson},
  {Giallongo}, {Giavalisco}, {Maiolino}, {Moorwood}, {Paris}, \&
  {Santini}}]{PL11}
{Pentericci}, L., {Fontana}, A., {Vanzella}, E., {et~al.} 2011, \apj, 743, 132,
  \dodoi{10.1088/0004-637X/743/2/132}

\bibitem[{{Pentericci} {et~al.}(2016){Pentericci}, {Carniani}, {Castellano},
  {Fontana}, {Maiolino}, {Guaita}, {Vanzella}, {Grazian}, {Santini}, {Yan},
  {Cristiani}, {Conselice}, {Giavalisco}, {Hathi}, \& {Koekemoer}}]{PL16}
{Pentericci}, L., {Carniani}, S., {Castellano}, M., {et~al.} 2016, \apjl, 829,
  L11, \dodoi{10.3847/2041-8205/829/1/L11}

\bibitem[{{Petrosian} {et~al.}(1972){Petrosian}, {Silk}, \&
  {Field}}]{petrosian74}
{Petrosian}, V., {Silk}, J., \& {Field}, G.~B. 1972, \apjl, 177, L69,
  \dodoi{10.1086/181054}

\bibitem[{{Pety} {et~al.}(2004){Pety}, {Beelen}, {Cox}, {Downes}, {Omont},
  {Bertoldi}, \& {Carilli}}]{PJ04}
{Pety}, J., {Beelen}, A., {Cox}, P., {et~al.} 2004, \aap, 428, L21,
  \dodoi{10.1051/0004-6361:200400096}

\bibitem[{{Pilyugin} {et~al.}(2003){Pilyugin}, {Thuan}, \&
  {V{\'\i}lchez}}]{pilyugin03}
{Pilyugin}, L.~S., {Thuan}, T.~X., \& {V{\'\i}lchez}, J.~M. 2003, \aap, 397,
  487, \dodoi{10.1051/0004-6361:20021458}

\bibitem[{{Pineda} {et~al.}(2013){Pineda}, {Langer}, {Velusamy}, \&
  {Goldsmith}}]{pineda13}
{Pineda}, J.~L., {Langer}, W.~D., {Velusamy}, T., \& {Goldsmith}, P.~F. 2013,
  \aap, 554, A103, \dodoi{10.1051/0004-6361/201321188}

\bibitem[{{Piqueras L{\'o}pez} {et~al.}(2016){Piqueras L{\'o}pez}, {Colina},
  {Arribas}, {Pereira-Santaella}, \& {Alonso-Herrero}}]{piqueras16}
{Piqueras L{\'o}pez}, J., {Colina}, L., {Arribas}, S., {Pereira-Santaella}, M.,
  \& {Alonso-Herrero}, A. 2016, \aap, 590, A67,
  \dodoi{10.1051/0004-6361/201527671}

\bibitem[{{Pope} {et~al.}(2017){Pope}, {Monta{\~n}a}, {Battisti}, {Limousin},
  {Marchesini}, {Wilson}, {Alberts}, {Aretxaga}, {Avila-Reese}, {Ram{\'o}n
  Bermejo-Climent}, {Brammer}, {Bravo-Alfaro}, {Calzetti}, {Chary}, {Cybulski},
  {Giavalisco}, {Hughes}, {Kado-Fong}, {Keller}, {Kirkpatrick}, {Labbe},
  {Lange-Vagle}, {Lowenthal}, {Murphy}, {Oesch}, {Rosa Gonzalez},
  {S{\'a}nchez-Arg{\"u}elles}, {Shipley}, {Stefanon}, {Vega}, {Whitaker},
  {Williams}, {Yun}, {Zavala}, \& {Zeballos}}]{PA17}
{Pope}, A., {Monta{\~n}a}, A., {Battisti}, A., {et~al.} 2017, \apj, 838, 137,
  \dodoi{10.3847/1538-4357/aa6573}

\bibitem[{{Pope} {et~al.}(2023){Pope}, {McKinney}, {Kamieneski}, {Battisti},
  {Aretxaga}, {Brammer}, {Diego}, {Hughes}, {Keller}, {Marchesini}, {Mizener},
  {Monta{\~n}a}, {Murphy}, {Whitaker}, {Wilson}, \& {Yun}}]{PA23a}
{Pope}, A., {McKinney}, J., {Kamieneski}, P., {et~al.} 2023, \apjl, 951, L46,
  \dodoi{10.3847/2041-8213/acdf5a}

\bibitem[{{Popesso} {et~al.}(2023){Popesso}, {Concas}, {Cresci}, {Belli},
  {Rodighiero}, {Inami}, {Dickinson}, {Ilbert}, {Pannella}, \&
  {Elbaz}}]{popesso23}
{Popesso}, P., {Concas}, A., {Cresci}, G., {et~al.} 2023, \mnras, 519, 1526,
  \dodoi{10.1093/mnras/stac3214}

\bibitem[{{Popping}(2023)}]{PG23}
{Popping}, G. 2023, \aap, 669, L8, \dodoi{10.1051/0004-6361/202244831}

\bibitem[{{Posses} {et~al.}(2024){Posses}, {Aravena}, {Gonz{\'a}lez-L{\'o}pez},
  {F{\"o}rster Schreiber}, {Liu}, {Lee}, {Solimano}, {D{\'\i}az-Santos},
  {Assef}, {Barcos-Mu{\~n}oz}, {Bovino}, {Bowler}, {Calistro Rivera}, {da
  Cunha}, {Davies}, {Killi}, {De Looze}, {Ferrara}, {Fisher}, {Herrera-Camus},
  {Ikeda}, {Lambert}, {Li}, {Lutz}, {Mitsuhashi}, {Palla}, {Rela{\~n}o},
  {Spilker}, {Naab}, {Tadaki}, {Telikova}, {{\"U}bler}, {van der Giessen}, \&
  {Villanueva}}]{PA24}
{Posses}, A., {Aravena}, M., {Gonz{\'a}lez-L{\'o}pez}, J., {et~al.} 2024, arXiv
  e-prints, arXiv:2403.03379, \dodoi{10.48550/arXiv.2403.03379}

\bibitem[{{Posses} {et~al.}(2023){Posses}, {Aravena}, {Gonz{\'a}lez-L{\'o}pez},
  {Assef}, {Lambert}, {Jones}, {Bouwens}, {Brisbin}, {D{\'\i}az-Santos},
  {Herrera-Camus}, {Ricci}, \& {Smit}}]{PA23b}
{Posses}, A.~C., {Aravena}, M., {Gonz{\'a}lez-L{\'o}pez}, J., {et~al.} 2023,
  \aap, 669, A46, \dodoi{10.1051/0004-6361/202243399}

\bibitem[{{Rawle} {et~al.}(2014){Rawle}, {Egami}, {Bussmann}, {Gurwell},
  {Ivison}, {Boone}, {Combes}, {Danielson}, {Rex}, {Richard}, {Smail},
  {Swinbank}, {Altieri}, {Blain}, {Clement}, {Dessauges-Zavadsky}, {Edge},
  {Fazio}, {Jones}, {Kneib}, {Omont}, {P{\'e}rez-Gonz{\'a}lez}, {Schaerer},
  {Valtchanov}, {van der Werf}, {Walth}, {Zamojski}, \& {Zemcov}}]{RT14}
{Rawle}, T.~D., {Egami}, E., {Bussmann}, R.~S., {et~al.} 2014, \apj, 783, 59,
  \dodoi{10.1088/0004-637X/783/1/59}

\bibitem[{{Ren} {et~al.}(2023){Ren}, {Fudamoto}, {Inoue}, {Sugahara},
  {Tokuoka}, {Tamura}, {Matsuo}, {Kohno}, {Umehata}, {Hashimoto}, {Bouwens},
  {Smit}, {Kashikawa}, {Okamoto}, {Shibuya}, \& {Shimizu}}]{RY23}
{Ren}, Y.~W., {Fudamoto}, Y., {Inoue}, A.~K., {et~al.} 2023, \apj, 945, 69,
  \dodoi{10.3847/1538-4357/acb8ab}

\bibitem[{{Reuter} {et~al.}(2020){Reuter}, {Vieira}, {Spilker}, {Weiss},
  {Aravena}, {Archipley}, {B{\'e}thermin}, {Chapman}, {De Breuck}, {Dong},
  {Everett}, {Fu}, {Greve}, {Hayward}, {Hill}, {Hezaveh}, {Jarugula}, {Litke},
  {Malkan}, {Marrone}, {Narayanan}, {Phadke}, {Stark}, \& {Strandet}}]{RC20}
{Reuter}, C., {Vieira}, J.~D., {Spilker}, J.~S., {et~al.} 2020, \apj, 902, 78,
  \dodoi{10.3847/1538-4357/abb599}

\bibitem[{{Richard} {et~al.}(2011{\natexlab{a}}){Richard}, {Jones}, {Ellis},
  {Stark}, {Livermore}, \& {Swinbank}}]{RJ11b}
{Richard}, J., {Jones}, T., {Ellis}, R., {et~al.} 2011{\natexlab{a}}, \mnras,
  413, 643, \dodoi{10.1111/j.1365-2966.2010.18161.x}

\bibitem[{{Richard} {et~al.}(2011{\natexlab{b}}){Richard}, {Kneib}, {Ebeling},
  {Stark}, {Egami}, \& {Fiedler}}]{RJ11a}
{Richard}, J., {Kneib}, J.-P., {Ebeling}, H., {et~al.} 2011{\natexlab{b}},
  \mnras, 414, L31, \dodoi{10.1111/j.1745-3933.2011.01050.x}

\bibitem[{{Riechers} {et~al.}(2013){Riechers}, {Bradford}, {Clements},
  {Dowell}, {P{\'e}rez-Fournon}, {Ivison}, {Bridge}, {Conley}, {Fu}, {Vieira},
  {Wardlow}, {Calanog}, {Cooray}, {Hurley}, {Neri}, {Kamenetzky}, {Aguirre},
  {Altieri}, {Arumugam}, {Benford}, {B{\'e}thermin}, {Bock}, {Burgarella},
  {Cabrera-Lavers}, {Chapman}, {Cox}, {Dunlop}, {Earle}, {Farrah}, {Ferrero},
  {Franceschini}, {Gavazzi}, {Glenn}, {Solares}, {Gurwell}, {Halpern},
  {Hatziminaoglou}, {Hyde}, {Ibar}, {Kov{\'a}cs}, {Krips}, {Lupu}, {Maloney},
  {Martinez-Navajas}, {Matsuhara}, {Murphy}, {Naylor}, {Nguyen}, {Oliver},
  {Omont}, {Page}, {Petitpas}, {Rangwala}, {Roseboom}, {Scott}, {Smith},
  {Staguhn}, {Streblyanska}, {Thomson}, {Valtchanov}, {Viero}, {Wang},
  {Zemcov}, \& {Zmuidzinas}}]{RD13}
{Riechers}, D.~A., {Bradford}, C.~M., {Clements}, D.~L., {et~al.} 2013, \nat,
  496, 329, \dodoi{10.1038/nature12050}

\bibitem[{{Riechers} {et~al.}(2014){Riechers}, {Carilli}, {Capak}, {Scoville},
  {Smol{\v{c}}i{\'c}}, {Schinnerer}, {Yun}, {Cox}, {Bertoldi}, {Karim}, \&
  {Yan}}]{RD14}
{Riechers}, D.~A., {Carilli}, C.~L., {Capak}, P.~L., {et~al.} 2014, \apj, 796,
  84, \dodoi{10.1088/0004-637X/796/2/84}

\bibitem[{{Riechers} {et~al.}(2020){Riechers}, {Hodge}, {Pavesi}, {Daddi},
  {Decarli}, {Ivison}, {Sharon}, {Smail}, {Walter}, {Aravena}, {Capak},
  {Carilli}, {Cox}, {Cunha}, {Dannerbauer}, {Dickinson}, {Neri}, \&
  {Wagg}}]{RD20}
{Riechers}, D.~A., {Hodge}, J.~A., {Pavesi}, R., {et~al.} 2020, \apj, 895, 81,
  \dodoi{10.3847/1538-4357/ab8c48}

\bibitem[{{Rigby} {et~al.}(2008){Rigby}, {Marcillac}, {Egami}, {Rieke},
  {Richard}, {Kneib}, {Fadda}, {Willmer}, {Borys}, {van der Werf},
  {P{\'e}rez-Gonz{\'a}lez}, {Knudsen}, \& {Papovich}}]{RJ08}
{Rigby}, J.~R., {Marcillac}, D., {Egami}, E., {et~al.} 2008, \apj, 675, 262,
  \dodoi{10.1086/525273}

\bibitem[{{Rigopoulou} {et~al.}(2018){Rigopoulou}, {Pereira-Santaella},
  {Magdis}, {Cooray}, {Farrah}, {Marques-Chaves}, {Perez-Fournon}, \&
  {Riechers}}]{RD18}
{Rigopoulou}, D., {Pereira-Santaella}, M., {Magdis}, G.~E., {et~al.} 2018,
  \mnras, 473, 20, \dodoi{10.1093/mnras/stx2311}

\bibitem[{{Rizzo} {et~al.}(2021){Rizzo}, {Vegetti}, {Fraternali}, {Stacey}, \&
  {Powell}}]{RF21}
{Rizzo}, F., {Vegetti}, S., {Fraternali}, F., {Stacey}, H.~R., \& {Powell}, D.
  2021, \mnras, 507, 3952, \dodoi{10.1093/mnras/stab2295}

\bibitem[{{Rizzo} {et~al.}(2020){Rizzo}, {Vegetti}, {Powell}, {Fraternali},
  {McKean}, {Stacey}, \& {White}}]{RF20}
{Rizzo}, F., {Vegetti}, S., {Powell}, D., {et~al.} 2020, \nat, 584, 201,
  \dodoi{10.1038/s41586-020-2572-6}

\bibitem[{{Rojas-Ruiz} {et~al.}(2021){Rojas-Ruiz}, {Ba{\~n}ados}, {Neeleman},
  {Connor}, {Eilers}, {Venemans}, {Khusanova}, {Carilli}, {Mazzucchelli},
  {Decarli}, {Momjian}, \& {Novak}}]{RS21}
{Rojas-Ruiz}, S., {Ba{\~n}ados}, E., {Neeleman}, M., {et~al.} 2021, \apj, 920,
  150, \dodoi{10.3847/1538-4357/ac1a13}

\bibitem[{{Roman-Oliveira} {et~al.}(2023){Roman-Oliveira}, {Fraternali}, \&
  {Rizzo}}]{RF23}
{Roman-Oliveira}, F., {Fraternali}, F., \& {Rizzo}, F. 2023, \mnras, 521, 1045,
  \dodoi{10.1093/mnras/stad530}

\bibitem[{{Romano} {et~al.}(2020){Romano}, {Franchini}, {Grisoni}, {Spitoni},
  {Matteucci}, \& {Morossi}}]{romano20}
{Romano}, D., {Franchini}, M., {Grisoni}, V., {et~al.} 2020, \aap, 639, A37,
  \dodoi{10.1051/0004-6361/202037972}

\bibitem[{{Rooney} {et~al.}(2025){Rooney}, {Peng}, {Vishwas}, {Stacey},
  {Nikola}, {Lamarche}, {Ball}, {Ferkinhoff}, {Brisbin}, \&
  {Hailey-Dunsheath}}]{RC25}
{Rooney}, C., {Peng}, B., {Vishwas}, A., {et~al.} 2025, \apj, 987, 61,
  \dodoi{10.3847/1538-4357/add9a1}

\bibitem[{{Rosenberg} {et~al.}(2015){Rosenberg}, {van der Werf}, {Aalto},
  {Armus}, {Charmandaris}, {D{\'\i}az-Santos}, {Evans}, {Fischer}, {Gao},
  {Gonz{\'a}lez-Alfonso}, {Greve}, {Harris}, {Henkel}, {Israel}, {Isaak},
  {Kramer}, {Meijerink}, {Naylor}, {Sanders}, {Smith}, {Spaans}, {Spinoglio},
  {Stacey}, {Veenendaal}, {Veilleux}, {Walter}, {Wei{\ss}}, {Wiedner}, {van der
  Wiel}, \& {Xilouris}}]{R15}
{Rosenberg}, M.~J.~F., {van der Werf}, P.~P., {Aalto}, S., {et~al.} 2015, \apj,
  801, 72, \dodoi{10.1088/0004-637X/801/2/72}

\bibitem[{{Rowland} {et~al.}(2024){Rowland}, {Hodge}, {Bouwens}, {Mancera
  Pi{\~n}a}, {Hygate}, {Algera}, {Aravena}, {Bowler}, {da Cunha}, {Dayal},
  {Ferrara}, {Herard-Demanche}, {Inami}, {van Leeuwen}, {de Looze}, {Oesch},
  {Pallottini}, {Phillips}, {Rybak}, {Schouws}, {Smit}, {Sommovigo},
  {Stefanon}, \& {van der Werf}}]{RL24}
{Rowland}, L.~E., {Hodge}, J., {Bouwens}, R., {et~al.} 2024, \mnras, 535, 2068,
  \dodoi{10.1093/mnras/stae2217}

\bibitem[{{Rowland} {et~al.}(2025){Rowland}, {Stefanon}, {Bouwens}, {Hodge},
  {Algera}, {Fisher}, {Dayal}, {Pallottini}, {Stark}, {Heintz}, {Aravena},
  {Bowler}, {Cescon}, {Endsley}, {Ferrara}, {Gonzalez}, {Graziani}, {Gulis},
  {Herard-Demanche}, {Inami}, {Laza-Ramos}, {van Leeuwen}, {de Looze},
  {Nanayakkara}, {Oesch}, {Ormerod}, {Sartorio}, {Schouws}, {Smit},
  {Sommovigo}, {Toft}, {Weaver}, \& {van der Werf}}]{RL25}
{Rowland}, L.~E., {Stefanon}, M., {Bouwens}, R., {et~al.} 2025, arXiv e-prints,
  arXiv:2501.10559, \dodoi{10.48550/arXiv.2501.10559}

\bibitem[{{Rujopakarn} {et~al.}(2012){Rujopakarn}, {Rieke}, {Papovich},
  {Weiner}, {Rigby}, {Rex}, {Bian}, {Kuhn}, \& {Thompson}}]{RW12}
{Rujopakarn}, W., {Rieke}, G.~H., {Papovich}, C.~J., {et~al.} 2012, \apj, 755,
  168, \dodoi{10.1088/0004-637X/755/2/168}

\bibitem[{{Rybak} {et~al.}(2020{\natexlab{a}}){Rybak}, {Hodge}, {Vegetti}, {van
  der Werf}, {Andreani}, {Graziani}, \& {McKean}}]{RM20a}
{Rybak}, M., {Hodge}, J.~A., {Vegetti}, S., {et~al.} 2020{\natexlab{a}},
  \mnras, 494, 5542, \dodoi{10.1093/mnras/staa879}

\bibitem[{{Rybak} {et~al.}(2020{\natexlab{b}}){Rybak}, {Zavala}, {Hodge},
  {Casey}, \& {Werf}}]{RM20b}
{Rybak}, M., {Zavala}, J.~A., {Hodge}, J.~A., {Casey}, C.~M., \& {Werf}, P.
  v.~d. 2020{\natexlab{b}}, \apjl, 889, L11, \dodoi{10.3847/2041-8213/ab63de}

\bibitem[{{Rybak} {et~al.}(2019){Rybak}, {Calistro Rivera}, {Hodge}, {Smail},
  {Walter}, {van der Werf}, {da Cunha}, {Chen}, {Dannerbauer}, {Ivison},
  {Karim}, {Simpson}, {Swinbank}, \& {Wardlow}}]{RM19}
{Rybak}, M., {Calistro Rivera}, G., {Hodge}, J.~A., {et~al.} 2019, \apj, 876,
  112, \dodoi{10.3847/1538-4357/ab0e0f}

\bibitem[{{Rybak} {et~al.}(2023){Rybak}, {Lemsom}, {Lundgren}, {Zavala},
  {Hodge}, {de Breuck}, {Casey}, {Decarli}, {Torstensson}, {Wardlow}, \& {van
  der Werf}}]{RM23}
{Rybak}, M., {Lemsom}, L., {Lundgren}, A., {et~al.} 2023, Research Notes of the
  American Astronomical Society, 7, 188, \dodoi{10.3847/2515-5172/acf579}

\bibitem[{{Saintonge} {et~al.}(2013){Saintonge}, {Lutz}, {Genzel}, {Magnelli},
  {Nordon}, {Tacconi}, {Baker}, {Bandara}, {Berta}, {F{\"o}rster Schreiber},
  {Poglitsch}, {Sturm}, {Wuyts}, \& {Wuyts}}]{SA13}
{Saintonge}, A., {Lutz}, D., {Genzel}, R., {et~al.} 2013, \apj, 778, 2,
  \dodoi{10.1088/0004-637X/778/1/2}

\bibitem[{{Sanders} {et~al.}(2024){Sanders}, {Shapley}, {Topping}, {Reddy}, \&
  {Brammer}}]{sanders24}
{Sanders}, R.~L., {Shapley}, A.~E., {Topping}, M.~W., {Reddy}, N.~A., \&
  {Brammer}, G.~B. 2024, \apj, 962, 24, \dodoi{10.3847/1538-4357/ad15fc}

\bibitem[{{Sargsyan} {et~al.}(2012){Sargsyan}, {Lebouteiller}, {Weedman},
  {Spoon}, {Bernard-Salas}, {Engels}, {Stacey}, {Houck}, {Barry}, {Miles}, \&
  {Samsonyan}}]{sargsyan12}
{Sargsyan}, L., {Lebouteiller}, V., {Weedman}, D., {et~al.} 2012, \apj, 755,
  171, \dodoi{10.1088/0004-637X/755/2/171}

\bibitem[{{Sarkar} {et~al.}(2025){Sarkar}, {Chakraborty}, {Vogelsberger},
  {McDonald}, {Torrey}, {Garcia}, {Khullar}, {Ferland}, {Forman}, {Wolk},
  {Schneider}, {Bautz}, {Miller}, {Grant}, \& {ZuHone}}]{sarkar25}
{Sarkar}, A., {Chakraborty}, P., {Vogelsberger}, M., {et~al.} 2025, \apj, 978,
  136, \dodoi{10.3847/1538-4357/ad8f32}

\bibitem[{{Sawamura} {et~al.}(2025){Sawamura}, {Izumi}, {Nakanishi}, {Okuda},
  {Strauss}, {Imanishi}, {Matsuoka}, {Toba}, {Umehata}, {Hashimoto}, {Baba},
  {Goto}, {Kawaguchi}, {Kohno}, {Salak}, {Kawamuro}, {Iwasawa}, {Onoue}, {Lee},
  \& {Lee}}]{SM25a}
{Sawamura}, M., {Izumi}, T., {Nakanishi}, K., {et~al.} 2025, arXiv e-prints,
  arXiv:2502.16858, \dodoi{10.48550/arXiv.2502.16858}

\bibitem[{{Schaerer} {et~al.}(2015{\natexlab{a}}){Schaerer}, {Boone},
  {Zamojski}, {Staguhn}, {Dessauges-Zavadsky}, {Finkelstein}, \&
  {Combes}}]{SD15b}
{Schaerer}, D., {Boone}, F., {Zamojski}, M., {et~al.} 2015{\natexlab{a}}, \aap,
  574, A19, \dodoi{10.1051/0004-6361/201424649}

\bibitem[{{Schaerer} {et~al.}(2015{\natexlab{b}}){Schaerer}, {Boone}, {Jones},
  {Dessauges-Zavadsky}, {Sklias}, {Zamojski}, {Cava}, {Richard}, {Ellis},
  {Rawle}, {Egami}, \& {Combes}}]{SD15a}
{Schaerer}, D., {Boone}, F., {Jones}, T., {et~al.} 2015{\natexlab{b}}, \aap,
  576, L2, \dodoi{10.1051/0004-6361/201425542}

\bibitem[{{Schaerer} {et~al.}(2020){Schaerer}, {Ginolfi}, {B{\'e}thermin},
  {Fudamoto}, {Oesch}, {Le F{\`e}vre}, {Faisst}, {Capak}, {Cassata},
  {Silverman}, {Yan}, {Jones}, {Amorin}, {Bardelli}, {Boquien}, {Cimatti},
  {Dessauges-Zavadsky}, {Giavalisco}, {Hathi}, {Fujimoto}, {Ibar}, {Koekemoer},
  {Lagache}, {Lemaux}, {Loiacono}, {Maiolino}, {Narayanan}, {Morselli},
  {M{\'e}ndez-Hern{\`a}ndez}, {Pozzi}, {Riechers}, {Talia}, {Toft}, {Vallini},
  {Vergani}, {Zamorani}, \& {Zucca}}]{schaerer20}
{Schaerer}, D., {Ginolfi}, M., {B{\'e}thermin}, M., {et~al.} 2020, \aap, 643,
  A3, \dodoi{10.1051/0004-6361/202037617}

\bibitem[{{Schenker} {et~al.}(2012){Schenker}, {Stark}, {Ellis}, {Robertson},
  {Dunlop}, {McLure}, {Kneib}, \& {Richard}}]{SM12}
{Schenker}, M.~A., {Stark}, D.~P., {Ellis}, R.~S., {et~al.} 2012, \apj, 744,
  179, \dodoi{10.1088/0004-637X/744/2/179}

\bibitem[{{Scholtz} {et~al.}(2025){Scholtz}, {Curti}, {D'Eugenio}, {{\"U}bler},
  {Maiolino}, {Marconcini}, {Smit}, {Perna}, {Witstok}, {Arribas}, {B{\"o}ker},
  {Bunker}, {Carniani}, {Charlot}, {Cresci}, {Lamperti}, {Parlanti},
  {P{\'e}rez-Gonz{\'a}lez}, {Rodr{\'\i}guez Del Pino}, \& {Venturi}}]{SJ25}
{Scholtz}, J., {Curti}, M., {D'Eugenio}, F., {et~al.} 2025, \mnras, 539, 2463,
  \dodoi{10.1093/mnras/staf518}

\bibitem[{{Schouws} {et~al.}(2022){Schouws}, {Stefanon}, {Bouwens}, {Smit},
  {Hodge}, {Labb{\'e}}, {Algera}, {Boogaard}, {Carniani}, {Fudamoto},
  {Holwerda}, {Illingworth}, {Maiolino}, {Maseda}, {Oesch}, \& {van der
  Werf}}]{SS22}
{Schouws}, S., {Stefanon}, M., {Bouwens}, R., {et~al.} 2022, \apj, 928, 31,
  \dodoi{10.3847/1538-4357/ac4605}

\bibitem[{{Schouws} {et~al.}(2023){Schouws}, {Bouwens}, {Smit}, {Hodge},
  {Stefanon}, {Witstok}, {Hilhorst}, {Labb{\'e}}, {Algera}, {Boogaard},
  {Maseda}, {Oesch}, {R{\"o}ttgering}, \& {van der Werf}}]{SS23}
{Schouws}, S., {Bouwens}, R., {Smit}, R., {et~al.} 2023, \apj, 954, 103,
  \dodoi{10.3847/1538-4357/ace10c}

\bibitem[{{Schouws} {et~al.}(2025{\natexlab{a}}){Schouws}, {Bouwens}, {Algera},
  {Smit}, {Kumari}, {Rowland}, {van Leeuwen}, {Sommovigo}, {Ferrara}, {Oesch},
  {Ormerod}, {Stefanon}, {Herard-Demanche}, {Hodge}, {Fudamoto},
  {R{\"o}ttgering}, \& {van der Werf}}]{SS25a}
{Schouws}, S., {Bouwens}, R.~J., {Algera}, H., {et~al.} 2025{\natexlab{a}},
  arXiv e-prints, arXiv:2502.01610, \dodoi{10.48550/arXiv.2502.01610}

\bibitem[{{Schouws} {et~al.}(2025{\natexlab{b}}){Schouws}, {Bouwens},
  {Ormerod}, {Smit}, {Algera}, {Sommovigo}, {Hodge}, {Ferrara}, {Oesch},
  {Rowland}, {van Leeuwen}, {Stefanon}, {Herard-Demanche}, {Fudamoto},
  {R{\"o}ttgering}, \& {van der Werf}}]{SS25b}
{Schouws}, S., {Bouwens}, R.~J., {Ormerod}, K., {et~al.} 2025{\natexlab{b}},
  \apj, 988, 19, \dodoi{10.3847/1538-4357/adbf1b}

\bibitem[{{Schreiber} {et~al.}(2018){Schreiber}, {Labb{\'e}}, {Glazebrook},
  {Bekiaris}, {Papovich}, {Costa}, {Elbaz}, {Kacprzak}, {Nanayakkara}, {Oesch},
  {Pannella}, {Spitler}, {Straatman}, {Tran}, \& {Wang}}]{SC18}
{Schreiber}, C., {Labb{\'e}}, I., {Glazebrook}, K., {et~al.} 2018, \aap, 611,
  A22, \dodoi{10.1051/0004-6361/201731917}

\bibitem[{{Schreiber} {et~al.}(2021){Schreiber}, {Glazebrook}, {Papovich},
  {D{\'\i}az-Santos}, {Verma}, {Elbaz}, {Kacprzak}, {Nanayakkara}, {Oesch},
  {Pannella}, {Spitler}, {Straatman}, {Tran}, \& {Wang}}]{SC21}
{Schreiber}, C., {Glazebrook}, K., {Papovich}, C., {et~al.} 2021, \aap, 646,
  A68, \dodoi{10.1051/0004-6361/201936460}

\bibitem[{{Shao} {et~al.}(2017){Shao}, {Wang}, {Jones}, {Carilli}, {Walter},
  {Fan}, {Riechers}, {Bertoldi}, {Wagg}, {Strauss}, {Omont}, {Cox}, {Jiang},
  {Narayanan}, \& {Menten}}]{SY17}
{Shao}, Y., {Wang}, R., {Jones}, G.~C., {et~al.} 2017, \apj, 845, 138,
  \dodoi{10.3847/1538-4357/aa826c}

\bibitem[{{Shao} {et~al.}(2019){Shao}, {Wang}, {Carilli}, {Wagg}, {Walter},
  {Li}, {Fan}, {Jiang}, {Riechers}, {Bertoldi}, {Strauss}, {Cox}, {Omont}, \&
  {Menten}}]{SY19}
{Shao}, Y., {Wang}, R., {Carilli}, C.~L., {et~al.} 2019, \apj, 876, 99,
  \dodoi{10.3847/1538-4357/ab133d}

\bibitem[{{Shao} {et~al.}(2022){Shao}, {Wang}, {Weiss}, {Wagg}, {Carilli},
  {Strauss}, {Walter}, {Cox}, {Fan}, {Menten}, {Narayanan}, {Riechers},
  {Bertoldi}, {Omont}, \& {Jiang}}]{SY22}
{Shao}, Y., {Wang}, R., {Weiss}, A., {et~al.} 2022, \aap, 668, A121,
  \dodoi{10.1051/0004-6361/202244610}

\bibitem[{{Sharon} {et~al.}(2019){Sharon}, {Tagore}, {Baker}, {Rivera},
  {Keeton}, {Lutz}, {Genzel}, {Wilner}, {Hicks}, {Allam}, \& {Tucker}}]{SC19}
{Sharon}, C.~E., {Tagore}, A.~S., {Baker}, A.~J., {et~al.} 2019, \apj, 879, 52,
  \dodoi{10.3847/1538-4357/ab22b9}

\bibitem[{{Simpson} {et~al.}(2020){Simpson}, {Smail},
  {Dudzevi{\v{c}}i{\={u}}t{\.{e}}}, {Matsuda}, {Hsieh}, {Wang}, {Swinbank},
  {Stach}, {An}, {Birkin}, {Ao}, {Bunker}, {Chapman}, {Chen}, {Coppin},
  {Ikarashi}, {Ivison}, {Mitsuhashi}, {Saito}, {Umehata}, {Wang}, \&
  {Zhao}}]{SJ20}
{Simpson}, J.~M., {Smail}, I., {Dudzevi{\v{c}}i{\={u}}t{\.{e}}}, U., {et~al.}
  2020, \mnras, 495, 3409, \dodoi{10.1093/mnras/staa1345}

\bibitem[{{Smit} {et~al.}(2018){Smit}, {Bouwens}, {Carniani}, {Oesch},
  {Labb{\'e}}, {Illingworth}, {van der Werf}, {Bradley}, {Gonzalez}, {Hodge},
  {Holwerda}, {Maiolino}, \& {Zheng}}]{SR18}
{Smit}, R., {Bouwens}, R.~J., {Carniani}, S., {et~al.} 2018, \nat, 553, 178,
  \dodoi{10.1038/nature24631}

\bibitem[{{Smol{\v{c}}i{\'c}} {et~al.}(2015){Smol{\v{c}}i{\'c}}, {Karim},
  {Miettinen}, {Novak}, {Magnelli}, {Riechers}, {Schinnerer}, {Capak}, {Bondi},
  {Ciliegi}, {Aravena}, {Bertoldi}, {Bourke}, {Banfield}, {Carilli}, {Civano},
  {Ilbert}, {Intema}, {Le F{\`e}vre}, {Finoguenov}, {Hallinan}, {Kl{\"o}ckner},
  {Koekemoer}, {Laigle}, {Masters}, {McCracken}, {Mooley}, {Murphy},
  {Navarette}, {Salvato}, {Sargent}, {Sheth}, {Toft}, \& {Zamorani}}]{SV15}
{Smol{\v{c}}i{\'c}}, V., {Karim}, A., {Miettinen}, O., {et~al.} 2015, \aap,
  576, A127, \dodoi{10.1051/0004-6361/201424996}

\bibitem[{{Solimano} {et~al.}(2024){Solimano}, {Gonz{\'a}lez-L{\'o}pez},
  {Aravena}, {Herrera-Camus}, {De Looze}, {F{\"o}rster Schreiber}, {Spilker},
  {Tadaki}, {Assef}, {Barcos-Mu{\~n}oz}, {Davies}, {D{\'\i}az-Santos},
  {Ferrara}, {Fisher}, {Guaita}, {Ikeda}, {Johnston}, {Lutz}, {Mitsuhashi},
  {Moya-Sierralta}, {Rela{\~n}o}, {Naab}, {Posses}, {Telikova}, {{\"U}bler},
  {van der Giessen}, {Veilleux}, \& {Villanueva}}]{SM24}
{Solimano}, M., {Gonz{\'a}lez-L{\'o}pez}, J., {Aravena}, M., {et~al.} 2024,
  \aap, 689, A145, \dodoi{10.1051/0004-6361/202449192}

\bibitem[{{Solimano} {et~al.}(2025){Solimano}, {Gonz{\'a}lez-L{\'o}pez},
  {Aravena}, {Alcalde Pampliega}, {Assef}, {B{\'e}thermin}, {Boquien},
  {Bovino}, {Casey}, {Cassata}, {da Cunha}, {Davies}, {De Looze}, {Ding},
  {D{\'\i}az-Santos}, {Faisst}, {Ferrara}, {Fisher}, {F{\"o}rster-Schreiber},
  {Fujimoto}, {Ginolfi}, {Gruppioni}, {Guaita}, {Hathi}, {Herrera-Camus},
  {Ibar}, {Inami}, {Jones}, {Koekemoer}, {Lee}, {Li}, {Liu}, {Liu}, {Molina},
  {Ogle}, {Posses}, {Pozzi}, {Rela{\~n}o}, {Riechers}, {Romano}, {Spilker},
  {Sulzenauer}, {Telikova}, {Vallini}, {Vasan}, {Veilleux}, {Vergani},
  {Villanueva}, {Wang}, {Yan}, \& {Zamorani}}]{SM25b}
---. 2025, \aap, 693, A70, \dodoi{10.1051/0004-6361/202451551}

\bibitem[{{Sommariva} {et~al.}(2012){Sommariva}, {Mannucci}, {Cresci},
  {Maiolino}, {Marconi}, {Nagao}, {Baroni}, \& {Grazian}}]{SV12}
{Sommariva}, V., {Mannucci}, F., {Cresci}, G., {et~al.} 2012, \aap, 539, A136,
  \dodoi{10.1051/0004-6361/201118134}

\bibitem[{{Sommovigo} {et~al.}(2022){Sommovigo}, {Ferrara}, {Pallottini},
  {Dayal}, {Bouwens}, {Smit}, {da Cunha}, {De Looze}, {Bowler}, {Hodge},
  {Inami}, {Oesch}, {Endsley}, {Gonzalez}, {Schouws}, {Stark}, {Stefanon},
  {Aravena}, {Graziani}, {Riechers}, {Schneider}, {van der Werf}, {Algera},
  {Barrufet}, {Fudamoto}, {Hygate}, {Labb{\'e}}, {Li}, {Nanayakkara}, \&
  {Topping}}]{SL22}
{Sommovigo}, L., {Ferrara}, A., {Pallottini}, A., {et~al.} 2022, \mnras, 513,
  3122, \dodoi{10.1093/mnras/stac302}

\bibitem[{{Speagle} {et~al.}(2014){Speagle}, {Steinhardt}, {Capak}, \&
  {Silverman}}]{speagle14}
{Speagle}, J.~S., {Steinhardt}, C.~L., {Capak}, P.~L., \& {Silverman}, J.~D.
  2014, \apjs, 214, 15, \dodoi{10.1088/0067-0049/214/2/15}

\bibitem[{{Spinoglio} {et~al.}(2015){Spinoglio}, {Pereira-Santaella}, {Dasyra},
  {Calzoletti}, {Malkan}, {Tommasin}, \& {Busquet}}]{S15}
{Spinoglio}, L., {Pereira-Santaella}, M., {Dasyra}, K.~M., {et~al.} 2015, \apj,
  799, 21, \dodoi{10.1088/0004-637X/799/1/21}

\bibitem[{{Spinoglio} {et~al.}(2022){Spinoglio}, {Fern{\'a}ndez-Ontiveros},
  {Malkan}, {Kumar}, {Pereira-Santaella}, {P{\'e}rez-D{\'\i}az},
  {P{\'e}rez-Montero}, {Krabbe}, {Vacca}, {Colditz}, \& {Fischer}}]{S22}
{Spinoglio}, L., {Fern{\'a}ndez-Ontiveros}, J.~A., {Malkan}, M.~A., {et~al.}
  2022, \apj, 926, 55, \dodoi{10.3847/1538-4357/ac37b7}

\bibitem[{{Stacey} {et~al.}(1991){Stacey}, {Geis}, {Genzel}, {Lugten},
  {Poglitsch}, {Sternberg}, \& {Townes}}]{stacey91}
{Stacey}, G.~J., {Geis}, N., {Genzel}, R., {et~al.} 1991, \apj, 373, 423,
  \dodoi{10.1086/170062}

\bibitem[{{Stacey} {et~al.}(2010){Stacey}, {Hailey-Dunsheath}, {Ferkinhoff},
  {Nikola}, {Parshley}, {Benford}, {Staguhn}, \& {Fiolet}}]{SG10}
{Stacey}, G.~J., {Hailey-Dunsheath}, S., {Ferkinhoff}, C., {et~al.} 2010, \apj,
  724, 957, \dodoi{10.1088/0004-637X/724/2/957}

\bibitem[{{Stacey} {et~al.}(2018){Stacey}, {McKean}, {Robertson}, {Ivison},
  {Isaak}, {Schleicher}, {van der Werf}, {Baan}, {Berciano Alba}, {Garrett}, \&
  {Loenen}}]{SH18}
{Stacey}, H.~R., {McKean}, J.~P., {Robertson}, N.~C., {et~al.} 2018, \mnras,
  476, 5075, \dodoi{10.1093/mnras/sty458}

\bibitem[{{Stark} {et~al.}(2015){Stark}, {Richard}, {Charlot}, {Cl{\'e}ment},
  {Ellis}, {Siana}, {Robertson}, {Schenker}, {Gutkin}, \& {Wofford}}]{SD15c}
{Stark}, D.~P., {Richard}, J., {Charlot}, S., {et~al.} 2015, \mnras, 450, 1846,
  \dodoi{10.1093/mnras/stv688}

\bibitem[{{Stiavelli} {et~al.}(2023){Stiavelli}, {Morishita}, {Chiaberge},
  {Grillo}, {Leethochawalit}, {Rosati}, {Schuldt}, {Trenti}, \& {Treu}}]{SM23}
{Stiavelli}, M., {Morishita}, T., {Chiaberge}, M., {et~al.} 2023, \apjl, 957,
  L18, \dodoi{10.3847/2041-8213/ad0159}

\bibitem[{{Strandet} {et~al.}(2016){Strandet}, {Weiss}, {Vieira}, {de Breuck},
  {Aguirre}, {Aravena}, {Ashby}, {B{\'e}thermin}, {Bradford}, {Carlstrom},
  {Chapman}, {Crawford}, {Everett}, {Fassnacht}, {Furstenau}, {Gonzalez},
  {Greve}, {Gullberg}, {Hezaveh}, {Kamenetzky}, {Litke}, {Ma}, {Malkan},
  {Marrone}, {Menten}, {Murphy}, {Nadolski}, {Rotermund}, {Spilker}, {Stark},
  \& {Welikala}}]{SM16}
{Strandet}, M.~L., {Weiss}, A., {Vieira}, J.~D., {et~al.} 2016, \apj, 822, 80,
  \dodoi{10.3847/0004-637X/822/2/80}

\bibitem[{{Strandet} {et~al.}(2017){Strandet}, {Weiss}, {De Breuck}, {Marrone},
  {Vieira}, {Aravena}, {Ashby}, {B{\'e}thermin}, {Bothwell}, {Bradford},
  {Carlstrom}, {Chapman}, {Cunningham}, {Chen}, {Fassnacht}, {Gonzalez},
  {Greve}, {Gullberg}, {Hayward}, {Hezaveh}, {Litke}, {Ma}, {Malkan}, {Menten},
  {Miller}, {Murphy}, {Narayanan}, {Phadke}, {Rotermund}, {Spilker}, \&
  {Sreevani}}]{SM17}
{Strandet}, M.~L., {Weiss}, A., {De Breuck}, C., {et~al.} 2017, \apjl, 842,
  L15, \dodoi{10.3847/2041-8213/aa74b0}

\bibitem[{{Sturm} {et~al.}(2010){Sturm}, {Verma}, {Graci{\'a}-Carpio},
  {Hailey-Dunsheath}, {Contursi}, {Fischer}, {Gonz{\'a}lez-Alfonso},
  {Poglitsch}, {Sternberg}, {Genzel}, {Lutz}, {Tacconi}, {Christopher}, \& {de
  Jong}}]{SE10}
{Sturm}, E., {Verma}, A., {Graci{\'a}-Carpio}, J., {et~al.} 2010, \aap, 518,
  L36, \dodoi{10.1051/0004-6361/201014560}

\bibitem[{{Sugahara} {et~al.}(2021){Sugahara}, {Inoue}, {Hashimoto},
  {Yamanaka}, {Fujimoto}, {Tamura}, {Matsuo}, {Binggeli}, \&
  {Zackrisson}}]{SY21}
{Sugahara}, Y., {Inoue}, A.~K., {Hashimoto}, T., {et~al.} 2021, \apj, 923, 5,
  \dodoi{10.3847/1538-4357/ac2a36}

\bibitem[{{Sugahara} {et~al.}(2025){Sugahara}, {{\'A}lvarez-M{\'a}rquez},
  {Hashimoto}, {Colina}, {Inoue}, {Costantin}, {Fudamoto}, {Mawatari}, {Ren},
  {Arribas}, {Bakx}, {Blanco-Prieto}, {Ceverino}, {Crespo G{\'o}mez},
  {Hagimoto}, {Hashigaya}, {Marques-Chaves}, {Matsuo}, {Nakazato},
  {Pereira-Santaella}, {Tamura}, {Usui}, \& {Yoshida}}]{SY25}
{Sugahara}, Y., {{\'A}lvarez-M{\'a}rquez}, J., {Hashimoto}, T., {et~al.} 2025,
  \apj, 981, 135, \dodoi{10.3847/1538-4357/adb02a}

\bibitem[{{Sun} {et~al.}(2024){Sun}, {Helton}, {Egami}, {Hainline}, {Rieke},
  {Willmer}, {Eisenstein}, {Johnson}, {Rieke}, {Robertson}, {Tacchella},
  {Alberts}, {Baker}, {Bhatawdekar}, {Boyett}, {Bunker}, {Charlot}, {Chen},
  {Chevallard}, {Curtis-Lake}, {Danhaive}, {DeCoursey}, {Ji}, {Lyu},
  {Maiolino}, {Rujopakarn}, {Sandles}, {Shivaei}, {{\"U}bler}, {Willott}, \&
  {Witstok}}]{SF24}
{Sun}, F., {Helton}, J.~M., {Egami}, E., {et~al.} 2024, \apj, 961, 69,
  \dodoi{10.3847/1538-4357/ad07e3}

\bibitem[{{Sun} {et~al.}(2025){Sun}, {Yang}, {Wang}, {Eisenstein}, {Decarli},
  {Fan}, {Rieke}, {Ba{\~n}ados}, {Bosman}, {Cai}, {Champagne}, {Colina},
  {D'Eugenio}, {Fudamoto}, {Li}, {Lin}, {Liu}, {Lyu}, {Mazzucchelli}, {Jin},
  {Jun}, {Wu}, \& {Zhang}}]{SF25}
{Sun}, F., {Yang}, J., {Wang}, F., {et~al.} 2025, arXiv e-prints,
  arXiv:2506.06418, \dodoi{10.48550/arXiv.2506.06418}

\bibitem[{{Sutter} {et~al.}(2019){Sutter}, {Dale}, {Croxall}, {Pelligrini},
  {Smith}, {Appleton}, {Beir{\~a}o}, {Bolatto}, {Calzetti}, {Crocker}, {De
  Looze}, {Draine}, {Galametz}, {Groves}, {Helou}, {Herrera-Camus}, {Hunt},
  {Kennicutt}, {Roussel}, \& {Wolfire}}]{S19}
{Sutter}, J., {Dale}, D.~A., {Croxall}, K.~V., {et~al.} 2019, \apj, 886, 60,
  \dodoi{10.3847/1538-4357/ab4da5}

\bibitem[{{Sutter} {et~al.}(2021){Sutter}, {Dale}, {Sandstrom}, {Smith},
  {Bolatto}, {Boquien}, {Calzetti}, {Croxall}, {De Looze}, {Galametz},
  {Groves}, {Helou}, {Herrera-Camus}, {Hunt}, {Kennicutt}, {Pelligrini},
  {Wilson}, \& {Wolfire}}]{sutter21}
{Sutter}, J., {Dale}, D.~A., {Sandstrom}, K., {et~al.} 2021, \mnras, 503, 911,
  \dodoi{10.1093/mnras/stab490}

\bibitem[{{Swinbank} {et~al.}(2012){Swinbank}, {Karim}, {Smail}, {Hodge},
  {Walter}, {Bertoldi}, {Biggs}, {de Breuck}, {Chapman}, {Coppin}, {Cox},
  {Danielson}, {Dannerbauer}, {Ivison}, {Greve}, {Knudsen}, {Menten},
  {Simpson}, {Schinnerer}, {Wardlow}, {Wei{\ss}}, \& {van der Werf}}]{SA12}
{Swinbank}, A.~M., {Karim}, A., {Smail}, I., {et~al.} 2012, \mnras, 427, 1066,
  \dodoi{10.1111/j.1365-2966.2012.22048.x}

\bibitem[{{Symeonidis} \& {Page}(2018)}]{symeonidis18}
{Symeonidis}, M., \& {Page}, M.~J. 2018, \mnras, 479, L91,
  \dodoi{10.1093/mnrasl/sly105}

\bibitem[{{Tacchella} {et~al.}(2023){Tacchella}, {Eisenstein}, {Hainline},
  {Johnson}, {Baker}, {Helton}, {Robertson}, {Suess}, {Chen}, {Nelson},
  {Pusk{\'a}s}, {Sun}, {Alberts}, {Egami}, {Hausen}, {Rieke}, {Rieke},
  {Shivaei}, {Williams}, {Willmer}, {Bunker}, {Cameron}, {Carniani}, {Charlot},
  {Curti}, {Curtis-Lake}, {Looser}, {Maiolino}, {Maseda}, {Rawle}, {Rix},
  {Smit}, {{\"U}bler}, {Willott}, {Witstok}, {Baum}, {Bhatawdekar}, {Boyett},
  {Danhaive}, {de Graaff}, {Endsley}, {Ji}, {Lyu}, {Sandles}, {Saxena},
  {Scholtz}, {Topping}, \& {Whitler}}]{TS23}
{Tacchella}, S., {Eisenstein}, D.~J., {Hainline}, K., {et~al.} 2023, \apj, 952,
  74, \dodoi{10.3847/1538-4357/acdbc6}

\bibitem[{{Tadaki} {et~al.}(2019){Tadaki}, {Iono}, {Hatsukade}, {Kohno}, {Lee},
  {Matsuda}, {Michiyama}, {Nakanishi}, {Nagao}, {Saito}, {Tamura}, {Ueda}, \&
  {Umehata}}]{TK19}
{Tadaki}, K.-i., {Iono}, D., {Hatsukade}, B., {et~al.} 2019, \apj, 876, 1,
  \dodoi{10.3847/1538-4357/ab1415}

\bibitem[{{Tadaki} {et~al.}(2022){Tadaki}, {Tsujita}, {Tamura}, {Kohno},
  {Hatsukade}, {Iono}, {Lee}, {Matsuda}, {Michiyama}, {Nagao}, {Nakanishi},
  {Nishimura}, {Saito}, {Umehata}, \& {Zavala}}]{TK22}
{Tadaki}, K.-i., {Tsujita}, A., {Tamura}, Y., {et~al.} 2022, \pasj, 74, L9,
  \dodoi{10.1093/pasj/psac018}

\bibitem[{{Tamura} {et~al.}(2020){Tamura}, {Mawatari}, {Hashimoto}, {Inoue},
  {Zackrissonm}, {Christensen}, {Binggeli}, {Matsuda}, {Matsuo}, {Takeuchi},
  {Asano}, {Sunaga}, {Shimizu}, {Okamoto}, {Yoshida}, {Lee}, {Shibuya},
  {Taniguchi}, {Umehata}, {Hatsukade}, {Kohno}, \& {Ota}}]{TY20}
{Tamura}, Y., {Mawatari}, K., {Hashimoto}, T., {et~al.} 2020, in IAU Symposium,
  Vol. 341, Panchromatic Modelling with Next Generation Facilities, ed.
  M.~{Boquien}, E.~{Lusso}, C.~{Gruppioni}, \& P.~{Tissera}, 211--215,
  \dodoi{10.1017/S1743921319002436}

\bibitem[{{Tamura} {et~al.}(2023){Tamura}, {C. Bakx}, {Inoue}, {Hashimoto},
  {Tokuoka}, {Imamura}, {Hatsukade}, {Lee}, {Moriwaki}, {Okamoto}, {Ota},
  {Umehata}, {Yoshida}, {Zackrisson}, {Hagimoto}, {Matsuo}, {Shimizu},
  {Sugahara}, \& {Takeuchi}}]{TY23}
{Tamura}, Y., {C. Bakx}, T. J.~L., {Inoue}, A.~K., {et~al.} 2023, \apj, 952, 9,
  \dodoi{10.3847/1538-4357/acd637}

\bibitem[{{Teplitz} {et~al.}(2000){Teplitz}, {McLean}, {Becklin}, {Figer},
  {Gilbert}, {Graham}, {Larkin}, {Levenson}, \& {Wilcox}}]{TH00}
{Teplitz}, H.~I., {McLean}, I.~S., {Becklin}, E.~E., {et~al.} 2000, \apjl, 533,
  L65, \dodoi{10.1086/312595}

\bibitem[{{Tielens} \& {Hollenbach}(1985)}]{tielens85}
{Tielens}, A.~G.~G.~M., \& {Hollenbach}, D. 1985, \apj, 291, 722,
  \dodoi{10.1086/163111}

\bibitem[{{Tokuoka} {et~al.}(2022){Tokuoka}, {Inoue}, {Hashimoto}, {Ellis},
  {Laporte}, {Sugahara}, {Matsuo}, {Tamura}, {Fudamoto}, {Moriwaki},
  {Roberts-Borsani}, {Shimizu}, {Yamanaka}, {Yoshida}, {Zackrisson}, \&
  {Zheng}}]{TT22}
{Tokuoka}, T., {Inoue}, A.~K., {Hashimoto}, T., {et~al.} 2022, \apjl, 933, L19,
  \dodoi{10.3847/2041-8213/ac7447}

\bibitem[{{Tripodi} {et~al.}(2023{\natexlab{a}}){Tripodi}, {Lelli}, {Feruglio},
  {Fiore}, {Fontanot}, {Bischetti}, \& {Maiolino}}]{TR23a}
{Tripodi}, R., {Lelli}, F., {Feruglio}, C., {et~al.} 2023{\natexlab{a}}, \aap,
  671, A44, \dodoi{10.1051/0004-6361/202245202}

\bibitem[{{Tripodi} {et~al.}(2022){Tripodi}, {Feruglio}, {Fiore}, {Bischetti},
  {D'Odorico}, {Carniani}, {Cristiani}, {Gallerani}, {Maiolino}, {Marconi},
  {Pallottini}, {Piconcelli}, {Vallini}, \& {Zana}}]{TR22}
{Tripodi}, R., {Feruglio}, C., {Fiore}, F., {et~al.} 2022, \aap, 665, A107,
  \dodoi{10.1051/0004-6361/202243920}

\bibitem[{{Tripodi} {et~al.}(2023{\natexlab{b}}){Tripodi}, {Feruglio},
  {Kemper}, {Civano}, {Costa}, {Elvis}, {Bischetti}, {Carniani}, {Di Mascia},
  {D'Odorico}, {Fiore}, {Gallerani}, {Ginolfi}, {Maiolino}, {Piconcelli},
  {Valiante}, \& {Zappacosta}}]{TR23b}
{Tripodi}, R., {Feruglio}, C., {Kemper}, F., {et~al.} 2023{\natexlab{b}},
  \apjl, 946, L45, \dodoi{10.3847/2041-8213/acc58d}

\bibitem[{{Tsukui} {et~al.}(2023){Tsukui}, {Wisnioski}, {Krumholz}, \&
  {Battisti}}]{TT23}
{Tsukui}, T., {Wisnioski}, E., {Krumholz}, M.~R., \& {Battisti}, A. 2023,
  \mnras, 523, 4654, \dodoi{10.1093/mnras/stad1464}

\bibitem[{{{\"U}bler} {et~al.}(2023){{\"U}bler}, {Maiolino}, {Curtis-Lake},
  {P{\'e}rez-Gonz{\'a}lez}, {Curti}, {Perna}, {Arribas}, {Charlot}, {Marshall},
  {D'Eugenio}, {Scholtz}, {Bunker}, {Carniani}, {Ferruit}, {Jakobsen}, {Rix},
  {Rodr{\'\i}guez Del Pino}, {Willott}, {Boeker}, {Cresci}, {Jones}, {Kumari},
  \& {Rawle}}]{UH23}
{{\"U}bler}, H., {Maiolino}, R., {Curtis-Lake}, E., {et~al.} 2023, \aap, 677,
  A145, \dodoi{10.1051/0004-6361/202346137}

\bibitem[{{{\"U}bler} {et~al.}(2024){{\"U}bler}, {Maiolino},
  {P{\'e}rez-Gonz{\'a}lez}, {D'Eugenio}, {Perna}, {Curti}, {Arribas}, {Bunker},
  {Carniani}, {Charlot}, {Rodr{\'\i}guez Del Pino}, {Baker}, {B{\"o}ker},
  {Cresci}, {Dunlop}, {Grogin}, {Jones}, {Kumari}, {Lamperti}, {Laporte},
  {Marshall}, {Mazzolari}, {Parlanti}, {Rawle}, {Scholtz}, {Venturi}, \&
  {Witstok}}]{UH24a}
{{\"U}bler}, H., {Maiolino}, R., {P{\'e}rez-Gonz{\'a}lez}, P.~G., {et~al.}
  2024, \mnras, 531, 355, \dodoi{10.1093/mnras/stae943}

\bibitem[{{Uematsu} {et~al.}(2023){Uematsu}, {Ueda}, {Kohno}, {Yamada}, {Toba},
  {Fujimoto}, {Hatsukade}, {Umehata}, {Espada}, {Sun}, {Magdis}, {Kokorev}, \&
  {Ao}}]{uematsu23}
{Uematsu}, R., {Ueda}, Y., {Kohno}, K., {et~al.} 2023, \apj, 945, 121,
  \dodoi{10.3847/1538-4357/acb4e9}

\bibitem[{{Umehata} {et~al.}(2017){Umehata}, {Matsuda}, {Tamura}, {Kohno},
  {Smail}, {Ivison}, {Steidel}, {Chapman}, {Geach}, {Hayes}, {Nagao}, {Ao},
  {Kawabe}, {Yun}, {Hatsukade}, {Kubo}, {Kato}, {Saito}, {Ikarashi},
  {Nakanishi}, {Lee}, {Izumi}, {Mori}, \& {Ouchi}}]{UH17}
{Umehata}, H., {Matsuda}, Y., {Tamura}, Y., {et~al.} 2017, \apjl, 834, L16,
  \dodoi{10.3847/2041-8213/834/2/L16}

\bibitem[{{Umehata} {et~al.}(2021){Umehata}, {Smail}, {Steidel}, {Hayes},
  {Scott}, {Swinbank}, {Ivison}, {Nagao}, {Kubo}, {Nakanishi}, {Matsuda},
  {Ikarashi}, {Tamura}, \& {Geach}}]{UH21}
{Umehata}, H., {Smail}, I., {Steidel}, C.~C., {et~al.} 2021, \apj, 918, 69,
  \dodoi{10.3847/1538-4357/ac1106}

\bibitem[{{Uzgil} {et~al.}(2016){Uzgil}, {Bradford}, {Hailey-Dunsheath},
  {Maloney}, \& {Aguirre}}]{UB16}
{Uzgil}, B.~D., {Bradford}, C.~M., {Hailey-Dunsheath}, S., {Maloney}, P.~R., \&
  {Aguirre}, J.~E. 2016, \apj, 832, 209, \dodoi{10.3847/0004-637X/832/2/209}

\bibitem[{{Valentino} {et~al.}(2022){Valentino}, {Brammer}, {Fujimoto},
  {Heintz}, {Weaver}, {Strait}, {Gould}, {Mason}, {Watson}, {Laursen}, \&
  {Toft}}]{VF22}
{Valentino}, F., {Brammer}, G., {Fujimoto}, S., {et~al.} 2022, \apjl, 929, L9,
  \dodoi{10.3847/2041-8213/ac62cc}

\bibitem[{{Valtchanov} {et~al.}(2011){Valtchanov}, {Virdee}, {Ivison},
  {Swinyard}, {van der Werf}, {Rigopoulou}, {da Cunha}, {Lupu}, {Benford},
  {Riechers}, {Smail}, {Jarvis}, {Pearson}, {Gomez}, {Hopwood}, {Altieri},
  {Birkinshaw}, {Coia}, {Conversi}, {Cooray}, {de Zotti}, {Dunne}, {Frayer},
  {Leeuw}, {Marston}, {Negrello}, {Portal}, {Scott}, {Thompson}, {Vaccari},
  {Baes}, {Clements}, {Micha{\l}owski}, {Dannerbauer}, {Serjeant}, {Auld},
  {Buttiglione}, {Cava}, {Dariush}, {Dye}, {Eales}, {Fritz}, {Ibar}, {Maddox},
  {Pascale}, {Pohlen}, {Rigby}, {Rodighiero}, {Smith}, {Temi}, {Carpenter},
  {Bolatto}, {Gurwell}, \& {Vieira}}]{VI11}
{Valtchanov}, I., {Virdee}, J., {Ivison}, R.~J., {et~al.} 2011, \mnras, 415,
  3473, \dodoi{10.1111/j.1365-2966.2011.18959.x}

\bibitem[{{van Leeuwen} {et~al.}(2024){van Leeuwen}, {Bouwens}, {van der Werf},
  {Hodge}, {Schouws}, {Stefanon}, {Algera}, {Aravena}, {Boogaard}, {Bowler},
  {da Cunha}, {Dayal}, {Decarli}, {Gonzalez}, {Inami}, {de Looze}, {Sommovigo},
  {Venemans}, {Walter}, {Barrufet}, {Ferrara}, {Graziani}, {Hygate}, {Oesch},
  {Palla}, {Rowland}, \& {Schneider}}]{vI24}
{van Leeuwen}, I.~F., {Bouwens}, R.~J., {van der Werf}, P.~P., {et~al.} 2024,
  \mnras, 534, 2062, \dodoi{10.1093/mnras/stae2171}

\bibitem[{{Vanzella} {et~al.}(2011){Vanzella}, {Pentericci}, {Fontana},
  {Grazian}, {Castellano}, {Boutsia}, {Cristiani}, {Dickinson}, {Gallozzi},
  {Giallongo}, {Giavalisco}, {Maiolino}, {Moorwood}, {Paris}, \&
  {Santini}}]{VE11}
{Vanzella}, E., {Pentericci}, L., {Fontana}, A., {et~al.} 2011, \apjl, 730,
  L35, \dodoi{10.1088/2041-8205/730/2/L35}

\bibitem[{{Venemans} {et~al.}(2019){Venemans}, {Neeleman}, {Walter}, {Novak},
  {Decarli}, {Hennawi}, \& {Rix}}]{VB19}
{Venemans}, B.~P., {Neeleman}, M., {Walter}, F., {et~al.} 2019, \apjl, 874,
  L30, \dodoi{10.3847/2041-8213/ab11cc}

\bibitem[{{Venemans} {et~al.}(2016){Venemans}, {Walter}, {Zschaechner},
  {Decarli}, {De Rosa}, {Findlay}, {McMahon}, \& {Sutherland}}]{VB16}
{Venemans}, B.~P., {Walter}, F., {Zschaechner}, L., {et~al.} 2016, \apj, 816,
  37, \dodoi{10.3847/0004-637X/816/1/37}

\bibitem[{{Venemans} {et~al.}(2012){Venemans}, {McMahon}, {Walter}, {Decarli},
  {Cox}, {Neri}, {Hewett}, {Mortlock}, {Simpson}, \& {Warren}}]{VB12}
{Venemans}, B.~P., {McMahon}, R.~G., {Walter}, F., {et~al.} 2012, \apjl, 751,
  L25, \dodoi{10.1088/2041-8205/751/2/L25}

\bibitem[{{Venemans} {et~al.}(2017{\natexlab{a}}){Venemans}, {Walter},
  {Decarli}, {Ba{\~n}ados}, {Hodge}, {Hewett}, {McMahon}, {Mortlock}, \&
  {Simpson}}]{VB17a}
{Venemans}, B.~P., {Walter}, F., {Decarli}, R., {et~al.} 2017{\natexlab{a}},
  \apj, 837, 146, \dodoi{10.3847/1538-4357/aa62ac}

\bibitem[{{Venemans} {et~al.}(2017{\natexlab{b}}){Venemans}, {Walter},
  {Decarli}, {Ba{\~n}ados}, {Carilli}, {Winters}, {Schuster}, {da Cunha},
  {Fan}, {Farina}, {Mazzucchelli}, {Rix}, \& {Weiss}}]{VB17b}
---. 2017{\natexlab{b}}, \apjl, 851, L8, \dodoi{10.3847/2041-8213/aa943a}

\bibitem[{{Venemans} {et~al.}(2020){Venemans}, {Walter}, {Neeleman}, {Novak},
  {Otter}, {Decarli}, {Ba{\~n}ados}, {Drake}, {Farina}, {Kaasinen},
  {Mazzucchelli}, {Carilli}, {Fan}, {Rix}, \& {Wang}}]{VB20}
{Venemans}, B.~P., {Walter}, F., {Neeleman}, M., {et~al.} 2020, \apj, 904, 130,
  \dodoi{10.3847/1538-4357/abc563}

\bibitem[{{Venturi} {et~al.}(2024){Venturi}, {Carniani}, {Parlanti},
  {Kohandel}, {Curti}, {Pallottini}, {Vallini}, {Arribas}, {Bunker}, {Cameron},
  {Castellano}, {Ferrara}, {Fontana}, {Gallerani}, {Gelli}, {Maiolino},
  {Ntormousi}, {Pacifici}, {Pentericci}, {Salvadori}, \& {Vanzella}}]{VG24}
{Venturi}, G., {Carniani}, S., {Parlanti}, E., {et~al.} 2024, \aap, 691, A19,
  \dodoi{10.1051/0004-6361/202449855}

\bibitem[{{Villa-V{\'e}lez} {et~al.}(2024){Villa-V{\'e}lez}, {Godard},
  {Guillard}, \& {Pineau des For{\^e}ts}}]{villa24}
{Villa-V{\'e}lez}, J.~A., {Godard}, B., {Guillard}, P., \& {Pineau des
  For{\^e}ts}, G. 2024, \aap, 688, A96, \dodoi{10.1051/0004-6361/202449212}

\bibitem[{{Villanueva} {et~al.}(2024){Villanueva}, {Herrera-Camus},
  {Gonz{\'a}lez-L{\'o}pez}, {Aravena}, {Assef}, {Baeza-Garay},
  {Barcos-Mu{\~n}oz}, {Bovino}, {Bowler}, {da Cunha}, {De Looze},
  {Diaz-Santos}, {Ferrara}, {F{\"o}rster Schreiber}, {Algera}, {Ikeda},
  {Killi}, {Mitsuhashi}, {Naab}, {Relano}, {Spilker}, {Solimano}, {Palla},
  {Price}, {Posses}, {Tadaki}, {Telikova}, \& {{\"U}bler}}]{VV24}
{Villanueva}, V., {Herrera-Camus}, R., {Gonz{\'a}lez-L{\'o}pez}, J., {et~al.}
  2024, \aap, 691, A133, \dodoi{10.1051/0004-6361/202451490}

\bibitem[{{Vizgan} {et~al.}(2022){Vizgan}, {Heintz}, {Greve}, {Narayanan},
  {Dav{\'e}}, {Olsen}, {Popping}, \& {Watson}}]{vizgan22}
{Vizgan}, D., {Heintz}, K.~E., {Greve}, T.~R., {et~al.} 2022, \apjl, 939, L1,
  \dodoi{10.3847/2041-8213/ac982c}

\bibitem[{{Wagg} {et~al.}(2010){Wagg}, {Carilli}, {Wilner}, {Cox}, {De Breuck},
  {Menten}, {Riechers}, \& {Walter}}]{WJ10}
{Wagg}, J., {Carilli}, C.~L., {Wilner}, D.~J., {et~al.} 2010, \aap, 519, L1,
  \dodoi{10.1051/0004-6361/201015424}

\bibitem[{{Wagg} {et~al.}(2012){Wagg}, {Wiklind}, {Carilli}, {Espada}, {Peck},
  {Riechers}, {Walter}, {Wootten}, {Aravena}, {Barkats}, {Cortes}, {Hills},
  {Hodge}, {Impellizzeri}, {Iono}, {Leroy}, {Mart{\'\i}n}, {Rawlings},
  {Maiolino}, {McMahon}, {Scott}, {Villard}, \& {Vlahakis}}]{WJ12}
{Wagg}, J., {Wiklind}, T., {Carilli}, C.~L., {et~al.} 2012, \apjl, 752, L30,
  \dodoi{10.1088/2041-8205/752/2/L30}

\bibitem[{{Walter} {et~al.}(2009){Walter}, {Riechers}, {Cox}, {Neri},
  {Carilli}, {Bertoldi}, {Weiss}, \& {Maiolino}}]{WF09}
{Walter}, F., {Riechers}, D., {Cox}, P., {et~al.} 2009, \nat, 457, 699,
  \dodoi{10.1038/nature07681}

\bibitem[{{Walter} {et~al.}(2012){Walter}, {Decarli}, {Carilli}, {Bertoldi},
  {Cox}, {da Cunha}, {Daddi}, {Dickinson}, {Downes}, {Elbaz}, {Ellis}, {Hodge},
  {Neri}, {Riechers}, {Weiss}, {Bell}, {Dannerbauer}, {Krips}, {Krumholz},
  {Lentati}, {Maiolino}, {Menten}, {Rix}, {Robertson}, {Spinrad}, {Stark}, \&
  {Stern}}]{WF12}
{Walter}, F., {Decarli}, R., {Carilli}, C., {et~al.} 2012, \nat, 486, 233,
  \dodoi{10.1038/nature11073}

\bibitem[{{Walter} {et~al.}(2018){Walter}, {Riechers}, {Novak}, {Decarli},
  {Ferkinhoff}, {Venemans}, {Ba{\~n}ados}, {Bertoldi}, {Carilli}, {Fan},
  {Farina}, {Mazzucchelli}, {Neeleman}, {Rix}, {Strauss}, {Uzgil}, \&
  {Wang}}]{WF18}
{Walter}, F., {Riechers}, D., {Novak}, M., {et~al.} 2018, \apjl, 869, L22,
  \dodoi{10.3847/2041-8213/aaf4fa}

\bibitem[{{Walter} {et~al.}(2022){Walter}, {Neeleman}, {Decarli}, {Venemans},
  {Meyer}, {Weiss}, {Ba{\~n}ados}, {Bosman}, {Carilli}, {Fan}, {Riechers},
  {Rix}, \& {Thompson}}]{WF22}
{Walter}, F., {Neeleman}, M., {Decarli}, R., {et~al.} 2022, \apj, 927, 21,
  \dodoi{10.3847/1538-4357/ac49e8}

\bibitem[{{Wang} {et~al.}(2019{\natexlab{a}}){Wang}, {Wang}, {Fan}, {Wu},
  {Yang}, {Neri}, \& {Yue}}]{WF19a}
{Wang}, F., {Wang}, R., {Fan}, X., {et~al.} 2019{\natexlab{a}}, \apj, 880, 2,
  \dodoi{10.3847/1538-4357/ab2717}

\bibitem[{{Wang} {et~al.}(2019{\natexlab{b}}){Wang}, {Yang}, {Fan}, {Wu},
  {Yue}, {Li}, {Bian}, {Jiang}, {Ba{\~n}ados}, {Schindler}, {Findlay},
  {Davies}, {Decarli}, {Farina}, {Green}, {Hennawi}, {Huang}, {Mazzuccheli},
  {McGreer}, {Venemans}, {Walter}, {Dye}, {Lyke}, {Myers}, \& {Nunez}}]{WF19b}
{Wang}, F., {Yang}, J., {Fan}, X., {et~al.} 2019{\natexlab{b}}, \apj, 884, 30,
  \dodoi{10.3847/1538-4357/ab2be5}

\bibitem[{{Wang} {et~al.}(2021{\natexlab{a}}){Wang}, {Yang}, {Fan}, {Hennawi},
  {Barth}, {Banados}, {Bian}, {Boutsia}, {Connor}, {Davies}, {Decarli},
  {Eilers}, {Farina}, {Green}, {Jiang}, {Li}, {Mazzucchelli}, {Nanni},
  {Schindler}, {Venemans}, {Walter}, {Wu}, \& {Yue}}]{WF21a}
---. 2021{\natexlab{a}}, \apjl, 907, L1, \dodoi{10.3847/2041-8213/abd8c6}

\bibitem[{{Wang} {et~al.}(2021{\natexlab{b}}){Wang}, {Fan}, {Yang},
  {Mazzucchelli}, {Wu}, {Li}, {Ba{\~n}ados}, {Farina}, {Nanni}, {Ai}, {Bian},
  {Davies}, {Decarli}, {Hennawi}, {Schindler}, {Venemans}, \& {Walter}}]{WF21b}
{Wang}, F., {Fan}, X., {Yang}, J., {et~al.} 2021{\natexlab{b}}, \apj, 908, 53,
  \dodoi{10.3847/1538-4357/abcc5e}

\bibitem[{{Wang} {et~al.}(2024{\natexlab{a}}){Wang}, {Yang}, {Fan}, {Venemans},
  {Decarli}, {Ba{\~n}ados}, {Walter}, {Barth}, {Bian}, {Davies}, {Eilers},
  {Farina}, {Hennawi}, {Li}, {Mazzucchelli}, {Wang}, {Wu}, \& {Yue}}]{WF24a}
{Wang}, F., {Yang}, J., {Fan}, X., {et~al.} 2024{\natexlab{a}}, \apj, 968, 9,
  \dodoi{10.3847/1538-4357/ad3fb4}

\bibitem[{{Wang} {et~al.}(2016{\natexlab{a}}){Wang}, {Koribalski}, {Serra},
  {van der Hulst}, {Roychowdhury}, {Kamphuis}, \& {Chengalur}}]{wang16}
{Wang}, J., {Koribalski}, B.~S., {Serra}, P., {et~al.} 2016{\natexlab{a}},
  \mnras, 460, 2143, \dodoi{10.1093/mnras/stw1099}

\bibitem[{{Wang} {et~al.}(2013){Wang}, {Wagg}, {Carilli}, {Walter}, {Lentati},
  {Fan}, {Riechers}, {Bertoldi}, {Narayanan}, {Strauss}, {Cox}, {Omont},
  {Menten}, {Knudsen}, {Neri}, \& {Jiang}}]{WR13}
{Wang}, R., {Wagg}, J., {Carilli}, C.~L., {et~al.} 2013, \apj, 773, 44,
  \dodoi{10.1088/0004-637X/773/1/44}

\bibitem[{{Wang} {et~al.}(2016{\natexlab{b}}){Wang}, {Wu}, {Neri}, {Fan},
  {Walter}, {Carilli}, {Momjian}, {Bertoldi}, {Strauss}, {Li}, {Wang},
  {Riechers}, {Jiang}, {Omont}, {Wagg}, \& {Cox}}]{WR16}
{Wang}, R., {Wu}, X.-B., {Neri}, R., {et~al.} 2016{\natexlab{b}}, \apj, 830,
  53, \dodoi{10.3847/0004-637X/830/1/53}

\bibitem[{{Wang} {et~al.}(2019{\natexlab{c}}){Wang}, {Shao}, {Carilli},
  {Jones}, {Walter}, {Fan}, {Riechers}, {Decarli}, {Bertoldi}, {Wagg},
  {Strauss}, {Omont}, {Cox}, {Jiang}, {Narayanan}, {Menten}, \&
  {Venemans}}]{WR19}
{Wang}, R., {Shao}, Y., {Carilli}, C.~L., {et~al.} 2019{\natexlab{c}}, \apj,
  887, 40, \dodoi{10.3847/1538-4357/ab4d4b}

\bibitem[{{Wang} {et~al.}(2024{\natexlab{b}}){Wang}, {Wylezalek}, {De Breuck},
  {Vernet}, {Rupke}, {Zakamska}, {Vayner}, {Lehnert}, {Nesvadba}, \&
  {Stern}}]{WW24}
{Wang}, W., {Wylezalek}, D., {De Breuck}, C., {et~al.} 2024{\natexlab{b}},
  \aap, 683, A169, \dodoi{10.1051/0004-6361/202348531}

\bibitem[{{Wang} {et~al.}(2025){Wang}, {De Breuck}, {Wylezalek}, {Vernet},
  {Lehnert}, {Stern}, {Rupke}, {Nesvadba}, {Vayner}, {Zakamska}, {Lin},
  {Kukreti}, {Dall'Agnol de Oliveira}, \& {Groth}}]{WW25}
{Wang}, W., {De Breuck}, C., {Wylezalek}, D., {et~al.} 2025, \aap, 696, A88,
  \dodoi{10.1051/0004-6361/202553668}

\bibitem[{{Ward} {et~al.}(2024){Ward}, {de la Vega}, {Mobasher}, {McGrath},
  {Iyer}, {Calabr{\`o}}, {Costantin}, {Dickinson}, {Holwerda},
  {Huertas-Company}, {Hirschmann}, {Lucas}, {Pandya}, {Wilkins}, {Yung},
  {Arrabal Haro}, {Bagley}, {Finkelstein}, {Kartaltepe}, {Koekemoer},
  {Papovich}, \& {Pirzkal}}]{ward24}
{Ward}, E., {de la Vega}, A., {Mobasher}, B., {et~al.} 2024, \apj, 962, 176,
  \dodoi{10.3847/1538-4357/ad20ed}

\bibitem[{{Wardlow} {et~al.}(2013){Wardlow}, {Cooray}, {De Bernardis},
  {Amblard}, {Arumugam}, {Aussel}, {Baker}, {B{\'e}thermin}, {Blundell},
  {Bock}, {Boselli}, {Bridge}, {Buat}, {Burgarella}, {Bussmann},
  {Cabrera-Lavers}, {Calanog}, {Carpenter}, {Casey}, {Castro-Rodr{\'\i}guez},
  {Cava}, {Chanial}, {Chapin}, {Chapman}, {Clements}, {Conley}, {Cox},
  {Dowell}, {Dye}, {Eales}, {Farrah}, {Ferrero}, {Franceschini}, {Frayer},
  {Frazer}, {Fu}, {Gavazzi}, {Glenn}, {Gonz{\'a}lez Solares}, {Griffin},
  {Gurwell}, {Harris}, {Hatziminaoglou}, {Hopwood}, {Hyde}, {Ibar}, {Ivison},
  {Kim}, {Lagache}, {Levenson}, {Marchetti}, {Marsden}, {Martinez-Navajas},
  {Negrello}, {Neri}, {Nguyen}, {O'Halloran}, {Oliver}, {Omont}, {Page},
  {Panuzzo}, {Papageorgiou}, {Pearson}, {P{\'e}rez-Fournon}, {Pohlen},
  {Riechers}, {Rigopoulou}, {Roseboom}, {Rowan-Robinson}, {Schulz}, {Scott},
  {Scoville}, {Seymour}, {Shupe}, {Smith}, {Streblyanska}, {Strom},
  {Symeonidis}, {Trichas}, {Vaccari}, {Vieira}, {Viero}, {Wang}, {Xu}, {Yan},
  \& {Zemcov}}]{WJ13}
{Wardlow}, J.~L., {Cooray}, A., {De Bernardis}, F., {et~al.} 2013, \apj, 762,
  59, \dodoi{10.1088/0004-637X/762/1/59}

\bibitem[{{Wardlow} {et~al.}(2017){Wardlow}, {Cooray}, {Osage}, {Bourne},
  {Clements}, {Dannerbauer}, {Dunne}, {Dye}, {Eales}, {Farrah}, {Furlanetto},
  {Ibar}, {Ivison}, {Maddox}, {Micha{\l}owski}, {Riechers}, {Rigopoulou},
  {Scott}, {Smith}, {Wang}, {van der Werf}, {Valiante}, {Valtchanov}, \&
  {Verma}}]{WJ17}
{Wardlow}, J.~L., {Cooray}, A., {Osage}, W., {et~al.} 2017, \apj, 837, 12,
  \dodoi{10.3847/1538-4357/837/1/12}

\bibitem[{{Watson} {et~al.}(2015){Watson}, {Christensen}, {Knudsen}, {Richard},
  {Gallazzi}, \& {Micha{\l}owski}}]{WD15}
{Watson}, D., {Christensen}, L., {Knudsen}, K.~K., {et~al.} 2015, \nat, 519,
  327, \dodoi{10.1038/nature14164}

\bibitem[{{Wei{\ss}} {et~al.}(2007){Wei{\ss}}, {Downes}, {Neri}, {Walter},
  {Henkel}, {Wilner}, {Wagg}, \& {Wiklind}}]{WA07}
{Wei{\ss}}, A., {Downes}, D., {Neri}, R., {et~al.} 2007, \aap, 467, 955,
  \dodoi{10.1051/0004-6361:20066117}

\bibitem[{{Wei{\ss}} {et~al.}(2003){Wei{\ss}}, {Henkel}, {Downes}, \&
  {Walter}}]{WA03}
{Wei{\ss}}, A., {Henkel}, C., {Downes}, D., \& {Walter}, F. 2003, \aap, 409,
  L41, \dodoi{10.1051/0004-6361:20031337}

\bibitem[{{Wei{\ss}} {et~al.}(2013){Wei{\ss}}, {De Breuck}, {Marrone},
  {Vieira}, {Aguirre}, {Aird}, {Aravena}, {Ashby}, {Bayliss}, {Benson},
  {B{\'e}thermin}, {Biggs}, {Bleem}, {Bock}, {Bothwell}, {Bradford}, {Brodwin},
  {Carlstrom}, {Chang}, {Chapman}, {Crawford}, {Crites}, {de Haan}, {Dobbs},
  {Downes}, {Fassnacht}, {George}, {Gladders}, {Gonzalez}, {Greve},
  {Halverson}, {Hezaveh}, {High}, {Holder}, {Holzapfel}, {Hoover}, {Hrubes},
  {Husband}, {Keisler}, {Lee}, {Leitch}, {Lueker}, {Luong-Van}, {Malkan},
  {McIntyre}, {McMahon}, {Mehl}, {Menten}, {Meyer}, {Murphy}, {Padin},
  {Plagge}, {Reichardt}, {Rest}, {Rosenman}, {Ruel}, {Ruhl}, {Schaffer},
  {Shirokoff}, {Spilker}, {Stalder}, {Staniszewski}, {Stark}, {Story},
  {Vanderlinde}, {Welikala}, \& {Williamson}}]{WA13}
{Wei{\ss}}, A., {De Breuck}, C., {Marrone}, D.~P., {et~al.} 2013, \apj, 767,
  88, \dodoi{10.1088/0004-637X/767/1/88}

\bibitem[{{Werner}(1970)}]{werner70}
{Werner}, M.~W. 1970, \aplett, 6, 81

\bibitem[{{Willott} {et~al.}(2015{\natexlab{a}}){Willott}, {Bergeron}, \&
  {Omont}}]{WC15b}
{Willott}, C.~J., {Bergeron}, J., \& {Omont}, A. 2015{\natexlab{a}}, \apj, 801,
  123, \dodoi{10.1088/0004-637X/801/2/123}

\bibitem[{{Willott} {et~al.}(2017){Willott}, {Bergeron}, \& {Omont}}]{WC17}
---. 2017, \apj, 850, 108, \dodoi{10.3847/1538-4357/aa921b}

\bibitem[{{Willott} {et~al.}(2015{\natexlab{b}}){Willott}, {Carilli}, {Wagg},
  \& {Wang}}]{WC15a}
{Willott}, C.~J., {Carilli}, C.~L., {Wagg}, J., \& {Wang}, R.
  2015{\natexlab{b}}, \apj, 807, 180, \dodoi{10.1088/0004-637X/807/2/180}

\bibitem[{{Willott} {et~al.}(2013){Willott}, {Omont}, \& {Bergeron}}]{WC13}
{Willott}, C.~J., {Omont}, A., \& {Bergeron}, J. 2013, \apj, 770, 13,
  \dodoi{10.1088/0004-637X/770/1/13}

\bibitem[{{Willott} {et~al.}(2007){Willott}, {Delorme}, {Omont}, {Bergeron},
  {Delfosse}, {Forveille}, {Albert}, {Reyl{\'e}}, {Hill}, {Gully-Santiago},
  {Vinten}, {Crampton}, {Hutchings}, {Schade}, {Simard}, {Sawicki}, {Beelen},
  \& {Cox}}]{WC07}
{Willott}, C.~J., {Delorme}, P., {Omont}, A., {et~al.} 2007, \aj, 134, 2435,
  \dodoi{10.1086/522962}

\bibitem[{{Wilson} {et~al.}(2023){Wilson}, {Heintz}, {Jakobsson}, {Madden},
  {Watson}, {Magdis}, {Valentino}, {Greve}, \& {Vizgan}}]{wilson23}
{Wilson}, S.~N., {Heintz}, K.~E., {Jakobsson}, P., {et~al.} 2023, arXiv
  e-prints, arXiv:2305.05213, \dodoi{10.48550/arXiv.2305.05213}

\bibitem[{{Witstok} {et~al.}(2022){Witstok}, {Smit}, {Maiolino}, {Kumari},
  {Aravena}, {Boogaard}, {Bouwens}, {Carniani}, {Hodge}, {Jones}, {Stefanon},
  {van der Werf}, \& {Schouws}}]{WJ22}
{Witstok}, J., {Smit}, R., {Maiolino}, R., {et~al.} 2022, \mnras, 515, 1751,
  \dodoi{10.1093/mnras/stac1905}

\bibitem[{{Wolfire} {et~al.}(2022){Wolfire}, {Vallini}, \&
  {Chevance}}]{wolfire22}
{Wolfire}, M.~G., {Vallini}, L., \& {Chevance}, M. 2022, \araa, 60, 247,
  \dodoi{10.1146/annurev-astro-052920-010254}

\bibitem[{{Wong} {et~al.}(2022{\natexlab{a}}){Wong}, {Wang}, {Hashimoto},
  {Takagi}, {Goto}, {Kim}, {Wu}, {On}, {Santos}, {Lu}, {Kilerci-Eser}, {Ho}, \&
  {Hsiao}}]{WY22b}
{Wong}, Y. H.~V., {Wang}, P., {Hashimoto}, T., {et~al.} 2022{\natexlab{a}},
  \apj, 929, 161, \dodoi{10.3847/1538-4357/ac5cc7}

\bibitem[{{Wong} {et~al.}(2022{\natexlab{b}}){Wong}, {Wang}, {Hashimoto},
  {Takagi}, {Goto}, {Kim}, {Wu}, {On}, {Santos}, {Lu}, {Kilerci-Eser}, {Ho}, \&
  {Hsiao}}]{WY22a}
---. 2022{\natexlab{b}}, arXiv e-prints, arXiv:2202.13613,
  \dodoi{10.48550/arXiv.2202.13613}

\bibitem[{{Wu} {et~al.}(2021){Wu}, {Cai}, {Neeleman}, {Finlator}, {Zhang},
  {Prochaska}, {Wang}, {Emonts}, {Fan}, {Keating}, {Wang}, {Yang}, {Hennawi},
  \& {Wang}}]{WY21}
{Wu}, Y., {Cai}, Z., {Neeleman}, M., {et~al.} 2021, Nature Astronomy, 5, 1110,
  \dodoi{10.1038/s41550-021-01471-4}

\bibitem[{{Wuyts} {et~al.}(2012{\natexlab{a}}){Wuyts}, {Rigby}, {Gladders},
  {Gilbank}, {Sharon}, {Gralla}, \& {Bayliss}}]{WE12b}
{Wuyts}, E., {Rigby}, J.~R., {Gladders}, M.~D., {et~al.} 2012{\natexlab{a}},
  \apj, 745, 86, \dodoi{10.1088/0004-637X/745/1/86}

\bibitem[{{Wuyts} {et~al.}(2012{\natexlab{b}}){Wuyts}, {Rigby}, {Sharon}, \&
  {Gladders}}]{WE12a}
{Wuyts}, E., {Rigby}, J.~R., {Sharon}, K., \& {Gladders}, M.~D.
  2012{\natexlab{b}}, \apj, 755, 73, \dodoi{10.1088/0004-637X/755/1/73}

\bibitem[{{Xiao} {et~al.}(2024){Xiao}, {Oesch}, {Elbaz}, {Bing}, {Nelson},
  {Weibel}, {Illingworth}, {van Dokkum}, {Naidu}, {Daddi}, {Bouwens},
  {Matthee}, {Wuyts}, {Chisholm}, {Brammer}, {Dickinson}, {Magnelli}, {Leroy},
  {Schaerer}, {Herard-Demanche}, {Lim}, {Barrufet}, {Endsley}, {Fudamoto},
  {G{\'o}mez-Guijarro}, {Gottumukkala}, {Labb{\'e}}, {Magee}, {Marchesini},
  {Maseda}, {Qin}, {Reddy}, {Shapley}, {Shivaei}, {Shuntov}, {Stefanon},
  {Whitaker}, \& {Wyithe}}]{XM24}
{Xiao}, M., {Oesch}, P.~A., {Elbaz}, D., {et~al.} 2024, \nat, 635, 311,
  \dodoi{10.1038/s41586-024-08094-5}

\bibitem[{{Yang} {et~al.}(2019{\natexlab{a}}){Yang}, {Wang}, {Fan}, {Yue},
  {Wu}, {Li}, {Bian}, {Jiang}, {Ba{\~n}ados}, \& {Beletsky}}]{YJ19b}
{Yang}, J., {Wang}, F., {Fan}, X., {et~al.} 2019{\natexlab{a}}, \aj, 157, 236,
  \dodoi{10.3847/1538-3881/ab1be1}

\bibitem[{{Yang} {et~al.}(2019{\natexlab{b}}){Yang}, {Venemans}, {Wang}, {Fan},
  {Novak}, {Decarli}, {Walter}, {Yue}, {Momjian}, {Keeton}, {Wang},
  {Zabludoff}, {Wu}, \& {Bian}}]{YJ19a}
{Yang}, J., {Venemans}, B., {Wang}, F., {et~al.} 2019{\natexlab{b}}, \apj, 880,
  153, \dodoi{10.3847/1538-4357/ab2a02}

\bibitem[{{Yang} {et~al.}(2020){Yang}, {Wang}, {Fan}, {Hennawi}, {Davies},
  {Yue}, {Banados}, {Wu}, {Venemans}, {Barth}, {Bian}, {Boutsia}, {Decarli},
  {Farina}, {Green}, {Jiang}, {Li}, {Mazzucchelli}, \& {Walter}}]{YJ20}
{Yang}, J., {Wang}, F., {Fan}, X., {et~al.} 2020, \apjl, 897, L14,
  \dodoi{10.3847/2041-8213/ab9c26}

\bibitem[{{Yang} {et~al.}(2021){Yang}, {Wang}, {Fan}, {Barth}, {Hennawi},
  {Nanni}, {Bian}, {Davies}, {Farina}, {Schindler}, {Ba{\~n}ados}, {Decarli},
  {Eilers}, {Green}, {Guo}, {Jiang}, {Li}, {Venemans}, {Walter}, {Wu}, \&
  {Yue}}]{YJ21}
---. 2021, \apj, 923, 262, \dodoi{10.3847/1538-4357/ac2b32}

\bibitem[{{Yi} {et~al.}(2014){Yi}, {Wang}, {Wu}, {Yang}, {Bai}, {Fan},
  {Brandt}, {Ho}, {Zuo}, {Kim}, {Wang}, {Yang}, {Zhang}, {Wang}, {Wang}, {Ai},
  {Fan}, {Chang}, {Wang}, {Lun}, \& {Xin}}]{YW14}
{Yi}, W.-M., {Wang}, F., {Wu}, X.-B., {et~al.} 2014, \apjl, 795, L29,
  \dodoi{10.1088/2041-8205/795/2/L29}

\bibitem[{{Yue} {et~al.}(2021){Yue}, {Yang}, {Fan}, {Wang}, {Spilker},
  {Georgiev}, {Keeton}, {Litke}, {Marrone}, {Walter}, {Wang}, {Wu}, {Venemans},
  \& {Zabludoff}}]{YM21}
{Yue}, M., {Yang}, J., {Fan}, X., {et~al.} 2021, \apj, 917, 99,
  \dodoi{10.3847/1538-4357/ac0af4}

\bibitem[{{Yun} {et~al.}(2015){Yun}, {Aretxaga}, {Gurwell}, {Hughes},
  {Monta{\~n}a}, {Narayanan}, {Rosa-Gonz{\'a}lez}, {S{\'a}nchez-Arg{\"u}elles},
  {Schloerb}, {Snell}, {Vega}, {Wilson}, {Zeballos}, {Chavez}, {Cybulski},
  {D{\'\i}az-Santos}, {De La Luz}, {Erickson}, {Ferrusca}, {Gim}, {Heyer},
  {Iono}, {Pope}, {Rogstad}, {Scott}, {Souccar}, {Terlevich}, {Terlevich},
  {Wilner}, \& {Zavala}}]{YM15}
{Yun}, M.~S., {Aretxaga}, I., {Gurwell}, M.~A., {et~al.} 2015, \mnras, 454,
  3485, \dodoi{10.1093/mnras/stv1963}

\bibitem[{{Zamora} {et~al.}(2024){Zamora}, {Venturi}, {Carniani}, {Bertola},
  {Parlanti}, {Perna}, {Arribas}, {B{\"o}ker}, {Bunker}, {Charlot},
  {D'Eugenio}, {Maiolino}, {Rodr{\'\i}guez Del Pino}, {{\"U}bler}, {Cresci},
  {Jones}, \& {Lamperti}}]{ZS24}
{Zamora}, S., {Venturi}, G., {Carniani}, S., {et~al.} 2024, arXiv e-prints,
  arXiv:2412.02751, \dodoi{10.48550/arXiv.2412.02751}

\bibitem[{{Zanella} {et~al.}(2018){Zanella}, {Daddi}, {Magdis}, {Diaz Santos},
  {Cormier}, {Liu}, {Cibinel}, {Gobat}, {Dickinson}, {Sargent}, {Popping},
  {Madden}, {Bethermin}, {Hughes}, {Valentino}, {Rujopakarn}, {Pannella},
  {Bournaud}, {Walter}, {Wang}, {Elbaz}, \& {Coogan}}]{ZA18}
{Zanella}, A., {Daddi}, E., {Magdis}, G., {et~al.} 2018, \mnras, 481, 1976,
  \dodoi{10.1093/mnras/sty2394}

\bibitem[{{Zanella} {et~al.}(2024){Zanella}, {Iani}, {Dessauges-Zavadsky},
  {Richard}, {De Breuck}, {Vernet}, {Kohandel}, {Arrigoni Battaia},
  {Bolamperti}, {Calura}, {Chen}, {Devereaux}, {Ferrara}, {Mainieri},
  {Pallottini}, {Rodighiero}, {Vallini}, \& {Vanzella}}]{ZA24}
{Zanella}, A., {Iani}, E., {Dessauges-Zavadsky}, M., {et~al.} 2024, \aap, 685,
  A80, \dodoi{10.1051/0004-6361/202349074}

\bibitem[{{Zavala} {et~al.}(2015){Zavala}, {Yun}, {Aretxaga}, {Hughes},
  {Wilson}, {Geach}, {Egami}, {Gurwell}, {Wilner}, {Smail}, {Blain}, {Chapman},
  {Coppin}, {Dessauges-Zavadsky}, {Edge}, {Monta{\~n}a}, {Nakajima}, {Rawle},
  {S{\'a}nchez-Arg{\"u}elles}, {Swinbank}, {Webb}, \& {Zeballos}}]{ZJ15}
{Zavala}, J.~A., {Yun}, M.~S., {Aretxaga}, I., {et~al.} 2015, \mnras, 452,
  1140, \dodoi{10.1093/mnras/stv1351}

\bibitem[{{Zavala} {et~al.}(2018){Zavala}, {Monta{\~n}a}, {Hughes}, {Yun},
  {Ivison}, {Valiante}, {Wilner}, {Spilker}, {Aretxaga}, {Eales},
  {Avila-Reese}, {Ch{\'a}vez}, {Cooray}, {Dannerbauer}, {Dunlop}, {Dunne},
  {G{\'o}mez-Ruiz}, {Micha{\l}owski}, {Narayanan}, {Nayyeri}, {Oteo}, {Rosa
  Gonz{\'a}lez}, {S{\'a}nchez-Arg{\"u}elles}, {Schloerb}, {Serjeant}, {Smith},
  {Terlevich}, {Vega}, {Villalba}, {van der Werf}, {Wilson}, \&
  {Zeballos}}]{ZJ18}
{Zavala}, J.~A., {Monta{\~n}a}, A., {Hughes}, D.~H., {et~al.} 2018, Nature
  Astronomy, 2, 56, \dodoi{10.1038/s41550-017-0297-8}

\bibitem[{{Zavala} {et~al.}(2024){Zavala}, {Bakx}, {Mitsuhashi}, {Castellano},
  {Calabro}, {Akins}, {Buat}, {Casey}, {Fernandez-Arenas}, {Franco}, {Fontana},
  {Hatsukade}, {Ho}, {Ikeda}, {Kartaltepe}, {Koekemoer}, {McKinney},
  {Napolitano}, {P{\'e}rez-Gonz{\'a}lez}, {Santini}, {Serjeant}, {Terlevich},
  {Terlevich}, \& {Yung}}]{ZJ24}
{Zavala}, J.~A., {Bakx}, T., {Mitsuhashi}, I., {et~al.} 2024, \apjl, 977, L9,
  \dodoi{10.3847/2041-8213/ad8f38}

\bibitem[{{Zavala} {et~al.}(2025){Zavala}, {Castellano}, {Akins}, {Bakx},
  {Burgarella}, {Casey}, {Ch{\'a}vez Ortiz}, {Dickinson}, {Finkelstein},
  {Mitsuhashi}, {Nakajima}, {P{\'e}rez-Gonz{\'a}lez}, {Arrabal Haro},
  {Bergamini}, {Buat}, {Backhaus}, {Calabr{\`o}}, {Cleri},
  {Fern{\'a}ndez-Arenas}, {Fontana}, {Franco}, {Grillo}, {Giavalisco},
  {Grogin}, {Hathi}, {Hirschmann}, {Ikeda}, {Jung}, {Kartaltepe}, {Koekemoer},
  {Larson}, {McKinney}, {Papovich}, {Rosati}, {Saito}, {Santini}, {Terlevich},
  {Terlevich}, {Treu}, \& {Yung}}]{ZJ25}
{Zavala}, J.~A., {Castellano}, M., {Akins}, H.~B., {et~al.} 2025, Nature
  Astronomy, 9, 155, \dodoi{10.1038/s41550-024-02397-3}

\bibitem[{{Zhang} {et~al.}(2018){Zhang}, {Ivison}, {George}, {Zhao}, {Dunne},
  {Herrera-Camus}, {Lewis}, {Liu}, {Naylor}, {Oteo}, {Riechers}, {Smail},
  {Yang}, {Eales}, {Hopwood}, {Maddox}, {Omont}, \& {van der Werf}}]{ZZ18}
{Zhang}, Z.-Y., {Ivison}, R.~J., {George}, R.~D., {et~al.} 2018, \mnras, 481,
  59, \dodoi{10.1093/mnras/sty2082}

\bibitem[{{Zhao} {et~al.}(2016){Zhao}, {Lu}, {Xu}, {Gao}, {Lord},
  {Charmandaris}, {Diaz-Santos}, {Evans}, {Howell}, {Petric}, {van der Werf},
  \& {Sanders}}]{Z16}
{Zhao}, Y., {Lu}, N., {Xu}, C.~K., {et~al.} 2016, \apj, 819, 69,
  \dodoi{10.3847/0004-637X/819/1/69}

\end{thebibliography}
\bibliographystyle{aasjournal}


\end{document}